\documentclass[%
 reprint,
superscriptaddress,
nofootinbib,
 amsmath,amssymb,
 aps,
]{revtex4-1}

\usepackage{graphicx}% Include figure files
\usepackage{epstopdf}
\usepackage{dcolumn}% Align table columns on decimal point
\usepackage{bm}% bold math
\usepackage{color}
\usepackage{amsmath}
\usepackage[caption=false]{subfig}
\usepackage{xspace}
\usepackage[normalem]{ulem}
\usepackage{epsfig}
\usepackage{multirow}
\usepackage[]{units}
\usepackage{hyperref}% add hypertext capabilities
\usepackage[mathlines]{lineno}% Enable numbering of text and display math
\usepackage{gensymb}
\usepackage{float}

\RequirePackage{xspace}
\usepackage{accents}

\def\nupot {7.482$\times10^{20}$\xspace}
\def\nubpot {7.471$\times10^{20}$\xspace}

\def\stot        {\ensuremath{\textrm{sin}^2(2\theta_{13})}\xspace}

\def\stt        {\ensuremath{\textrm{sin}^2\theta_{23}}\xspace} 
\def\sto       {\ensuremath{\textrm{sin}^2\theta_{13}}\xspace}

\def\dmsq         {\ensuremath{\Delta m^2_{32}}\xspace}

\def\thint   {\ensuremath{\textrm \,\theta_{13}}\xspace}

\def\dcp      {\ensuremath{\delta_{\mathrm{CP}}}\xspace}

\def\erec {\ensuremath{E_\nu^{rec}}\xspace}

\newcommand{\nd}{ND280\xspace}

\newcommand{\ccpip}{CC1$\pi^+$\xspace}

\newcommand{\nova}{NO$\nu$A\xspace}

\newcommand{\bplus}[1]{\accentset{(+)}{#1}}
\newcommand{\bminus}[1]{\accentset{(-)}{#1}}

\def\p      {\ensuremath{\rm \pi}\xspace}

\def\mup        {\ensuremath{\mu^+}\xspace}
\def\nunu       {\ensuremath{\nu}\xspace}
\def\nub        {\ensuremath{\overline{\nu}}\xspace}

\def\nue        {\ensuremath{\nu_e}\xspace}
\def\nueb       {\ensuremath{\nub_e}\xspace}

\def\num        {\ensuremath{\nu_\mu}\xspace}
\def\numb       {\ensuremath{\nub_\mu}\xspace}

\def\pipi   {\ensuremath{\pi}\xspace}
\def\piz   {\ensuremath{\pi^0}\xspace}

\def\pip   {\ensuremath{\pi^+}\xspace}

\def\Kbar  {\kern 0.2em\overline{\kern -0.2em K}{}\xspace}

\def\Kz    {\ensuremath{K^0}\xspace}
\def\Kzb   {\ensuremath{\Kbar^0}\xspace}
\def\KzKzb {\ensuremath{\Kz \kern -0.16em \Kzb}\xspace}
\def\Kp    {\ensuremath{K^+}\xspace}
\def\Km    {\ensuremath{K^-}\xspace}

\def\KpKm  {\ensuremath{\Kp \kern -0.16em \Km}\xspace}

\newcommand{\tev}{\ensuremath{\mathrm{\,Te\kern -0.1em V}}\xspace}
\newcommand{\gev}{\ensuremath{\mathrm{\,Ge\kern -0.1em V}}\xspace}
\newcommand{\mev}{\ensuremath{\mathrm{\,Me\kern -0.1em V}}\xspace}
\newcommand{\kev}{\ensuremath{\mathrm{\,ke\kern -0.1em V}}\xspace}
\newcommand{\ev}{\ensuremath{\mathrm{\,e\kern -0.1em V}}\xspace}
\newcommand{\gevc}{\ensuremath{{\mathrm{\,Ge\kern -0.1em V\!/}c}}\xspace}
\newcommand{\mevc}{\ensuremath{{\mathrm{\,Me\kern -0.1em V\!/}c}}\xspace}
\newcommand{\gevcc}{\ensuremath{{\mathrm{\,Ge\kern -0.1em V\!/}c^2}}\xspace}
\newcommand{\mevcc}{\ensuremath{{\mathrm{\,Me\kern -0.1em V\!/}c^2}}\xspace}
\newcommand{\evcc}{\ensuremath{{\mathrm{\,e\kern -0.1em V\!/}c^2}}\xspace}
\def\mus  {\ensuremath{\rm \,\mus}\xspace}
\def\ns   {\ensuremath{\rm \,ns}\xspace}

\def\mus        {\ensuremath{\,\mu{\rm s}}\xspace}    %% microsecond
\def\ns         {\ensuremath{{\rm \,ns}}\xspace}      %% nanosecond
\newcommand{\pf}{\ensuremath{p_{\mathrm{F}}}\xspace}
\newcommand{\eb}{\ensuremath{E_{\mathrm{b}}}\xspace}
\newcommand{\ma}{\ensuremath{M_{\textrm{A}}}\xspace}
\newcommand{\maqe}{\ensuremath{M_{\textrm{A}}^{\mbox{\scriptsize{QE}}}}\xspace}

\def\qq        {\ensuremath{Q^{2}}\xspace}

\def\bracketbar{\hbox{\kern-8pt\raise1pt%
         \hbox{{\tiny(}{\lower1.4pt\hbox{\bf--}}{\tiny)}}}}
\def\numormb{\ensuremath{\nu^{\bracketbar}_\mu}\xspace}
\def\nueoreb{\ensuremath{\nu^{\bracketbar}_e}\xspace}

\begin{document}

\preprint{XXX}

%\title[fragile]{Measurement of neutrino and antineutrino oscillations by the T2K experiment  including an additional sample of charged-current \nue interactions at the far detector with one charged pion in the final state}
\title[fragile]{Measurement of neutrino and antineutrino oscillations by the T2K experiment  including a new additional sample of \nue interactions at the far detector}
%Measurement of neutrino and antineutrino oscillations in appearance and disappearance channels by the T2K experiment with $7.5\times10^{20}$ proton on target in neutrino and antineutrino mode}%mode and $7.5\times10^{20}$ in antineutrino mode}% Force line breaks with \\

%%%%%%%%%%%%%%%%%%%%%%%%%%%%%%%%%%%%%%%%%%%%%%%%%%%%%%%%%%%%%%
% T2K author list generated on Mon, 26 Jun 2017 21:57:13 +0900
% setting: extra = 0 revtex = 1 ptep = 0 simple = 0 xml = 0 yearrule = 1 shiftrule = 1
%         author list from archive (starting 2017/04/21 until now)
%         exemption(s) granted to: prodrigues
% Number of authors = 295
%%%%%%%%%%%%%%%%%%%%%%%%%%%%%%%%%%%%%%%%%%%%%%%%%%%%%%%%%%%%%%

\newcommand{\INSTEE}{\affiliation{University of Bern, Albert Einstein Center for Fundamental Physics, Laboratory for High Energy Physics (LHEP), Bern, Switzerland}}
\newcommand{\INSTFE}{\affiliation{Boston University, Department of Physics, Boston, Massachusetts, U.S.A.}}
\newcommand{\INSTD}{\affiliation{University of British Columbia, Department of Physics and Astronomy, Vancouver, British Columbia, Canada}}
\newcommand{\INSTGA}{\affiliation{University of California, Irvine, Department of Physics and Astronomy, Irvine, California, U.S.A.}}
\newcommand{\INSTI}{\affiliation{IRFU, CEA Saclay, Gif-sur-Yvette, France}}
\newcommand{\INSTGB}{\affiliation{University of Colorado at Boulder, Department of Physics, Boulder, Colorado, U.S.A.}}
\newcommand{\INSTFG}{\affiliation{Colorado State University, Department of Physics, Fort Collins, Colorado, U.S.A.}}
\newcommand{\INSTFH}{\affiliation{Duke University, Department of Physics, Durham, North Carolina, U.S.A.}}
\newcommand{\INSTBA}{\affiliation{Ecole Polytechnique, IN2P3-CNRS, Laboratoire Leprince-Ringuet, Palaiseau, France }}
\newcommand{\INSTEG}{\affiliation{University of Geneva, Section de Physique, DPNC, Geneva, Switzerland}}
\newcommand{\INSTDG}{\affiliation{H. Niewodniczanski Institute of Nuclear Physics PAN, Cracow, Poland}}
\newcommand{\INSTCB}{\affiliation{High Energy Accelerator Research Organization (KEK), Tsukuba, Ibaraki, Japan}}
\newcommand{\INSTED}{\affiliation{Institut de Fisica d'Altes Energies (IFAE), The Barcelona Institute of Science and Technology, Campus UAB, Bellaterra (Barcelona) Spain}}
\newcommand{\INSTEC}{\affiliation{IFIC (CSIC \& University of Valencia), Valencia, Spain}}
\newcommand{\INSTEI}{\affiliation{Imperial College London, Department of Physics, London, United Kingdom}}
\newcommand{\INSTGF}{\affiliation{INFN Sezione di Bari and Universit\`a e Politecnico di Bari, Dipartimento Interuniversitario di Fisica, Bari, Italy}}
\newcommand{\INSTBE}{\affiliation{INFN Sezione di Napoli and Universit\`a di Napoli, Dipartimento di Fisica, Napoli, Italy}}
\newcommand{\INSTBF}{\affiliation{INFN Sezione di Padova and Universit\`a di Padova, Dipartimento di Fisica, Padova, Italy}}
\newcommand{\INSTBD}{\affiliation{INFN Sezione di Roma and Universit\`a di Roma ``La Sapienza'', Roma, Italy}}
\newcommand{\INSTEB}{\affiliation{Institute for Nuclear Research of the Russian Academy of Sciences, Moscow, Russia}}
\newcommand{\INSTHA}{\affiliation{Kavli Institute for the Physics and Mathematics of the Universe (WPI), The University of Tokyo Institutes for Advanced Study, University of Tokyo, Kashiwa, Chiba, Japan}}
\newcommand{\INSTCC}{\affiliation{Kobe University, Kobe, Japan}}
\newcommand{\INSTCD}{\affiliation{Kyoto University, Department of Physics, Kyoto, Japan}}
\newcommand{\INSTEJ}{\affiliation{Lancaster University, Physics Department, Lancaster, United Kingdom}}
\newcommand{\INSTFC}{\affiliation{University of Liverpool, Department of Physics, Liverpool, United Kingdom}}
\newcommand{\INSTFI}{\affiliation{Louisiana State University, Department of Physics and Astronomy, Baton Rouge, Louisiana, U.S.A.}}
\newcommand{\INSTHB}{\affiliation{Michigan State University, Department of Physics and Astronomy,  East Lansing, Michigan, U.S.A.}}
\newcommand{\INSTCE}{\affiliation{Miyagi University of Education, Department of Physics, Sendai, Japan}}
\newcommand{\INSTDF}{\affiliation{National Centre for Nuclear Research, Warsaw, Poland}}
\newcommand{\INSTFJ}{\affiliation{State University of New York at Stony Brook, Department of Physics and Astronomy, Stony Brook, New York, U.S.A.}}
\newcommand{\INSTGJ}{\affiliation{Okayama University, Department of Physics, Okayama, Japan}}
\newcommand{\INSTCF}{\affiliation{Osaka City University, Department of Physics, Osaka, Japan}}
\newcommand{\INSTGG}{\affiliation{Oxford University, Department of Physics, Oxford, United Kingdom}}
\newcommand{\INSTBB}{\affiliation{UPMC, Universit\'e Paris Diderot, CNRS/IN2P3, Laboratoire de Physique Nucl\'eaire et de Hautes Energies (LPNHE), Paris, France}}
\newcommand{\INSTGC}{\affiliation{University of Pittsburgh, Department of Physics and Astronomy, Pittsburgh, Pennsylvania, U.S.A.}}
\newcommand{\INSTFA}{\affiliation{Queen Mary University of London, School of Physics and Astronomy, London, United Kingdom}}
\newcommand{\INSTE}{\affiliation{University of Regina, Department of Physics, Regina, Saskatchewan, Canada}}
\newcommand{\INSTGD}{\affiliation{University of Rochester, Department of Physics and Astronomy, Rochester, New York, U.S.A.}}
\newcommand{\INSTHC}{\affiliation{Royal Holloway University of London, Department of Physics, Egham, Surrey, United Kingdom}}
\newcommand{\INSTBC}{\affiliation{RWTH Aachen University, III. Physikalisches Institut, Aachen, Germany}}
\newcommand{\INSTFB}{\affiliation{University of Sheffield, Department of Physics and Astronomy, Sheffield, United Kingdom}}
\newcommand{\INSTDI}{\affiliation{University of Silesia, Institute of Physics, Katowice, Poland}}
\newcommand{\INSTEH}{\affiliation{STFC, Rutherford Appleton Laboratory, Harwell Oxford,  and  Daresbury Laboratory, Warrington, United Kingdom}}
\newcommand{\INSTCH}{\affiliation{University of Tokyo, Department of Physics, Tokyo, Japan}}
\newcommand{\INSTBJ}{\affiliation{University of Tokyo, Institute for Cosmic Ray Research, Kamioka Observatory, Kamioka, Japan}}
\newcommand{\INSTCG}{\affiliation{University of Tokyo, Institute for Cosmic Ray Research, Research Center for Cosmic Neutrinos, Kashiwa, Japan}}
\newcommand{\INSTGI}{\affiliation{Tokyo Metropolitan University, Department of Physics, Tokyo, Japan}}
\newcommand{\INSTF}{\affiliation{University of Toronto, Department of Physics, Toronto, Ontario, Canada}}
\newcommand{\INSTB}{\affiliation{TRIUMF, Vancouver, British Columbia, Canada}}
\newcommand{\INSTG}{\affiliation{University of Victoria, Department of Physics and Astronomy, Victoria, British Columbia, Canada}}
\newcommand{\INSTDJ}{\affiliation{University of Warsaw, Faculty of Physics, Warsaw, Poland}}
\newcommand{\INSTDH}{\affiliation{Warsaw University of Technology, Institute of Radioelectronics, Warsaw, Poland}}
\newcommand{\INSTFD}{\affiliation{University of Warwick, Department of Physics, Coventry, United Kingdom}}
\newcommand{\INSTGH}{\affiliation{University of Winnipeg, Department of Physics, Winnipeg, Manitoba, Canada}}
\newcommand{\INSTEA}{\affiliation{Wroclaw University, Faculty of Physics and Astronomy, Wroclaw, Poland}}
\newcommand{\INSTHE}{\affiliation{Yokohama National University, Faculty of Engineering, Yokohama, Japan}}
\newcommand{\INSTH}{\affiliation{York University, Department of Physics and Astronomy, Toronto, Ontario, Canada}}

\INSTEE
\INSTFE
\INSTD
\INSTGA
\INSTI
\INSTGB
\INSTFG
\INSTFH
\INSTBA
\INSTEG
\INSTDG
\INSTCB
\INSTED
\INSTEC
\INSTEI
\INSTGF
\INSTBE
\INSTBF
\INSTBD
\INSTEB
\INSTHA
\INSTCC
\INSTCD
\INSTEJ
\INSTFC
\INSTFI
\INSTHB
\INSTCE
\INSTDF
\INSTFJ
\INSTGJ
\INSTCF
\INSTGG
\INSTBB
\INSTGC
\INSTFA
\INSTE
\INSTGD
\INSTHC
\INSTBC
\INSTFB
\INSTDI
\INSTEH
\INSTCH
\INSTBJ
\INSTCG
\INSTGI
\INSTF
\INSTB
\INSTG
\INSTDJ
\INSTDH
\INSTFD
\INSTGH
\INSTEA
\INSTHE
\INSTH

\author{K.\,Abe}\INSTBJ
\author{J.\,Amey}\INSTEI
\author{C.\,Andreopoulos}\INSTEH\INSTFC
\author{M.\,Antonova}\INSTEB
\author{S.\,Aoki}\INSTCC
\author{A.\,Ariga}\INSTEE
\author{Y.\,Ashida}\INSTCD
\author{S.\,Ban}\INSTCD
\author{M.\,Barbi}\INSTE
\author{G.J.\,Barker}\INSTFD
% ...

\collaboration{The T2K Collaboration}\noaffiliation

\date{\today}% It is always \today, today,

\begin{abstract}
The T2K experiment reports an updated analysis of neutrino and antineutrino oscillations in appearance and disappearance channels. 
A sample of electron neutrino candidates at Super-Kamiokande in which a pion decay has been tagged is added to the four single-ring samples used in previous T2K oscillation analyses.
Through combined analyses of these five samples, simultaneous measurements of four oscillation parameters, $|\dmsq|$, \stt, \sto, and \dcp and of the mass ordering are made.  
A set of studies of simulated data indicates that the sensitivity to the oscillation parameters is not limited by neutrino interaction model uncertainty.
Multiple oscillation analyses are performed, and frequentist and Bayesian intervals are presented for combinations of the oscillation parameters with and without the inclusion of reactor constraints on \sto. When combined with reactor measurements, the hypothesis of CP conservation (\dcp$=0$ or \pipi) is excluded at 90\% confidence level. The 90\% confidence region for \dcp is [-2.95,-0.44] ([-1.47, -1.27]) for normal (inverted) ordering. The central values and 68\% confidence intervals for the other oscillation parameters for normal (inverted) ordering are $\Delta m^{2}_{32}=2.54\pm0.08$ ($2.51\pm0.08$) $\times 10^{-3}$ eV$^2 / c^4$ and $\stt = 0.55^{+0.05}_{-0.09}$ ($0.55^{+0.05}_{-0.08}$), compatible with maximal mixing. In the Bayesian analysis, the data weakly prefer normal ordering (Bayes factor 3.7) and the upper octant for \stt (Bayes factor 2.4). 

%\begin{description}

%\end{description}
\end{abstract}

\pacs{Valid PACS appear here}% PACS, the Physics and Astronomy
\maketitle

%\tableofcontents
%Goal: $\sim20-25$ pages paper, first draft by the end of the year?\\

\section{Introduction}
\label{sec:intro}
%Introduction, brief T2K description, data taking.
%Total length: 1 page.
%Editor: Claudio.

Neutrino oscillations have been firmly established by multiple experiments. Super-Kamiokande (SK) observed an energy and pathlength dependent deficit in the atmospheric muon neutrino flux~\cite{Fukuda:1998mi}; and Sudbury Neutrino Observatory (SNO) resolved the long-standing solar neutrino problem by demonstrating that the previously observed deficit of electron neutrinos from the Sun was due to flavor transitions~\cite{Ahmad:2002jz}.
%Neutrino oscillations were firmly established in the late 1990s with the observation by the Super-Kamiokande (SK) experiment that muon neutrinos, produced by cosmic rays in the atmosphere, change their flavor while traveling through the Earth~\cite{Fukuda:1998mi}. A few years later, the long-standing puzzle of the solar neutrino deficit was solved by the Sudbury Neutrino Observatory (SNO), which that revealed that neutrino oscillations were responsible for the deficit of electron neutrinos produced in the Sun~\cite{Ahmad:2002jz}.
These two experiments, together with accelerator-based (K2K~\cite{Ahn:2006zza}, MINOS~\cite{Adamson:2014vgd}) and reactor-based (KamLAND~\cite{Eguchi:2002dm}) long-baseline experiments measured the two mass-squared differences between mass eigenstates and two of the three mixing angles in the PMNS matrix.

The mixing angle, \thint, has been measured as nonzero by T2K~\cite{Abe:2011sj,Abe:2013hdq}, by reactor experiments~\cite{An:2012eh,Ahn:2012nd,Abe:2011fz}, and more recently by \nova~\cite{Adamson:2016tbq}. Establishing that all three mixing angles are nonzero opens a way to study CP violation in the leptonic sector through neutrino oscillations. 
CP violation in neutrino oscillations arises from \dcp, an irreducible CP-odd phase in the PMNS matrix. This phase introduces a difference in the appearance probability between neutrinos and antineutrinos. To investigate this phenomenon, after taking data with a beam predominantly composed of muon neutrinos in order to observe the appearance of electron neutrinos at the far detector, T2K has switched to taking data with a beam predominantly composed of muon antineutrinos. %This allow to maximize the sensitivity to
A direct measurement of CP violation can then be obtained by comparing $\num\rightarrow\nue$ and $\numb\rightarrow\nueb$ channels. 
%In the rest of this paper we will refer to the run in neutrino mode as FHC (Forward Horn Current) and to the one in antineutrino mode as RHC (Reversed Horn Current). 

To produce neutrinos, protons extracted from the J-PARC main ring strike a target producing hadrons which are then focused and selected by charge with a system of magnetic horns. The hadrons decay in flight, producing an intense neutrino beam. A beam predominantly composed of neutrinos or antineutrinos can be produced by choosing the direction of the current in the magnetic horn. T2K uses the so-called off-axis technique with the beam axis directed 2.5$^{\circ}$ away from SK in order to produce a narrow--band neutrino beam, peaked at an energy of 600~\mev, where the effect of neutrino oscillations is maximum for a baseline of 295~km. Neutrinos are also observed at a near detector complex, installed 280~m from the target, comprising an on-axis detector (INGRID) which provides day-to-day monitoring of the beam profile and direction, and a magnetized off-axis detector (ND280), at the same off-axis angle as SK, which measures neutrino interaction rates before oscillation.

The analyses described in this paper are based on an exposure of \nupot protons on target (POT) in neutrino mode (\nunu-mode) and  \nubpot POT in antineutrino mode (\nub-mode) collected at SK during seven physics runs as detailed in Tab.~\ref{tbl:run1to7_pot}. %The so-far collected POT correspond to the 20\% of the total T2K expected exposure.
The neutrino oscillation parameters are measured by combining \num and \numb disappearance channels with \nue and \nueb appearance channels, using the same analysis techniques described in Ref.~\cite{Abe:2017uxa}. The analyzed data set is the same as in Ref.~\cite{Abe:2017uxa}, but an additional SK sample is included in the oscillation analysis.  
%In this paper we will describe the most recent measurements of neutrino oscillations in T2K, obtained by combining measurement of \num and \numb disappearance and of \nue and \nueb appearance. 
Previously, for the appearance channel, only the SK single-ring e-like interactions without additional activity in the detector were used for the oscillation analysis. 
%The analysis presented in this paper also uses events with an additional electron ring, consistent with coming from the decays of $pi^+$'s produced in neutrino interactions. We refer to these electrons from the decay chain $pi^+\rightarrow mu^+ \rightarrow e^+$ as Michel electrons.
The analysis presented in this paper includes an additional sample enriched in \nue interactions in which the e-like ring is accompanied by delayed Michel electron due to the decay chain $\pip\rightarrow\mup\rightarrow e^+$ of \pip's produced in the neutrino interactions. 
This sample is currently only used in $\nu$-mode, and increases the statistics of the \nue sample in SK by roughly 10\%.

\begin{table}[h]
\caption{
T2K data-taking periods and collected POT used in the analyses
presented in this paper.
}
\label{tbl:run1to7_pot}
\begin{tabular}{ l c c c }
\hline\hline
Run & Dates & \(\nu\)-mode POT & \(\bar{\nu}\)-mode POT \\ 
Period & & (\(\times10^{20}\)) & (\(\times10^{20}\))   \\ \hline
Run 1 & Jan. 2010-Jun. 2010 & 0.323 & -- \\
Run 2 & Nov. 2010-Mar. 2011 & 1.108 & -- \\
Run 3 & Mar. 2012-Jun. 2012 & 1.579 & -- \\
Run 4 & Oct. 2012-May 2013 & 3.560 & -- \\
Run 5 & May 2014-Jun. 2014 & 0.242 & 0.506 \\
Run 6 & Nov. 2014-Jun. 2015 & 0.190 & 3.505 \\
Run 7 & Feb. 2016-May 2016 & 0.480 & 3.460 \\
\hline
Total & Jan. 2010-May 2016 & 7.482 &  7.471 \\
\hline \hline
\end{tabular}
\end{table}

The paper is organized as follows: the neutrino beam and the modeling of the neutrino fluxes are described in Sec.~\ref{sec:beam}. The neutrino interaction model developed for this analysis will then be described in Sec.~\ref{sec:niwg}, followed by the selection of neutrinos in the near detector complex in Sec.~\ref{sec:nd}. The neutrino flux and neutrino interaction inputs, and near detector selections are combined to reduce flux and cross-section uncertainties at the far detector as will be shown in Sec.~\ref{sec:banff}. The far detector selections are described in Sec.~\ref{sec:sk}. The neutrino oscillations and the T2K oscillation analyses frameworks are then described in Sec.~\ref{sec:oscprob} and in Sec.~\ref{sec:oamethod} respectively. Sec.~\ref{sec:fds} is dedicated to a description of the impact of the uncertainties of the neutrino interaction model on the T2K oscillation analyses. Finally, the results of the oscillation analyses are presented in Sec.~\ref{sec:nuebar} and in Sec.~\ref{sec:oaresults} and some concluding remarks are given in Sec.~\ref{sec:conclusions}.

%
%We will then describe the neutrino interactions model that we have developed for this analysis. Next we describe the ND280 complex and the neutrino and antineutrino selections at both, INGRID and ND280 and how these data are used in the oscillation analysis. Particular care is given to the understanding of the neutrino interaction models, for which a strategy based on fake data studies have been developed. We then describe the SK selection, introducing for the first time the CC1$\pi^+$ selection at the far detector. Finally we will give detailed description of the results of the joint analyses with the five samples used in the fit.
 \section{The T2K beam}
\label{sec:beam}

The neutrino beam is produced by the interaction of 30~GeV protons from the J-PARC 
main ring accelerator on a 1.9 interaction-length graphite target. Secondary hadrons, mainly pions and kaons,
leaving the target pass through three electromagnetic horns~\cite{SEKIGUCHI201557}, which are operated at a current of either +250~kA or \(-\)250~kA to focus positively or negatively charged particles respectively. The outgoing hadrons decay 
in a 96-m long decay volume, where a relatively pure beam of muon neutrinos is produced by 
the decay of positively charged hadrons in positive 
focusing mode (\(\nu\)-mode), and a beam mostly composed of muon antineutrinos is produced in negative focusing mode (\(\bar{
\nu}\)-mode).  Protons and undecayed hadrons are stopped in a beam dump,
while muons above 5 GeV pass through and are detected in a muon monitor (MUMON~\cite{doi:10.1093/ptep/ptv054}),
and are used to monitor the secondary beam stability.
%The T2K experiment makes use of an off-axis configuration, where the SK
%far detector is 2.5$\degree$~off-axis from the neutrino beam center, producing a
%neutrino flux which is relatively sharply peaked in energy around 0.6~GeV.
The T2K beamline hardware has been described in detail elsewhere \cite{Abe:2011ks}.  

\begin{figure*}[!htbp]
\includegraphics[trim=100 290 120 50,clip=true,width=0.4\textwidth]{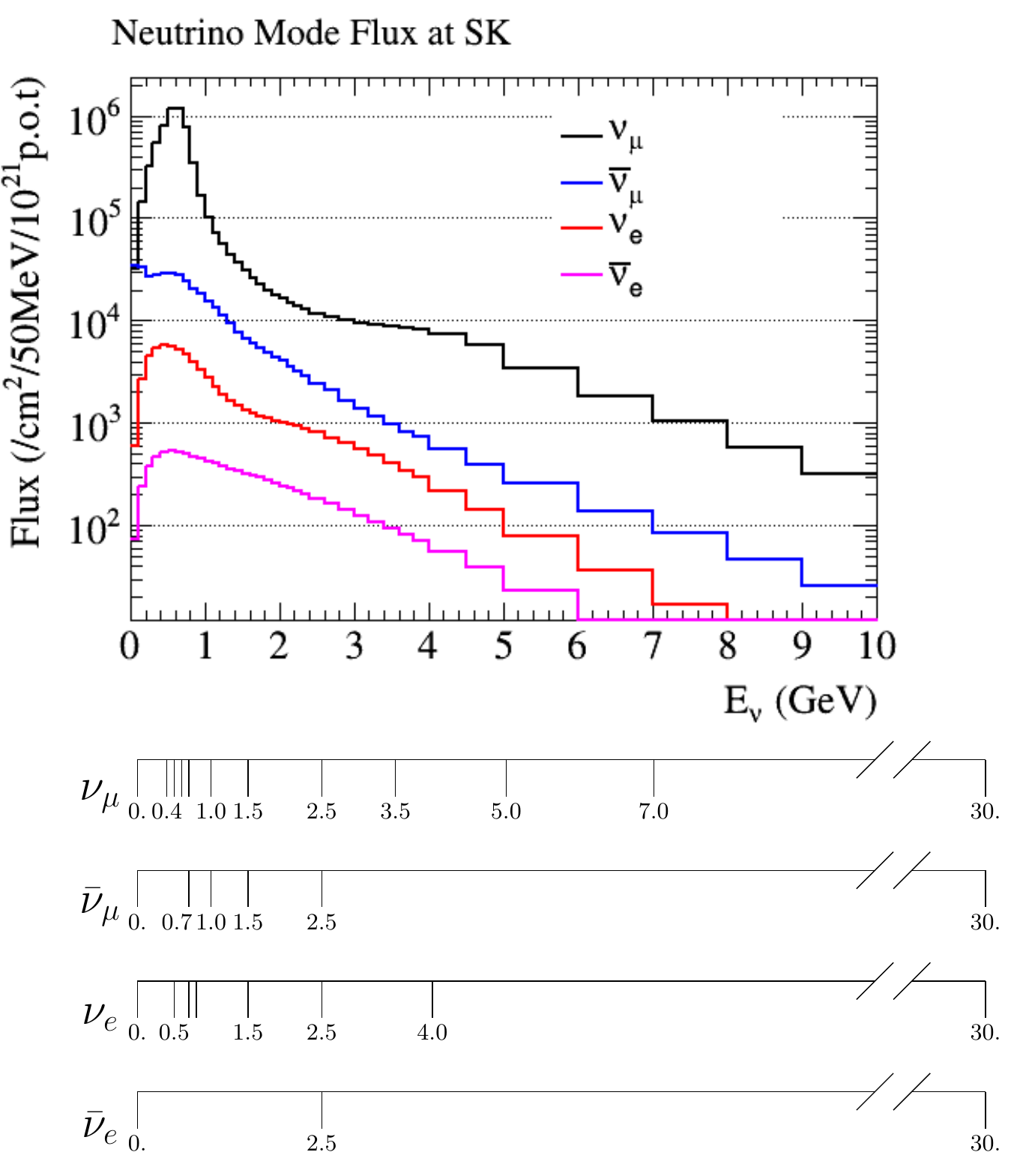}
~~
\includegraphics[trim=100 290 120 50,clip=true,width=0.4\textwidth]{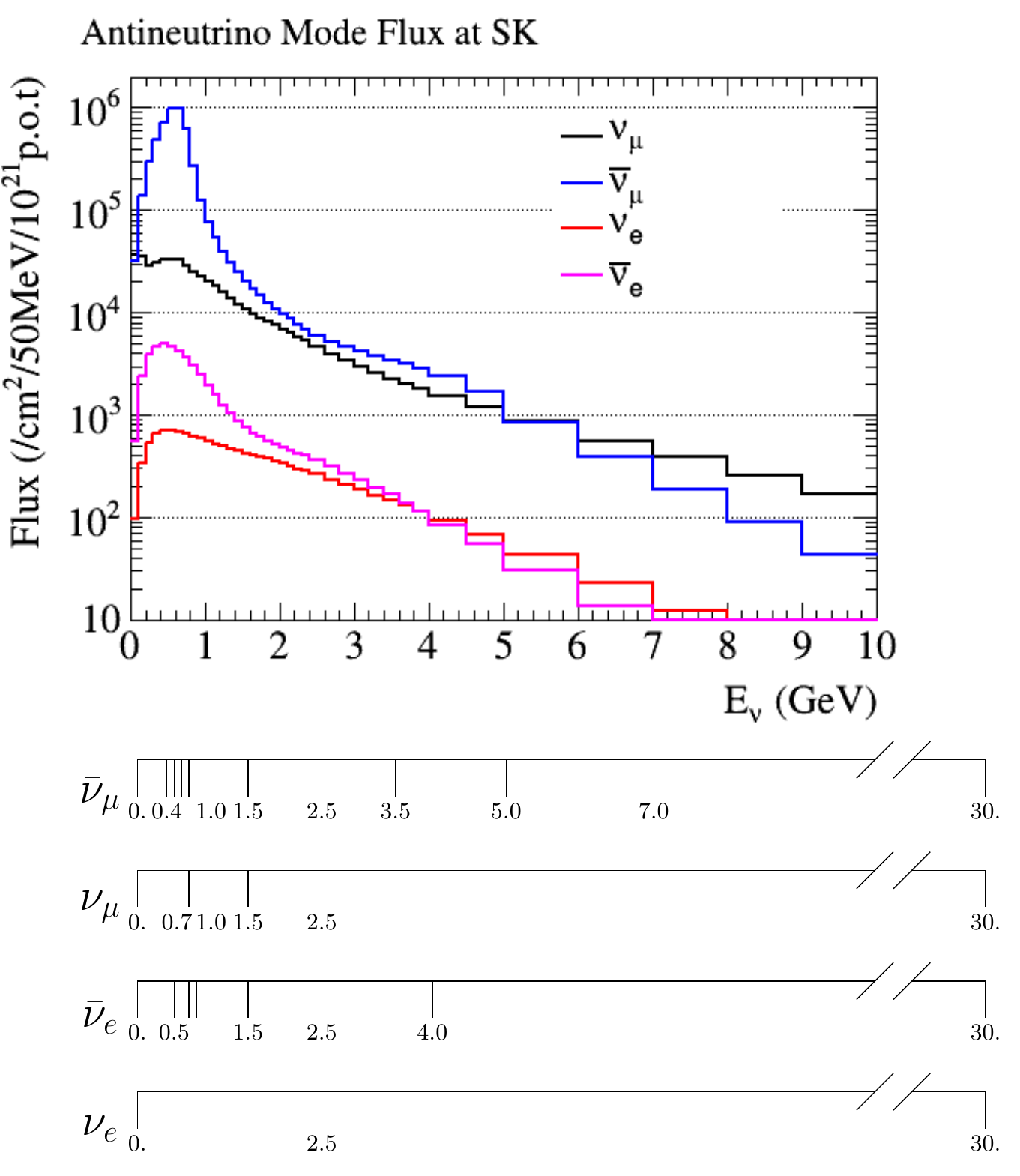}
\caption{The T2K unoscillated neutrino flux prediction at SK for \(\nu\)- (left) and 
\(\bar{\nu}\)- (right) modes.  The binning used for the flux systematic parameters is also shown.
%The flux in the range 8~GeV \(<\) E\(_\nu\) \(<\) 30~GeV is 
%also simulated but not shown.
}
\label{fig:skflux}
\end{figure*}

%\subsection{Neutrino Flux Prediction and Flux Errors}
The T2K neutrino flux at the near and far detectors in case of no neutrino oscillation 
is predicted by a simulation which has been described in detail in Ref.~\cite{Abe:2013fp}.   
Interactions of the primary proton beam, whose profile
is measured for each run period by a suite of proton beam monitors, %\red{and the INGRID (Sec.~\ref{sec:ingrid}) detector}
 as well as subsequently-produced pions and kaons, are simulated within 
the graphite target by the FLUKA 2011 package \cite{Ferrari:2005zk,Fluka:2014}.   The predicted 
hadron production rates inside and 
outside the target are then adjusted based on the results from the latest analysis of the full 2009 
thin-target dataset by the NA61/SHINE experiment 
\cite{Abgrall:2011ae,Abgrall:2011ts,Abgrall:2016fs}, 
%which measured hadron production from a 30~GeV beam incident on a graphite target with a thickness of 4\% of an interaction length, 
as well as other hadron production experiments \cite{eichten,allaby,e910}.  
Particles which exit the target and subsequently decay are tracked through the horns and decay volume
by a GEANT3 \cite{GEANT3} simulation using the GCALOR \cite{GCALOR} package.
The predicted (anti)neutrino fluxes at the far 
detector for T2K Run 1-7 is shown for both \(\nu\)- and \(\bar{\nu}\)-modes in Fig.~\ref{fig:skflux}.

Most of the ``right-sign'' \(\nu_\mu\) flux (i.e. \(\nu\)'s in \(\nu\)-mode and
\(\bar{\nu}\)'s in \(\bar{\nu}\)-mode) comes from 
mesons produced inside the target, dominantly ``right-sign'' (focused) pion and kaon 
production, which subsequently decay to produce muons.%, although secondary proton and neutron production also contributes.
Interactions producing ``right-sign'' \(\nu_e\)'s also predominantly come from interactions in the target, with
a larger fraction of \nue produced by kaon (rather than pion) decays.
% production interactions (with subsequent muon decays).
Interactions producing the ``wrong-sign'' \(\nu_\mu\) flux have a higher fractional rate of
out-of-target interactions, which are dominated by protons,
neutrons, and pions scattering in the horns and decay volume walls.   
Interactions producing the ``wrong-sign'' \(\nu_e\) flux have a significant fraction of \(K^0\)
production from proton or neutron interactions, as well as charged kaon production.

In general, the \(\nu\)- and \(\bar{\nu}\)-mode fluxes are similar at low
energy, although the ``right-sign'' \(\nu_\mu\) (and \(\nu_e\)) flux in
\(\nu\)-mode is \(\sim\)15\% higher around the flux peak than the ``right-sign''
\(\bar{\nu}_\mu\) (and \(\bar{\nu}_e\)) flux in \(\bar{\nu}\)-mode.  The ``wrong-sign'' background
\(\bar{\nu}\) flux is also lower in \(\nu\)-mode compared to the \(\nu\) flux in
\(\bar{\nu}\)-mode, especially at high energy.  These differences are due to the
higher production multiplicities of positive, rather than negative, parent particles.
%, since
%``right-sign''  neutrinos  in \(\nu\)-mode come from \(\pi^+\) production  at
%low  energy and \(K^+\) production at high energy, while ``right-sign''
%neutrinos in \(\bar{\nu}\)-mode are dominated by \(\pi^-\) production  at  low
%energy  and \(K^-\) and \(K^0_L\) production  at  high  energy, 

\begin{figure*}[htbp]
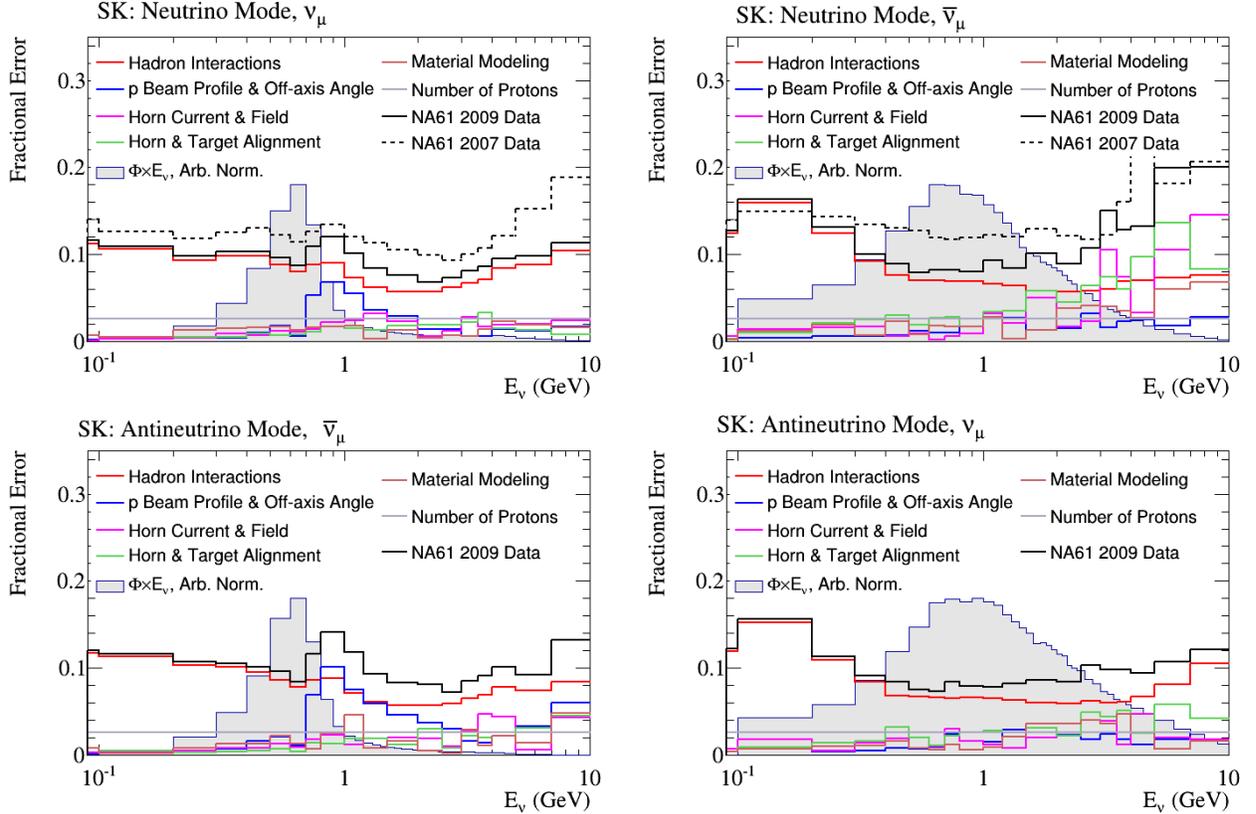

\includegraphics[width=0.45\textwidth]{fig_2a.pdf}
~~
\includegraphics[width=0.45\textwidth]{fig_2b.pdf}
\includegraphics[width=0.45\textwidth]{fig_2c.pdf}
~~
\includegraphics[width=0.45\textwidth]{fig_2d.pdf}
\caption{The T2K fractional systematic uncertainties on the SK flux arising from the
beamline configuration and hadron production prior to constraints from near detector
data. Uncertainties are given for \nunu's in a \(\nu\)-mode beam (top
left), \(\bar{\nu}\)'s in a \(\nu\)-mode beam (top right), \nub's in an
\(\bar{\nu}\)-mode beam (bottom left), and \nunu's in an \(\bar{\nu}\)-mode
beam (bottom right).  For the \(\nu\)-mode plots, the total current uncertainties (NA61 2009 data) are
compared to the total uncertainties estimated for the previous T2K results (NA61 2007
data)~\cite{Abe:2015awa}.}
\label{fig:skfluxerror}
\end{figure*}

%\subsection{Neutrino Flux Errors}
Uncertainties in the neutrino flux prediction arise from the hadron production
model, proton beam profile, off-axis angle, horn current, horn alignment, and
other factors. For each source of error, the underlying parameters
in the model are varied to evaluate the effect on the flux prediction in bins of neutrino
energy for each neutrino flavor as described in detail elsewhere
\cite{Abe:2013fp}.  The uncertainties on the unoscillated \(\nu_\mu\) and
\(\bar{\nu}_\mu\) beam fluxes at the far detector are shown in
Fig.~\ref{fig:skfluxerror} and are currently dominated by uncertainties on hadron production.  The uncertainties on the background \(\nu_e\) and
\(\bar{\nu}_e\) fluxes from the beam are %\(\sim\)
7--10\% in the relevant
region. %, but are not shown here.

\section{Neutrino interaction model}
\label{sec:niwg}
The neutrino interaction model used in this analysis is based on NEUT~\cite{neut} version 5.3.2, which includes many significant improvements over the old version 5.1.4.2, used in previous T2K oscillation analyses (described in detail in Ref.~\cite{Abe:2015awa}). This model is constrained where possible by external experiments that are used to provide initial cross section parameter uncertainties. Such uncertainties are reduced using ND280 data, as explained in Sec.~\ref{sec:banff}. Alternative models are used to build simulated datasets to test the robustness of the T2K analysis against model-dependent assumptions, as explained in Sec.~\ref{sec:fds}.
This section describes the updated NEUT interaction model and alternative models used for the oscillation analyses.

%Expected length: 3 pages.

%Editors: Callum and Sara.

\subsection{Neutrino interaction model used in the oscillation analyses}
\label{sec:niwgmodel}
%Editor: Callum
The interaction rate at T2K energies is dominated by charged current quasi-elastic (CCQE) events, $\nu_{l}n\rightarrow l^{-}p$ ($\nub_{l}p\rightarrow l^{+}n$). Because CCQE is a two-body process and the neutrino direction is known, the neutrino energy can be reconstructed from the outgoing lepton kinematics alone. However, nuclear effects and other processes, which have the same experimental signature of a single muon and no final state pions (CC$0\pi$, or CCQE-like), are indistinguishable from CCQE and can affect the reconstructed neutrino energy and thus the oscillation result if not accounted for~\cite{Ankowski:2016jdd, Martini:2012fa, Nieves:2012yz, coloma_erec, ankowski_erec, lalakulich_erec_2012}. The T2K cross-section modeling has been updated to include recent theoretical models of these processes (full details can be found in Ref.~\cite{Wilkinson:2016wmz}). In previous analyses, the CCQE model was based on the Llewellyn-Smith neutrino-nucleon scattering model~\cite{llewellyn-smith} with a dipole axial form factor and BBBA05 vector form factors~\cite{bbba05}, and used the Smith-Moniz Relativistic Fermi Gas (RFG) model~\cite{smith-moniz} to account for the fact that the nucleons are bound in a nucleus. The main improvements available in NEUT v5.3.2 are the inclusion of the Spectral Function (SF) model from Ref.~\cite{Benhar:1999bg}, which provides a more sophisticated description of the initial state of the nucleus than the RFG: the inclusion of the multi-nucleon interaction (2p2h) model from Refs.~\cite{Nieves:2011pp, nievesExtension}; and the implementation of the Random Phase Approximation (RPA) correction from Ref.~\cite{Nieves:2011pp}. The 2p2h model includes interactions with more than one nucleon bound within the nucleus, which contribute considerable strength to the CCQE-like cross section, and add significant smearing to the reconstructed neutrino energy distribution (as it is not a two-body process). RPA is a nuclear screening effect due to long-range nucleon-nucleon correlations which modifies the interaction strength as a function of four-momentum transfer, $Q^2$. The models make different physical assumptions, so they cannot be combined arbitrarily. Two candidate models were considered for the default model: SF and RFG+RPA+2p2h. RFG+RPA+2p2h was selected as the default because it was most consistently able to describe the available MiniBooNE~\cite{mb-ccqe-2010, mb-ccqe-antinu-2013} and MINERvA~\cite{minerva-nu-ccqe, minerva-antinu-ccqe} CCQE-like data (see Ref.~\cite{Wilkinson:2016wmz} for details).

Various parameters have been introduced to describe the theoretical uncertainties and approximations in the RFG+RPA+2p2h model, which are constrained using the near detector data. The variable parameters are the axial-mass, \ma, the Fermi momentum, \pf, the binding energy, \eb, and the 2p2h cross-section normalization. Given the poor agreement between MINERvA and MiniBooNE datasets~\cite{Wilkinson:2016wmz}, and slight inconsistencies between the signal definitions from the two experiments, no external constraints are applied on the variable parameters prior to the ND280 fit. Given the absence of firm predictions on the scaling of various nuclear effects with the nucleus mass number, the Fermi momentum, binding energy and 2p2h normalization are treated as uncorrelated between $^{12}$C and $^{16}$O. The 2p2h normalization is also considered to be uncorrelated between neutrino and antineutrino interactions. All of these parameters can be separately constrained by the T2K near detector data with the inclusion of samples with interactions on both $^{12}$C and on $^{16}$O, with $\nu$- and $\bar{\nu}$-mode data.

%Although constraints on the variable model parameters were also produced by the fits to external CCQE-like data, and indeed were used in previously reported analyses~\cite{Abe:2015ibe}, it was decided to drop these constraints for the current analysis and to allow the ND280 data to fully constrain the CCQE-like parameters. This decision was motivated by the poor agreement between MINERvA and MiniBooNE datasets, and slight inconsistencies between the signal definitions from the two experiments, which makes interpreting fits to them difficult. The variable parameters in the RFG+RPA+2p2h model are the axial-mass $M_{\mathrm{A}}$ for which the default NEUT value is used, the Fermi momentum ($p_{\mathrm{F}}$) and binding energy ($E_{\mathrm{b}}$) RFG parameters, which are treated separated for $^{12}$C and $^{16}$O and take their central values from fits to electron scattering data~\cite{moniz:fg_escat}, and 2p2h model normalization parameters for $^{12}$C and $^{16}$O, as well as a $\bar{\nu}$ 2p2h parameter which allows for difference in the neutrino and antineutrino 2p2h normalizations.

%% This section relies heavily on TN-197, for which Phil Rodrigues and Aaron Bercellie were primary authors. I imagine they aren't on the T2K author list anymore, but probably should be for this paper?
The NEUT model for resonant pion production is based on the Rein-Sehgal model~\cite{rein-sehgal} with updated nucleon form factors~\cite{sobczyk_pion_FFs}, and with the invariant hadronic mass restricted to be $W \leq 2$ GeV to avoid double-counting pions produced through Deep Inelastic Scattering (DIS). Three variable model parameters are considered: the resonant axial mass, $M_{\mathrm{A}}^{\mathrm{RES}}$; 
the value of the axial form factor at zero transferred 4-momentum,
%the normalization of the axial form factor for resonant pion production, 
$C^{\mathrm{A}}_{5}$; and the normalization of the isospin non-resonant component predicted in the Rein-Sehgal model, I$\frac{1}{2}$. Initial central values and uncertainties for these parameters are obtained in a fit to low energy neutrino--deuterium single pion production data from ANL~\cite{ANL_Radecky_1982} and BNL~\cite{BNL_Kitagaki_1986} for the resonant pion production channels $\nu_{\mu}p\rightarrow \mu^{-}p\pi^{+}$, $\nu_{\mu}n\rightarrow \mu^{-}p\pi^{0}$ and $\nu_{\mu}n\rightarrow \mu^{-}n\pi^{+}$. For the dominant production channel, $\nu_{\mu}p\rightarrow \mu^{-}p\pi^{+}$, the reanalyzed dataset from Ref.~\cite{anl_bnl_reanalysis} was used. Resonant kaon, photon and eta production is also modeled using the Rein-Sehgal resonance production amplitudes, with modified branching ratios to account for the decay of the resonances to kaons, photons, or eta, rather than to pions. External neutrino-nucleus and antineutrino-nucleus pion production data from MiniBooNE~\cite{AguilarArevalo:2009ww, AguilarArevalo:2010xt, AguilarArevalo:2010bm}, MINERvA~\cite{Eberly:2014mra}, SciBooNE~\cite{Kurimoto:2009wq} and K2K~\cite{Nakayama:2004dp} were used as a cross check to ensure that the broad features of all datasets were consistent with the uncertainties on the interaction level parameters ($M_{\mathrm{A}}^{\mathrm{RES}}$, $C^{\mathrm{A}}_{5}$ and I$\frac{1}{2}$ component) or the uncertainties on Final State Interactions (FSI, which will be described shortly). A full fit to the external data is difficult due to strong correlations between FSI parameters and the neutrino-nucleus interaction model parameters, and a lack of information on the correlations between external data points.

% Fits to neutrino--nucleus single pion production data were used in previous T2K analyses (see Ref.~\cite{Abe:2015awa} for details), but were not used for this analysis because it proved to be difficult to separate the effect of Final State Interaction (FSI) parameters from interaction level parameters, and problems with the data (highly correlated datapoints without correlations provided) made it difficult to trust the errors produced from the fit. In this analysis, external neutrino-nucleus pion production data from MiniBooNE, MINERvA and K2K was used as a cross check to ensure that the broad features of all datasets were consistent with the uncertainties on the interaction level parameters ($M_{\mathrm{A}}^{\mathrm{RES}}$, $C^{\mathrm{A}}_{5}$ and I $\frac{1}{2}$ background) or the FSI uncertainties (which will be described shortly). 

The coherent pion production model used is the Rein-Sehgal model described in Refs.~\cite{rein-coh, rein_2007}. However, recent results from MINERvA~\cite{minerva_coh_2014} are better described by the Berger-Sehgal model~\cite{Berger:2008xs}, so a rough reweighting of the coherent events as a function of the outgoing pion energy, $E_{\pi}$, is applied to approximate the Berger-Sehgal model using the weights and binning given in Tab.~\ref{tab:cccoh_weights}. Normalization uncertainties of 30\% are introduced separately for CC- and NC-coherent events, based on comparisons to the MINERvA data (after the weights in Table~\ref{tab:cccoh_weights} are applied), which are fully correlated between $^{12}$C and $^{16}$O. 
\begin{table}
  \centering      
\caption{Weights applied to coherent pion interactions as a function of the pion energy, $E_{\pi}$.}\label{tab:cccoh_weights}
      {\renewcommand{\arraystretch}{1.2}
        \begin{tabular}{>{\centering}p{2.5cm}|p{2.5cm}<{\centering}}
          \hline \hline
          $E_{\pi}$ (GeV) & Weight \\
          \hline
          0.00 - 0.25 & 0.135 \\
          0.25 - 0.50 & 0.400 \\
          0.50 - 0.75 & 0.294 \\
          0.75 - 1.00 & 1.206 \\
          \hline \hline
      \end{tabular}}
\end{table}

The Deep Inelastic Scattering (DIS) model is unchanged from previous analyses (described in Ref.~\cite{Abe:2015awa}). The DIS cross section is calculated for $W \geq 1.3$ GeV, using GRV98 structure functions~\cite{Gluck1998} with Bodek-Yang corrections~\cite{bodek-yang}. Single pion production through DIS is suppressed for $W \leq 2$ GeV to avoid double counting with the resonant pion production contributions, and uses a custom hadronization model described in Ref.~\cite{neut}. For $W > 2$ GeV, PYTHIA/Jetset~\cite{SJOSTRAND199474} is used for hadronization. 
%A CC-other shape parameter, $x^{\mathrm{CC-Other}}$ which scales the cross section by $(1 + x^{\mathrm{CC-Other}}/E_{\nu})$ was introduced to give flexibility to the CC-DIS contribution. The parameter was designed to give greater flexibility at low $E_{\nu}$, and takes an initial uncertainty from comparisons with MINOS CC-inclusive data~\cite{Adamson:2009ju}. CC-Other includes CC resonant kaon, photon and eta production, as well as CC-DIS events.
A CC-other shape parameter, $x^{\mathrm{CC-Other}}$ was introduced to give flexibility to the CC-DIS contribution. This parameter applies to CC resonant kaon, photon and eta production, as well as CC-DIS events, and it scales the cross section by $(1 + x^{\mathrm{CC-Other}}/E_{\nu})$. It was designed to give greater flexibility at low $E_{\nu}$, and the initial uncertainty was set by NEUT comparisons with MINOS CC-inclusive data~\cite{Adamson:2009ju}.

In addition to the previously described NC-coherent parameter, two other NC-specific parameters have been introduced in this analysis. A study in Ref.~\cite{Wang:2015ivq} showed that the NEUT neutral current single photon production (NC1$\gamma$) cross section prediction was approximately a factor of two smaller than a recent theoretical model~\cite{Wang:2013wva}. Because of this, the NC1$\gamma$ cross section has been set to be 200\% of the NEUT nominal prediction, with an uncertainty of 100\%. An NC-Other normalization parameter is applied to neutral current elastic, NC resonant kaon and eta production, as well as NC-DIS events, with an initial uncertainty set at 30\%.

Hadrons produced inside the nucleus may undergo FSI before leaving the nuclear environment, which changes the outgoing particle content and kinematics in the final state. NEUT models FSI for pions, kaons, etas and nucleons using a cascade model described in Ref.~\cite{neut}. Interactions are generated inside the nucleus according to a Woods-Saxon density distribution~\cite{woods-saxon}, and all outgoing hadrons are stepped through the nucleus with interaction probabilities calculated at each step until they leave the nucleus. Particles produced in DIS interactions are propagated some distance without interacting to allow for a formation zone, where the initial step size is based on results from the SKAT experiment~\cite{skat_1979}. The allowed pion interactions in the nucleus are: charge exchange, where the charge of the pion changes; absorption, where the pion is absorbed through two or three body processes; elastic scattering, where the pion only exchanges momentum and energy; and inelastic scattering, where additional pions are produced. If an interaction occurs, new and modified particles are also added to the cascade. For pion momenta $p_{\pi} \geq 500$ MeV, nucleons are treated as free particles, and separate high ($p_{\pi} \geq 500$ MeV) and low  ($p_{\pi} < 500$ MeV) energy scattering parameters are introduced for charge exchange and elastic scattering. Initial interaction uncertainties are obtained from fits to a large body of pion--nucleon and pion--nucleus scattering data for nuclei ranging from carbon to lead, as described in Ref.~\cite{Abe:2015awa}. The variable parameters included to vary the pion FSI cross section are summarized in Tab.~\ref{tab:xsec_parameters}. Pion FSI parameters are assumed to be fully correlated between $^{12}$C and $^{16}$O. Uncertainties on nucleon, kaon and eta FSI interaction probabilities are not considered in the current analysis. 

To account for effects which may potentially affect $\nu^{\bracketbar}_e$ but not $\nu^{\bracketbar}_{\mu}$ cross sections, such as radiative corrections or second class currents (see, for example, Ref.~\cite{day_nue_numu_2012}), which are not included in the NEUT cross section model, additional uncertainties which affect \nueoreb have been introduced. These include an uncorrelated 2\% uncertainty on the $\nu_{e}/\nu_{\mu}$ and $\bar{\nu}_{e}/\bar{\nu}_{\mu}$ cross section ratios to account for radiative corrections, and an additional 2\% uncertainty which is fully anticorrelated between $\nu_e$ and $\bar{\nu}_{e}$ to allow for second class currents.

The full list of cross-section uncertainties and their values before and after the ND280 data constraints is provided in Tab.~\ref{tab:xsec_parameters}.

\subsection{Alternative neutrino interaction models for studies of simulated data}
\label{sec:niwgfake}
%Editor: Sara.

Neutrino interactions with $^{12}$C and $^{16}$O nuclear targets at the near and far detectors may be affected 
by important nuclear effects which are not well understood. Various theoretical models are available to describe such effects,
which are based on different approximations and with different ranges of validity.
None of the available models are capable of describing all the available measurements of neutrino--nucleus 
cross sections from T2K and from other experiments.
It is therefore crucial to test that the T2K oscillation analysis is insensitive to reasonable modifications of the neutrino interaction model
described in Sec.\ref{sec:niwgmodel},
which will now be refered to as the ``reference model''. With this aim, various simulated datasets have been built based on alternative models. The following effects
have been considered: variations of the distribution of the momentum of the initial nucleons
in the nucleus and of the energy needed to extract the nucleons from the nucleus (the binding energy);
uncertainties on the long-range nuclear correlations modifying the cross section as a function of $Q^2$; and
modifications of the modeling of multi-nucleon interaction model (2p2h), including short-range nuclear correlations
and meson exchange currents.

To test the nuclear effects in the initial state two alternative models have been considered beyond the
RFG simulation used as reference: the SF developed in Ref.~\cite{Benhar:1999bg} 
and the Local Fermi Gas (LFG) model from Ref.~\cite{Nieves:2011pp}. The LFG model also differs from the reference model in the implementation of the binding energy.
In the latter, an effective value is considered, based on the average momentum of nucleons within the nucleus, 
while the LFG model considers the different state of the initial and final nucleus
after the nucleon ejection, naturally including a different binding energy for neutrino and antineutrino interactions. 
The simulated datasets built with this alternative model will be referred to as the ``alternative 1p1h model''.

\begin{figure}[htbp]
 \includegraphics[width=0.5\textwidth]{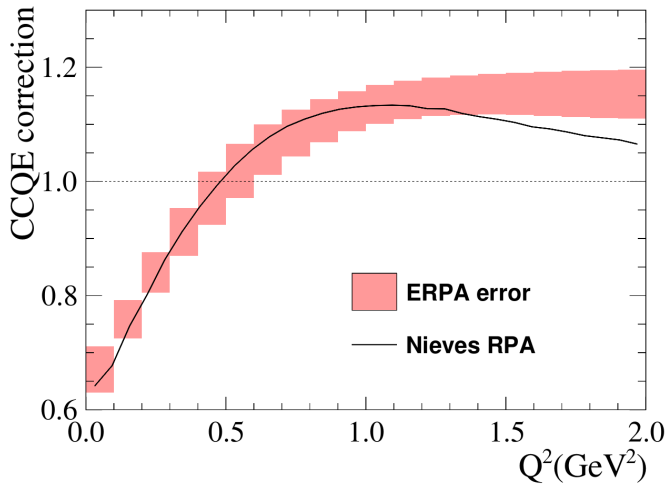}
 \caption{Effective-RPA with error band from the fit to external data compared with RPA corrections computed in Ref.~\cite{Nieves:2011pp}.}
 \label{fig:niwg_erpa}
\end{figure}

The correction to the CCQE cross section due to long-range nuclear correlations, described by RPA in the reference model,
has been parametrized as a function of $Q^2$ in terms of five free parameters. 
A joint fit to the MiniBooNE~\cite{mb-ccqe-2010, mb-ccqe-antinu-2013} and MINERvA~\cite{minerva-nu-ccqe, minerva-antinu-ccqe}
$\nu_\mu$ and $\bar{\nu}_{\mu}$ datasets has been performed 
to extract an alternative, data-driven RPA correction, labeled ``effective-RPA'' in the following. The effective-RPA correction
%agrees with the reference RPA at intermediate $Q^2$ but 
deviates from the reference RPA at high $Q^2$, as can be seen in Fig.~\ref{fig:niwg_erpa}.
%agrees with the reference RPA at intermediate $Q^2$ but has a $<$5\% smaller cross-section suppression at low and high $Q^2$ 
%($Q^2 \lesssim 0.2$ GeV$^2$ and $Q^2 \gtrsim 1.4$ GeV$^2$).

\begin{figure}[htbp]
 \includegraphics[width=0.5\textwidth]{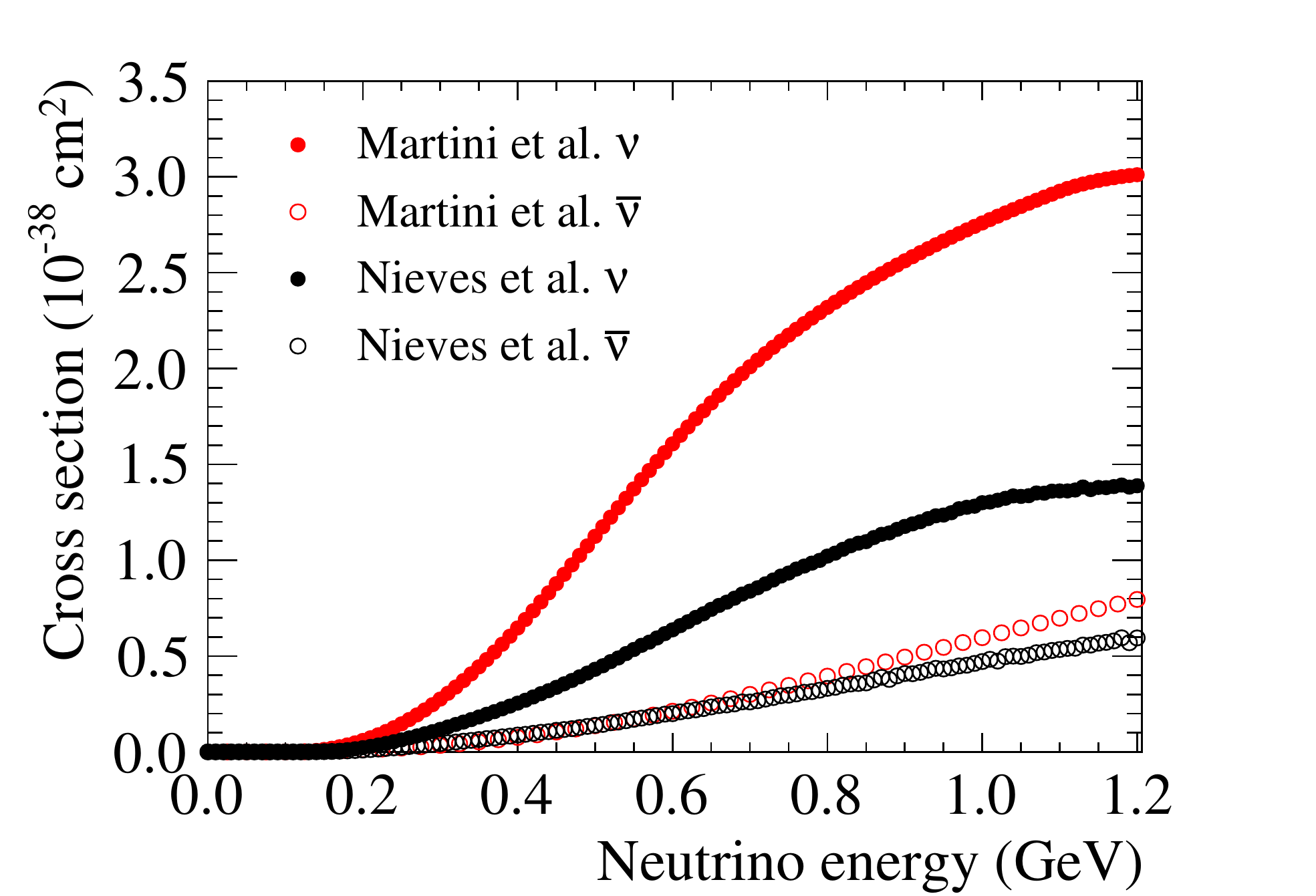}
 \caption{Multi-nucleon interactions (2p2h) cross section on $^{12}$C as a function of energy from the models of Nieves (reference model in the text)~\cite{Nieves:2011pp}
and Martini (alternative model in the text)~\cite{Martini:2011wp}.}
 \label{fig:niwg_2p2h}
\end{figure}

\begin{figure}[htbp]
 \includegraphics[width=0.5\textwidth]{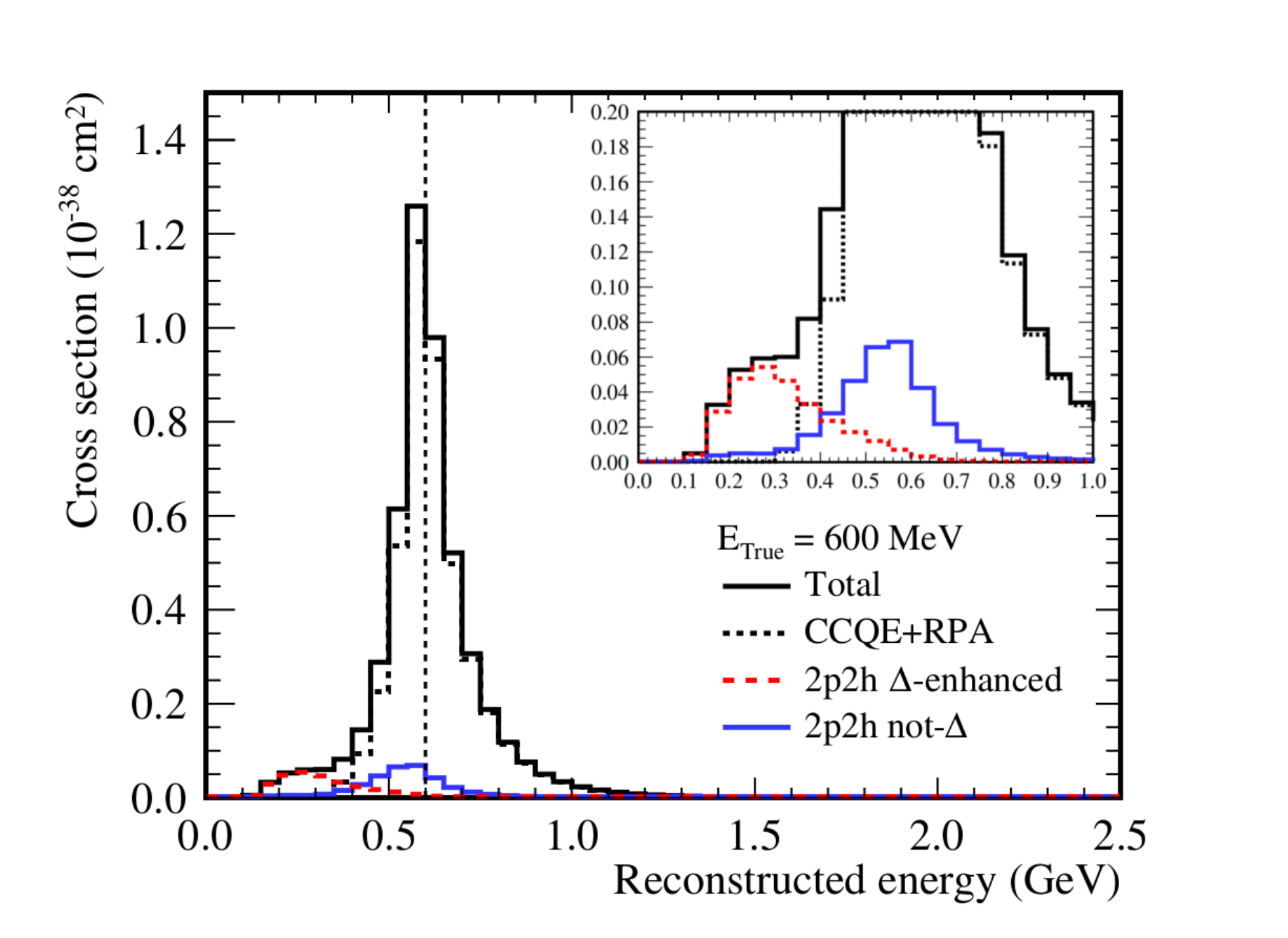}
 \caption{
 %Bias in neutrino energy estimation, using the CCQE two-body approximation, due to the different components of multi-nucleon interactions (2p2h) on $^{12}$C from the model in Ref.~\cite{Nieves:2011pp}.
Neutrino energy calculated with the CCQE two-body assumption for CCQE and 2p2h interactions of 600 MeV muon neutrinos on $^{12}$C simulated with the reference model. The different components of 2p2h show differing amounts of bias.}
 \label{fig:niwg_Delta_notDelta}
\end{figure}

The model in Ref.~\cite{Martini:2011wp} has been considered as an alternative 2p2h model, which differs from the reference model in many respects. The alternative 2p2h cross section is twice as large 
for neutrino interactions but has a similar strength for antineutrino interactions, except at high neutrino energies 
($E_\nu\gtrsim 1$~GeV) where it is about 30\% larger, as can be seen in Fig.~\ref{fig:niwg_2p2h}. 
The difference between the 2p2h normalization for neutrino and antineutrino interactions is constrained with the ND280 data 
in order to avoid biases in the CP asymmetry measurement in the oscillation analysis. 
The alternative model has also been used for one of the studies of simulated data.
Another important difference between the two models consists in the relative proportion of nucleon-nucleon correlations,
meson exchange currents and their interference, the first being strongly enhanced in the
alternative model. This difference affects the estimation of the neutrino energy from the
outgoing lepton kinematics. This estimation assumes the CCQE hypothesis, and it is well known that
the 2p2h contribution biases the neutrino energy reconstruction~\cite{Martini:2012fa,Nieves:2012yz} 
if not properly taken into account in the simulation. 
The reference model includes 2p2h events, and so this effect is included in the T2K neutrino oscillation analysis. Nevertheless,
%This effect is taken into account using the reference interaction model described above. Nevertheless, 
the different 2p2h components produce different biases in the neutrino energy estimation, as shown in Fig.~\ref{fig:niwg_Delta_notDelta},
therefore incorrectly estimating the relative proportions of nucleon--nucleon correlations and
meson exchange current can cause a residual bias in the neutrino energy estimation.
To address this, three simulated datasets have been built. In the first, the multi-nucleon
interactions have been reweighted as a function of neutrino energy, separately for neutrino and antineutrino, to
reproduce the alternative model (referred to as the ``alternative 2p2h model'' in the following). 
In the other two simulated datasets, the full 2p2h cross section has been assigned either to
meson exchange currents (``Delta-enhanced 2p2h'') 
or nucleon-nucleon correlations only (``not-Delta 2p2h'') 
by reweighting the muon kinematics as a function of muon angle, muon momentum and neutrino energy. 

The results obtained by considering all the alternative models, SF, alternative 1p1h, effective RPA, alternative 2p2h, Delta-enhanced 2p2h, not-Delta 2p2h, are shown in Sec.~\ref{sec:fds}.

\section{The ND280 complex}
\label{sec:nd}

The precise measurement of neutrino oscillations in T2K requires a good understanding of the neutrino beam properties and of neutrino interactions. The two previous sections 
%have described constraints on the neutrino flux and cross section models based on external measurements. 
have described the neutrino flux model and neutrino-nucleus interaction model, and constraints on those models based on external measurements.
As we will show in Sec.~\ref{sec:oasyst}, with only that information, the precision on the measurements of neutrino oscillations parameters would be limited. In order to reduce the model uncertainties a near detector complex has been built 280~m downstream of the production target. The goal of the near detectors is to directly measure the neutrino beam properties and the neutrino interaction rate. The near detector complex comprises an on-axis detector (INGRID) and an off-axis detector (ND280). INGRID is composed of a set of modules with sufficient target mass and transverse extent to monitor the beam direction and profile on a day-to-day basis. The ND280 is composed of a set of subdetectors, installed inside a magnet, and is able to measure the products of neutrino interactions in detail.

In this section, the methods used to select high purity samples of neutrino and antineutrino interactions in INGRID and ND280 will be described, and the results are compared with the predictions obtained from the beam line simulation and the interaction models. The use of the ND280 data to reduce the systematic uncertainties in the T2K oscillation analysis will be described in Sec.~\ref{sec:banff}.

\subsection{On-axis near detector}
\label{sec:ingrid}
%%%%%%%%%%%%%%%%%%%%%%%%%%%%%%%%%%%
%%%%%%%%%%Plan%%%%%%%%%%%%%%%%%%%%%
%1. Description of the detector%%%%
%2. Event selection%%%%%%%%%%%%%%%%
%3. Corrections%%%%%%%%%%%%%%%%%%%%
%4. Systematics%%%%%%%%%%%%%%%%%%%%
%5. Results on number of events &%%
%Beam position %%%%%%%%%%%%%%%%%%%%
%6. Impact on total uncertainty%%%%
%%%%%%%%%%%%%%%%%%%%%%%%%%%%%%%%%%%
The INGRID detector is used to monitor the neutrino beam rate, profile and center. Those parameters are used to determine the off-axis angle at SK. INGRID is centered on the neutrino beam axis and samples the neutrino beam with a transverse cross section of $10~\text{m} \times 10~\text{m}$ using 14 modules positioned in the shape of a cross. Each INGRID module holds 11 tracking segments built from pairs of orthogonally oriented scintillator planes interleaved with nine iron planes. There are also three veto planes, located on the top, bottom and one side of each module. The most upstream tracking plane is used as a front veto plane. The scintillator planes are built from 24 plastic scintillator bars instrumented with fibers connected to multi-pixel photon counters (MPPCs) to detect scintillation light. More details can be found in Ref.~\cite{Abe:2011xv}.

%%%%%%%%%%%%%%%%%%%%%%%%%%%%%%%%%%%
%%%%%%The scheme of INGRID%%%%%%%%%
%%%%%%%%%%%%%%%%%%%%%%%%%%%%%%%%%%%

\subsubsection{Event selection and corrections}
Neutrino and antineutrino interactions within INGRID modules are first reconstructed independently in the horizontal and vertical layers of scintillators. Pairs of tracks in the two different orientations are then matched by comparing the most upstream point 
%Z coordinates 
to form 3D tracks. The upstream edges of the different 3D tracks are then compared in the longitudinal and transverse direction with respect to the beam direction in order to construct a common vertex. The subsequent reconstructed event is rejected if the vertex is reconstructed out of the fiducial volume, if the external veto planes have hits within 8 cm from the upstream extrapolated position of a reconstructed track, or if the event timing deviates from more than $100~$ns to the expected event timing.

In order to reduce the systematic uncertainty on the track reconstruction in \nub-mode, the selection has been improved from the one used in Ref.~\cite{Abe:2015awa}. %in the antineutrino mode. 
To reduce the impact of MPPC dark noise, the reconstruction is only applied to events where two consecutive tracking planes have a hit coincidence on their horizontal and vertical planes. This condition was not used in Ref.~\cite{Abe:2015awa} and has been applied only to the \nub-mode in the analyses presented here. A total of $12.8 \times 10^{6}$ and $4.1 \times 10^{6}$ neutrino events are reconstructed respectively in $\nu$- and \nub-mode, with estimated purities of $99.6\%$ and $98.0\%$ respectively.

The selected number of events in each module is corrected to take into account the impact of the detector dead channels, the event loss due to non-reconstructed neutrino interactions caused by pile-up, the variation of the iron mass between the modules,  the time variation of the MPPC noise during the data taking and the contamination from external background as the previous INGRID analysis~\cite{Abe:2011xv, Abe:2015awa}.

\subsubsection{Systematic uncertainties}
The systematic uncertainties on the event selection are estimated using the simulation and control samples. The sources of error are the same as those identified in Ref.~\cite{Abe:2015awa} and include the neutrino target mass, the accidental coincidence with MPPC dark noise, the hit efficiency, the event pile-up, the cosmic and beam-induced backgrounds along with errors associated to event selection cuts. The method for estimating the uncertainty has not been changed since Ref.~\cite{Abe:2015awa} for $\nu$-mode, and is also applied here for \nub-mode data. The uncertainties are evaluated to be $0.9\%$ and $1.7\%$ for neutrino and antineutrino data respectively. The larger uncertainty in \nub-mode mainly arises from a discrepancy between data and simulation for interactions producing a track that cross less than four tracking planes.

\subsubsection{Results of neutrino beam measurement}
\begin{figure*}
  \centering
  \includegraphics[width=1.0\textwidth]{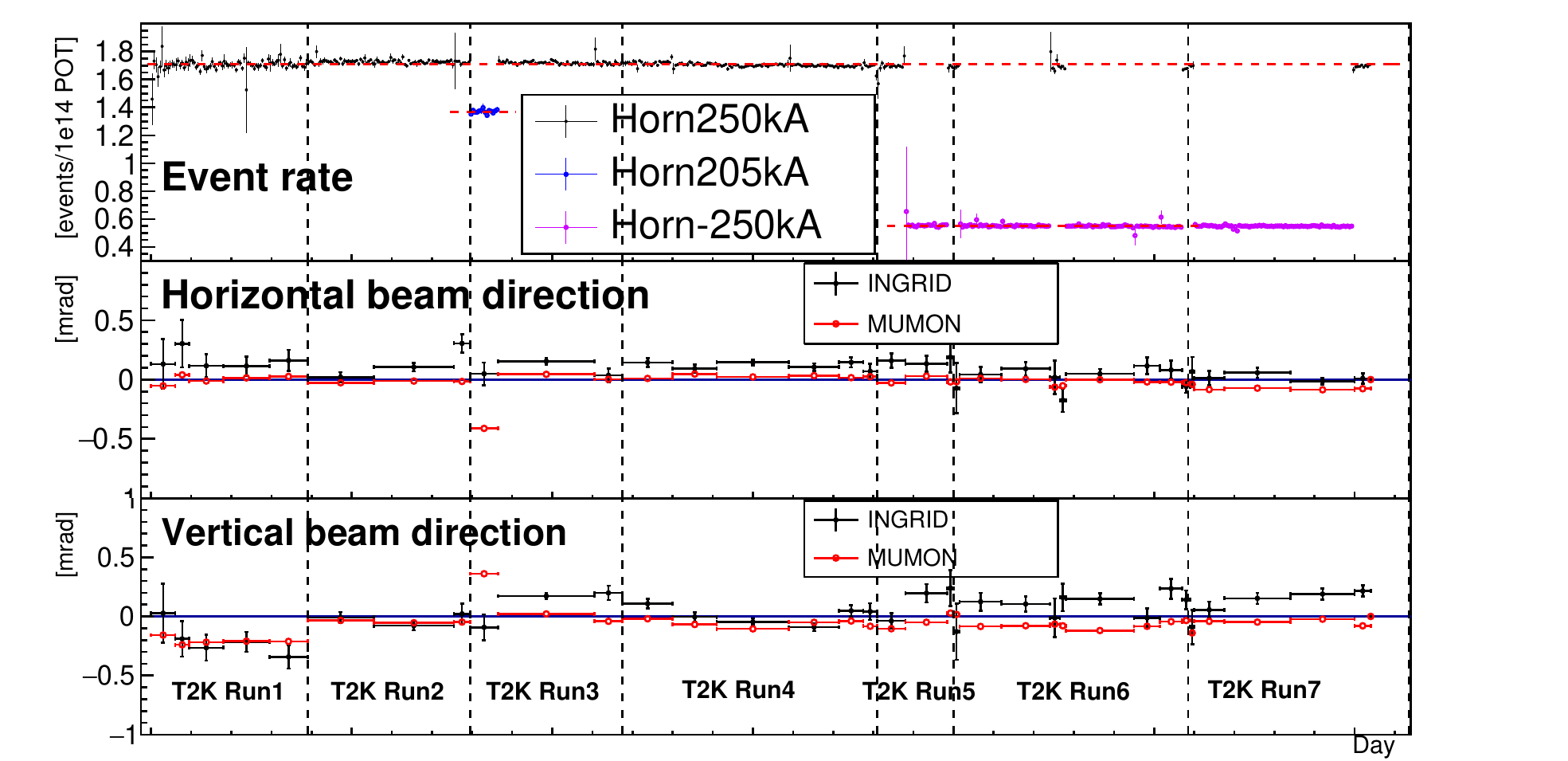}
    \caption{The event rate at INGRID as a function of time is shown in the top panel. The horizontal dashed redline corresponds to the best fit for the different horn currents with a constant function.
    %The horizontal dashed line re the average event rate for the different horn currents. 
    The central and bottom panels show the neutrino beam direction measured at INGRID and at MUMON along the horizontal and vertical transverse directions with respect to the beam as a function of time. The dashed vertical lines separate the seven T2K physics runs.}\label{Fig:INGRID:Rate}
\end{figure*}

The stability of the neutrino flux is monitored by measuring the event rate, that is the total number of selected events per protons on target.
%The intensity of the neutrino flux is estimated using the total number of selected events per protons on target. 
Fig.~\ref{Fig:INGRID:Rate} shows the intensity stability as a function of time for both $\nu$- and \nub-modes. Most of the data have been taken with the horn currents set to an absolute value of 250~kA, except for a small fraction of $\nu$-mode data taken during T2K run 3 in which horns were operated at 205~kA. 
%The averaged intensities for these two modes are respe
The average event rates are compared with the simulation, and the ratios are:
\begin{eqnarray}
  \begin{aligned}
    \frac{N^{\text{data, $\nu$}}_{250kA}}{N^{\text{MC, $\nu$}}_{250kA}} & = 1.010 \pm 0.001(stat.) \pm 0.009(syst.), \\
    \frac{N^{\text{data, $\nu$}}_{205kA}}{N^{\text{MC, $\nu$}}_{205kA}} & = 1.026 \pm 0.002(stat.) \pm 0.009(syst.), \\
    \frac{N^{\text{data, $\nub$}}_{-250kA}}{N^{\text{MC, $\nub$}}_{-250kA}} & = 0.984 \pm 0.001(stat.) \pm 0.017(syst.). \\
    \label{Eq:INGRID:Rate}
  \end{aligned}
\end{eqnarray}
%The intensities are in agreement in the simulation within the uncertainties.\\
The quoted systematic uncertainties do not include the uncertainties on flux and cross section model, but they only include INGRID detector systematic uncertainties. The numbers of expected events in the Monte Carlo are obtained with the cross section models described in Sect.~\ref{sec:niwgmodel}. 
The spatial spread of the neutrino beam is measured using the number of reconstructed events in each INGRID module. 
%The spread of the neutrino beam is estimated using the number of reconstructed events in each INGRID module. 
This produces a measurement of the number of events as function of the distance from the center in both vertical and horizontal directions. The two distributions are fit with a Gaussian, and the neutrino beam center and width are given by the mean and the sigma of the fit.

The measurement of the position of the beam center is crucial to determine the off-axis angle, and therefore, the neutrino beam energy at SK. A deviation of $1~$mrad of the beam direction would shift the peak neutrino energy by $2\%$. Fig.~\ref{Fig:INGRID:Rate} shows the beam direction stability for all data taking periods. The variations are well within the design goal of 1~mrad.
The average angles %of the measured beam center compared to nominal expectations 
are
\begin{eqnarray}
  \begin{aligned}
    \overline{\theta}^{\text{beam, $\nu$}}_{X} & = 0.027 \pm 0.010(stat.) \pm 0.095(syst.)~{\rm mrad} \\
    \overline{\theta}^{\text{beam, $\nu$}}_{Y} & = 0.036 \pm 0.011(stat.) \pm 0.105(syst.)~{\rm mrad} \\
    \label{Eq:INGRID:BeamPositionFHC}
  \end{aligned}
\end{eqnarray}
for $\nu$-mode, and
\begin{eqnarray}
  \begin{aligned}
    \overline{\theta}^{\text{beam, $\nub$}}_{X} & = -0.032 \pm 0.012(stat.) \pm 0.121(syst.)~{\rm mrad} \\
    \overline{\theta}^{\text{beam, $\nub$}}_{Y} & = 0.137 \pm 0.020(stat.) \pm 0.140(syst.)~{\rm mrad} \\
    \label{Eq:INGRID:BeamPositionFHC}
  \end{aligned}
\end{eqnarray}
for $\nub$-mode. All values are compatible with the expected beam direction.

%1. Intensity measurement.
%2. Beam center measurement + say it stays within 28mrad (stability plot!)
%3. Beam width measurement

%Total systematic error: different for neutrino and antineutrino mode.
%Neutrino mode is shown on OA2014: error of 0.9%
%Antineutrino mode: variation due to active plane selection. 1.7%

%The same selection is applied both for foward and reverse horn current modes. 

%Say that the numu selection has been extensively described in (Long OA Paper 2014). We will focus on antineutrino.
%The numubar selection
%What is new in the numubar selection? What is the purity?

%Width with stat/syst error for all runs!
%Beam intensity for all runs

%1. Intensity, beam center position in X/Y (in mrad) and widths for Run 4, 5 and 6 in FHC
%MC predicted width.

%Total number of events in FHC and \nub, efficiency of what and purity of what? numu events or numu+numubar? No, of all neutrino events, numu+numub+nue+nueb!
%2. Expected quantities by MC: 
%Neutrino250kA: 1.706, 0, 0, 432.0, 452.0
%Neutrino205kA: 1.317, 0, 0, 432.0, 452.0
%Antineutrino: 0.564, 0, 0, ?, ?

\subsection{The off-axis ND280 detector}
\label{sec:nd280}

The off-axis near detector ND280 measures the neutrino energy spectrum, flavor content, and interaction rates of the unoscillated beam. 
These measurements are crucial to reduce the uncertainties on neutrino flux and interaction models which affect the prediction on the number of expected events at the far detector.

The ND280 detector consists of a set of subdetectors installed inside the refurbished UA1/NOMAD magnet, 
which provides a 0.2 T field, used to measure the charge and the momentum of particles passing through ND280. 
For the analyses described in this paper, \num and \numb charged current interactions are selected in the tracker region of ND280, 
which consists of three Time Projection Chambers (TPC1, 2, 3) \cite{Abgrall:2010hi}, interleaved with two Fine-Grained Detectors (FGD1, 2) \cite{Amaudruz:2012agx}. %Thanks to the particle identification capability ND280 is also used to measure the \nue event rate, due to the intrinsic \nue contamination in the \num beam before the oscillations as described in~\cite{Abe:2014usb,Abe:2015mxf}.

The upstream FGD1 detector consists of fifteen polystyrene scintillator modules, while the downstream FGD2 contains seven polystyrene scintillator modules interleaved with six water modules. 
The FGDs provide target mass for neutrino interactions and track the charged particles coming from the interaction vertex, 
while the TPCs provide 3D tracking and determine the momentum and energy loss of each charged particle traversing them.  
The observed energy loss in the TPCs, combined with the measurement of the momentum, is used for particle identification of the charged tracks produced in neutrino interactions in order to measure exclusive CC event rates.
The major updates in the near detector analysis with respect to Ref.~\cite{Abe:2015awa} are the use of interactions in FGD2 and the inclusion of data taken with the \nub-mode beam. 

The charge and particle identification ability of the tracker is important because it provides separation between $\mu^+$ (produced by $\bar\nu_\mu$ CC interactions) and $\mu^-$ (produced by $\nu_\mu$ CC interactions) when T2K runs in $\bar\nu$-mode.
Moreover, by including both FGD1 and FGD2 samples, the properties of neutrino interactions on water can be effectively isolated from those on carbon, 
reducing the uncertainties related to extrapolating across differing nuclear targets in the near and far detectors. 
The near detector analysis described here uses a reduced dataset comprising $5.81\times10^{20}$ POT in $\nu$-mode and $2.84\times10^{20}$ POT in $\bar\nu$-mode, as shown in Tab. \ref{tbl:nd280_pot}. 
%which is a sub-sample of the total statistic collected by T2K and used in the oscillation analysis.

\begin{table}[h]
\caption{
Collected POT for each dataset used in the ND280 analysis.  
}
\label{tbl:nd280_pot}
\begin{tabular}{ l c c c }
\hline\hline
Run & Dates & \(\nu\)-Mode POT & \(\bar{\nu}\)-Mode POT \\ 
Period & & (\(\times10^{20}\)) & (\(\times10^{20}\))   \\ \hline
Run 2 & Nov. 2010-Mar. 2011 & 0.78 & -- \\
Run 3 & Mar. 2012-Jun. 2012 & 1.56 & -- \\
Run 4 & Oct. 2012-May 2013 & 3.47 & -- \\
Run 5 & Jun. 2014 & -- & 0.43 \\
Run 6 & Nov. 2014-Apr. 2015 & -- & 2.41 \\
\hline
Total & Nov. 2010-Apr. 2015 & 5.81 &  2.84 \\
\hline \hline
\end{tabular}
\end{table}

\subsubsection{ND280 \num CC selection in $\nu$-mode}
\label{sec:nd280numu}

The event selection in $\nu$-mode beam is unchanged since the previous analysis described in Ref.~\cite{Abe:2015awa}. 
Muon-neutrino-induced CC interactions are selected by identifying the
$\mu^-$ produced in the final state as the highest-momentum, negative-curvature track in each event
with a vertex in FGD1 (FGD2) Fiducial Volume (FV) and crossing the middle (last) TPC. 
The energy lost by the selected track in the TPC must be consistent with a muon.

\begin{figure*}[htbp]
    \centering
    \includegraphics[height=6cm]{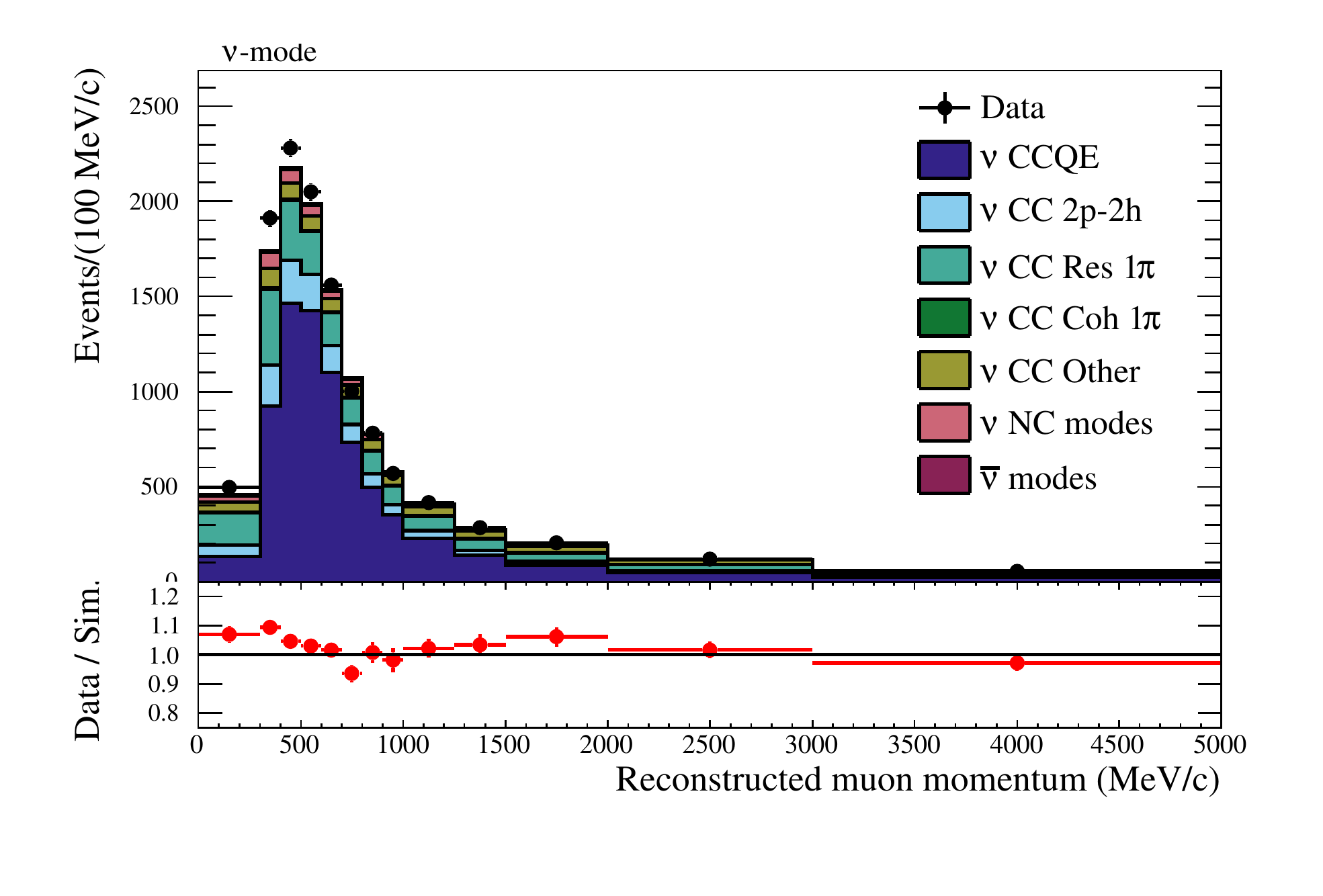}
    \hfill
    \includegraphics[height=6cm]{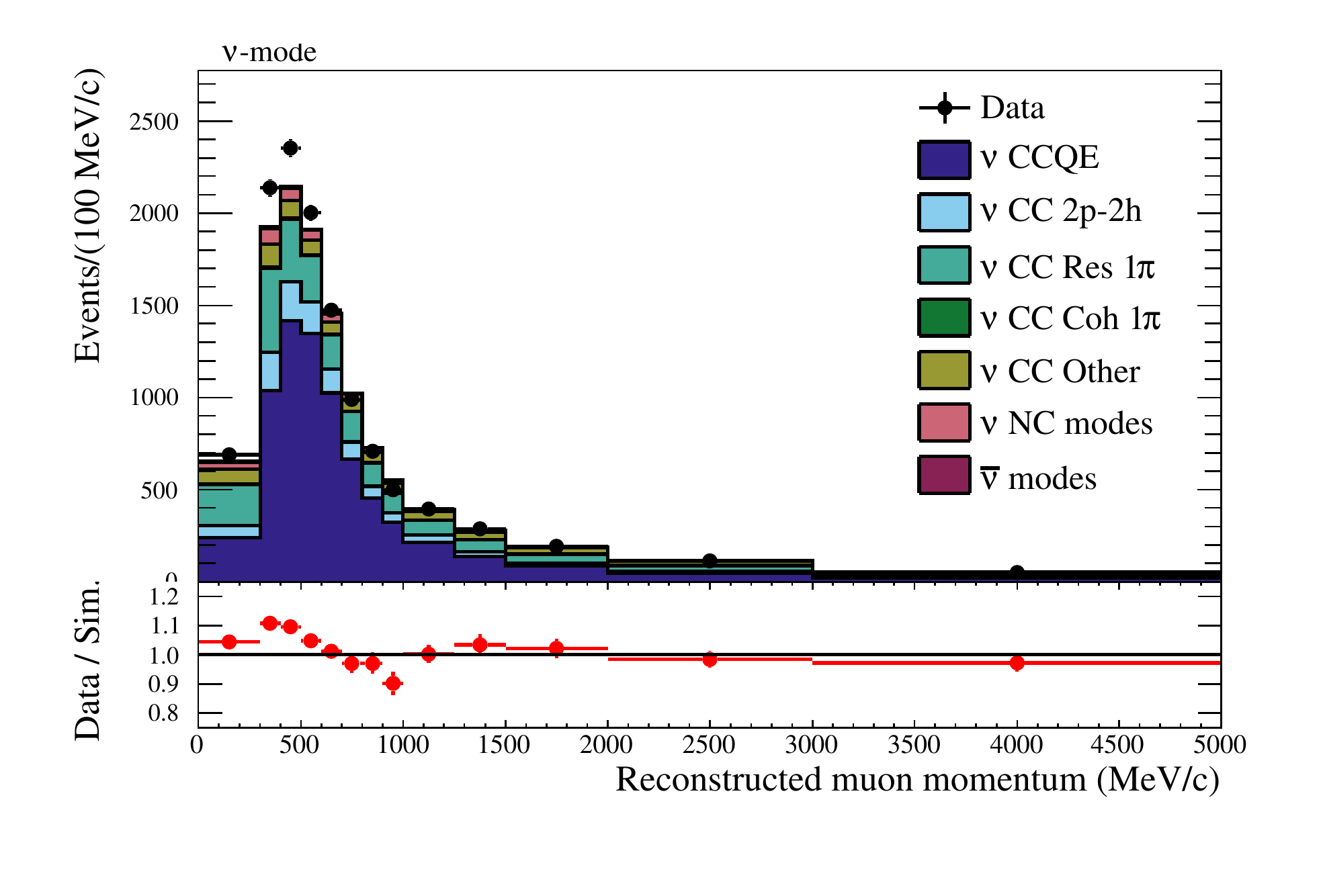}
    \hfill
        \includegraphics[height=6cm]{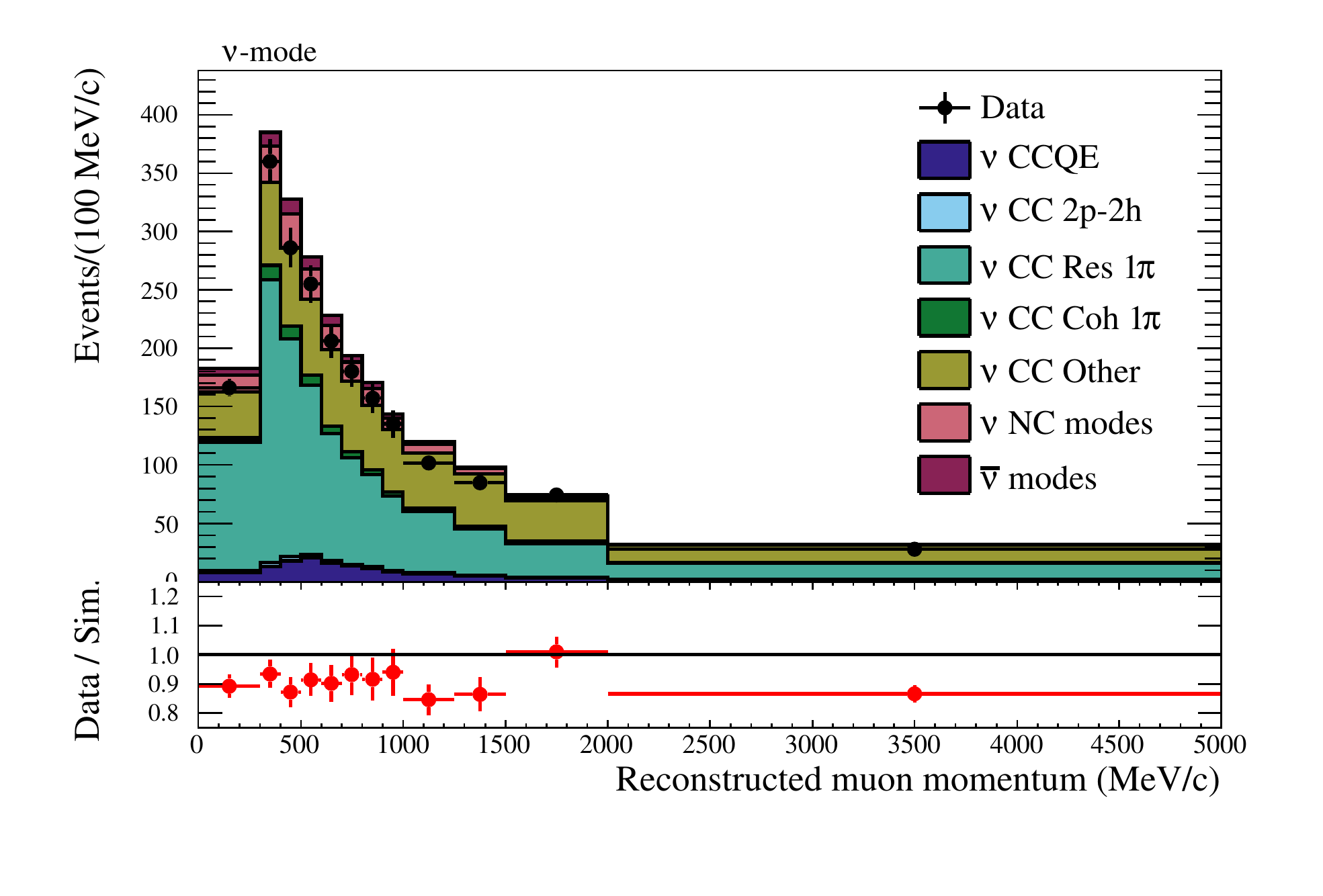}
    \hfill
    \includegraphics[height=6cm]{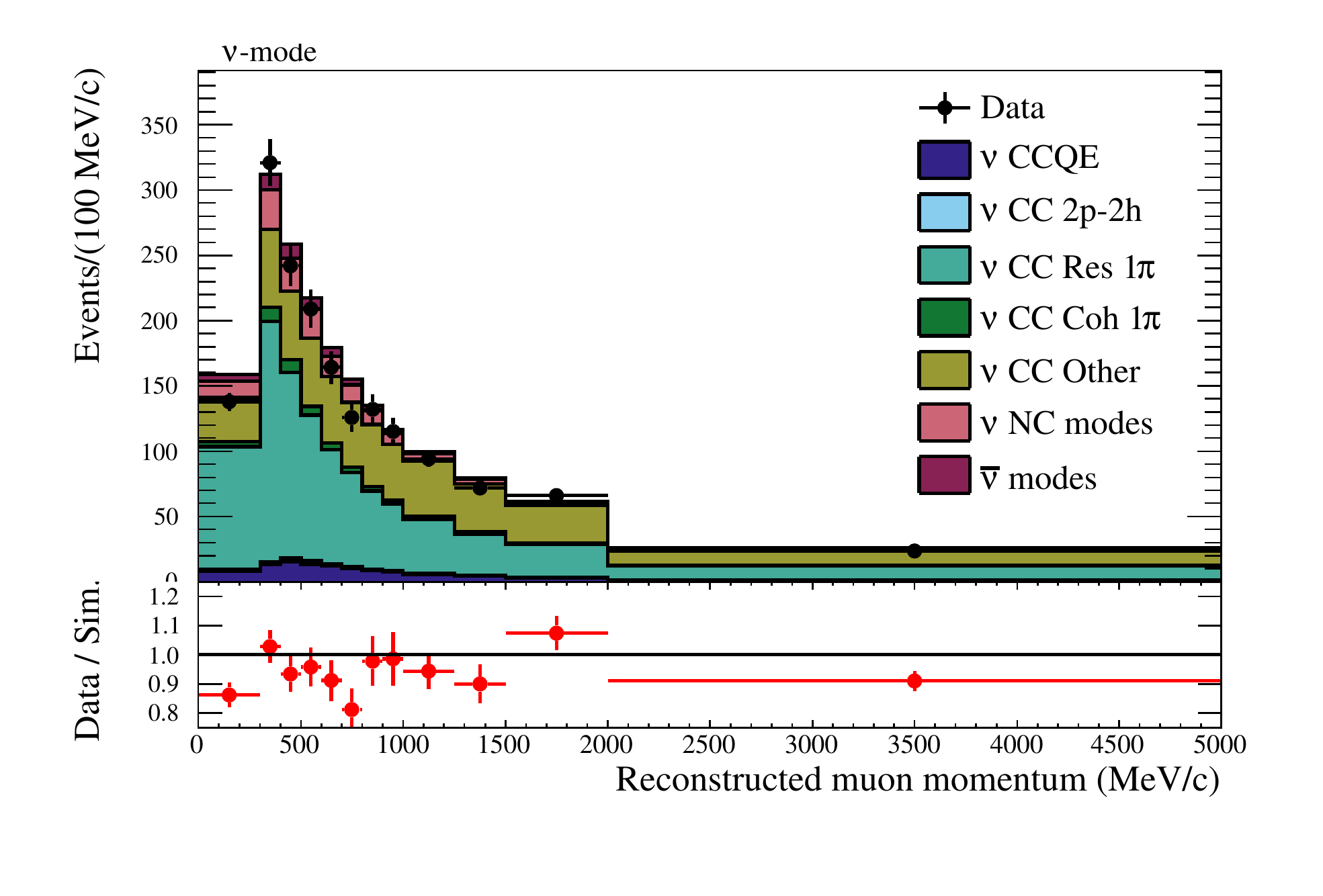}
    \caption{Top: muon momentum distributions of the $\nu$-mode $\nu_\mu$ CC-0$\pi$ samples in FGD1 (left) and FGD2 (right). 
    Bottom: muon momentum distributions of the $\nu$-mode $\nu_\mu$ CC-1$\pi^+$ samples in FGD1 (left) and FGD2 (right). 
    All distributions are shown prior to the ND280 fit.
    %The simulation is broken-down by neutrino reaction type.
    }
    \label{fig:nd280_fhc_cc0pi}
\end{figure*}

All the events generated upstream of FGD1 are rejected by excluding events with a track in the first TPC.
The selected $\nu_\mu$ CC candidates are then divided into three subsamples, according to the number of identified pions in the event:
CC-$0\pi$, CC-$1\pi^+$ and CC-Other, which are dominated by ``quasi-elastic'' (CCQE), CC resonant pion production,
and DIS interactions, respectively. 
Pions are selected in different ways according to their charge. A $\pi^+$ can be identified in three ways:
an FGD+TPC track with positive-curvature and an energy loss in the TPC consistent with a pion;
an FGD-contained track with a charge deposition consistent with a pion; 
or a delayed energy deposit in the FGD due to a decay electron from stopped $\pi^+\rightarrow\mu^+$.
In this analysis, $\pi^{-}$'s are only identified by selecting negative-curvature FGD+TPC tracks, 
while $\pi^0$'s are identified by looking for tracks in the TPC with charge depositions consistent with an electron from a $\gamma$ conversion.
The output of the $\nu$-mode tracker selection are six samples, three per FGD.
The selected CC-0$\pi$ and CC-$1\pi^+$ samples in both FGDs before the ND280 fit are shown in Fig. \ref{fig:nd280_fhc_cc0pi}.
For each of the selected samples, the number of observed and predicted events are shown in Tab.~\ref{tbl:nd_fhc_samp}.
\begin{table}[h!]
\begin{centering}
\caption{Observed and predicted event rates for different ND280 samples collected in $\nu$-mode beam. Before the ND280 fit that will be described in Sec.~\ref{sec:banff}, uncertainties of $\sim$20\% on the event rates are expected.}
\begin{tabular}{  l c  c }
\hline \hline
FGD1 sample & Data & Prediction\\
\hline
$\nu_\mu$ CC-0$\pi$ & 17354 & 16951\\
$\nu_\mu$ CC-1$\pi^+$ & 3984 & 4460\\
$\nu_\mu$ CC-Other & 4220 & 4010 \\
\hline \hline
FGD2 sample & Data & Prediction\\
\hline
$\nu_\mu$ CC-0$\pi$ & 17650 & 17212\\
$\nu_\mu$ CC-1$\pi^+$ & 3383 & 3617\\
$\nu_\mu$ CC-Other & 4118 & 3627\\
\hline \hline
\end{tabular}
\label{tbl:nd_fhc_samp}
\end{centering}
\end{table}

\subsubsection{ND280 \numb and \num CC selections in \nub-mode}
\label{sec:nd280numubar}

The main difference between $\nu$- and \nub-modes is the increase in the number of interactions produced by ``wrong-sign" neutrinos. Once differences in the flux and the cross section are taken into account, the wrong-sign contamination in \nub-mode is expected to be approximately 30\%, while the wrong-sign contamination in \nunu-mode is approximately 4\%.
%While SK and INGRID cannot distinguish the charge of the particles, the magnetic field at the ND280 allows the separation of \num and \numb CC interactions as has previously been remarked.
% allowing a precise measurement of the neutrino background.

%For this reason, since SK is not immersed in a magnetic field and thus it cannot distinguish the charge of particles coming from an antineutrino or a neutrino interaction when T2K runs in $\bar\nu$-mode,
%both $\bar\nu_\mu$ and $\nu_\mu$ CC interactions are selected at the near detector, 
%in order to have a precise understanding of the neutrino background.

The lepton selection criteria of $\bar\nu_\mu$ ( $\nu_\mu$) CC interactions is similar to the one used in the neutrino beam mode, except for the condition 
that the highest-momentum, positively (negatively) charged particles must also be the highest momentum track in the event. 
This additional cut is essential to reduce the background due to $\pi^+$ ($\pi^-$) generated in neutrino (antineutrino) interactions that can be misidentified as the muon candidate. The selected $\bar\nu_\mu$ CC ($\nu_\mu$ CC) candidate events are divided in two subsamples: 
CC-1-Track, dominated by CCQE-like interactions; and CC-N-Tracks (N$>$1), a mixture of resonant production and DIS.
These two subsamples are defined by the number of reconstructed tracks crossing the TPC.
For these selections the CC candidates are not divided into three subsamples as in Sec. \ref{sec:nd280numu}, according to the number of identified pions in the event in order to avoid samples with low statistics. 
%The low statistic of CC events in $\bar\nu$-mode is due to the poor statistic collected in this beam configuration with respect to $\nu$-mode and to the small antineutrino cross section with respect to the neutrino one. 

\begin{figure*}[htbp]
    \centering
    \includegraphics[height=6cm]{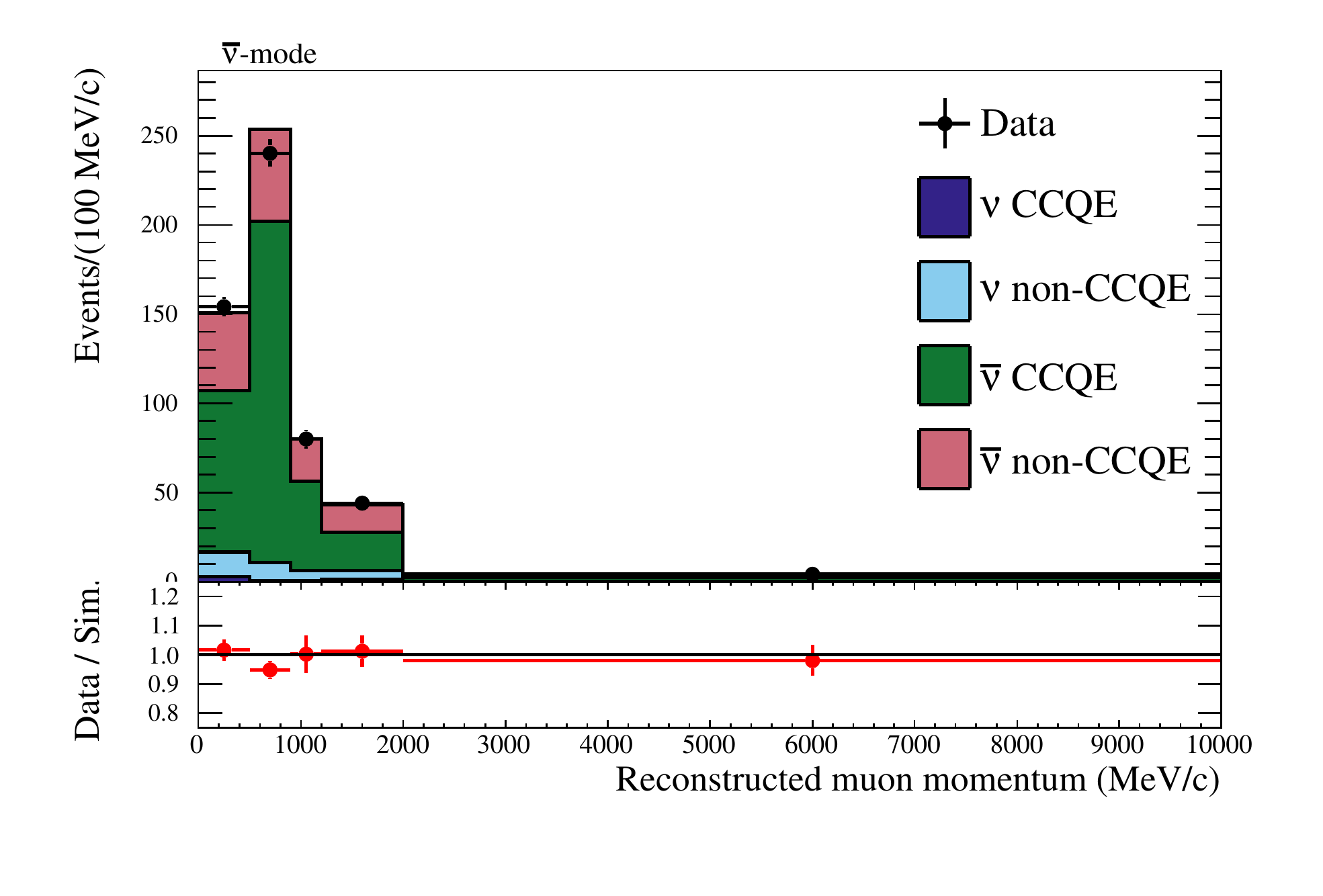}
    \hfill
    \includegraphics[height=6cm]{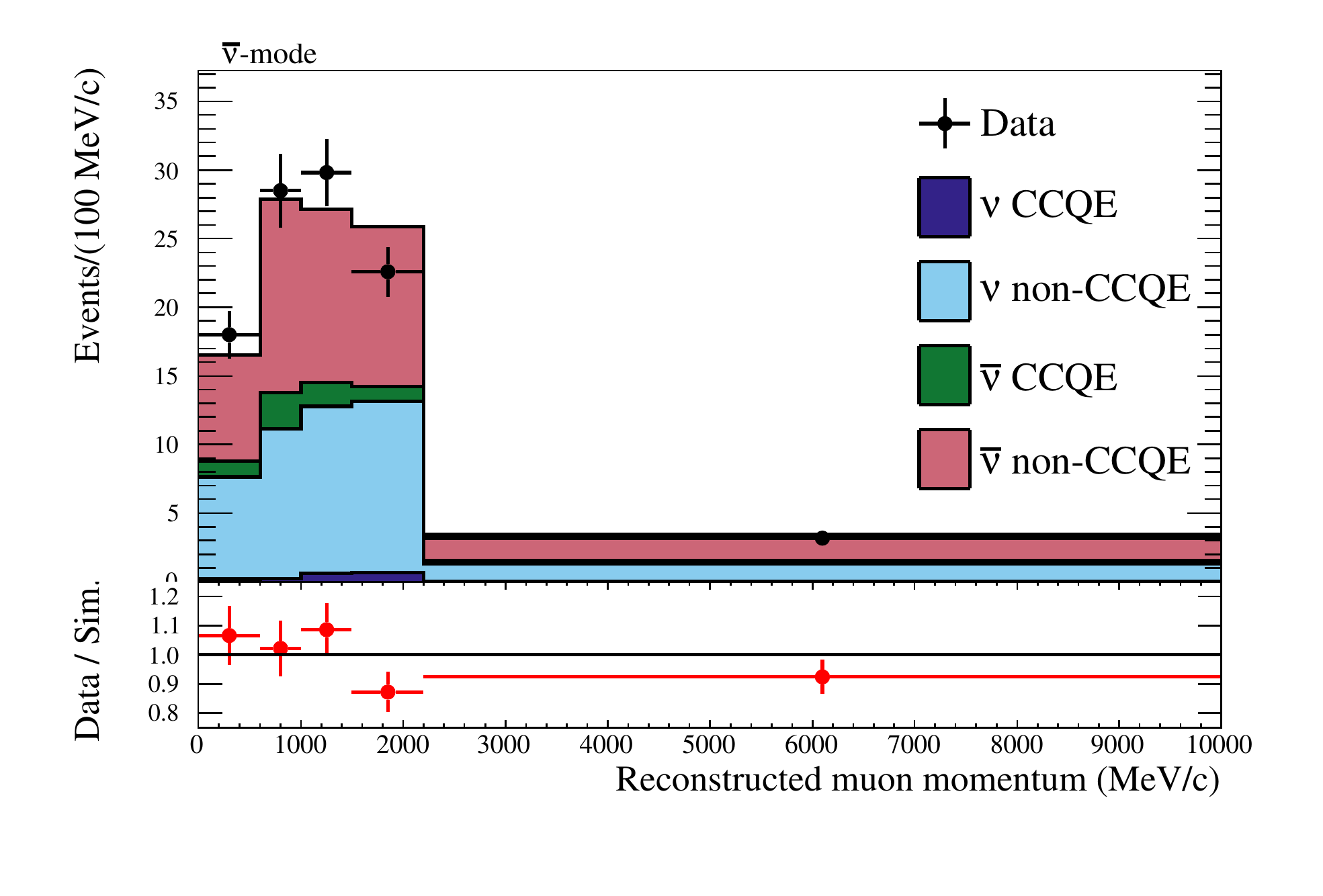}
    \hfill
    \includegraphics[height=6cm]{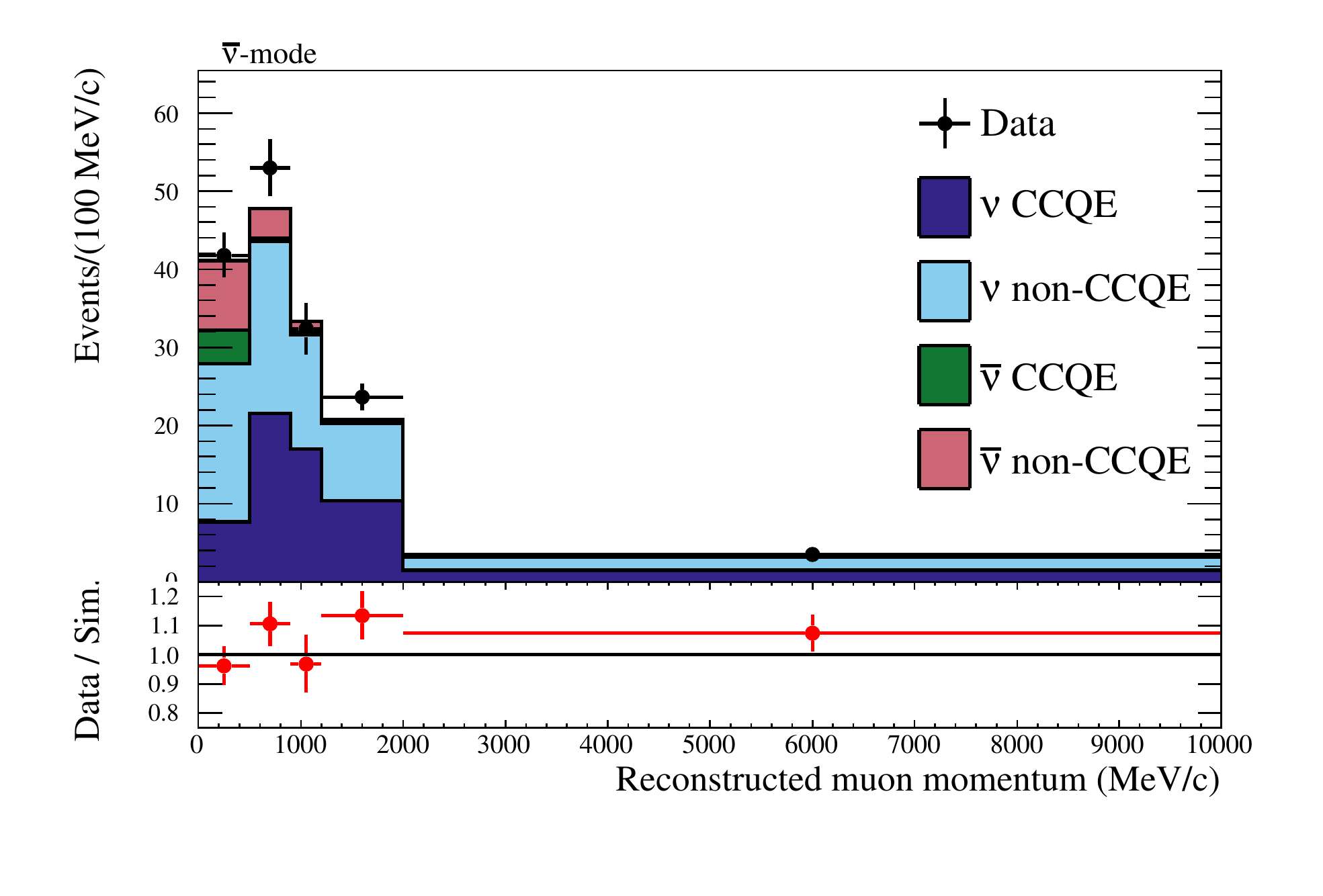}
    \hfill
    \includegraphics[height=6cm]{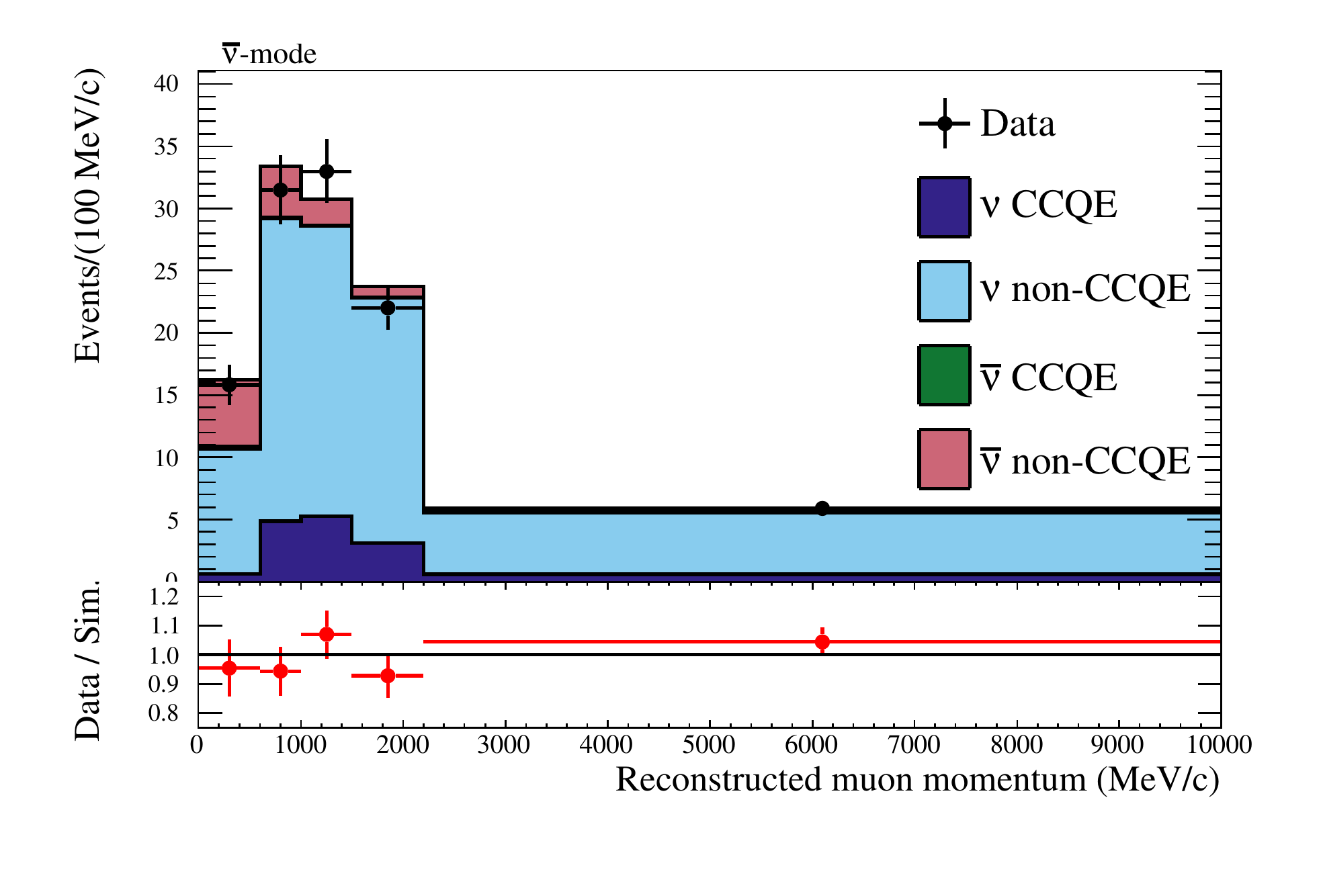}
    \caption{Top: muon momentum distributions for the $\bar\nu$-mode $\bar\nu_\mu$ CC-1-Track (left) and CC-N-Tracks (right) samples. %The simulation is broken-down by neutrino reaction type.
    Bottom: muon momentum distributions for the $\bar\nu$-mode $\nu_\mu$ CC-1-Track (left) and CC-N-Tracks (right) samples. 
    All distributions are shown prior to the ND280 fit.
    }
    \label{fig:nd280_rhc}
\end{figure*}

The output of the \nub-mode tracker selection are eight samples, four per FGD. 
For each of the selected sample, the number of predicted events and the ones observed in data are shown in Tab. \ref{tab:nd_rhc_samp}.
The four selected samples in FGD1, before the ND280 fit, are shown in Fig. \ref{fig:nd280_rhc}.

\begin{table}[h!]
\begin{centering}
\caption{Observed and predicted event rates for different ND280 samples collected in $\bar\nu$-mode beam. Before the ND280 fit that will be described in Sec.~\ref{sec:banff} uncertainties of $~$20\% on the event rates are expected.}
\begin{tabular}{  l  c  c }
\hline \hline
FGD1 sample & Data & Prediction\\
\hline
$\bar\nu_\mu$ CC-1-Track & 2663 & 2709\\
$\bar\nu_\mu$ CC-N-Tracks & 775 & 798\\
$\nu_\mu$ CC-1-Track & 989 & 938\\
$\nu_\mu$ CC-N-Tracks & 1001 & 995 \\
\hline \hline
FGD2 sample & Data & Prediction\\
\hline
$\bar\nu_\mu$ CC-1-Track & 2762 & 2730\\
$\bar\nu_\mu$ CC-N-Tracks & 737 & 804\\
$\nu_\mu$ CC-1-Track & 980 & 944\\
$\nu_\mu$ CC-N-Tracks & 936 &  917\\
\hline \hline
\end{tabular}
\label{tab:nd_rhc_samp}
\end{centering}
\end{table}

\subsection{ND280 detector systematic uncertainties}
\label{sec:nd280syst}

In order to assess systematic uncertainties related to the ND280 detector modeling, various different control samples are used, as described in Ref.~\cite{Abe:2015awa}. 
The control samples include muons produced in neutrino interactions outside ND280, cosmic muons, interactions upstream of TPC1, and stopping muons. 
All control samples are independent of the samples used for the ND280 analyses described earlier.
The method to propagate the systematic uncertainties in the near detector analysis is also unchanged with respect to Ref.~\cite{Abe:2015awa}: a vector of systematic parameters $\vec{d}$ scales the expected number of events in bins of $p_{\mu}$ and $\cos\theta_{\mu}$. The covariance of $\vec{d}$, $V_{d}$, is evaluated by varying each systematic parameter.
%by simultaneously varying each systematic to correctly account for non-linearities in their effects.

The difference with respect to the previous analysis are the inclusion of a time of flight (ToF) systematic and new methods used to evaluate charge misidentification and FGD tracking efficiency uncertainties and uncertainties due to interactions outside the fiducial volume. The ToF between FGD1 and FGD2 is used to select events with a backward muon candidate in the FGD2 samples.
The ToF systematic uncertainty is $\sim$0.1\% for the $\nu$-mode FGD2 samples, and $\sim$0.01\% for the $\bar\nu$-mode FGD2 samples.
The ToF uncertainty is smaller in $\bar\nu$-mode because fewer backward-going $\mu^+$ are produced in \numb interactions than backward-going $\mu^-$ in \num interactions.

%The overall improvements in the systematic error reduction are mainly due to a better reconstruction. 
%Nevertheless few systematic uncertainties are decreased since currently treated in a different way: charge mis-identification, 
%FGD tracking efficiency and out of fiducial volume background.
The charge mis-identification uncertainty is parametrized as a function of the momentum resolution in the TPCs.

The FGD tracking efficiency for CC-events, where either a short pion or proton track is also produced, is estimated using a hybrid data-MC sample. This sample uses events with a long FGD-TPC matched muon candidate track with the addition of an FGD-isolated track generated via particle gun with a common vertex.

The method used to estimate the number of out-of-fiducial volume (OOFV) events has been refined by estimating the number of events, and the error, separately for each detector in which the OOFV events occur, rather than averaging over the number of OOFV events produced in all of the detectors outside the tracker as previously.

Most sources of systematic error are common between $\nu$- and \nub-modes because the selection criteria are similar, as described in Sec.~\ref{sec:nd280numubar}. However, as \nub-mode data are divided into CC 1-Track and CC N-Track samples, only based on the number of reconstructed FGD-TPC matched tracks,
%, rather than categorizing events based on the number of tagged pions, 
most uncertainties relating to the FGD reconstruction are not relevant. The exceptions are the FGD-TPC matching and ToF uncertainties, which apply to both modes. Other differences between modes arise because some errors change with the beam conditions (sand muons, pile-up and OOFV) and are evaluated independently for each run period.

The total systematic uncertainties are shown in Tab.~\ref{tbl:nd_syst}.

\begin{table}[h]
\begin{centering}
\caption{Systematic uncertainty on the total event rate affecting the near detector samples.}
\begin{tabular}{  l  c }
\hline \hline
ND280 sample & Total systematic \\
 & uncertainty (\%)\\
\hline \hline
\multicolumn{2}{c}{$\nu$-mode}\\
\hline
FGD1 $\nu_\mu$ CC-0$\pi$ & 1.7 \\
FGD1 $\nu_\mu$ CC-1$\pi^+$ & 3.3\\
FGD1 $\nu_\mu$ CC-Other & 6.5\\
\hline
FGD2 $\nu_\mu$ CC-0$\pi$ & 1.7 \\
FGD2 $\nu_\mu$ CC-1$\pi^+$ & 3.9\\
FGD2 $\nu_\mu$ CC-Other & 5.9\\
\hline
\hline
\multicolumn{2}{c}{$\bar\nu$-mode}\\
\hline
FGD1 $\bar\nu_\mu$ CC-1-Track & 5.4\\
FGD1 $\bar\nu_\mu$ CC-N-Tracks & 10.4\\
FGD1 $\nu_\mu$ CC-1-Track & 2.5\\
FGD1 $\nu_\mu$ CC-N-Tracks & 4.8\\
\hline
FGD2 $\bar\nu_\mu$ CC-1-Track & 3.5\\
FGD2 $\bar\nu_\mu$ CC-N-Tracks & 7.3\\
FGD2 $\nu_\mu$ CC-1-Track & 2.0\\
FGD2 $\nu_\mu$ CC-N-Tracks & 4.0\\
\hline \hline
\end{tabular}
\label{tbl:nd_syst}
\end{centering}
\end{table}

The dominant source of uncertainty for all ND280 samples comes from the pion re-interaction model, used to estimate the rate of pion interactions in the FGDs. This is due to differences between the GEANT4 model, used to simulate pion re-interactions outside the nucleus, and the available experimental data. For example, the systematic uncertainty related to pion interactions affecting the FGD1 $\nu_\mu$ CC-0$\pi$ ($\bar\nu_\mu$ CC-1-Track) sample is 1.4\% (4.9\%), with a total error of 1.7\% (5.4\%). The pion re-interaction uncertainty is larger for \nub-mode samples than for $\nu$-mode samples because $\pi^-$ interactions on carbon and water are less well understood than  $\pi^+$ interactions at the relevant energies, and because the fraction of $\pi^+$ from wrong-sign contamination in \nub-mode misidentified as $\mu^+$ candidate is larger than the fraction of $\pi^-$ misidentified as $\mu^-$ in the \nunu-mode.

\section{Near Detector data analysis}
\label{sec:banff}
The predicted event rates at both the ND280 and SK are based on parametrized neutrino flux and interaction models, described in Sec.~\ref{sec:beam} and Sec.~\ref{sec:niwgmodel}.
These models are fit to the precisely measured, high statistics data at the ND280, producing both a better central prediction of the SK event rate and reducing the systematic uncertainties associated with the flux and interaction models.
The near detector analysis uses event samples from both FGD1 and FGD2, and from the $\nu$-mode and \nub-mode data, giving 14 samples in total.
These, along with their associated systematic uncertainty, were described in Sec.~\ref{sec:nd280}.

\subsection{Near detector likelihood and fitting methods}
\label{sec:banfffit}
The form of the ND280 likelihood and the fitting method are the same as described in Ref.~\cite{Abe:2015awa}. 
The 14 event samples are binned in $p_{\mu}$ and $\cos\theta_{\mu}$, giving 1062 fit bins in total, though only the $p_{\mu}$ projection is shown for clarity. %, as shown in Fig.~\ref{fig:XXX}, 
The likelihood assumes that the observed number of events in each bin follows a Poisson distribution, with an expectation calculated according to the flux, cross-section and detector systematic parameters discussed above.
A multivariate Gaussian likelihood function is used to constrain these parameters in the fit, with the initial constraints that are described in Secs.~\ref{sec:beam},~\ref{sec:niwgmodel} and~\ref{sec:nd280}.
The near detector systematic and near detector flux parameters are treated as nuisance parameters, as are the cross-section systematic parameters governing neutral current and electron neutrino interactions.
The fitted neutrino cross-section and unoscillated SK flux parameters are passed to the oscillation analysis, using a covariance matrix to describe their uncertainties.

One significant difference with respect to Ref.~\cite{Abe:2015awa} is that, as discussed in Sec.~\ref{sec:niwgmodel}, the CCQE cross-section parameters (except the nucleon binding energy, \eb) have no external constraint.
These parameters are constrained solely by the ND280 data. In addition, in order to alleviate possible biases on the estimation of the oscillation parameters (see Sec.~\ref{sec:fds} for more details), the differences between the reference model and the alternative model for the 1p1h component of the neutrino-nucleus interaction cross section described in Sec.~\ref{sec:niwg} are taken into account in the likelihood. This is done by adding the difference in the expected number of events between the two models in each $p_{\mu}$ and $\cos\theta_{\mu}$ bin to the diagonal of the ND280 detector covariance matrix $V_d$.  
Finally, another significant difference is the inclusion of event samples from FGD2, which contains a water target, and \nub-mode data samples.
 
\subsection{Fit results}
\label{sec:banffresults}
\begin{figure*}[htbp]
    \centering
    \includegraphics[height=6cm]{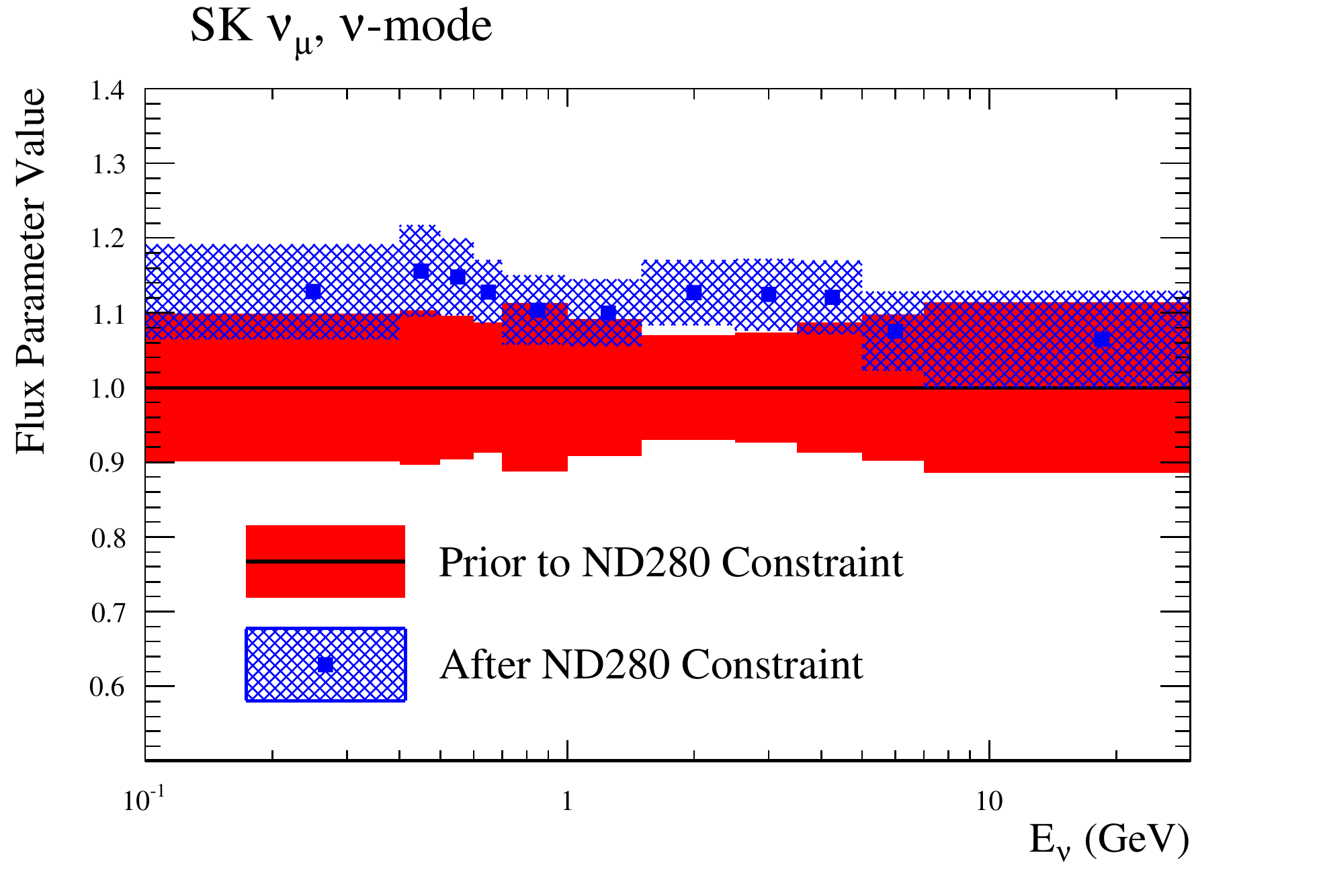}
    \hfill
    \includegraphics[height=6cm]{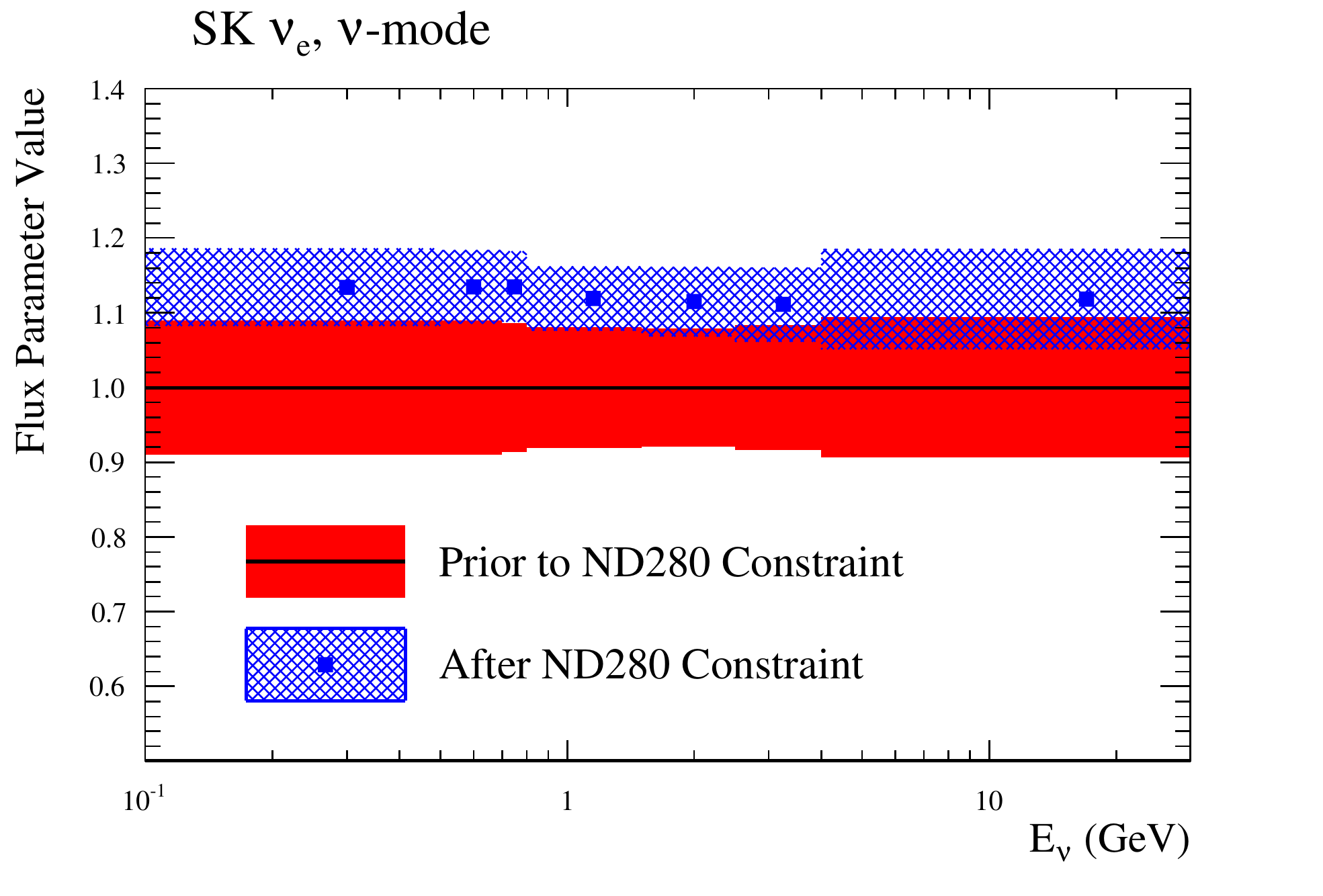}
    \hfill
    \includegraphics[height=6cm]{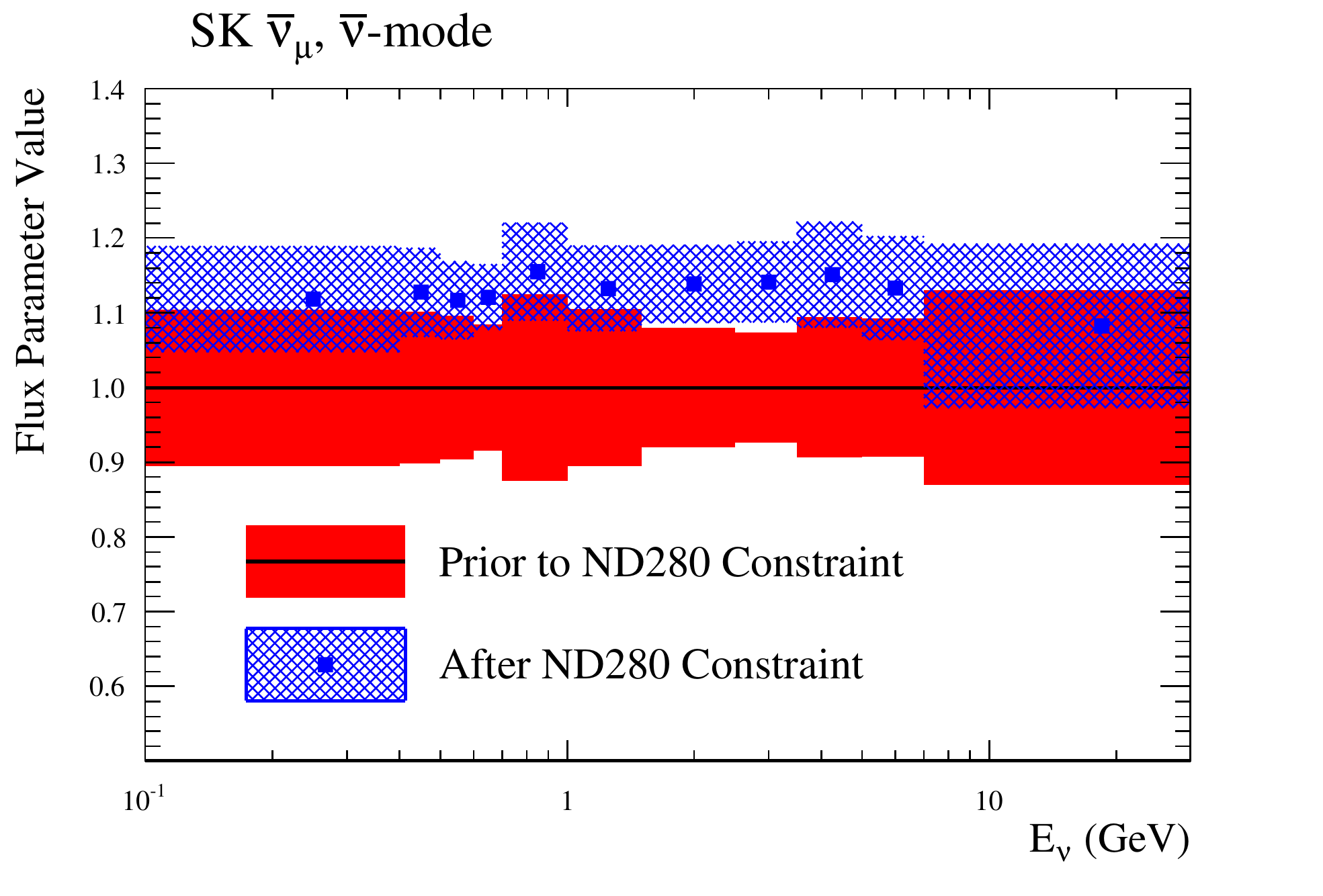}
    \hfill
    \includegraphics[height=6cm]{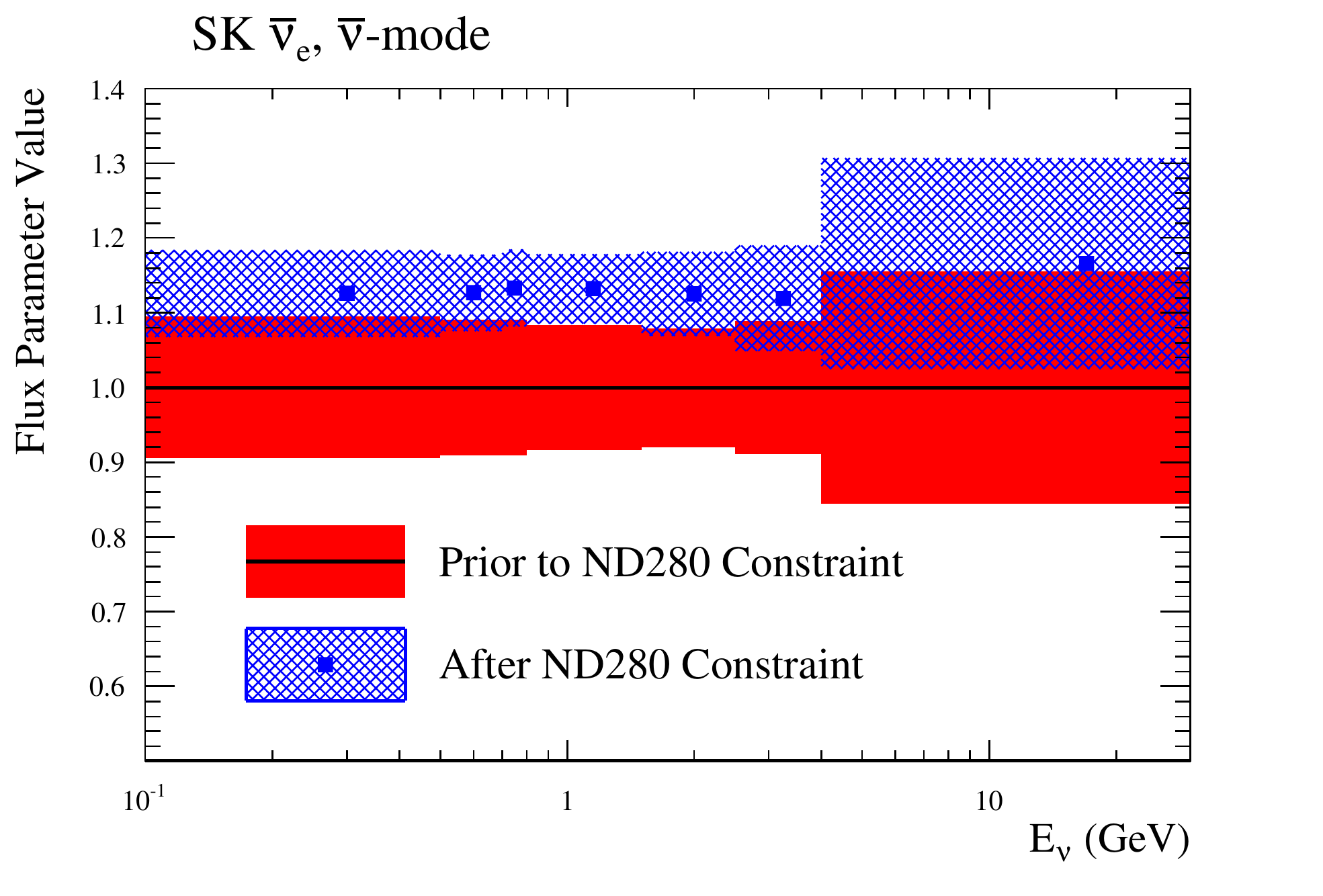}
    \caption{The SK flux parameters for the $\nu^{\bracketbar}_{\mu}$ (left) and $\nu^{\bracketbar}_{e}$ (right) neutrino species in $\nu$- (top) and \nub-modes (bottom), shown as a fraction of the nominal value.  The bands indicate the 1$\sigma$ uncertainty on the parameters before (solid, red) and after (hatched, blue) the near detector fit.}
    \label{fig:banffparamsflux}
\end{figure*}

\begin{figure*}[htbp]
    \centering
    \includegraphics[height=10cm]{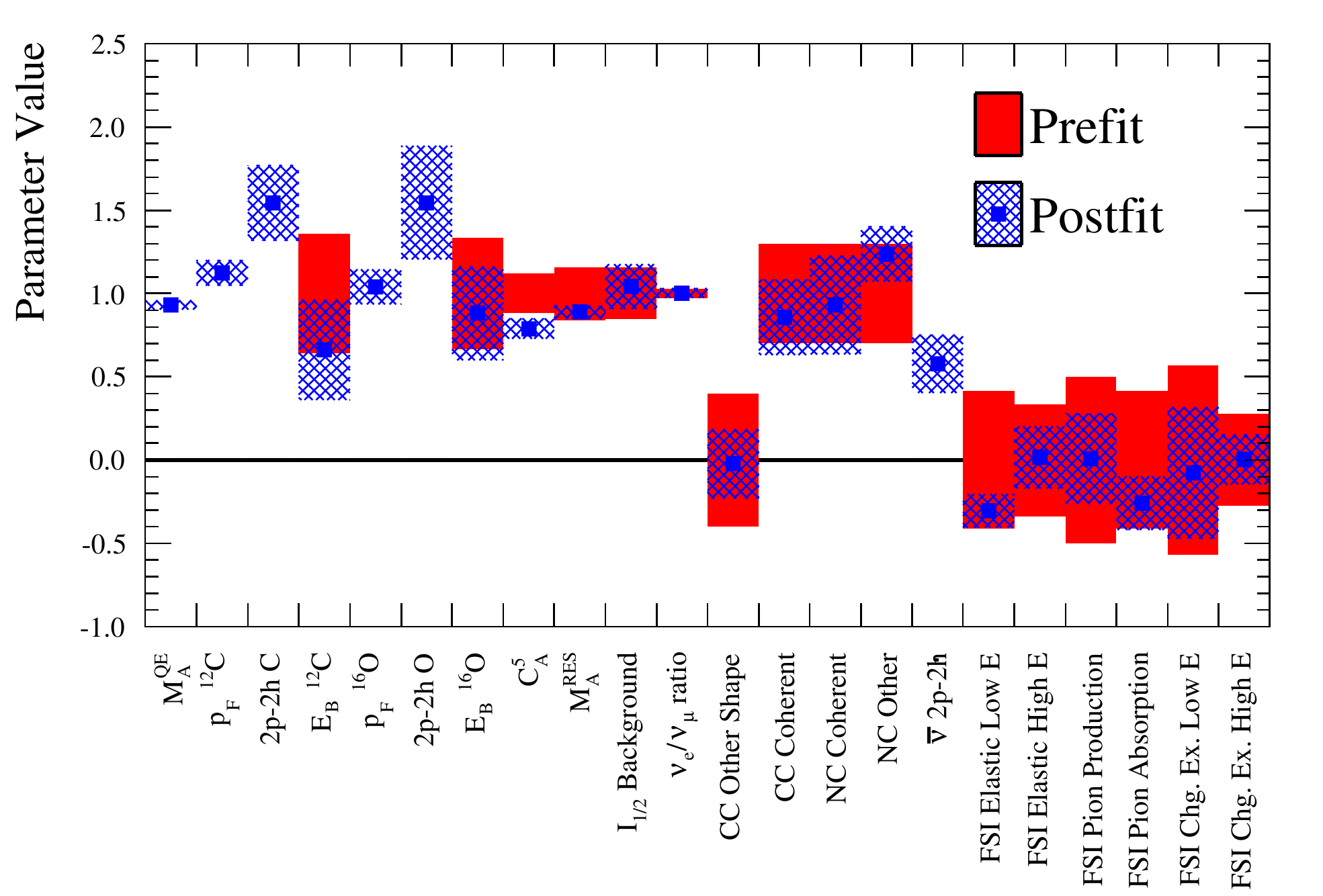}
    \caption{Cross-section parameters before (solid, red) and after (hatched, blue) the near detector fit, shown as a fraction of the nominal value (given in Tab.~\ref{tab:xsec_parameters}).  The extent of the colored band shows the 1$\sigma$ uncertainty. }
    \label{fig:banffparamsxsec}
\end{figure*}

The fit produces central values for the flux, cross-section and detector systematic parameters along with a covariance.
Fig.~\ref{fig:banffparamsflux} shows the values of the unoscillated SK flux parameters and Fig.~\ref{fig:banffparamsxsec} the cross-section parameters before and after the fit as a fraction of the nominal value, along with their prior constraints.
These parameter values are listed in Tables~\ref{tab:sk_fhc_flux_parameters}, \ref{tab:sk_rhc_flux_parameters} and \ref{tab:xsec_parameters}, showing the best-fit point for each along with its uncertainty, calculated as the square root of the diagonal of the covariance.

\begin{table}
    \begin{centering}
        \caption{Prefit and postfit values for the SK $\nu$-mode flux parameters}
        \begin{tabular}{  c | c | c }
            \hline \hline
            \nunu-mode & Prefit & \nd~postfit \\
            flux parameter (GeV) & & \\ 
             \hline
          SK $\nu_{\mu}$ [0.0--0.4] & 1.000 $\pm$ 0.099 & 1.128 $\pm$ 0.064 \\% \hline
          SK $\nu_{\mu}$ [0.4--0.5] & 1.000 $\pm$ 0.103 & 1.156 $\pm$ 0.061 \\% \hline
          SK $\nu_{\mu}$ [0.5--0.6] & 1.000 $\pm$ 0.096 & 1.148 $\pm$ 0.051 \\% \hline
          SK $\nu_{\mu}$ [0.6--0.7] & 1.000 $\pm$ 0.087 & 1.128 $\pm$ 0.043 \\% \hline
          SK $\nu_{\mu}$ [0.7--1.0] & 1.000 $\pm$ 0.113 & 1.104 $\pm$ 0.047 \\% \hline
          SK $\nu_{\mu}$ [1.0--1.5] & 1.000 $\pm$ 0.092 & 1.100 $\pm$ 0.045 \\% \hline
          SK $\nu_{\mu}$ [1.5--2.5] & 1.000 $\pm$ 0.070 & 1.127 $\pm$ 0.044 \\% \hline
          SK $\nu_{\mu}$ [2.5--3.5] & 1.000 $\pm$ 0.074 & 1.124 $\pm$ 0.048 \\% \hline
          SK $\nu_{\mu}$ [3.5--5.0] & 1.000 $\pm$ 0.087 & 1.121 $\pm$ 0.049 \\% \hline
          SK $\nu_{\mu}$ [5.0--7.0] & 1.000 $\pm$ 0.098 & 1.075 $\pm$ 0.053 \\% \hline
          SK $\nu_{\mu}$ $>$7.0 & 1.000 $\pm$ 0.114 & 1.064 $\pm$ 0.065 \\% \hline
   SK $\bar{\nu}_{\mu}$ [0.0--0.7]  & 1.000 $\pm$ 0.103 & 1.100 $\pm$ 0.081 \\% \hline
   SK $\bar{\nu}_{\mu}$ [0.7--1.0]  & 1.000 $\pm$ 0.079 & 1.112 $\pm$ 0.048 \\% \hline
   SK $\bar{\nu}_{\mu}$ [1.0--1.5]  & 1.000 $\pm$ 0.084 & 1.111 $\pm$ 0.060 \\% \hline
   SK $\bar{\nu}_{\mu}$ [1.5--2.5]  & 1.000 $\pm$ 0.086 & 1.116 $\pm$ 0.070 \\% \hline
   SK $\bar{\nu}_{\mu}$ $>$2.5  & 1.000 $\pm$ 0.086 & 1.162 $\pm$ 0.069 \\% \hline
           SK $\nu_{e}$ [0.0--0.5]  & 1.000 $\pm$ 0.090 & 1.134 $\pm$ 0.052 \\% \hline
           SK $\nu_{e}$ [0.5--0.7]  & 1.000 $\pm$ 0.090 & 1.135 $\pm$ 0.049 \\% \hline
           SK $\nu_{e}$ [0.7--0.8]  & 1.000 $\pm$ 0.086 & 1.135 $\pm$ 0.047 \\% \hline
           SK $\nu_{e}$ [0.8--1.5]  & 1.000 $\pm$ 0.081 & 1.119 $\pm$ 0.043 \\% \hline
           SK $\nu_{e}$ [1.5--2.5]  & 1.000 $\pm$ 0.079 & 1.115 $\pm$ 0.046 \\% \hline
           SK $\nu_{e}$ [2.5--4.0]  & 1.000 $\pm$ 0.084 & 1.111 $\pm$ 0.050 \\% \hline
           SK $\nu_{e}$ $>$4.0  & 1.000 $\pm$ 0.094 & 1.118 $\pm$ 0.067 \\% \hline
     SK $\bar{\nu}_{e}$ [0.0--2.5]  & 1.000 $\pm$ 0.074 & 1.121 $\pm$ 0.057 \\% \hline
     SK $\bar{\nu}_{e}$ $>$2.5 & 1.000 $\pm$ 0.128 & 1.153 $\pm$ 0.117 \\
            \hline \hline
        \end{tabular}
        \label{tab:sk_fhc_flux_parameters}
    \end{centering}
\end{table}

\begin{table}
    \begin{centering}
        \caption{Prefit and postfit values for the SK \nub-mode flux parameters}
        \begin{tabular}{  c | c | c }
          \hline \hline
            \nub-mode & Prefit & \nd~postfit \\
            flux parameter (GeV) & & \\ 
             \hline
          SK $\nu_{\mu}$ [0.0--0.7]  & 1.000 $\pm$ 0.094 & 1.098 $\pm$ 0.072 \\% \hline
          SK $\nu_{\mu}$ [0.7--1.0]  & 1.000 $\pm$ 0.079 & 1.121 $\pm$ 0.052 \\% \hline
          SK $\nu_{\mu}$ [1.0--1.5]  & 1.000 $\pm$ 0.077 & 1.130 $\pm$ 0.048 \\% \hline
          SK $\nu_{\mu}$ [1.5--2.5]  & 1.000 $\pm$ 0.081 & 1.155 $\pm$ 0.054 \\% \hline
          SK $\nu_{\mu}$ $>$2.5 & 1.000 $\pm$ 0.080 & 1.111 $\pm$ 0.055 \\% \hline
   SK $\bar{\nu}_{\mu}$ [0.0--0.4]  & 1.000 $\pm$ 0.104 & 1.118 $\pm$ 0.071 \\% \hline
   SK $\bar{\nu}_{\mu}$ [0.4--0.5]   & 1.000 $\pm$ 0.102 & 1.127 $\pm$ 0.060 \\% \hline
   SK $\bar{\nu}_{\mu}$ [0.5--0.6]   & 1.000 $\pm$ 0.096 & 1.117 $\pm$ 0.052 \\% \hline
   SK $\bar{\nu}_{\mu}$ [0.6--0.7]   & 1.000 $\pm$ 0.085 & 1.121 $\pm$ 0.044 \\% \hline
   SK $\bar{\nu}_{\mu}$ [0.7--1.0]   & 1.000 $\pm$ 0.125 & 1.155 $\pm$ 0.066 \\% \hline
   SK $\bar{\nu}_{\mu}$ [1.0--1.5]   & 1.000 $\pm$ 0.105 & 1.132 $\pm$ 0.057 \\% \hline
   SK $\bar{\nu}_{\mu}$ [1.5--2.5]   & 1.000 $\pm$ 0.078 & 1.139 $\pm$ 0.053 \\% \hline
   SK $\bar{\nu}_{\mu}$ [2.5--3.5]   & 1.000 $\pm$ 0.074 & 1.141 $\pm$ 0.054 \\% \hline
   SK $\bar{\nu}_{\mu}$ [3.5--5.0]   & 1.000 $\pm$ 0.094 & 1.151 $\pm$ 0.071 \\% \hline
   SK $\bar{\nu}_{\mu}$ [5.0--7.0]   & 1.000 $\pm$ 0.093 & 1.133 $\pm$ 0.070 \\% \hline
  SK $\bar{\nu}_{\mu}$ $>$7.0   & 1.000 $\pm$ 0.130 & 1.082 $\pm$ 0.110 \\% \hline
           SK $\nu_{e}$ [0.0--2.5]   & 1.000 $\pm$ 0.069 & 1.118 $\pm$ 0.051 \\% \hline
           SK $\nu_{e}$ $>$2.5  & 1.000 $\pm$ 0.085 & 1.112 $\pm$ 0.071 \\% \hline
     SK $\bar{\nu}_{e}$ [0.0--0.5]  & 1.000 $\pm$ 0.095 & 1.126 $\pm$ 0.058 \\% \hline
     SK $\bar{\nu}_{e}$ [0.5--0.7]  & 1.000 $\pm$ 0.091 & 1.127 $\pm$ 0.051 \\% \hline
     SK $\bar{\nu}_{e}$ [0.7--0.8]  & 1.000 $\pm$ 0.091 & 1.133 $\pm$ 0.052 \\% \hline
     SK $\bar{\nu}_{e}$ [0.8--1.5]  & 1.000 $\pm$ 0.084 & 1.132 $\pm$ 0.046 \\% \hline
     SK $\bar{\nu}_{e}$ [1.5--2.5] & 1.000 $\pm$ 0.080 & 1.125 $\pm$ 0.056 \\% \hline
     SK $\bar{\nu}_{e}$ [2.5--4.0]  & 1.000 $\pm$ 0.089 & 1.119 $\pm$ 0.071 \\% \hline
     SK $\bar{\nu}_{e}$ $>$4.0  & 1.000 $\pm$ 0.156 & 1.166 $\pm$ 0.141 \\
            \hline \hline
        \end{tabular}
        \label{tab:sk_rhc_flux_parameters}
    \end{centering}
\end{table}

\begin{table}
    \begin{centering}
          \caption{Prefit and postfit values for the cross-section parameters used in the oscillation fits.  If no prefit uncertainty is shown then the parameter had a flat prior assigned. If a parameter was not constrained by the ND280 fit this is noted in the postfit column.}
      \begin{tabular}{l|c|c}
	   \hline \hline
            Cross section & Prefit & \nd~postfit \\
            parameter & & \\ 
            \hline
            \\[-1em] 
            \maqe (GeV/$c^2$)                             & 1.20            & 1.12  $\pm$ 0.03 \\[2pt]
            \pf $^{12}$C (MeV/c)                          & 217.0           & 243.9 $\pm$ 16.6 \\[2pt]
            2p2h $^{12}$C                                 & 100.0           & 154.5 $\pm$ 22.7 \\[2pt]
            \eb $^{12}$C (MeV)                            & 25.0 $\pm$ 9.00 & 16.5  $\pm$ 7.53 \\[2pt]
            \pf $^{16}$O (MeV/c)                          & 225.0           & 234.2 $\pm$ 23.7 \\[2pt]
            2p2h $^{16}$O                                 & 100.0           & 154.6 $\pm$ 34.3 \\[2pt]
            \eb $^{16}$O (MeV)                            & 27.0 $\pm$ 9.00 & 23.8  $\pm$ 7.61 \\[2pt]
            $C_\mathrm{A}^{5}$                            & 1.01 $\pm$ 0.12 & 0.80  $\pm$ 0.06 \\[2pt]
            $\ma^{\mathrm{RES}}$  (GeV/$c^2$)             & 0.95 $\pm$ 0.15 & 0.84  $\pm$ 0.04 \\[2pt]
            I$_{\frac{1}{2}}$ background                  & 1.30 $\pm$ 0.20 & 1.36  $\pm$ 0.17 \\[2pt]
            CC other shape                                & 0.00 $\pm$ 0.40 & -0.02 $\pm$ 0.21 \\[2pt]
            CC coherent                                   & 1.00 $\pm$ 0.30 & 0.86  $\pm$ 0.23 \\[2pt]
            NC coherent                                   & 1.00 $\pm$ 0.30 & 0.93  $\pm$ 0.30 \\ [2pt]
	        2p2h $\bar{\nu}$                              & 1.00            & 0.58  $\pm$ 0.18 \\ [2pt]
            NC other                                      & 1.00 $\pm$ 0.30 & Not constrained \\ [2pt]
            NC 1-$\gamma$                                 & 1.00 $\pm$ 1.00 & Not constrained \\ [2pt]
            $\nu_{e}$/$\nu_{\mu}$ ratio                   & 1.00 $\pm$ 0.02 & Not constrained \\ [2pt]
            $\bar{\nu}_{e}$/$\bar{\nu}_{\mu}$ ratio       & 1.00 $\pm$ 0.02 & Not constrained \\ [2pt]
            FSI elastic low-E                             & 1.00 $\pm$ 0.41 & Not constrained \\ [2pt]
            FSI elastic high-E                            & 1.00 $\pm$ 0.34 & Not constrained \\ [2pt]
            FSI pion production                           & 1.00 $\pm$ 0.50 & Not constrained \\ [2pt]
            FSI pion absorption                           & 1.00 $\pm$ 0.41 & Not constrained \\ [2pt]
            FSI charge exchange low-E                     & 1.00 $\pm$ 0.57 & Not constrained \\ [2pt]
            FSI charge exchange high-E                    & 1.00 $\pm$ 0.28 & Not constrained \\ 
            \hline \hline
        \end{tabular}
        \label{tab:xsec_parameters}
    \end{centering}
\end{table}

Most noticeable in these results is the 10--15~\% increase in the neutrino flux, seen across all species and energies in both $\nu$- and \nub-modes.
%Given the high degree of correlation between the flux parameters, this is not a significant shift, as discussed in Sec.~\ref{sec:banffgofs}.

Small changes are seen in the central values of the CCQE cross-section parameters, with the fit increasing the Fermi momentum parameter while reducing the nucleon binding energy and axial mass parameters.
More interestingly, the 2p2h normalization is increased to approximately 1.5 times its nominal value, indicating that the fit is sensitive to differences in lepton kinematics between CCQE and 2p2h interactions.
The antineutrino 2p2h normalization is reduced compared to the neutrino parameter, highlighting a difference in the neutrino and antineutrino CC--0$\pi$ event rates that cannot be explained by flux or detector systematics. The fit also reduces the value of the charged current single pion parameters, as seen in the previous analysis~\cite{Abe:2015awa}. This accounts for the relative deficit observed in the CC--1$\pi$ sample compared to the CC--0$\pi$ sample.

%Sec.~\ref{sec:banffpostfit} shows a comparison of the data samples to the Monte Carlo prediction, both at the nominal systematic parameter values and using the best-fit values.
%From these comparisons it can be seen that better agreement is achieved between the data and MC after the fit.

\subsubsection{Goodness of fit and fit validation}
\label{sec:banffgofs}

\begin{figure}[htbp]
    \centering
    \includegraphics[height=5cm]{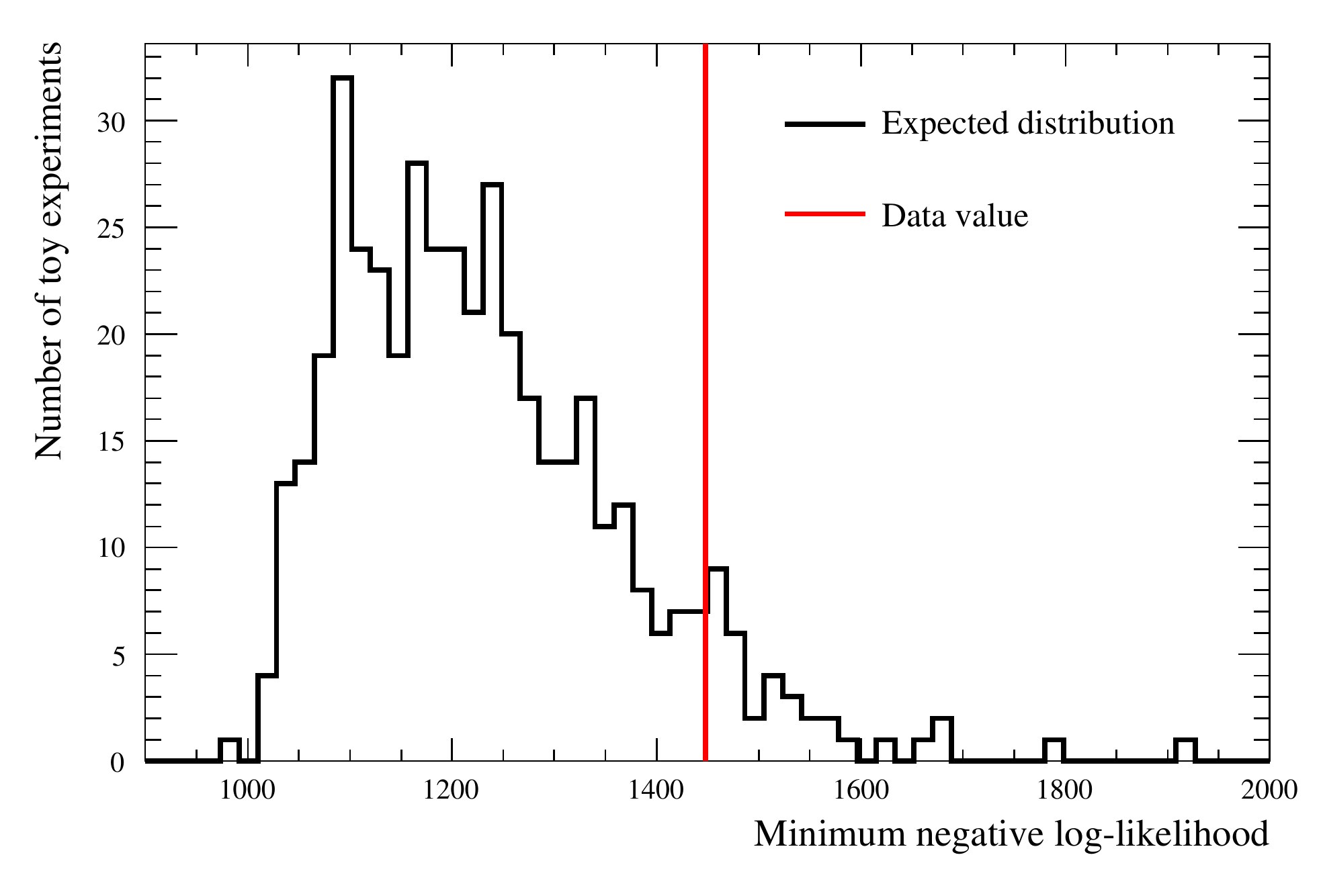}
%        \caption{Total $\Delta\chi^{2}$ distribution.}
    \caption{Distribution of the minimum negative log-likelihood values from fits to the mock datasets (black), with the value from the fit to the data superimposed in red.} 
    \label{fig:banffgof}
\end{figure}

\begin{figure}[htbp]
    \centering
    \includegraphics[height=5cm]{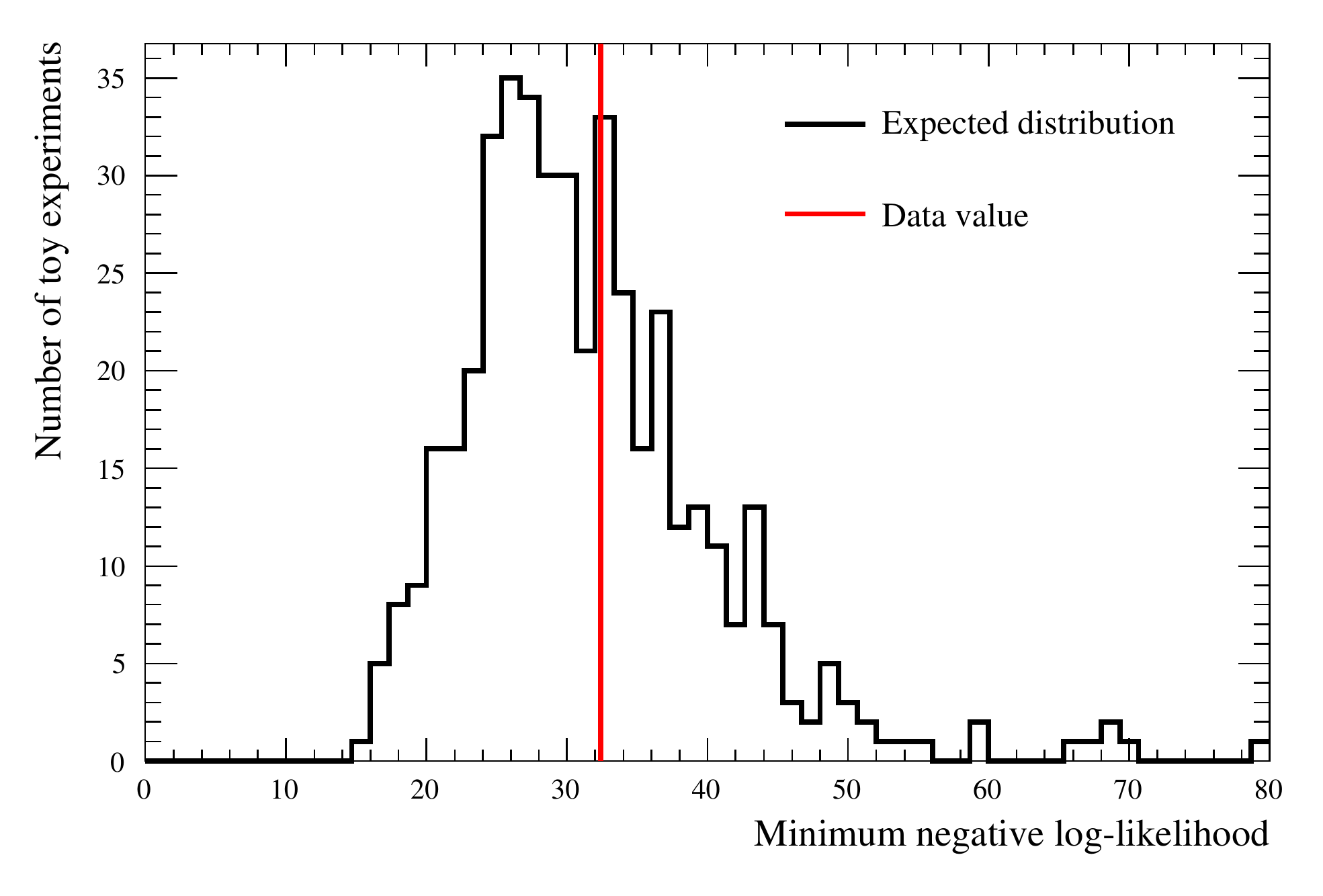}
\hfill
    \includegraphics[height=5cm]{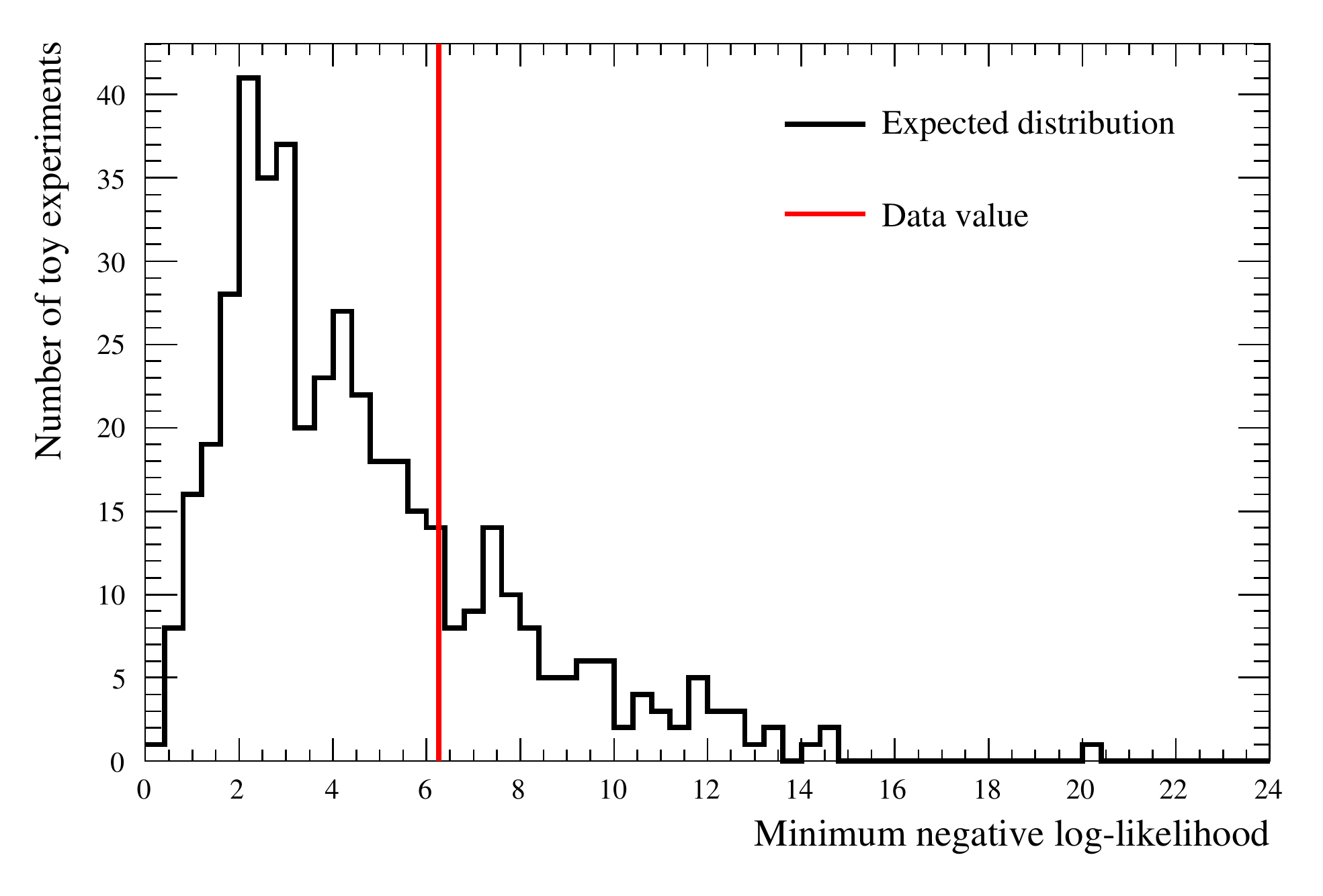}
    \caption{Distribution of the minimum negative log-likelihood values from fits to the mock datasets (black), with the value from the fit to the data superimposed in red. The distributions shown make up the contribution from the flux (top) and cross-section (bottom) prior terms.} 
    \label{fig:gofprior}
\end{figure}

The goodness of fit for the near detector analysis was estimated using mock datasets including statistical uncertainties.
Mock datasets are generated by simultaneously varying the systematic parameters in the fit according to their prior covariance then applying these to the nominal MC.
These were then fit and the minimum negative log-likelihood value found. The distribution of the minimum negative log-likelihood values is shown by the histogram in Fig.~\ref{fig:banffgof}, with the value from the data fit indicated with a red line.
The overall p-value for the fit is 8.6\%.
Figure~\ref{fig:gofprior} shows the same distribution for the flux and cross-section parameter priors, demonstrating that the fitted parameter values propagated to the oscillation analysis are reasonable.

\begin{figure*}[htbp]
    \centering
    \includegraphics[height=10cm]{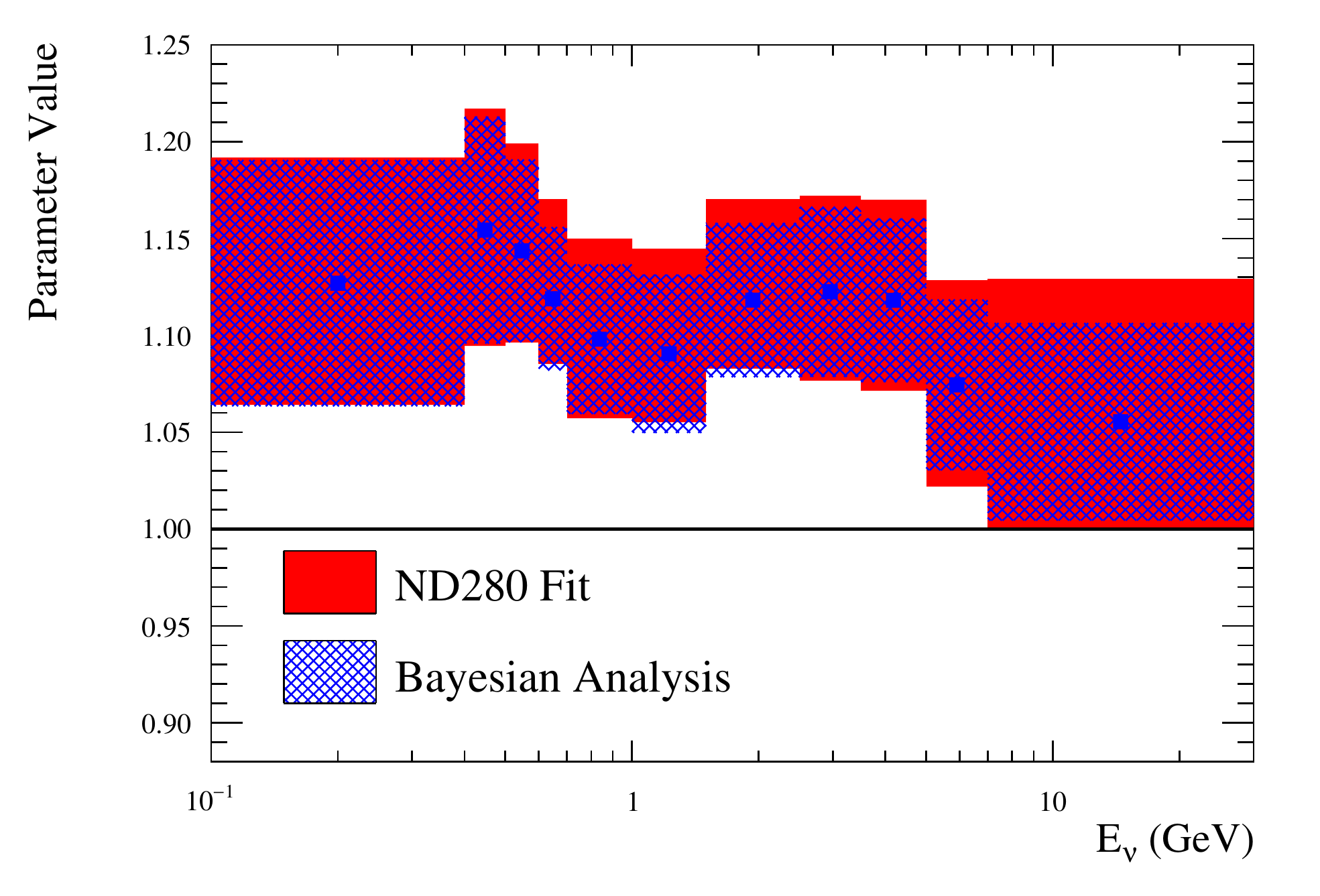}
    \hfill
    \includegraphics[height=10cm]{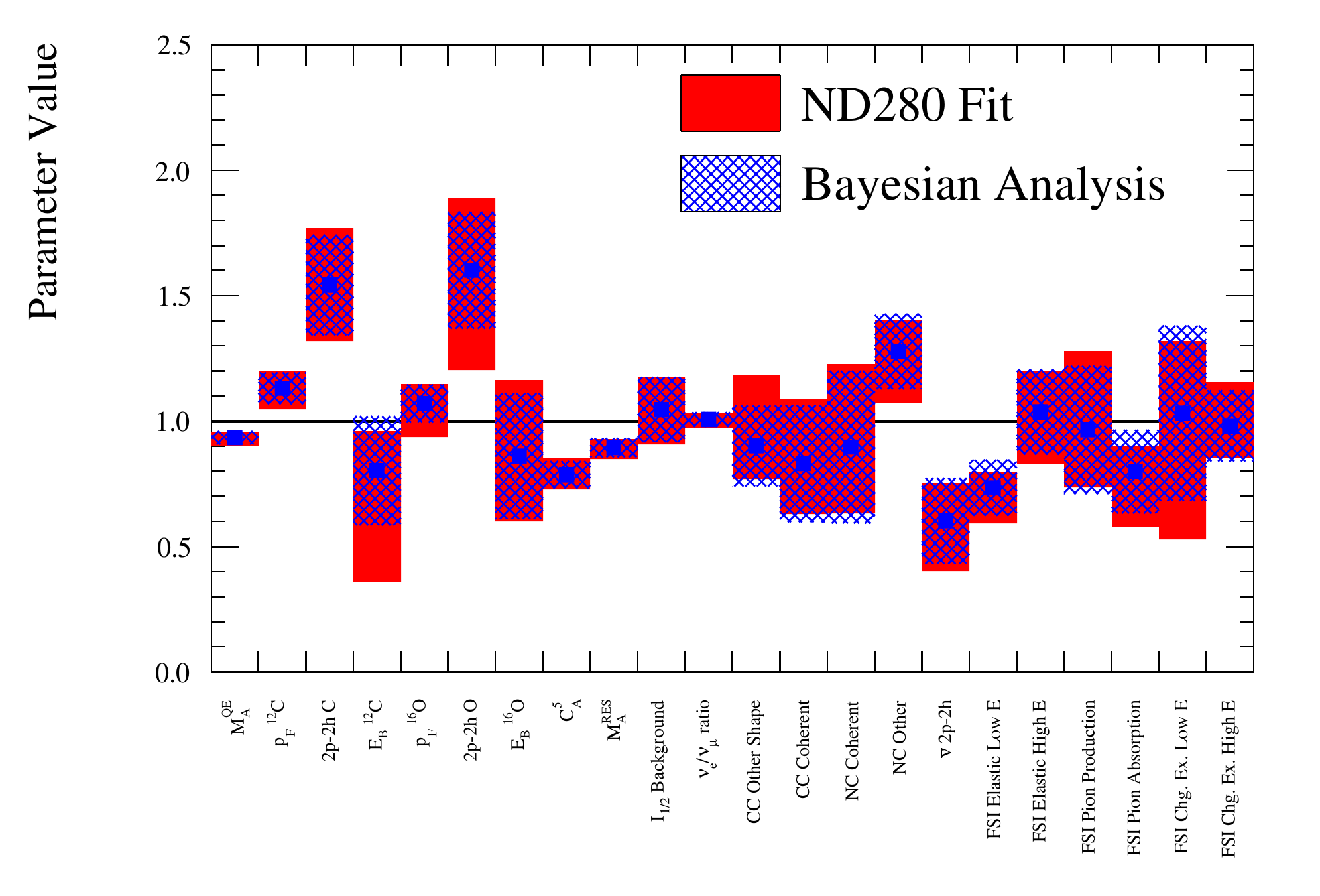}
    \caption{Comparison of the fitted SK $\nu$-mode flux parameters (top) and cross-section parameters (bottom) between the ND280 fit (red, solid) and the Bayesian analysis (blue, hashed).}
    \label{fig:mach3comp}
\end{figure*}

In addition, the Bayesian analysis which simultaneously fits both near and far detector samples, that will be described in Sec.~\ref{sec:oabayesian}, was used to cross-check the primary result by only fitting near detector data.
The results of this fit are compared to the best-fit parameters from the near detector analysis in Fig.~\ref{fig:mach3comp}, showing excellent agreement between the two.

\subsection{ND280 postfit distributions}
\label{sec:banffpostfit}
\begin{figure*}[htbp]
    \centering
    \includegraphics[height=6cm]{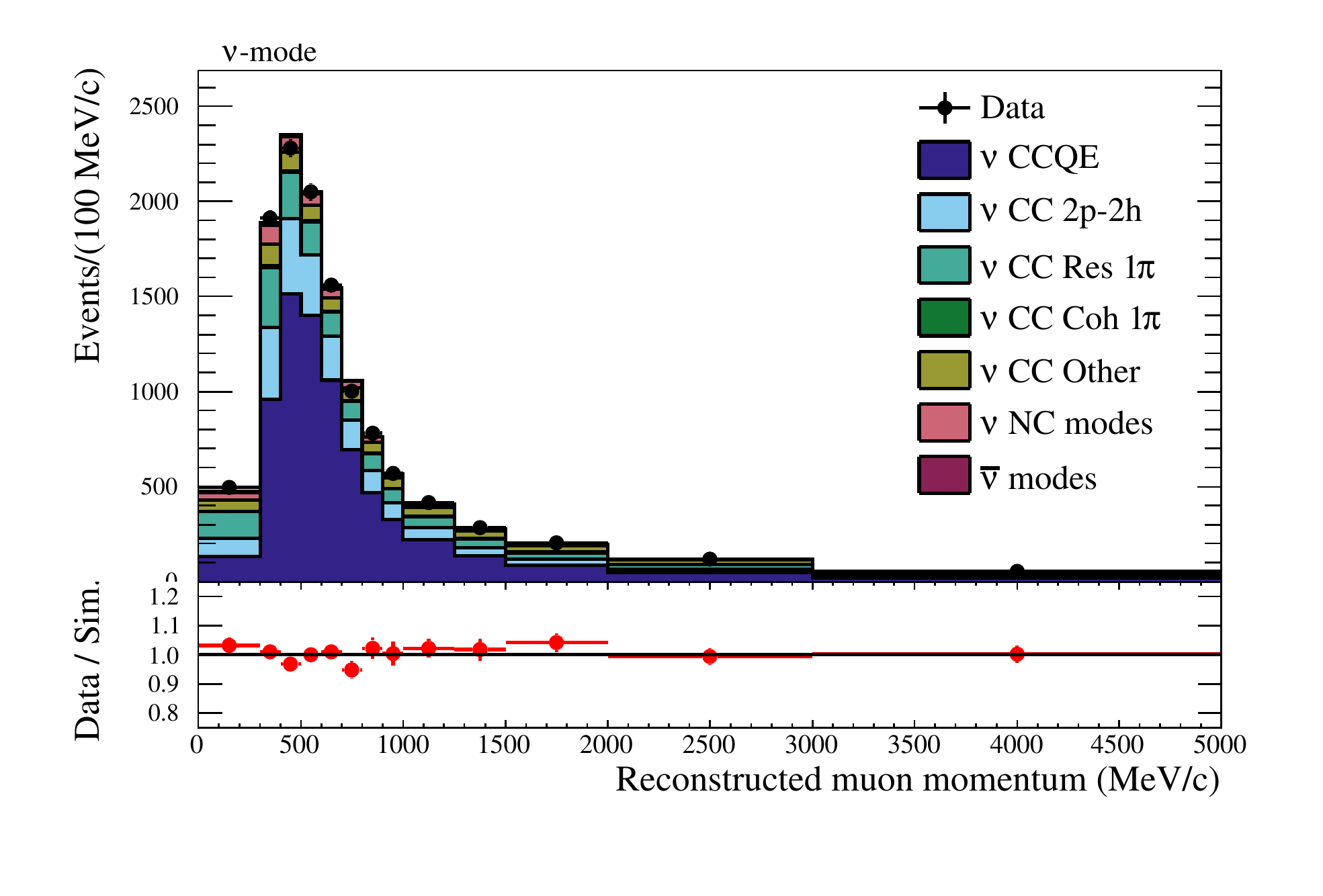}
    \hfill
    \includegraphics[height=6cm]{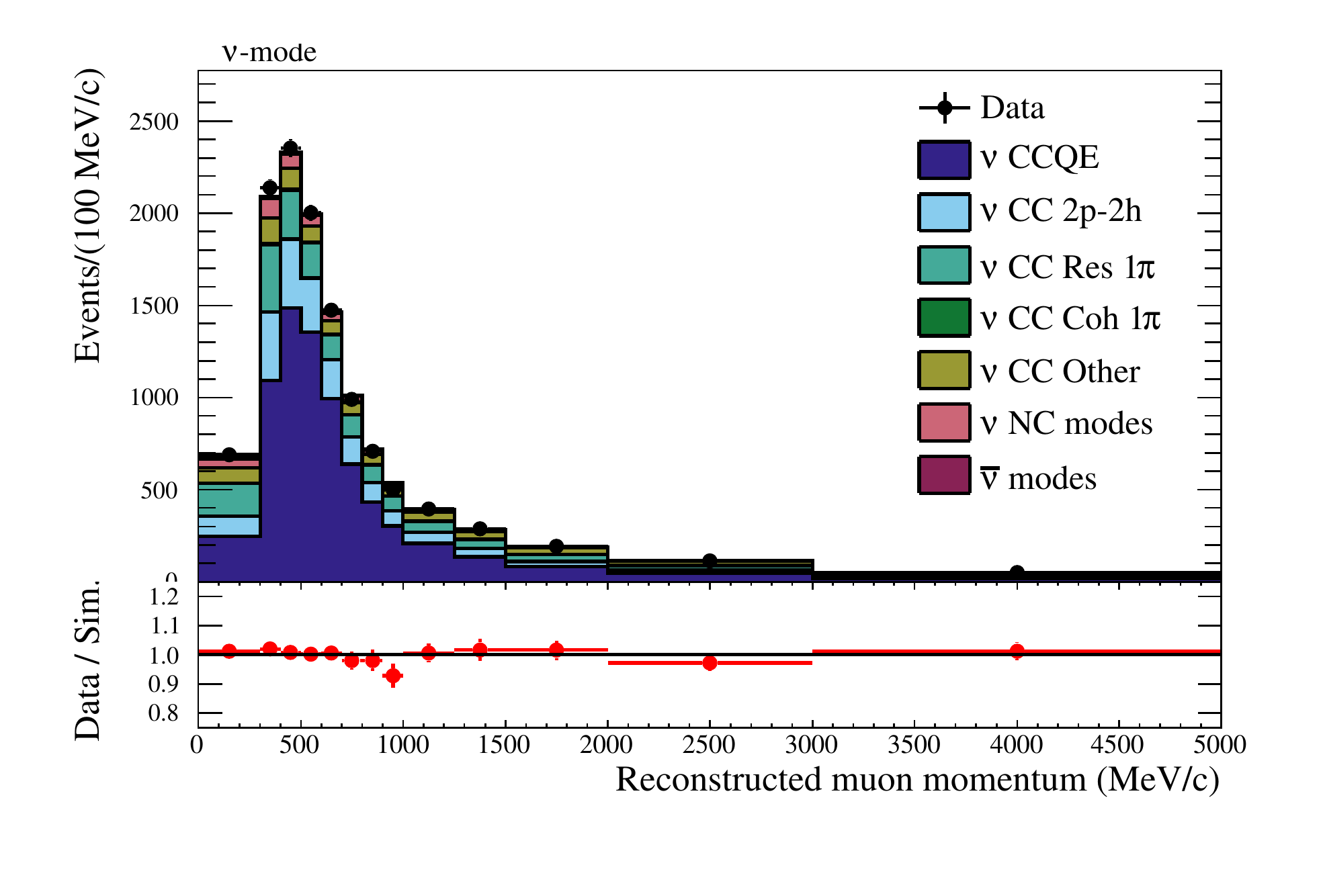}
    \hfill
    \includegraphics[height=6cm]{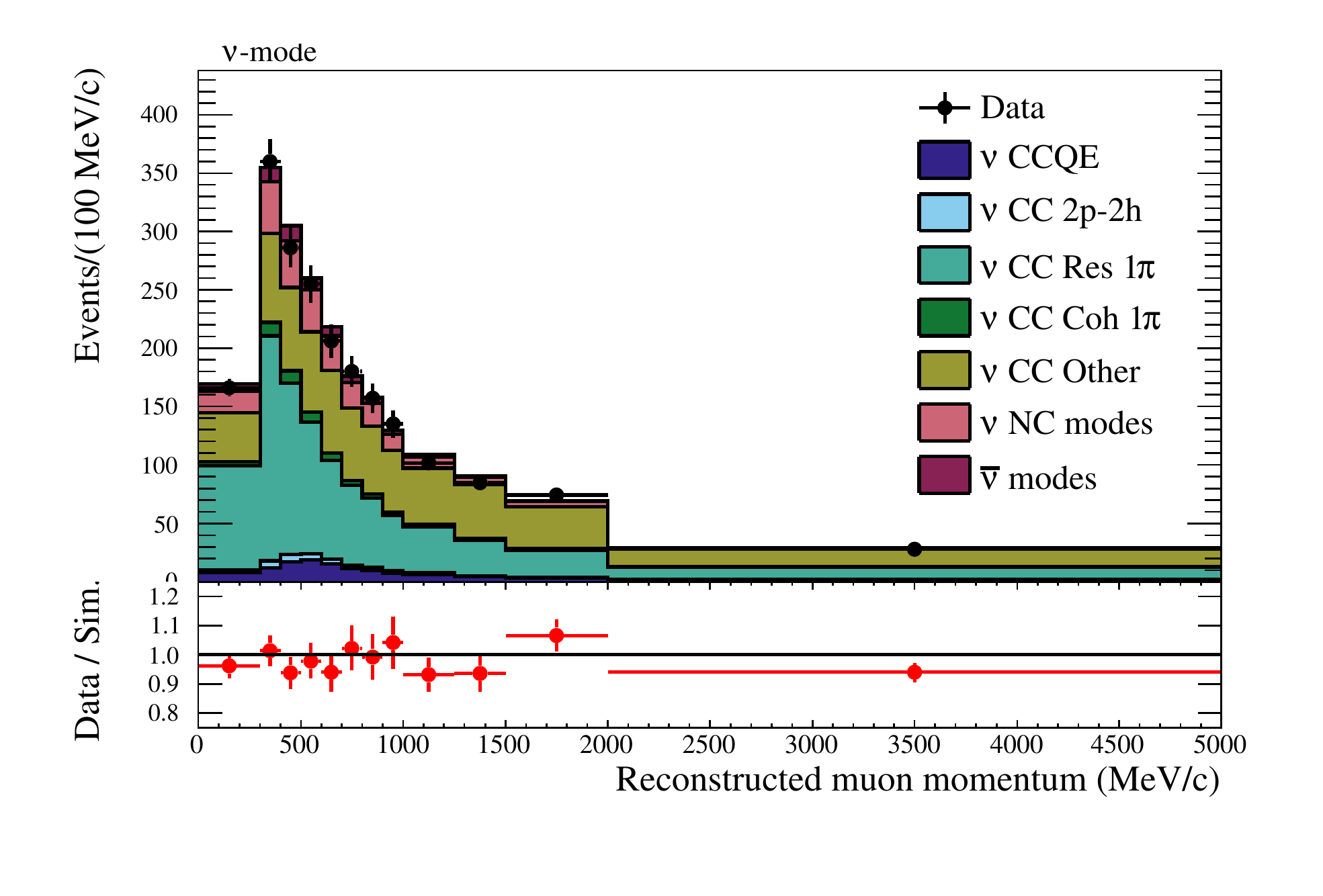}
    \hfill
    \includegraphics[height=6cm]{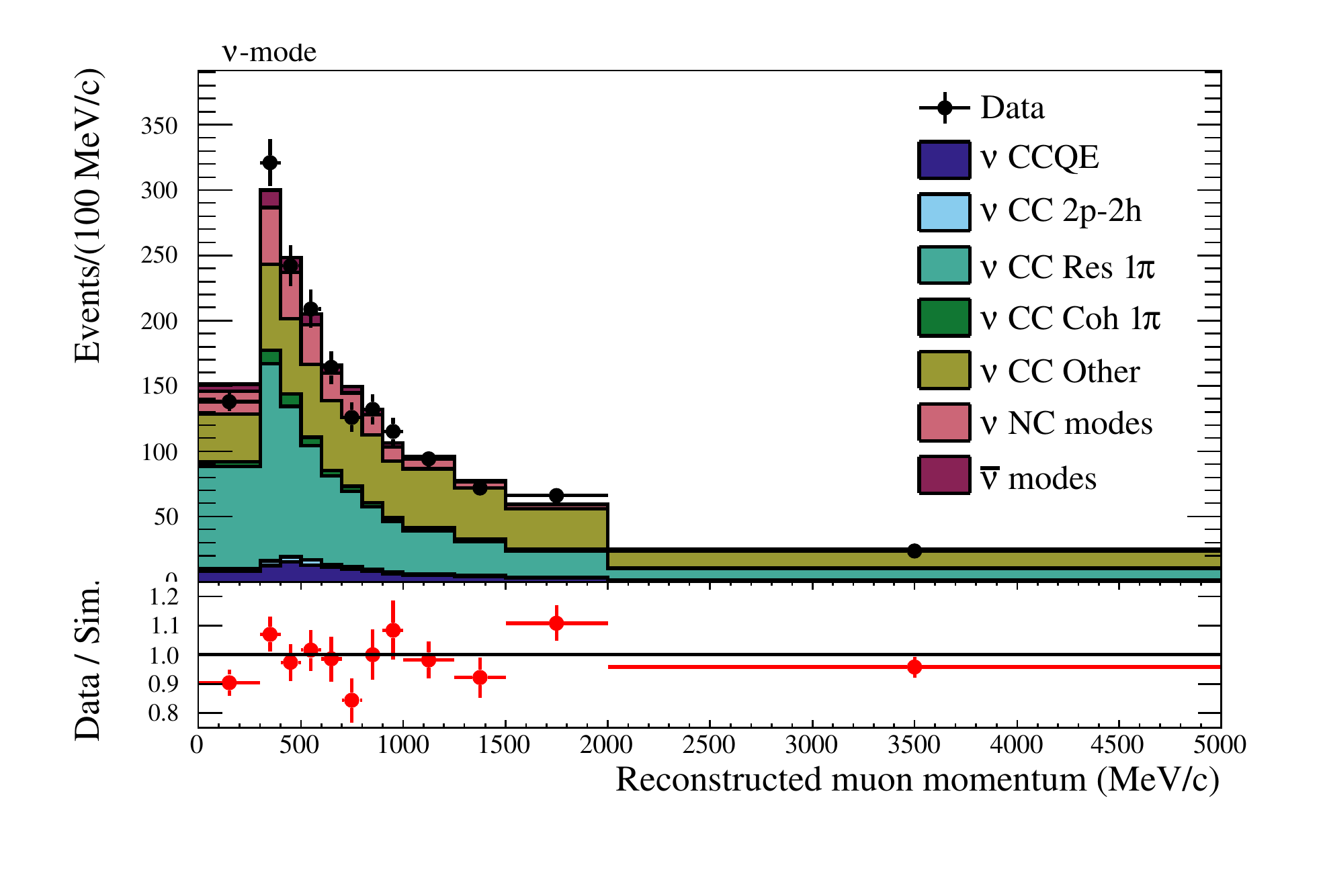}
    \caption{Data-MC comparison of $\nu$-mode $\nu_\mu$ CC-0$\pi$ (top) and CC-1$\pi^+$ (bottom) samples in FGD1 (left) and FGD2 (right) after the ND280 fit.
    The simulation is broken-down by neutrino reaction type.}
    \label{fig:nd280_fhc_cc0pi_banff}
\end{figure*}

\begin{figure*}[htbp]
    \centering
    \includegraphics[height=6cm]{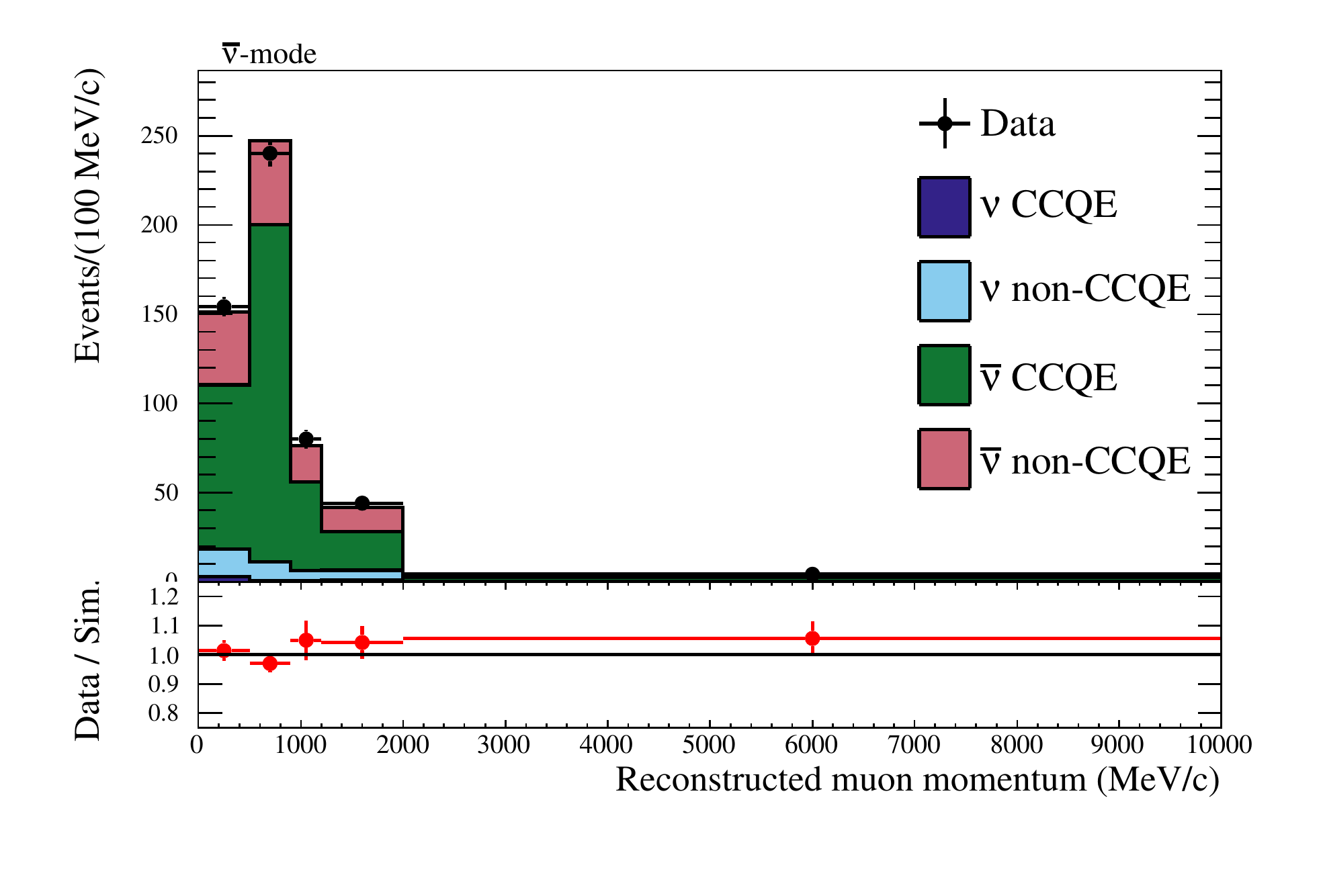}
    \hfill
    \includegraphics[height=6cm]{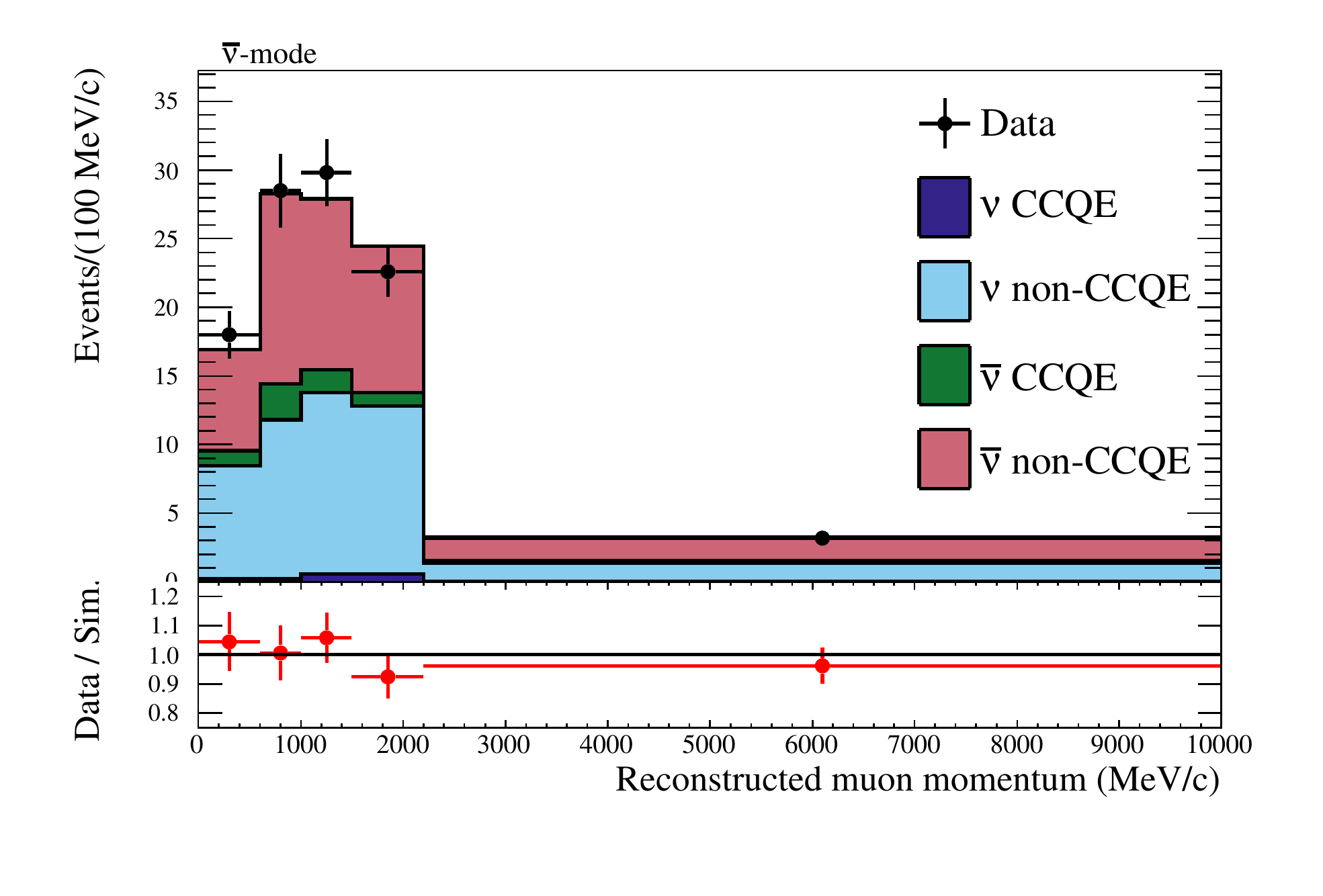}
    \hfill
    \includegraphics[height=6cm]{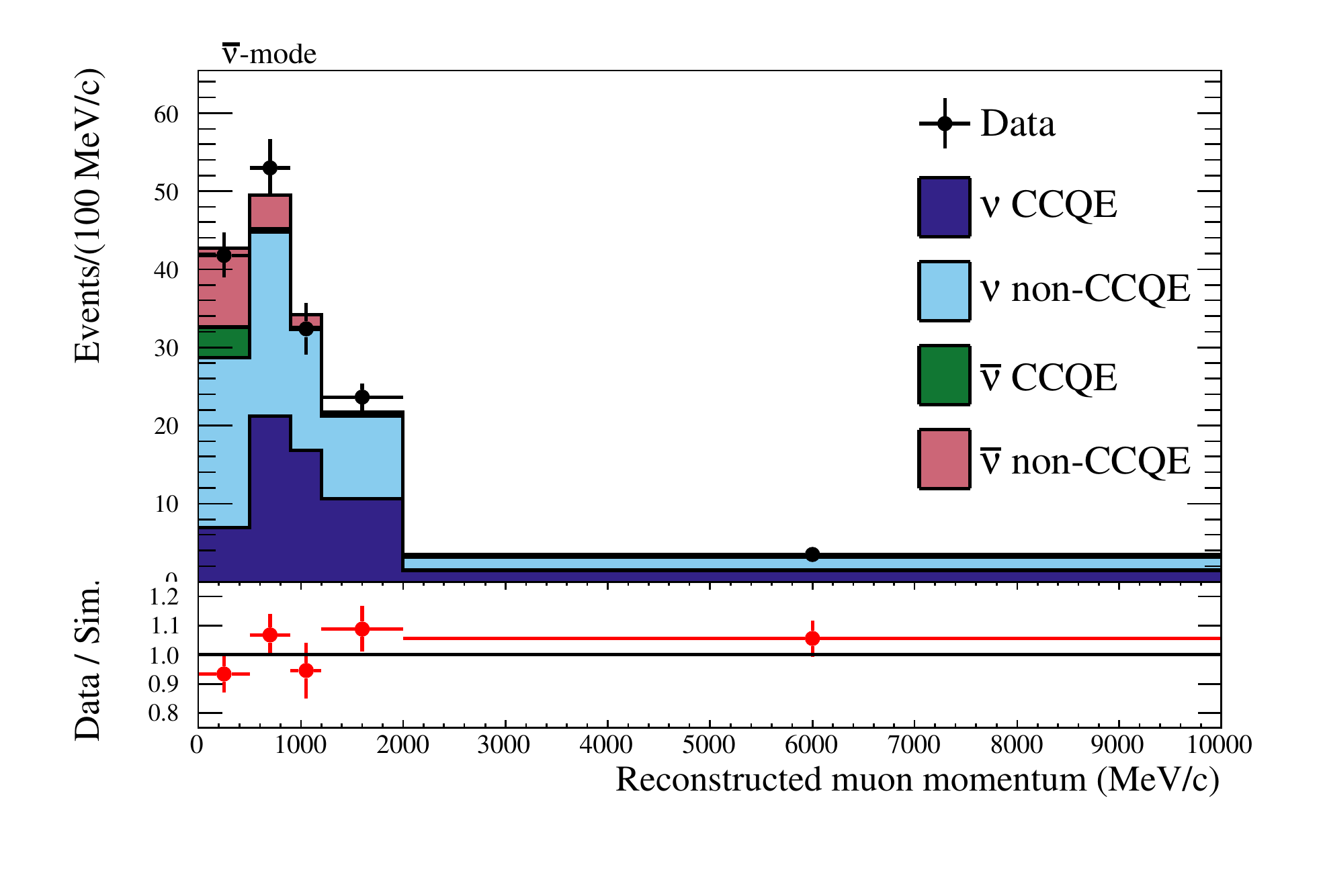}
    \hfill
    \includegraphics[height=6cm]{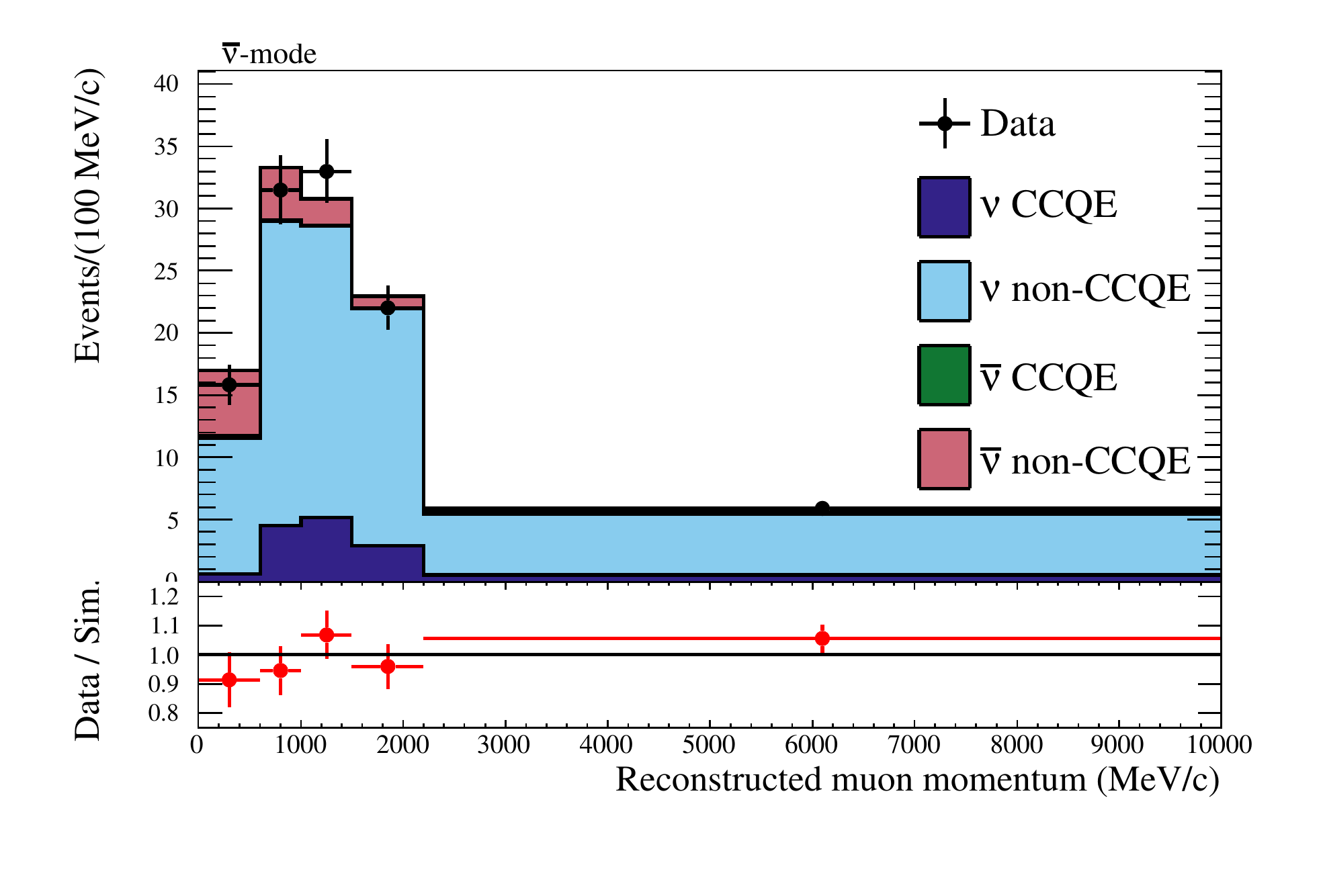}
    \caption{Top: data-MC comparison of $\bar\nu_\mu$ (right-sign) CC-1-Track (left) and CC-N-Tracks (right) samples after the ND280 fit. The simulation is broken-down by neutrino reaction type.
    Bottom: same muon momentum distributions for $\nu_\mu$ background samples.}
    \label{fig:nd280_rhc_banff}
\end{figure*}

The expected muon momentum spectrum after the ND280 fit for the CC-0\pipi and CC-1$\pi^+$ samples in \nunu-mode and the FGD1 samples in \nub-mode are shown in Fig.~\ref{fig:nd280_fhc_cc0pi_banff} and Fig.~\ref{fig:nd280_rhc_banff} respectively.
After the ND280 fit, the expected distributions show in general a better agreement with the data. 
The numbers of postfit predicted events for all the 14 samples are shown in Tab.~\ref{tab:nd_samp_banff}.
%for \nunu-mode and \nub-mode respectively.
%It is worth to notice the effect of the ND280 fit on each reaction code grouping in the simulation prediction.
The effects of the ND280 fit on the different neutrino interactions contributing to each ND280 sample are detailed in Tab.~\ref{tab:nd_fhc_reac} and Tab.~\ref{tab:nd_rhc_reac}.
%To better match the data, the ND280 fit has the tendency to increase the 2p2h and DIS contributions, while it decrease the resonant component.

\begin{table}[h!]
\begin{centering}
\caption{Observed and predicted events rates for different ND280 samples after the ND280 fit.}
\begin{tabular}{l | c  c | c  c }
\hline \hline
\nd~Sample   &  FGD1 & FGD1 & FGD2  & FGD2  \\
                & Data & Postfit & Data & Postfit \\

\hline
$\nu_\mu$ CC-0$\pi$ & 17354 & 17345 & 17650 & 17638\\
$\nu_\mu$ CC-1$\pi^+$ & 3984 & 4113 & 3383 & 3449\\
$\nu_\mu$ CC-Other & 4220 & 4150 & 4118 & 3965\\
\hline
$\bar\nu_\mu$ CC-1-Track & 2663 & 2639 & 2762 & 2728\\
$\bar\nu_\mu$ CC-N-Tracks & 775 & 785 & 737 & 814\\\
$\nu_\mu$ CC-1-Track & 989 & 966 & 980 & 987\\
$\nu_\mu$ CC-N-Tracks & 1001 & 989 & 936 &  937\\
\hline \hline
\end{tabular}
\label{tab:nd_samp_banff}
\end{centering}
\end{table}

%\begin{table}[h!]
%\begin{centering}
%\begin{tabular}{  l | c | c }
%\hline
%FGD1 Sample & Data & Postfit prediction\\
%\hline
%$\bar\nu_\mu$ CC-1-Track & 2663 & 2639.31\\
%$\bar\nu_\mu$ CC-N-Tracks & 775 & 784.72\\
%$\nu_\mu$ CC-1-Track & 989 & 966.18\\
%$\nu_\mu$ CC-N-Tracks & 1001 & 989.39 \\
%\hline
%FGD2 Sample & Data & Postfit prediction\\
%\hline
%$\bar\nu_\mu$ CC-1-Track & 2762 & 2728.34\\
%$\bar\nu_\mu$ CC-N-Tracks & 737 & 814.22\\
%$\nu_\mu$ CC-1-Track & 980 & 987.12\\
%$\nu_\mu$ CC-N-Tracks & 936 &  937.08\\
%\end{tabular}
%\caption{Observed and predicted events rates for different ND280 samples collected in $\bar\nu$-mode beam, after the ND280 fit.}
%\label{tab:nd_rhc_samp_banff}
%\end{centering}
%\end{table}

\begin{table}[h!]
\begin{centering}
\caption{Prefit and postfit expected fraction of events for each neutrino interactions for samples taken in \nunu-mode.}
\begin{tabular}{  l | c c | c c }
\hline \hline
\multicolumn{1}{c|}{}&\multicolumn{2}{c|}{FGD1 $\nu_\mu$ CC-0$\pi$}&\multicolumn{2}{c}{FGD2 $\nu_\mu$ CC-0$\pi$}\\
\hline
Reaction&Prefit & Postfit &Prefit & Postfit \\
\hline
$\nu$ CCQE &0.562&0.540&0.547&0.533\\
$\nu$ CC 2p2h&0.100&0.165&0.094&0.155\\
$\nu$ CC Res $1\pi$&0.202&0.152&0.219&0.164\\
$\nu$ CC Coh $1\pi$&0.003&0.003&0.003&0.003\\
$\nu$ CC Other &0.096&0.098&0.101&0.103\\
$\nu$ NC modes&0.032&0.037&0.032&0.037\\
$\bar\nu$ modes&0.003&0.003&0.004&0.004\\
%\end{tabular}

%\begin{tabular}{  l | c c | c c }
\hline \hline
\multicolumn{1}{c|}{}&\multicolumn{2}{c|}{FGD1 $\nu_\mu$ CC-1$\pi^+$}&\multicolumn{2}{c}{FGD2 $\nu_\mu$ CC-1$\pi^+$}\\
\hline
Reaction&Prefit & Postfit &Prefit & Postfit \\
\hline
$\nu$ CCQE &0.054&0.053&0.055&0.054\\
$\nu$ CC 2p2h&0.008&0.014&0.007&0.013\\
$\nu$ CC Res $1\pi$&0.485&0.409&0.476&0.401\\
$\nu$ CC Coh $1\pi$&0.025&0.025&0.026&0.025\\
$\nu$ CC Other &0.348&0.389&0.353&0.395\\
$\nu$ NC modes&0.059&0.087&0.60&0.089\\
$\bar\nu$ modes&0.022&0.024&0.022&0.024\\
%\end{tabular}
\hline \hline
%\begin{tabular}{  l | c c | c c }
\multicolumn{1}{c|}{}&\multicolumn{2}{c|}{FGD1 $\nu_\mu$ CC-Other}&\multicolumn{2}{c}{FGD2 $\nu_\mu$ CC-Other}\\
\hline
Reaction&Prefit & Postfit &Prefit & Postfit \\
\hline
$\nu$ CCQE &0.047&0.043&0.049&0.045\\
$\nu$ CC 2p2h&0.010&0.015&0.010&0.016\\
$\nu$ CC Res $1\pi$&0.148&0.112&0.154&0.117\\
$\nu$ CC Coh $1\pi$&0.004&0.003&0.003&0.003\\
$\nu$ CC Other &0.701&0.712&0.696&0.707\\
$\nu$ NC modes&0.078&0.102&0.075&0.099\\
$\bar\nu$ modes&0.013&0.014&0.11&0.013\\
\hline \hline
\end{tabular}
\label{tab:nd_fhc_reac}
\end{centering}
\end{table}

\begin{table}[h!]
\begin{centering}
\caption{Prefit and postfit expected fraction of events for each neutrino interactions for samples taken in \nub-mode.}

\begin{tabular}{  l | c c | c c }
\multicolumn{1}{c|}{}&\multicolumn{2}{c|}{FGD1 $\bar\nu_\mu$ CC-1-Track}&\multicolumn{2}{c}{FGD2 $\bar\nu_\mu$ CC-1-Track}\\
\hline \hline
Reaction&Prefit & Postfit &Prefit & Postfit \\
\hline
$\nu$ CCQE &0.011&0.012&0.015&0.014\\ %CCQE+MEC
$\nu$ non-CCQE &0.069&0.077&0.074&0.080\\
$\bar\nu$ CCQE & 0.733&0.751&0.731&0.751\\%CCQE+MEC
$\bar\nu$ non-CCQE & 0.186 & 0.161 & 0.181 & 0.156\\
\hline \hline
\multicolumn{1}{c|}{}&\multicolumn{2}{c|}{FGD1 $\bar\nu_\mu$ CC-N-Tracks}&\multicolumn{2}{c}{FGD2 $\bar\nu_\mu$ CC-N-Tracks}\\
\hline
Reaction&Prefit & Postfit &Prefit & Postfit \\
\hline
$\nu$ CCQE &0.017&0.017&0.021&0.021\\ %CCQE+MEC
$\nu$ non-CCQE &0.422&0.447&0.412&0.437\\
$\bar\nu$ CCQE & 0.085&0.083&0.088&0.087\\%CCQE+MEC
$\bar\nu$ non-CCQE & 0.474&0.452&0.478&0.453\\
\hline \hline
\multicolumn{1}{c|}{}&\multicolumn{2}{c|}{FGD1 $\nu_\mu$ CC-1-Track}&\multicolumn{2}{c}{FGD2 $\nu_\mu$ CC-1-Track}\\
\hline
Reaction&Prefit & Postfit &Prefit & Postfit \\
\hline
$\nu$ CCQE & 0.480 & 0.516 & 0.466 & 0.499\\ %CCQE+MEC
$\nu$ non-CCQE & 0.412 & 0.372&0.411&0.363\\
$\bar\nu$ CCQE & 0.037 & 0.033& 0.042&0.038\\%CCQE+MEC
$\bar\nu$ non-CCQE & 0.071&0.079&0.081&0.100\\
\hline \hline
\multicolumn{1}{c|}{}&\multicolumn{2}{c|}{FGD1 $\nu_\mu$ CC-N-Tracks}&\multicolumn{2}{c}{FGD2 $\nu_\mu$ CC-N-Tracks}\\
\hline
Reaction&Prefit & Postfit &Prefit & Postfit \\
\hline
$\nu$ CCQE & 0.150&0.163&0.143&0.154\\ %CCQE+MEC
$\nu$ non-CCQE &0.771&0.756&0.777&0.764\\
$\bar\nu$ CCQE & 0.003 & 0.003 &0.004 &0.003\\%CCQE+MEC
$\bar\nu$ non-CCQE & 0.076&0.078&0.077&0.078\\
\hline \hline
\end{tabular}
\label{tab:nd_rhc_reac}
\end{centering}
\end{table}

%Show comparison between data, prefit, and postfit distributions.

\section{Far detector event selection and systematics}
\label{sec:sk}
%%%%%%%%%%%%%%%%%%%%%%%%%%%%%%%%%%%%%%%%%%%%%%%%%%%%%%%%%%%%%%%%%%%%%%%%%%%%%%%%%%%%%%%%%%%
%% Description of the standard single-ring $\mu$ and e selection for FHC and RHC. 

%% Introduction of the CC1pi sample. 

%% SK detector systematics, describing in particular the work done to study systematic
%% uncertainties for the CC1pi sample.

%% Expected length 3 pages.
 
%% Editors: Ko and James.
%%%%%%%%%%%%%%%%%%%%%%%%%%%%%%%%%%%%%%%%%%%%%%%%%%%%%%%%%%%%%%%%%%%%%%%%%%%%%%%%%%%%%%%%%%%

The T2K far detector is Super-Kamiokande (SK)~\cite{FUKUDA2003418}, a 50~kton water Cherenkov detector located in the Kamioka Observatory, Gifu, Japan. SK is divided into two concentric cylinders, defining an inner detector (ID) instrumented with 11,129 20--inch photomultiplier tubes (PMT) and an outer detector (OD) instrumented with 1885 PMTs. The outer detector is mainly used as a veto for entering backgrounds while neutrino interactions are selected in a fiducial volume inside the ID. 

In order to precisely measure neutrino oscillations parameters, together with the large target volume, high acceptance and efficient discrimination is necessary to distinguish the leptons produced in \num and \nue interactions. Vertex, momentum reconstruction, and Particle identification (PID) in SK are done by observing the Cherenkov radiation produced by charged particles traversing the detector. These particles produce ring patterns that are recorded by the PMTs and are the primary tool used for the PID. Muons produced by \numormb CC interactions are usually unscattered thanks to their large mass and produce a clear ring pattern. In contrast, electrons from \nueoreb CC interactions produce electromagnetic showers resulting in diffuse ring edges. In addition to the shape of the Cherenkov ring, the opening angle also helps to distinguish between electrons and muons. At the typical energies of leptons produced by neutrinos from the T2K beam, the probability to misidentify a single electron (muon) as a muon (electron) is 0.7\% (0.8\%). In SK it is not possible to distinguish neutrinos from antineutrinos event-by-event since the charge of the outgoing leptons cannot be reconstructed. For this reason the selection that will be described in this section is identical for data taken in $\nu$-mode and in \nub-mode. 
  
%% Introduction
T2K data are extracted from the incoming stream in $\pm500\,\mus$ windows centered on the
beam trigger.
A scan of each window recovers individual events which are then classified.

%%FC

Events in which Cherenkov light is deposited exclusively in the ID comprise the fully contained (FC) sample.
% events whose Cherenkov light is deposited exclusively in the ID.%A sample of fully contained (FC) events is defined via pre-selection criteria.  
PMTs in the OD
that register light are grouped into clusters.  If the largest such cluster contains
more than 15 PMTs the event is moved to an OD sample.  Low energy (LE) events are
separated into a dedicated sample by requiring that the total charge from the ID PMT
hits in a 300 ns window be greater than 200 photoelectrons (p.e.), which corresponds
to the charge observed from a 20\,MeV electromagnetic shower.  Events are also designated
as LE if a single ID PMT hit constitutes more than half of the total p.e. observed to reject
events due to noise.  LE events are not included in the analysis presented here.

\begin{figure}[htbp]
\begin{center}
\includegraphics[width=0.45\textwidth]{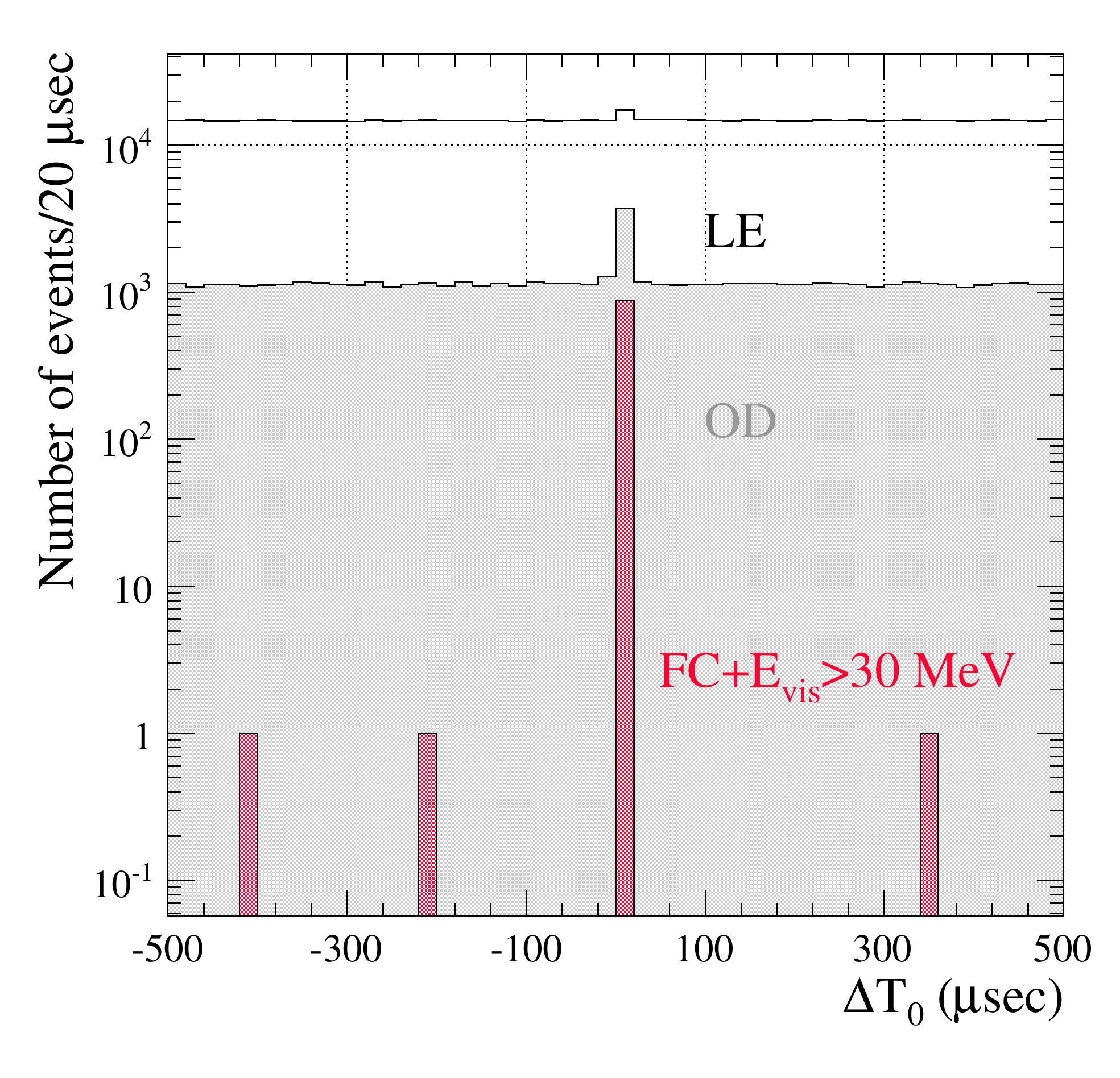}
\caption{$\Delta T_0$ distribution of all FC, OD and LE events within $\pm$500\mus of the 
expected beam arrival time observed during T2K Run 1-7.  The histograms are stacked in 
that order.}
\label{fig:SK_dT0_offtiming}
\end{center}
\end{figure}

\begin{figure}[htbp]
\begin{center}
\includegraphics[width=0.45\textwidth]{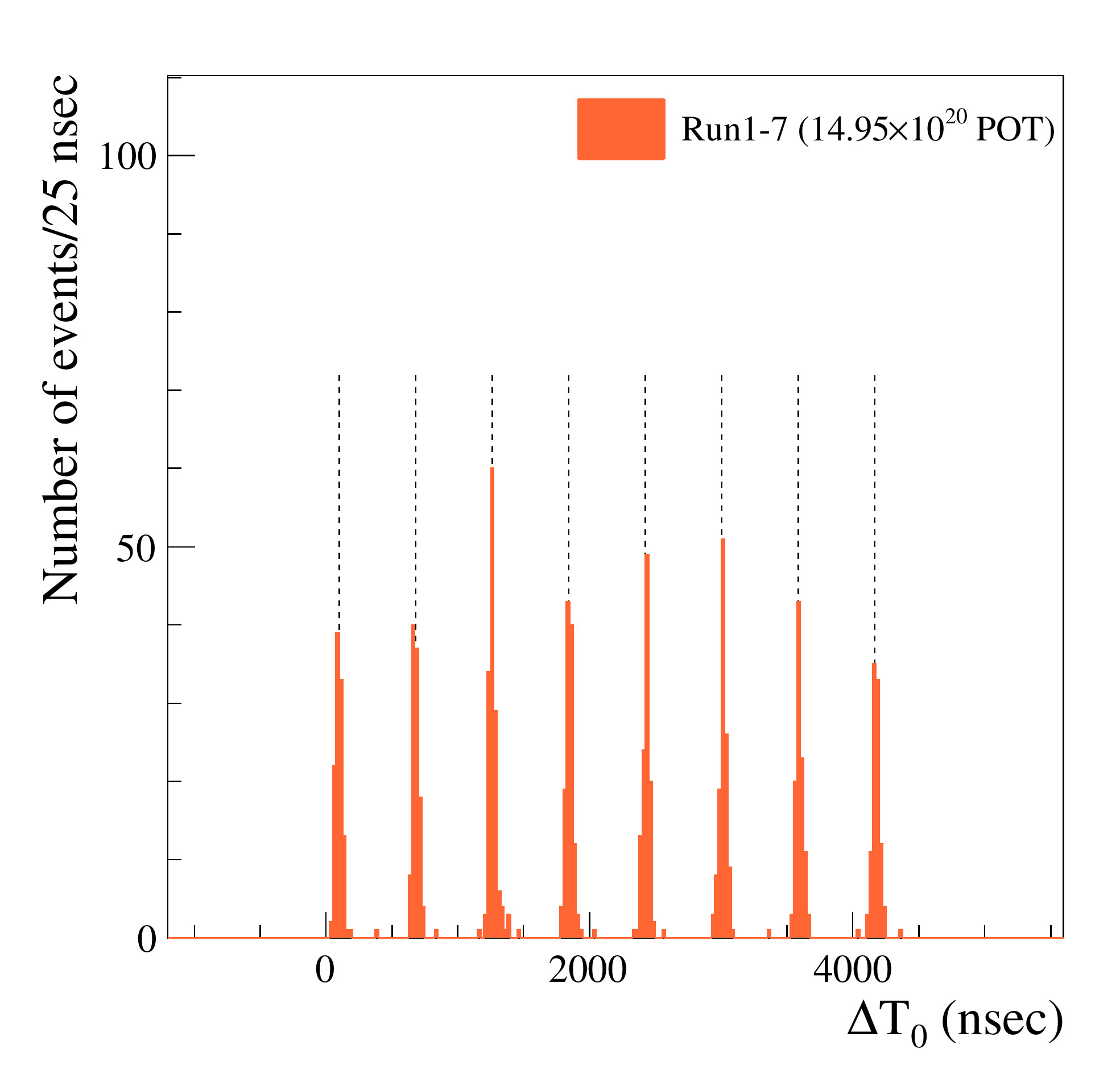}
\caption{$\Delta T_0$ distribution all FC events observed during T2K Run 1-7 zoomed in on 
the expected beam arrival time.}
\label{fig:SK_dT0_ontiming}
\end{center}
\end{figure}

%%Event Timing
Events at the far detector are timed with respect to the leading edge of the beam spill,
taking into account the neutrino time of flight, the Cherenkov photon propagation time
and delays in the DAQ electronics.
Fig.~\ref{fig:SK_dT0_offtiming} shows the event timing ($\Delta T_0$) distribution for all 
OD, LE, and FC events within $\pm500\mus$ of the beam arrival time. 
For FC events a visible energy ($E_{\mathrm{vis}}$) greater than 30\,MeV is also requested.  
A clear peak is observed around $\Delta T_0 = 0$ in the FC sample.
For an event to be incorporated into the analysis, $\Delta T_0$ must lie 
%between $-2$--$10\,\mus$
in the interval [-2,10]~\mus.  A total of 4 events have been observed outside this range.
The expected number of such events is 4.17, estimated by using data taken with no beam.
FC events within the spill window 
can be seen in Fig.~\ref{fig:SK_dT0_ontiming} where the beam structure with eight bunches 
is clearly visible.  The dotted lines represent the fitted bunch center times with 
a fixed bunch interval of 581\ns.

%FCFV
A fiducial volume is defined within the ID, 2\,m away from the detector wall, with a
fiducial mass of 22.5\,kt.  Events whose vertex is reconstructed within this volume 
%and with  $E_{\mathrm{vis}}$ greater than 30\,MeV 
are selected into the fully contained
fiducial volume sample (FCFV).  Visible energy is defined as the energy of an electromagnetic
shower that produces the observed amount of Cherenkov light. In total, 608 events are classified
as FCFV.  The expected number of background events from non-beam related sources in accidental
coincidence is estimated to be 0.0145.

\subsection{SK charged-current quasi-elastic selection}
\label{sec:skstandard}

\begin{figure}[htbp]
\begin{center}
\includegraphics[width=0.45\textwidth]{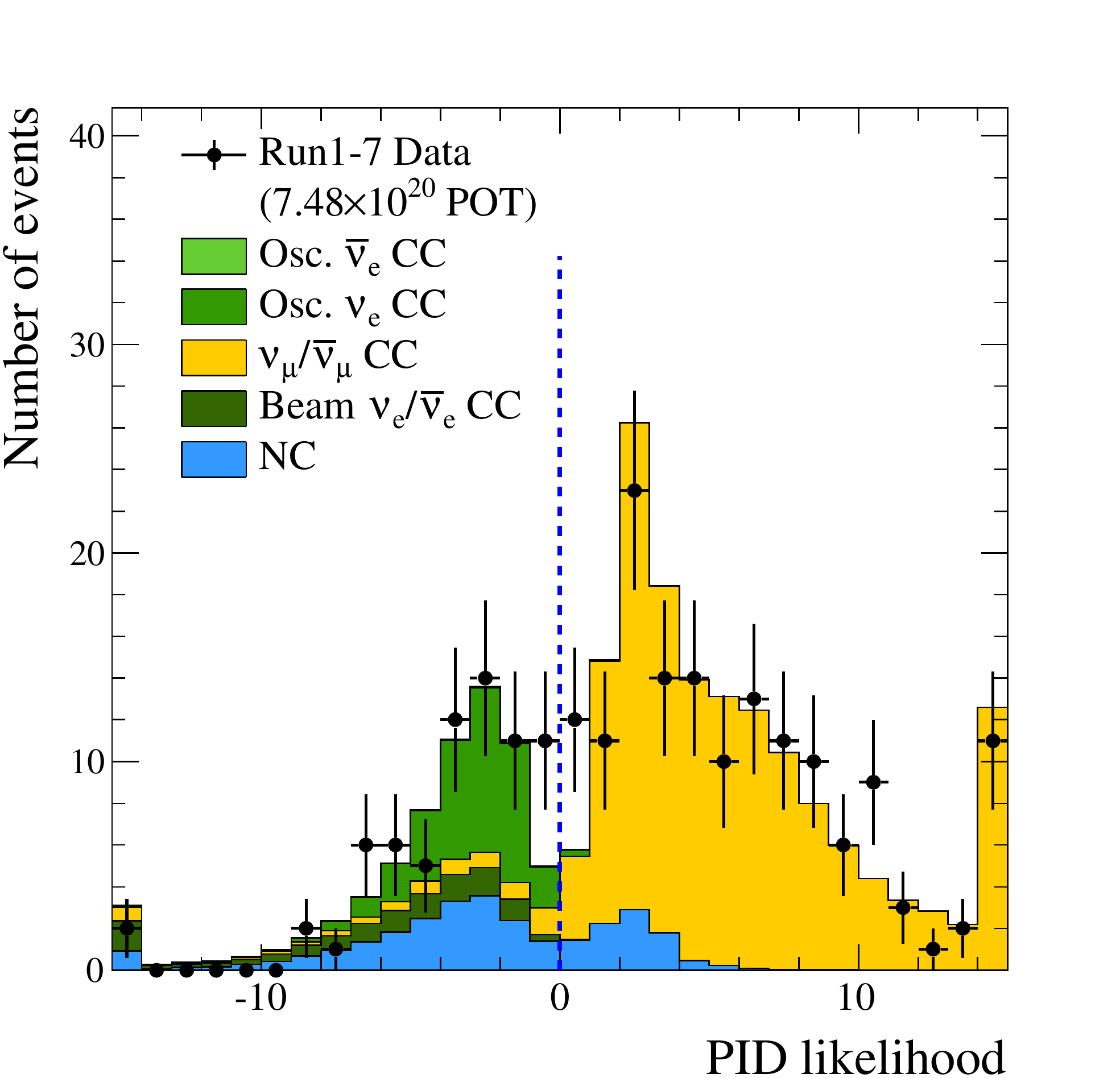}
\caption{The PID likelihood distribution 
of the observed \nunu-mode CC event samples after FCFV and single-ring cuts have been applied.  The data are shown as points with statistical error bars and the shaded, stacked 
histograms are the MC predictions.
The expectation is based on the parameters of Tab.~\ref{tab:nomosc_par}.}
\label{fig:SK_pid_nue_plots}
\end{center}
\end{figure}

%Candidate Samples Selections - CCQE
Charged-current interactions $(\;\nu^{\bracketbar} + N \rightarrow l^{\pm} + X)$ in the narrow energy range of the T2K 
beam most commonly produce single-ring events at SK because most of the resulting particles, except for the 
primary lepton, do not escape the nucleus, or are below detection threshold.  The energy 
of the incoming neutrino can be calculated assuming the kinematics of a CCQE interaction 
and neglecting Fermi motion:% as shown in equation \ref{eq:SK_Erec}.
%% CW: I rewrote this to include both nu and nubar modes
\begin{equation}
\label{eq:SK_Erec}
\erec = \frac{m^2_f - (m_i')^2 - m^2_l + 2m_i' E_l}{2(m_i' - E_l + p_l\cos\theta_l)}
\end{equation}
\noindent where \erec is the reconstructed neutrino energy, $m_i$ and $m_f$ are the initial and final nucleon masses respectively, and $m_i' = m_i - \eb$, where $\eb=27$\mev is the binding energy of a nucleon inside $^{16}\mathrm{O}$ nuclei.
$E_l$, $p_l$ and $\theta_l$ are the reconstructed lepton energy, momentum, and 
angle with respect to the beam, respectively.  
The selection criteria for both \nueoreb CC and \numormb CC events
were fixed using MC studies before being applied to data.  Events are determined 
to be $e$-like or $\mu$-like based on the PID of the brightest Cherenkov ring.  The PID of each ring 
is determined by a likelihood incorporating information on the charge distribution and the opening 
angle of the Cherenkov cone. The PID likelihood distribution for \nunu-mode FCFV single-ring events is shown in Fig.~\ref{fig:SK_pid_nue_plots}. The same criteria are applied to events observed for both \nunu- and \nub-mode data taking.

\begin{table}
  \centering      
\caption{Values of the oscillation parameters used for the Monte Carlo simulation at SK. The values of $\sin^2 \theta_{12}$, $\Delta m^2_{21}$ and $\sin^2 \theta_{13}$ are taken from Ref.~\cite{PDG2015},
while all the other oscillation parameters correspond to the most probable values obtained by the Bayesian analysis in Ref.~\cite{Abe:2015awa}.}
 %     {\renewcommand{\arraystretch}{1.2}
        \begin{tabular}{ c c}
  \hline \hline
          Parameter & Value \\
          \hline
           $\sin^2 2 \theta_{12}$ 	& 0.846   \\
          $\Delta m^2_{21}$   	& $7.53 \times 10^{-5}~\text{eV}^2/\text{c}^4$   \\
         $\sin^{2}\theta_{23}$ & 0.528 \\
          $\Delta m^2_{32}$ & $2.509\times10^{-3}\,\rm{eV}^{2}/\rm{c}^4$\\
          $\sin^{2}2\theta_{13}$ & 0.085 \\
          \dcp & -1.601 \\
          Mass ordering & normal \\
\hline \hline
        \end{tabular}
%      }
\label{tab:nomosc_par}
\end{table}

\nueoreb CC candidate events are selected using the criteria listed in Tab.~\ref{tab:SK_nue_events}.
The $E_{\mathrm{vis}}$ requirement removes low energy NC interactions and electrons from the 
decay of unseen parents that are below Cherenkov threshold or fall outside the 
beam time window.  The $\pi^0$-like event rejection uses an independent 
reconstruction algorithm which was introduced in previous analyses \cite{Abe:2015awa}.
The cut $\erec < 1.25\gev$ is required, as above this energy the intrinsic beam 
\nueoreb background is dominant.
The numbers of events remaining in the neutrino and antineutrino beam data after successive selection criteria for a simulation sample 
produced with the oscillation parameters of Tab.~\ref{tab:nomosc_par} are shown in Tab.~\ref{tab:SK_nue_events}.

\begin{figure*}[htbp]
   \centering
  \includegraphics[width=0.48\textwidth]{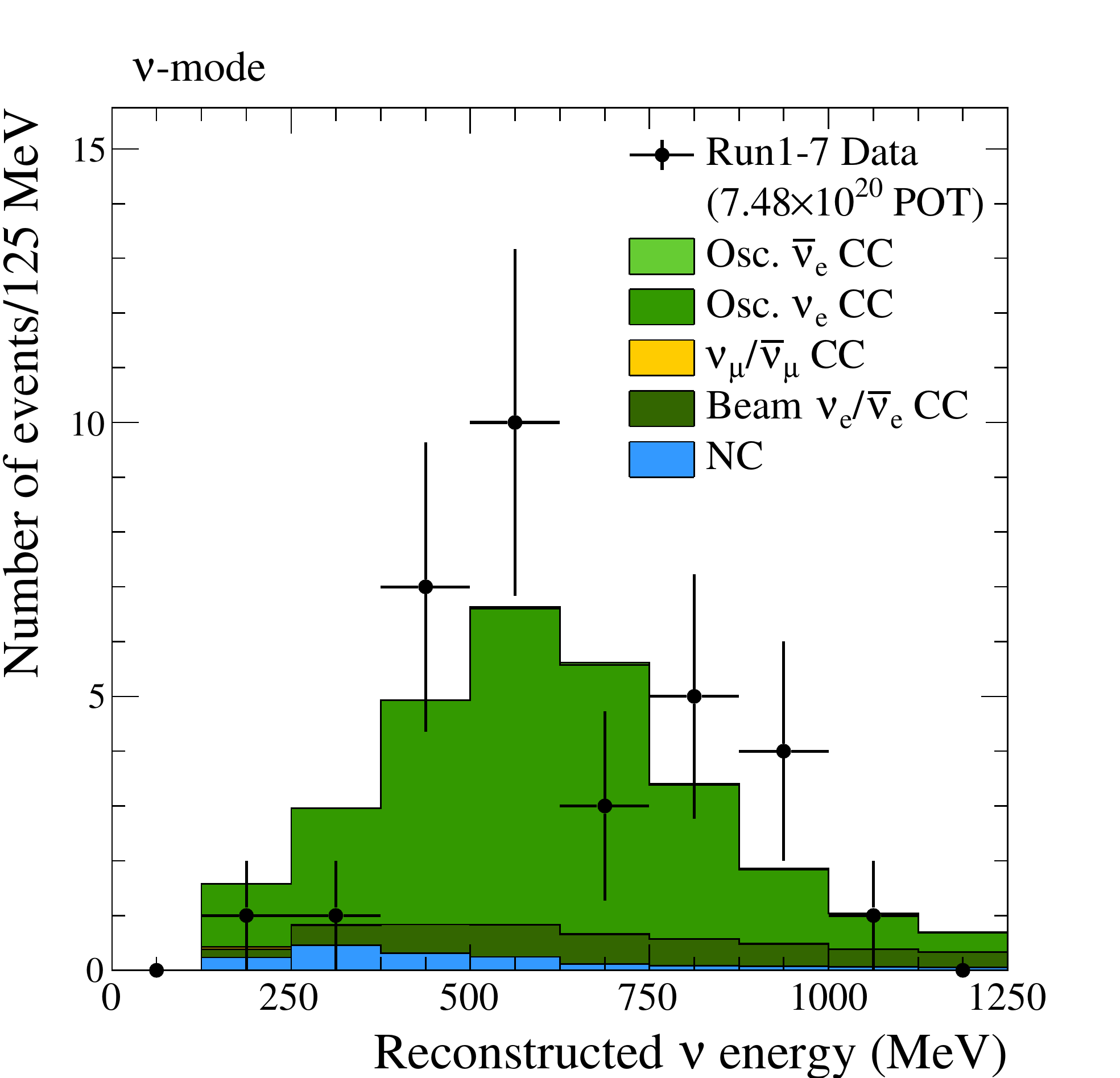} \hfill
  \includegraphics[width=0.48\textwidth]{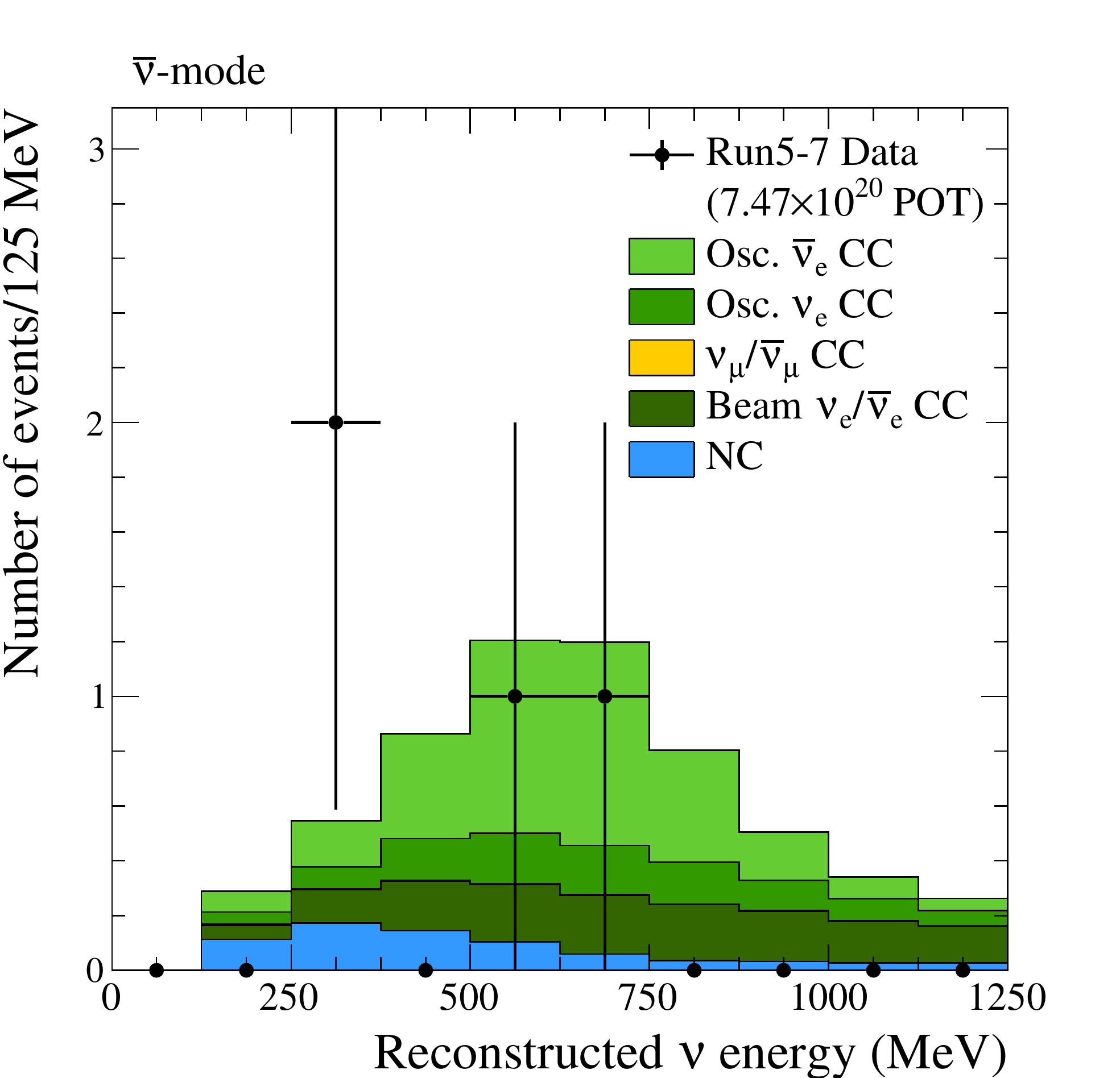}
\caption{The reconstructed energy spectra 
of the observed \nue in \nunu-mode (left) and \nueb in \nub-mode (right) CC candidate event samples assuming CCQE interaction kinematics.  
The data are shown as points with statistical error bars and the shaded, stacked 
histograms are the MC predictions.
The expectation is based on the parameters of Tab.~\ref{tab:nomosc_par}.}
\label{fig:SK_enurec_nue_plots}
%\end{center}
\end{figure*}

After all cuts, 32 events remain in the \nue CC candidate sample and 4 in the \nueb CC candidate sample, as shown in Fig.~\ref{fig:SK_enurec_nue_plots}.  Kolmogorov-Smirnov 
(KS) tests of the accumulated events as a function of POT %, shown in Fig. \ref{fig:SK_nue_KS}, 
are compatible with a constant rate, with p-values of 0.99 and 0.25 respectively.

\begin{figure*}[htbp]
\begin{center}
%\subfloat[]{
  %\includegraphics[width=0.23\textwidth]{figures/SK/nue_vtxxy_nu_run1-7c.eps}
  \includegraphics[width=0.48\textwidth]{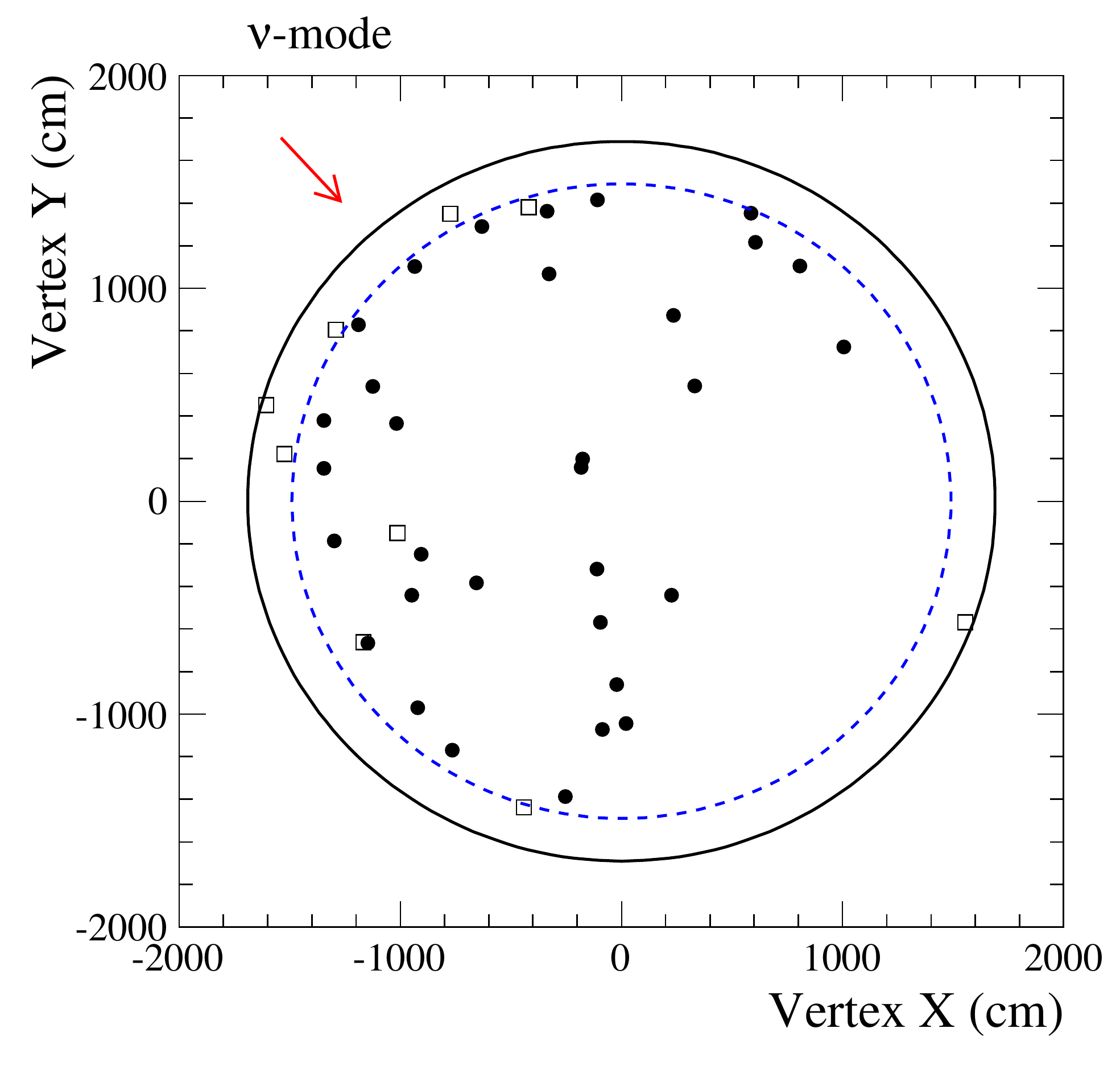}
  \includegraphics[width=0.48\textwidth]{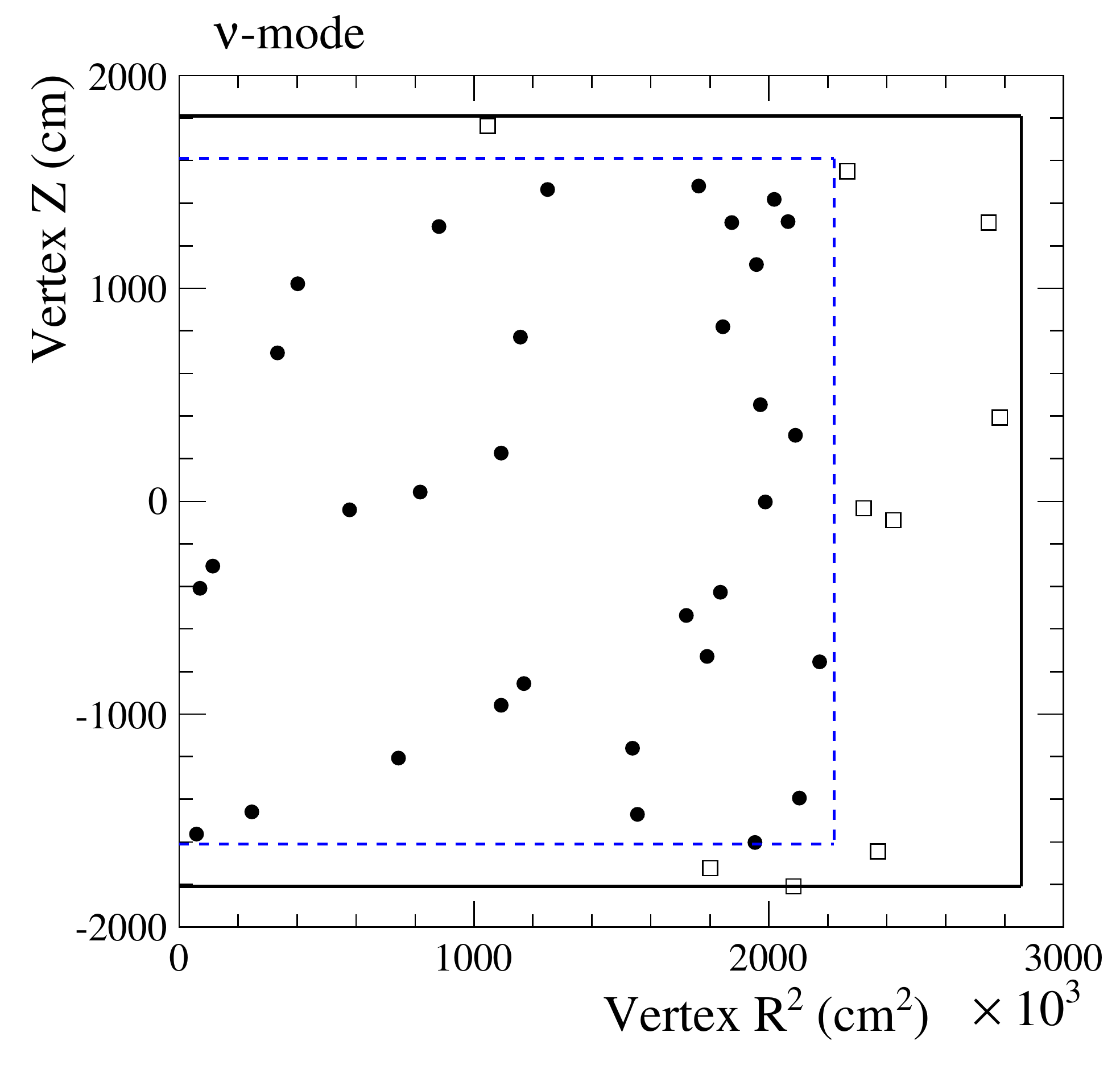}
%}
\\
%\subfloat[]{
%\includegraphics[width=0.23\textwidth]{figures/SK/nue_vtxr2_nu_run1-7c.eps}
  %\includegraphics[width=0.23\textwidth]{figures/SK/nue_vtxxy_antinu_run5-7.eps}
  \includegraphics[width=0.48\textwidth]{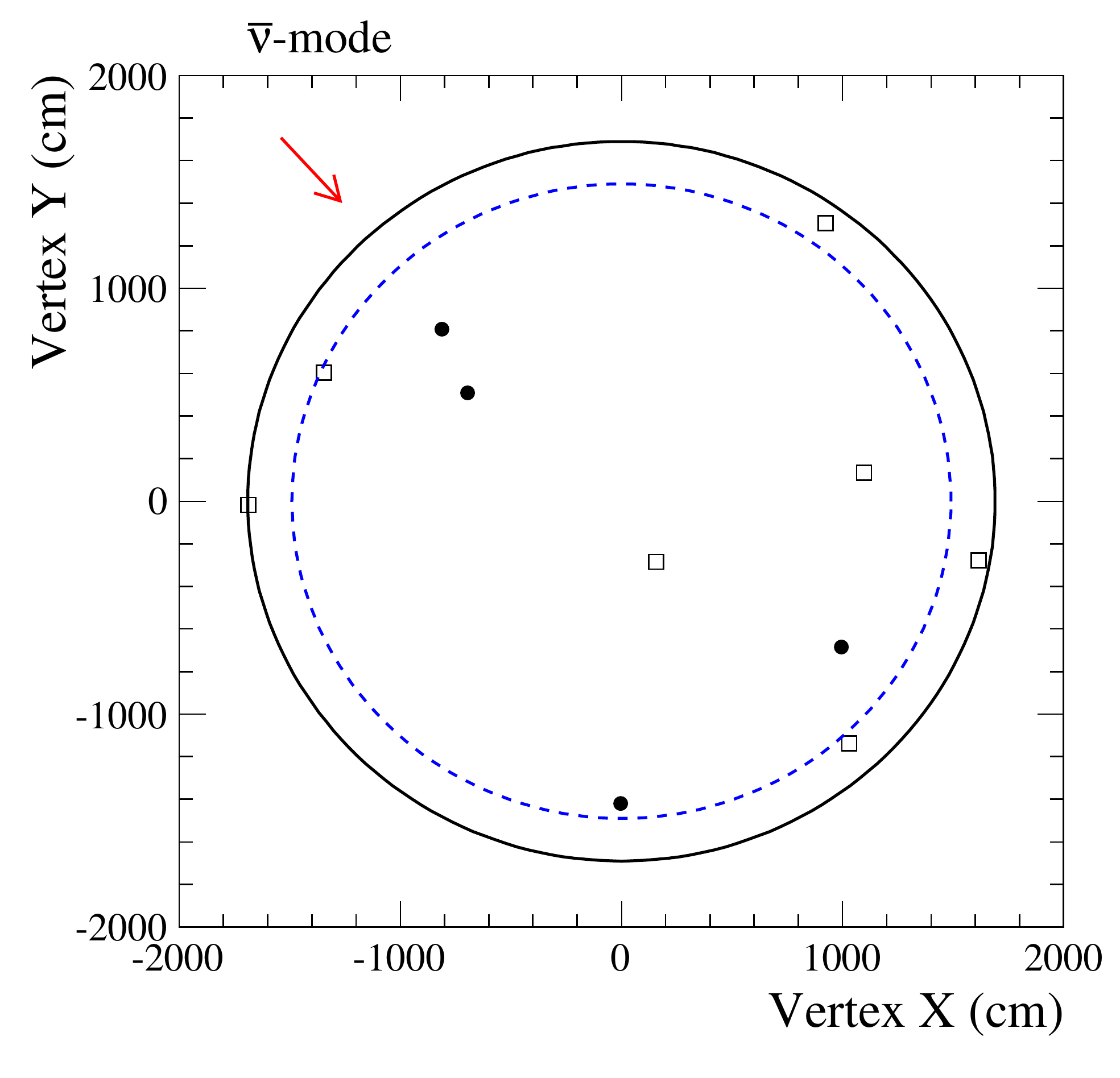}
  \includegraphics[width=0.48\textwidth]{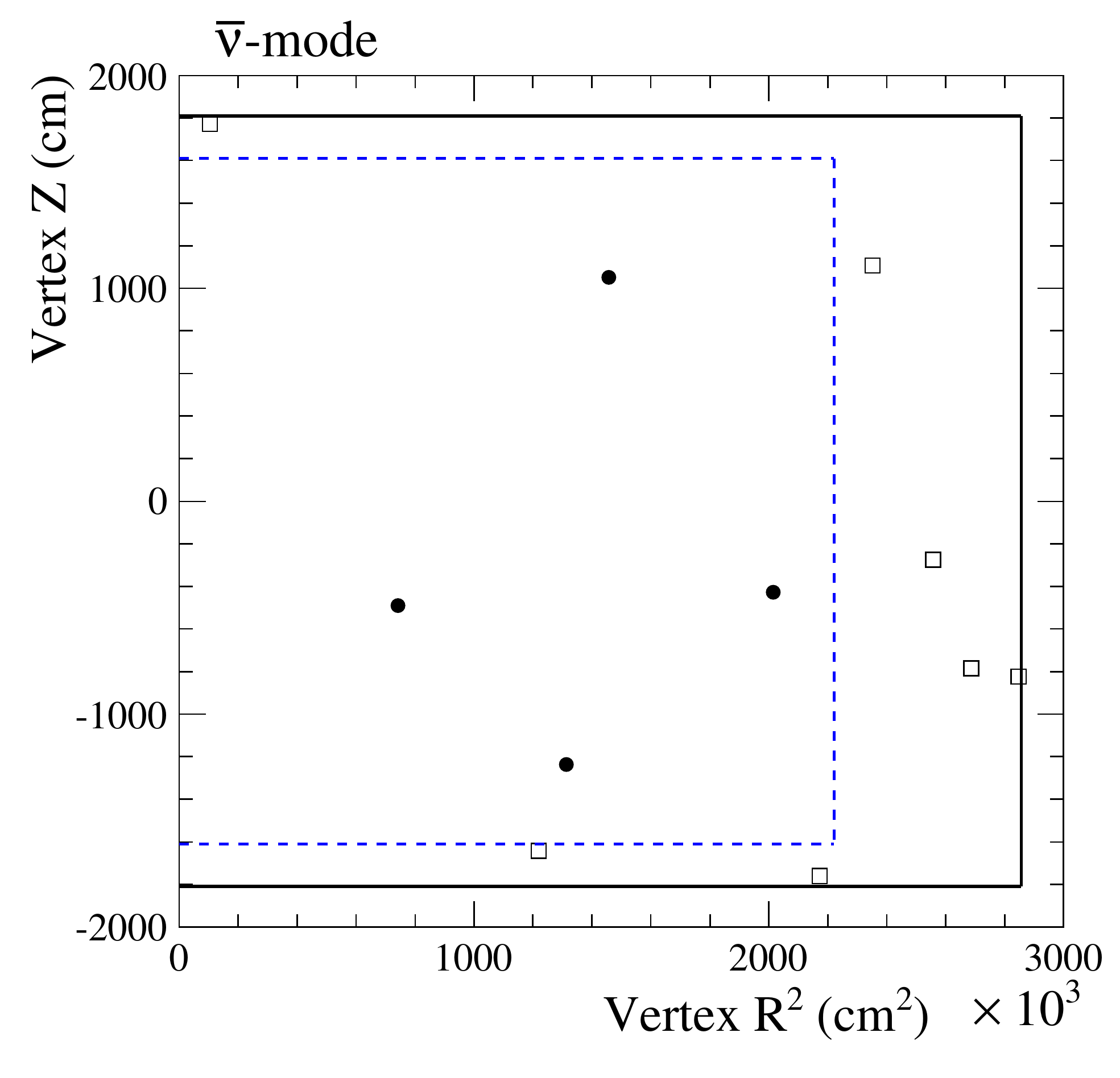}
%}
\caption{Two-dimensional vertex distributions of the observed \nue CC candidate events in 
$(X,Y)$ and $(R^2,Z)$ for $\nu$-mode (top) and $\nub$-mode (bottom). 
%and $(r^2 = x^2 + y^2,z)$.  
The arrow indicates the neutrino beam direction and 
the dashed line indicates the fiducial volume boundary.  
Events indicated 
by open square markers passed all of the \nue selection cuts except for the fiducial 
volume cut.}
\label{fig:SK_nue_vtx}
\end{center}
\end{figure*}

The vertex distributions of the candidate event samples are checked for 
signs of bias that might suggest background contamination.  Fig.~\ref{fig:SK_nue_vtx} 
shows the vertex distribution of the \nue CC candidate events 
in the SK tank coordinate system.  %No unexpected clustering is observed, and 
Combined KS tests for uniformity in $r^2$ and $z$ yield p-values of 0.1 and 0.6 for the \nue and \nueb samples respectively.

\begin{table*}[!t]
%\setlength\tabcolsep{3pt}
%{\footnotesize
\begin{center}
\caption{
Event reduction for the \nue CC selection at the far detector.  The numbers of 
expected MC events divided into five categories are shown after each 
selection criterion is applied.  The MC expectation is based upon three-neutrino 
oscillations with the parameters as shown in Tab.~\ref{tab:nomosc_par}.
}
\begin{ruledtabular}
\begin{tabular}{lccccccc}
%\hline
%\hline
&  & $\num+\numb$ & $\nue+\nueb$ & $\nu+\bar{\nu}$ & $\numb\rightarrow\nueb$ & $\num\rightarrow\nue$ &\\
$\nu$-beam mode & MC total & CC & CC & NC & CC & CC & Data\\
\hline
interactions in FV                                 & 744.89 & 364.32 & 18.55 & 326.16 & 0.39 & 35.47 &  - \\
FCFV                                               & 431.85 & 279.88 & 18.09 &  98.72 & 0.38 & 34.78 & 438\\
single ring\footnotemark[1]                        & 223.49 & 153.40 & 11.15 &  28.68 & 0.32 & 29.95 & 220\\
electron-like\footnotemark[2]                      &  66.94 &   6.46 & 11.06 &  19.53 & 0.31 & 29.57 & 70\\
$E_{\rm vis}>100{\rm \mev}$\footnotemark[3]          &  61.78 &   4.59 & 11.01 &  16.81 & 0.31 & 29.06 & 66\\
$N_{\rm Michel-e} = 0$\footnotemark[4]                &  50.60 &   0.97 &  8.97 &  14.24 & 0.31 & 26.11 & 51\\
$E_{\nu}^{\rm rec}<1250{\rm \mev}$\footnotemark[5]    &  40.71 &   0.25 &  4.26 &  10.85 & 0.22 & 25.14 & 46\\
not $\pi^{0}$-like\footnotemark[6]                  &  28.55 &   0.09 &  3.68 &   1.35 & 0.18 & 23.25 & 32\\
\hline
\hline
%% \end{tabular}
%% \vskip 6mm
%% \begin{tabular}{lcccccc}
%% \hline
%% \hline
%& & \ \ $\num+\numb$ \ \ & \ \ $\nue+\nueb$ \ \ & \ \ $\nu+\bar{\nu}$ \ \ & \ \ $\numb\rightarrow\nueb$ \ \ & \ \ $\num\rightarrow\nue$ \ \ \\
%& \ \ MC total\ \ & CC & CC & NC & CC & CC\\
$\bar{\nu}$-beam mode & & & & & & \\
\hline
interactions in FV                  & 312.38  & 164.04  & 9.00 & 132.75 &  4.30 & 2.29  & -   \\
FCFV                                     & 180.48  & 123.24  & 8.75 &  42.05 &  4.20 & 2.24  &170  \\
single ring                              &  96.06  &  73.21  & 5.51 &  11.87 &  3.74 & 1.73  & 94  \\
electron-like                           &  21.55  &   2.31  & 5.48 &   8.36   & 3.70 & 1.71 & 16  \\
$E_{\rm vis}>100{\rm \mev}$ &  20.05  &   1.83  & 5.46 &   7.39 &  3.68 & 1.69  & 14  \\
$N_{\rm Michel-e} = 0$           &  16.40  &   0.33  & 4.71 &   6.24 &  3.66 & 1.46  & 12  \\
$\erec<1250{\rm \mev}$        &  11.40  &   0.08 & 1.89 &   4.83 &  3.42 & 1.19  & 9 \\
not $\pi^{0}$-like                   &   6.28  &   0.02 & 1.58 &   0.60 & 3.04 &  1.05  &  4 \\

%\hline
%\hline
\end{tabular}
\end{ruledtabular}

\label{tab:SK_nue_events}
%\vspace{-1cm}
\footnotetext[1]{There is only one reconstructed Cherenkov ring}
\footnotetext[2]{The ring is $e$-like}
\footnotetext[3]{The visible energy, $E_{\mathrm{vis}}$, is greater than 100 \mev}
\footnotetext[4]{There is no reconstructed Michel electron}
\footnotetext[5]{The reconstructed energy, \erec, is less than 1.25 \gev}
\footnotetext[6]{The event is not consistent with a $\pi^0$ hypothesis}

\end{center}
%}
\end{table*}

%% \begin{enumerate}
%% \item There is only one reconstructed Cherenkov ring
%% \item The ring is $\mu$-like
%% \item The reconstructed momentum, $p_{\mu}$, is greater than 200 \mevc
%% \item There are less than two reconstructed Michel electrons
%% \end{enumerate}

\begin{table*}[!t]
\begin{center}
\caption{
Event reduction for the \num CC selection at the far detector.  The numbers of 
expected MC events divided into four categories are shown after each 
selection criterion is applied.  The MC expectation is based upon three-neutrino 
oscillations with the parameters as shown in Tab.~\ref{tab:nomosc_par}.
}
%% \begin{tabular}{ll}
%% (1) & There is only one reconstructed Cherenkov ring \\
%% (2) & The ring is $\mu$-like \\
%% (3) & The reconstructed momentum, $p_{\mu}$, is greater than 200 \mevc \\
%% (4) & There are less than two reconstructed Michel electrons \\
%% \end{tabular}
%% \vskip 5mm
\begin{ruledtabular}
\begin{tabular}{lccccccc}
%\hline
%\hline
 &             &  \num  &  \numb  & $\num+\numb$ &  $\nue+\nueb$  &  $\nu+\bar{\nu}$ & \\
$\nu$-beam mode & MC total & CCQE  & CCQE &  CC nonQE  & CC & NC & Data\\
\hline
interactions in FV                       &  744.89 & 100.17 & 6.45 & 257.70 & 54.41 & 326.16 &  - \\  
FCFV                                     &  431.85 &  78.75 & 4.85 & 196.28 & 53.25 &  98.72 & 438 \\  
single ring\footnotemark[7]              &  223.49 &  73.49 & 4.70 &  75.21 & 41.41 &  28.68 & 220 \\  
muon-like\footnotemark[8]                &  156.56 &  72.22 & 4.65 &  70.06 &  0.47 &   9.16 & 150 \\  
$p_{\mu}>200 \mevc$\footnotemark[9]       &  156.24 &  72.03 & 4.65 &  70.00 &  0.47 &   9.08 & 150 \\ 
$N_{\rm Michel-e} \leq 1$\footnotemark[10]  &  137.76 &  71.28 & 4.63 &  52.61 &  0.46 &   8.78 & 135 \\
\hline
\hline
%% \end{tabular}
%% \vskip 5mm
%% \begin{tabular}{lcccccc}
%% \hline
%% \hline
% &             & \ \ \num \ \ & \ \ \numb \ \ & $\num+\numb$ & \ \ $\nue+\nueb$ \ \ & \ \ $\nu+\bar{\nu}$ \ \ \\
%& MC total\ \ & CCQE  & CCQE & \ \ CC nonQE\ \  & CC & NC \\
$\bar{\nu}$-beam mode & & & & & & \\
\hline
interactions in FV      & 312.38 & 20.04 & 30.77 & 113.23 & 15.59 & 132.75 &  -  \\     
FCFV                    & 180.48 & 15.04 & 24.95 &  83.26 & 15.19 &  42.05 & 170 \\    
single ring             &  96.06 & 13.52 & 24.28 &  35.41 & 10.98 &  11.87 &  94 \\    
muon-like               &  74.52 & 13.40 & 23.96 &  33.56 &  0.09 &   3.52 &  78 \\    
$p_{\mu}>200 \mevc$      &  74.42 & 13.39 & 23.92 &  33.54 &  0.09 &   3.48 &  78 \\   
$N_{\rm Michel-e} \leq 1$  &  68.26 & 13.18 & 23.85 &  27.79 &  0.09 &   3.35 &  66 \\         
%\hline
%\hline
\end{tabular}
\end{ruledtabular}
\label{tab:SK_numu_events}
\footnotetext[7]{There is only one reconstructed Cherenkov ring}
\footnotetext[8]{The ring is $\mu$-like}
\footnotetext[9]{The reconstructed momentum, $p_{\mu}$, is greater than 200 \mevc}
\footnotetext[10]{There are less than two reconstructed Michel electrons}

\end{center}
\end{table*}

\begin{figure*}[htbp]
\centering
%\subfloat[]{
  %\includegraphics[width=0.45\textwidth]{figures/SK/numu_enurec_nu_run1-7c.eps}
  \includegraphics[width=0.48\textwidth]{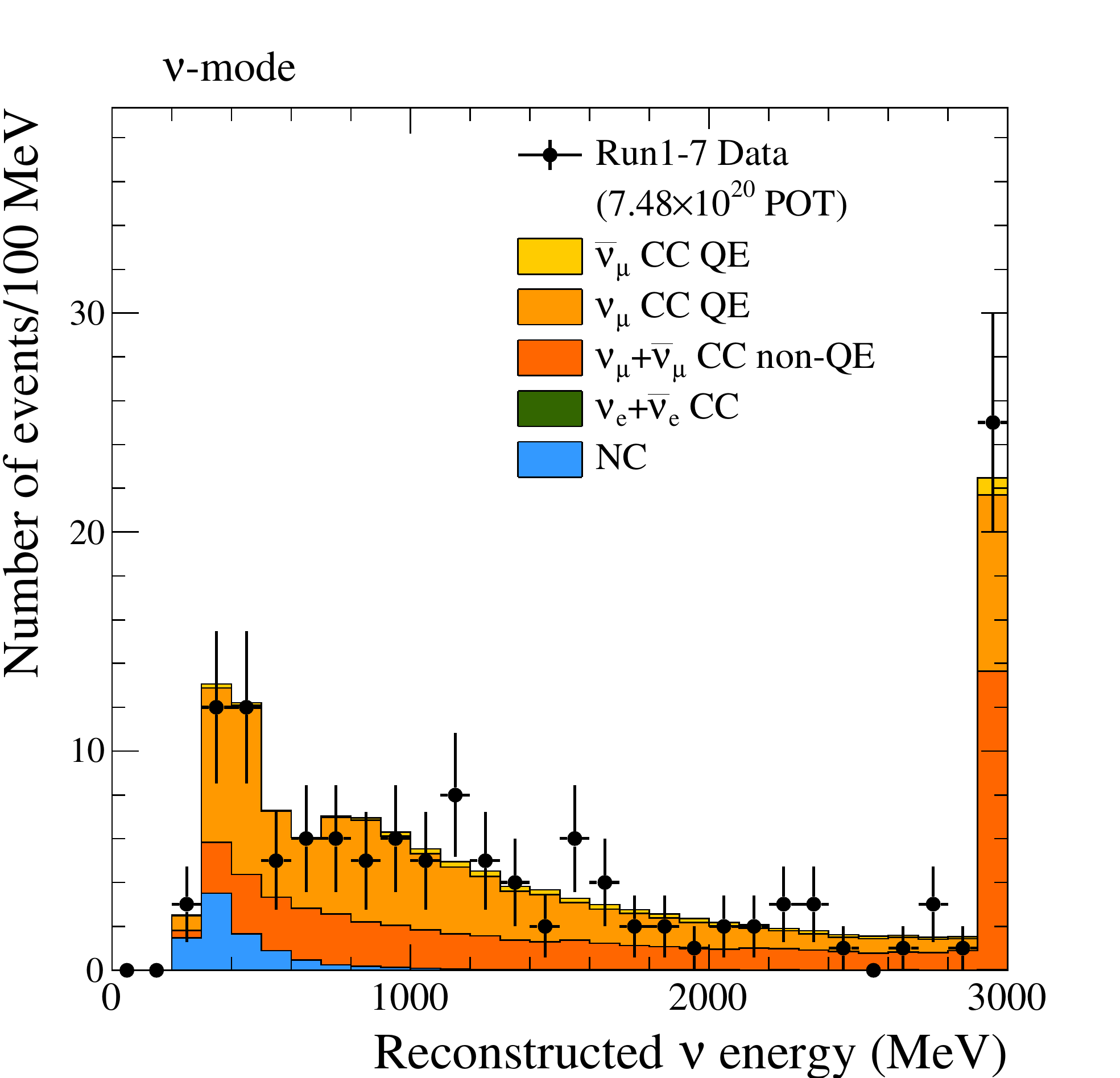} \hfill
%}
%\\
%\subfloat[]{
  %\includegraphics[width=0.45\textwidth]{figures/SK/numu_enurec_antinu_run5-7.eps}
    \includegraphics[width=0.48\textwidth]{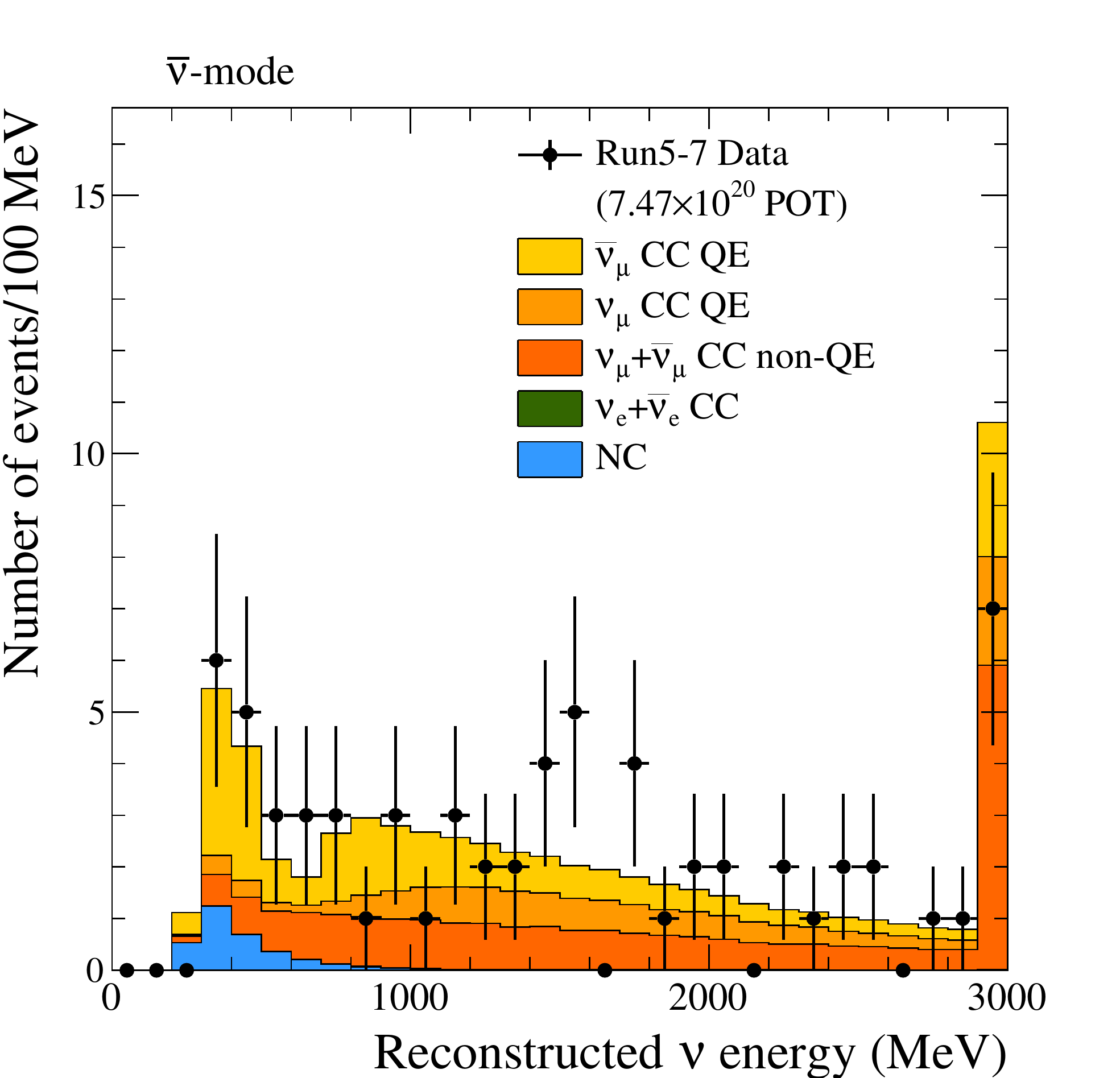}
%}
\caption{The reconstructed energy spectra 
of the observed \num in \nunu-mode (left) and \numb in \nub-mode (right) CC candidate event samples assuming CCQE interaction kinematics.  
The data are shown as points with statistical error bars and the shaded, stacked 
histograms are the MC predictions, and the rightmost bin includes overflow.
The expectation is based on the parameters of Tab.~\ref{tab:nomosc_par}.}
\label{fig:SK_enurec_plots}
%\end{center}
\end{figure*}

\numormb CC candidate events are selected using the criteria shown in Tab.~\ref{tab:SK_numu_events}.
The momentum cut rejects charged pions and misidentified electrons from the decay of unobserved muons 
and pions.  Fewer than two Michel electrons are required to reject events with additional unseen muons or pions.
After all cuts are applied, 135 events remain in the \num CC candidate sample and 66 in the \numb CC candidate sample
as shown in Fig.~\ref{fig:SK_enurec_plots}.

\subsection{SK charged-current single pion selection}
\label{sec:skcc1pi}

\begin{figure*}[htbp]
\begin{center}
  \includegraphics[width=0.48\textwidth]{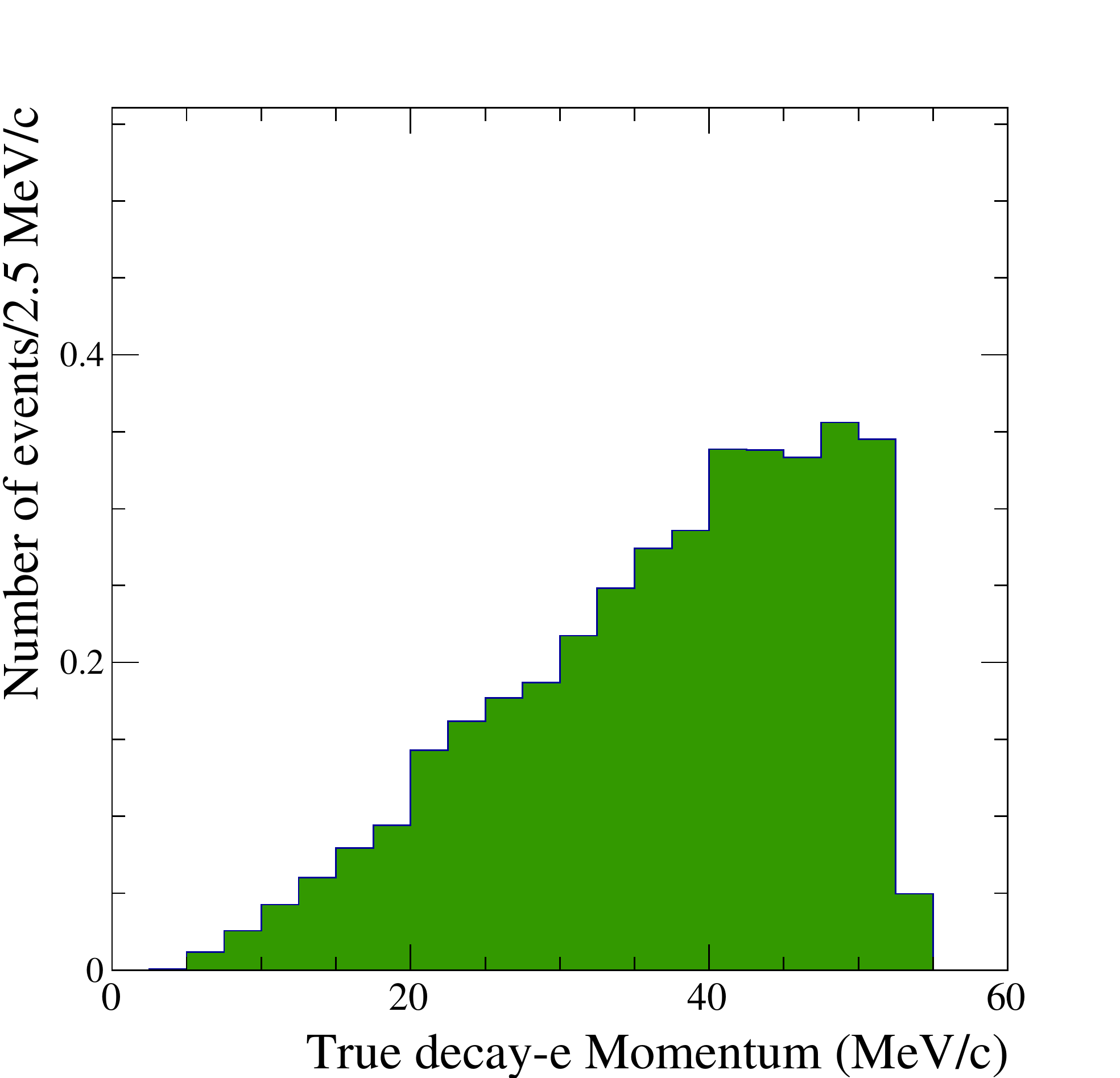}
  \includegraphics[width=0.48\textwidth]{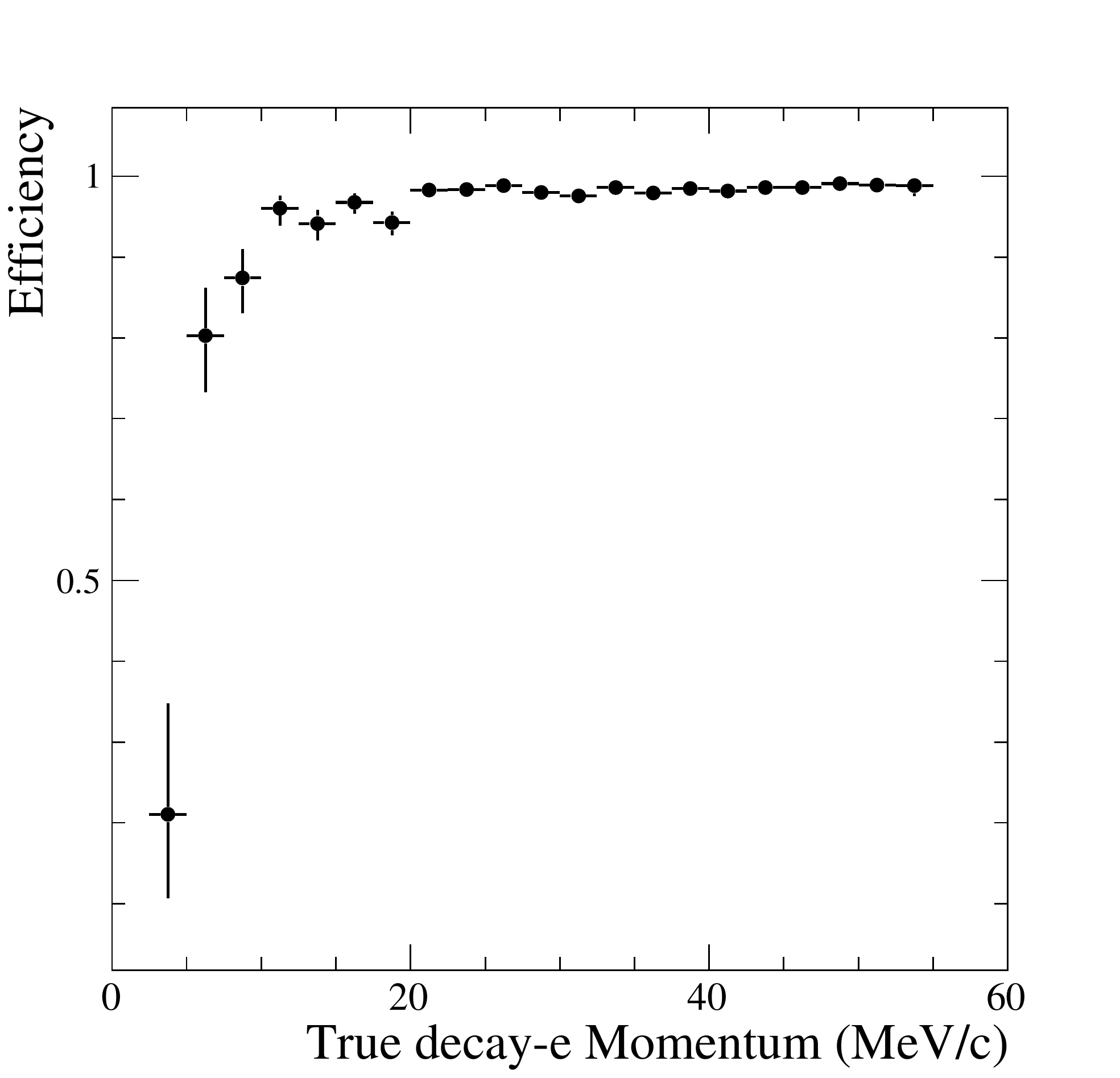}
  \caption{The true momentum distribution for tagged Michel electrons from true \ccpip simulated events reconstructed in the fiducial volume (left) and the tagging efficiency for these events (right).  The expectation is based on the parameters of Tab.~\ref{tab:nomosc_par}.}
\label{fig:SK_CC1pi_demom}
\end{center}
\end{figure*}

\begin{figure*}[htbp]
\begin{center}
  \includegraphics[width=0.48\textwidth]{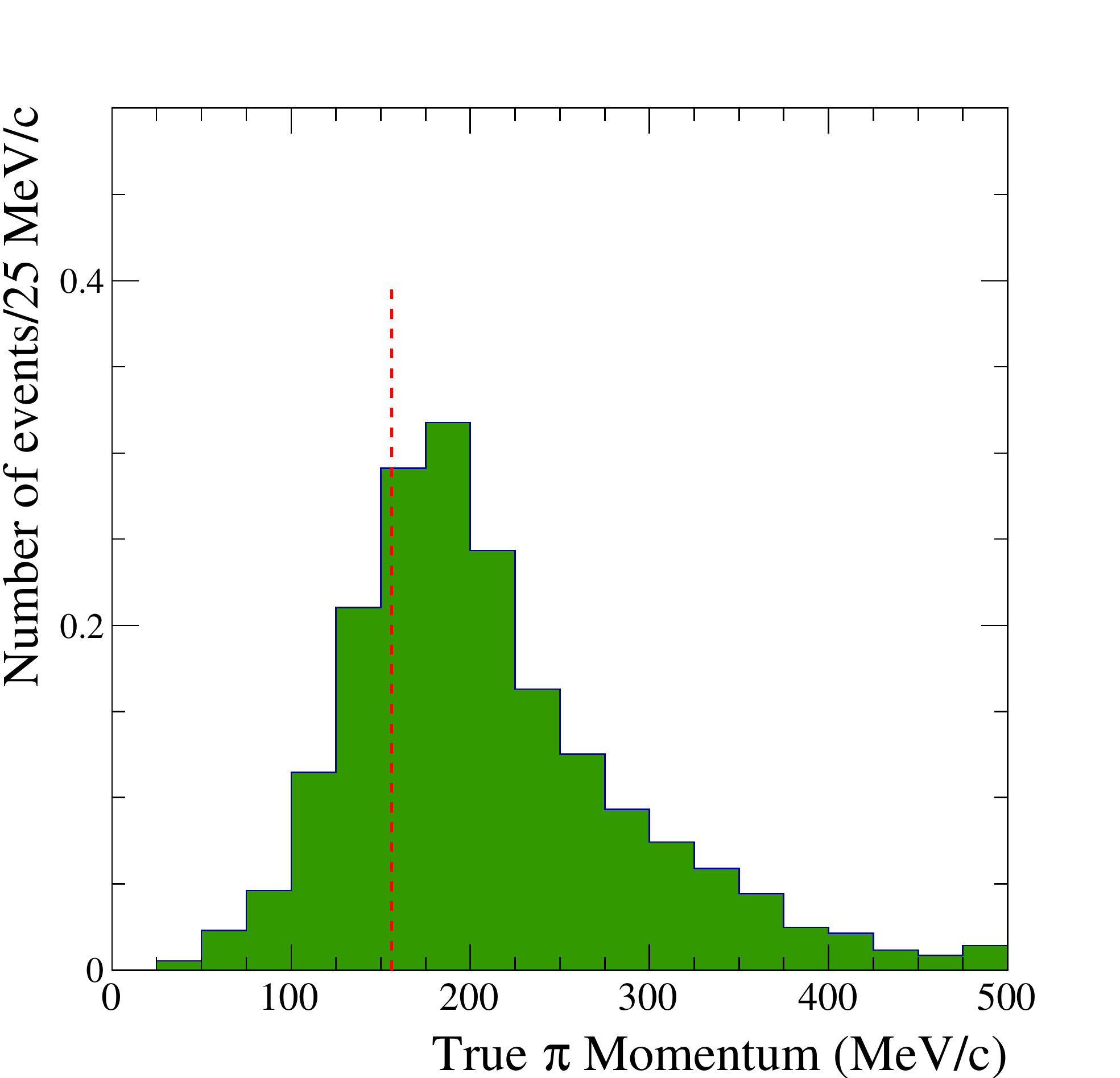}
  \includegraphics[width=0.48\textwidth]{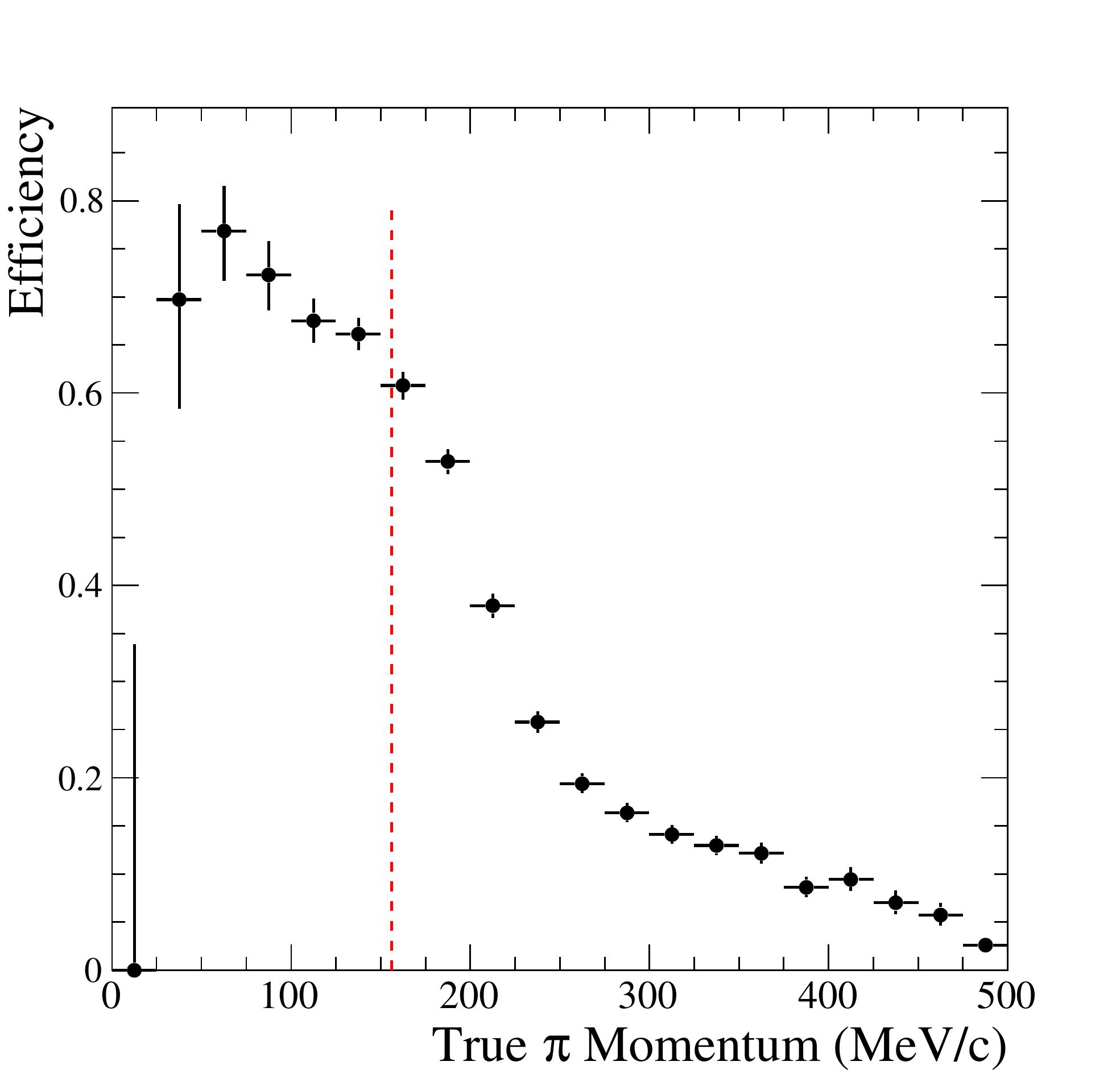}
  \caption{The true momentum distribution for selected simulated signal events in the \ccpip candidate sample (left) and the
    selection efficiency for these events (right).  The expectation is based on the parameters of Tab.~\ref{tab:nomosc_par}. The
  red dashed line indicates the Cherenkov threshold for charged pions.}
\label{fig:SK_CC1pi_pimom}
\end{center}
\end{figure*}

\begin{figure*}[htbp]
\centering
  \includegraphics[width=0.48\textwidth]{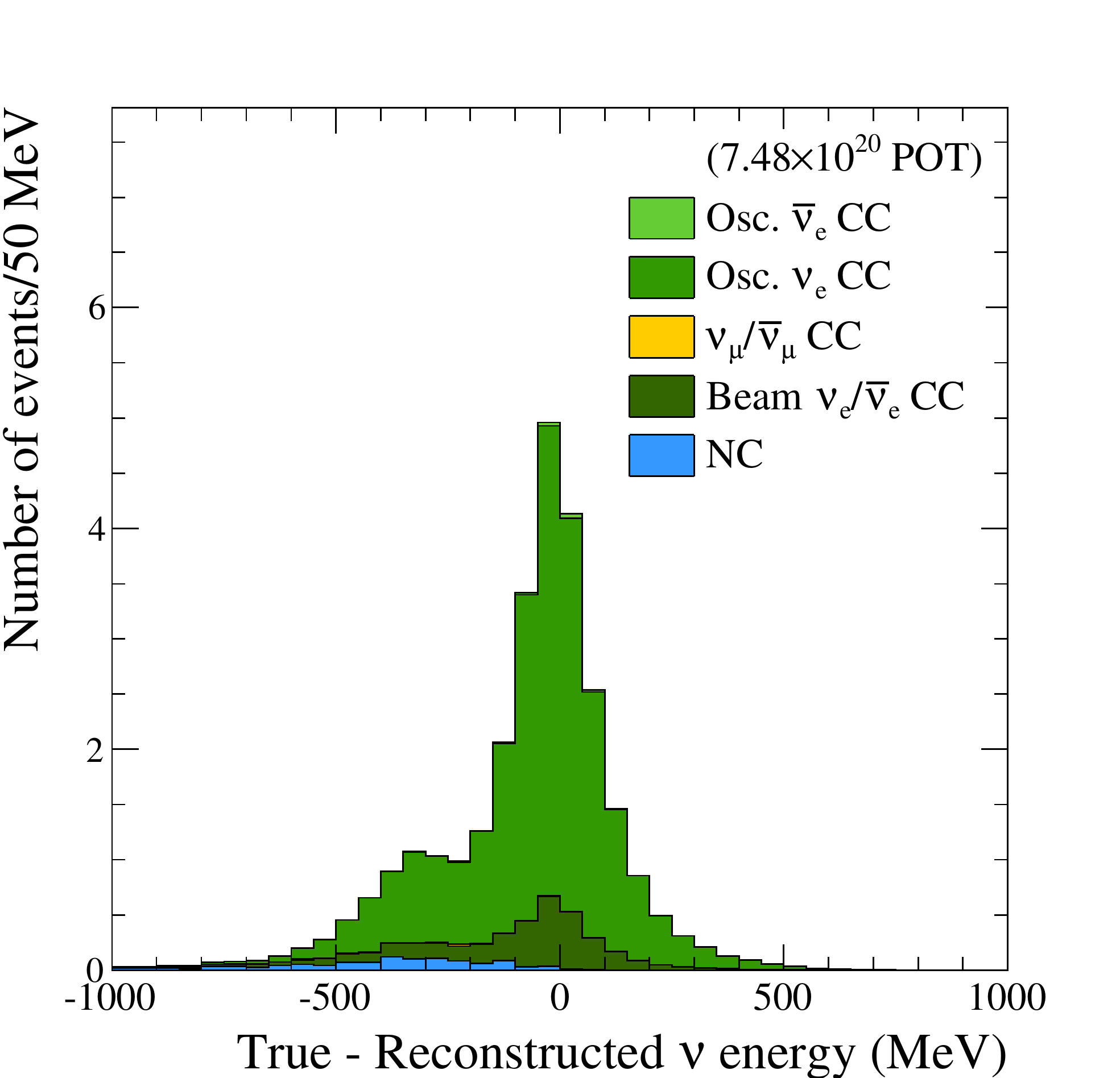} \hfill
  \includegraphics[width=0.48\textwidth]{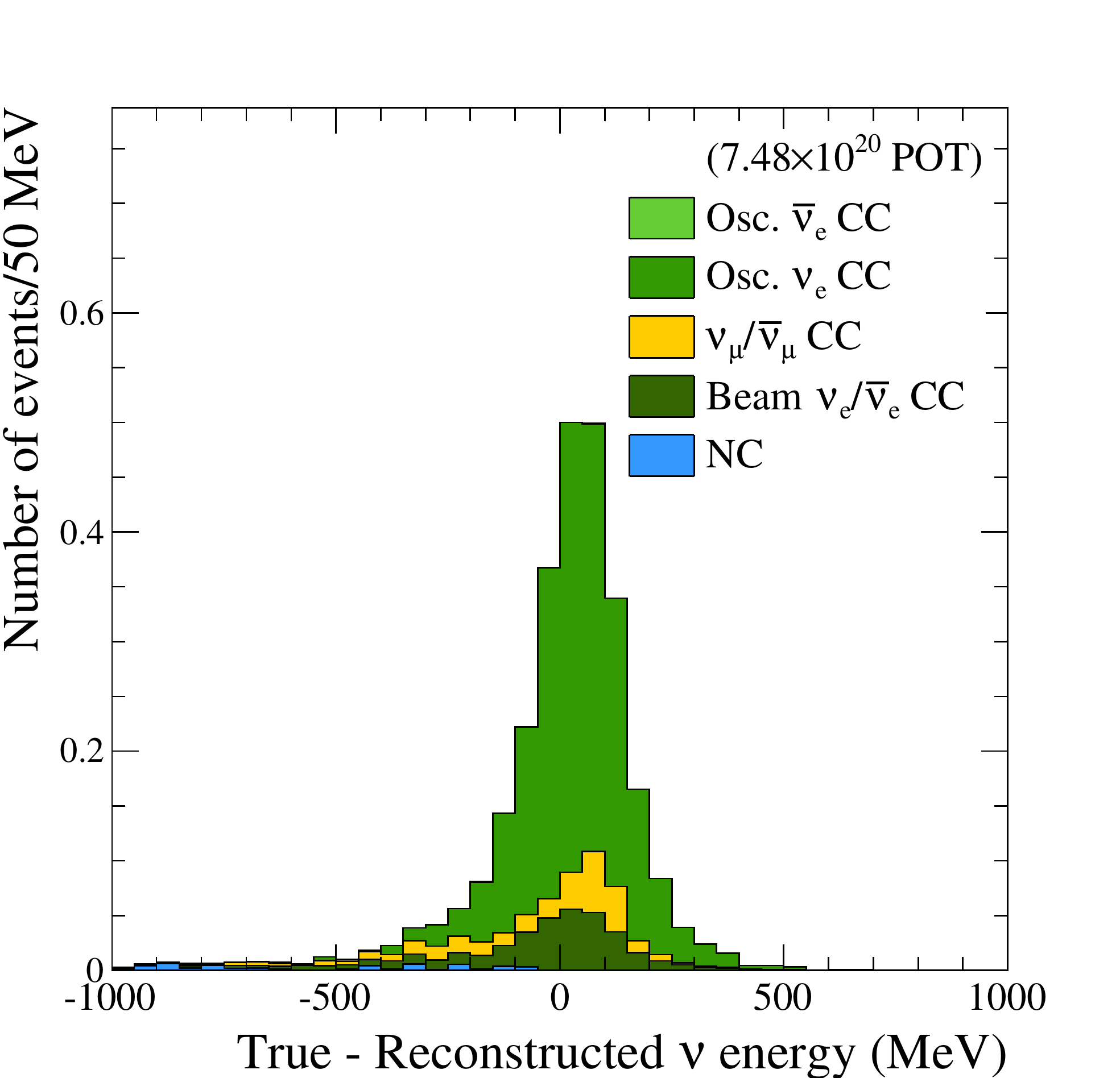}
\caption{Difference between true and reconstructed energy 
  of the \nue CCQE-like (left) and \ccpip (right) simulated samples.  The energy for the CCQE-like sample is reconstructed using
  Eq.~\ref{eq:SK_Erec}.  For the \ccpip sample the modified interaction kinematics in Eq.~\ref{eq:SK_CC1pi_Erec}
  are used.  
The expectation is based on the parameters of Tab.~\ref{tab:nomosc_par}.}
\label{fig:SK_CC1pi_enurec_res}
%\end{center}
\end{figure*}

\begin{figure}[htbp]
\begin{center}
  \includegraphics[width=0.48\textwidth]{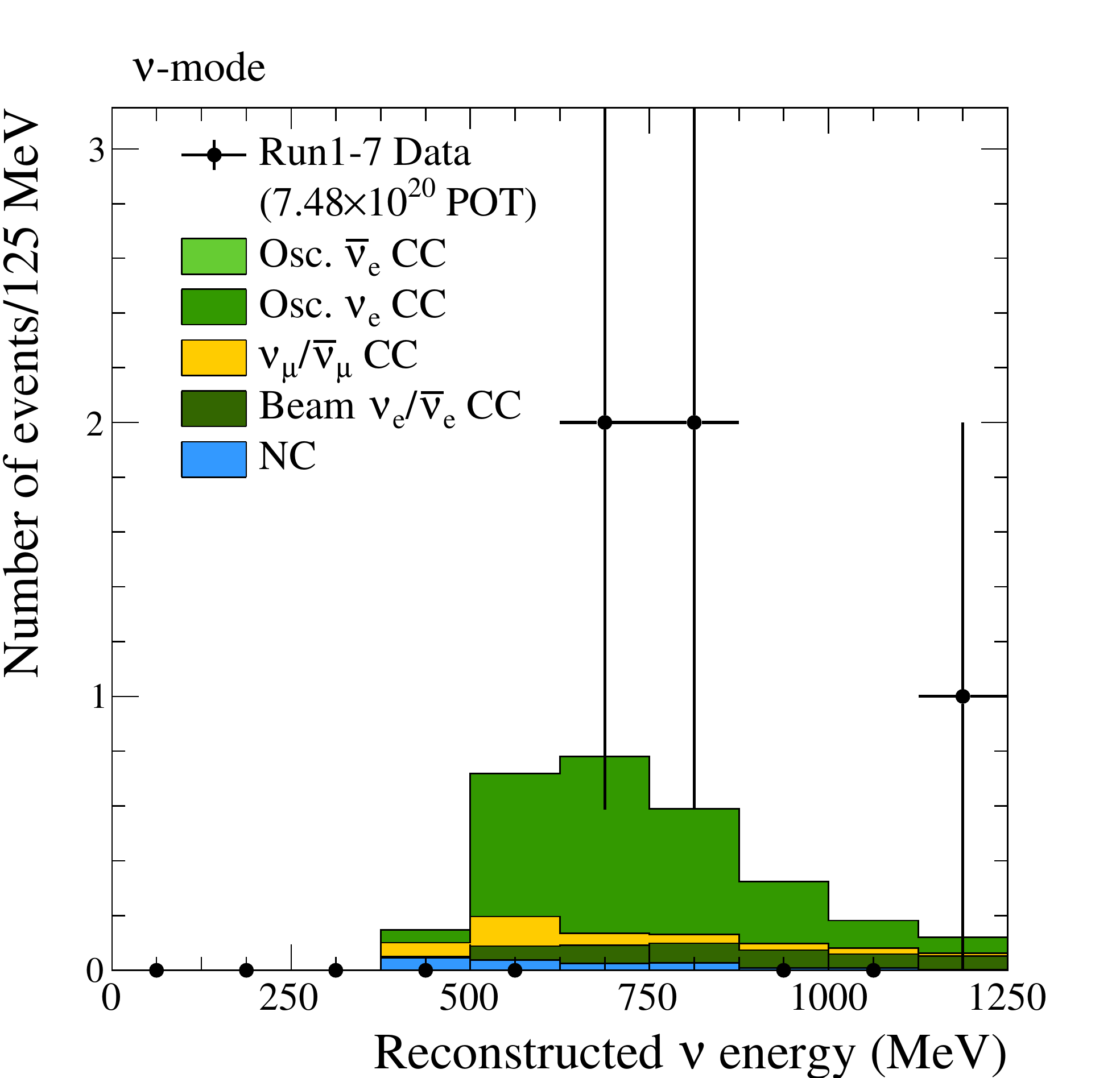}
\caption{The reconstructed energy spectrum 
of the observed \nue \ccpip candidate events assuming the modified interaction kinematics in equation \ref{eq:SK_CC1pi_Erec}.  
The data are shown as points with statistical error bars and the shaded, stacked 
histograms are the MC predictions.
The expectation is based on the parameters of Tab.~\ref{tab:nomosc_par}.}
\label{fig:SK_CC1pi_enurec_plots}
\end{center}
\end{figure}

%Candidate Samples Selections - CC1pi
A new far detector event sample has been included in the oscillation
analysis described here.
As mentioned previously, single-ring events produced by quasi-elastic interactions are the most common at T2K neutrino energies.  By modifying the
event selection criteria to include an additional Michel electron candidate, it is possible to
select events with a pion produced below Cherenkov threshold. Michel electrons are tagged at SK
by searching for secondary hit clusters within the same trigger window and in subsequent trigger windows
that pass time, charge and vertex goodness criteria. Note that this selection tags only positive $\pi^{+}$'s as $\p^{-}$'s are absorbed by nuclei before they decay when stopped in water. Fig.~\ref{fig:SK_CC1pi_demom} shows the true Michel electron momentum distribution
for true \ccpip simulated events reconstructed inside the fiducial volume.
Fig.~\ref{fig:SK_CC1pi_pimom} shows the true pion momentum distribution
for selected signal events and the selection efficiency.  The Cherenkov threshold for charged pions is also shown.
The efficiency falls above this threshold as the pion produces more light and the event-fitting algorithm is increasingly likely to find a second ring, thus disqualifying events from the sample.  The event reduction for the full selection is shown in Tab.~\ref{tab:SK_nueCCpi_events}.
There is a larger background from misidentified muons in this sample compared with the single-ring selection as
such events are more likely to contain a Michel electron candidate.
It can be seen in Tab.~\ref{tab:SK_nueCCpi_events} that there is an apparent difference between the data and the MC expectation after cut number five.  Applying an equivalent cut sequence to atmospheric neutrino data and MC (in this case the neutrino direction is not known so a cut on $E_{vis}$ is used instead of \erec) yielded no such discrepancy.
This single-ring \ccpip selection has been
implemented as a new sample for \nue appearance with neutrino beam data.
The reconstructed energy equation is modified from the CCQE case by recognizing that the outgoing
baryon is a $\Delta^{++}$ instead of a proton and neglecting nuclear effects:
\begin{equation}
\label{eq:SK_CC1pi_Erec}
\erec = \frac{m^2_{\Delta^{++}} - m^2_p - m^2_l + 2m_pE_l}{2(m_p - E_l + p_l\cos\theta_l)}
\end{equation}
where $m_{\Delta^{++}}$ is the mass of the $\Delta^{++}$ (1232.0 MeV/c$^2$).
Fig.~\ref{fig:SK_CC1pi_enurec_res} shows the difference in the true and reconstructed neutrino
energy for the final \ccpip candidate selection along with that for the single-ring selection for comparison.
Fig.~\ref{fig:SK_CC1pi_enurec_plots} shows the reconstructed energy distribution for the final
sample. Five \nue \ccpip candidates are reconstructed in the data while 3.1 events are expected for the oscillation parameters of Tab.~\ref{tab:nomosc_par}.

%% \begin{figure}[tbp]
%% \begin{center}
%% %\includegraphics[width=0.4\textwidth]{}
%% \caption{The difference between the true and reconstructed energy spectrum for MC events passing the \nue CC1$\pi$ sample selection 
%% using the modified interaction kinematics in equation \ref{eq:SK_CC1pi_Erec}.  
%% The expectation is based on the parameters of Tab.~\ref{tab:nomosc_par}.
%% }
%% \label{fig:SK_CC1pi_enurec_res}
%% \end{center}
%% \end{figure}

\begin{table*}[htbp]
\begin{center}
\caption{
Event reduction for the \nue \ccpip selection at the far detector.  The numbers of 
expected MC events divided into five categories are shown after each 
selection criterion is applied.  The MC expectation is based on the parameters of Tab.~\ref{tab:nomosc_par}.
}
%\vskip 6mm
\begin{ruledtabular}
\begin{tabular}{lccccccc}
%\hline
%\hline
&  & $\num+\numb$ & $\nue+\nueb$ & $\nu+\bar{\nu}$ & $\numb\rightarrow\nueb$ & $\num\rightarrow\nue$ &\\
$\nu$-beam mode & MC total & CC & CC & NC & CC & CC & Data\\
%% &  & \ \ $\num+\numb$ \ \ & \ \ $\nue+\nueb$ \ \ & \ \ $\nu+\bar{\nu}$ \ \ & \ \ $\numb\rightarrow\nueb$ \ \ & \ \ $\num\rightarrow\nue$ \ \ \\
%% & \ \ MC total\ \ & CC & CC & NC & CC & CC\\
\hline
interactions in FV                                  & 744.89 & 364.32 & 18.55 & 326.16 & 0.39 & 35.47  &  -  \\
FCFV                                                & 431.85 & 279.88 & 18.09 &  98.72 & 0.38 & 34.78  & 438 \\
(1) single ring\footnotemark[1]                     & 223.49 & 153.40 & 11.15 &  28.68 & 0.32 & 29.95  & 220 \\
(2) electron-like\footnotemark[2]                   &  66.94 &   6.46 & 11.06 &  19.53 & 0.31 & 29.57  &  70 \\
(3) $E_{\rm vis}>100{\rm \mev}$\footnotemark[3]       &  61.78 &   4.59 & 11.01 &  16.81 & 0.31 & 29.06  &  66 \\
(4) $N_{\rm Michel-e} = 1$\footnotemark[4]             &   9.36 &   2.42 &  1.87 &   2.14 & 0.01 &  2.92  &  14 \\
(5) $\erec<1250{\rm \mev}$\footnotemark[5]          &   4.66 &   0.70 &  0.50 &   0.78 & $<0.01$ &  2.66  &  11 \\
(6) not $\pi^{0}$-like\footnotemark[6]               &   3.14 &   0.29 &  0.39 &   0.15 & $<0.01$ &  2.31   &  5 \\
%\hline
%\hline
%\vspace{-1cm}

\end{tabular}
\end{ruledtabular}
\label{tab:SK_nueCCpi_events}
\footnotetext[1]{There is only one reconstructed Cherenkov ring}
\footnotetext[2]{The ring is $e$-like}
\footnotetext[3]{The visible energy, $E_{\mathrm{vis}}$, is greater than 100 \mev}
\footnotetext[4]{There is one reconstructed Michel electron}
\footnotetext[5]{The reconstructed energy, \erec, is less than 1.25 \gev}
\footnotetext[6]{The event is not consistent with a $\pi^0$ hypothesis}
\end{center}
\end{table*}

%plot of Etrue-Erec for CC1pi

\begin{figure*}[htbp]
\begin{center}
%\subfloat[]{
  %\includegraphics[width=0.23\textwidth]{figures/SK/nue_vtxxy_nu_run1-7c.eps}
  \includegraphics[width=0.48\textwidth]{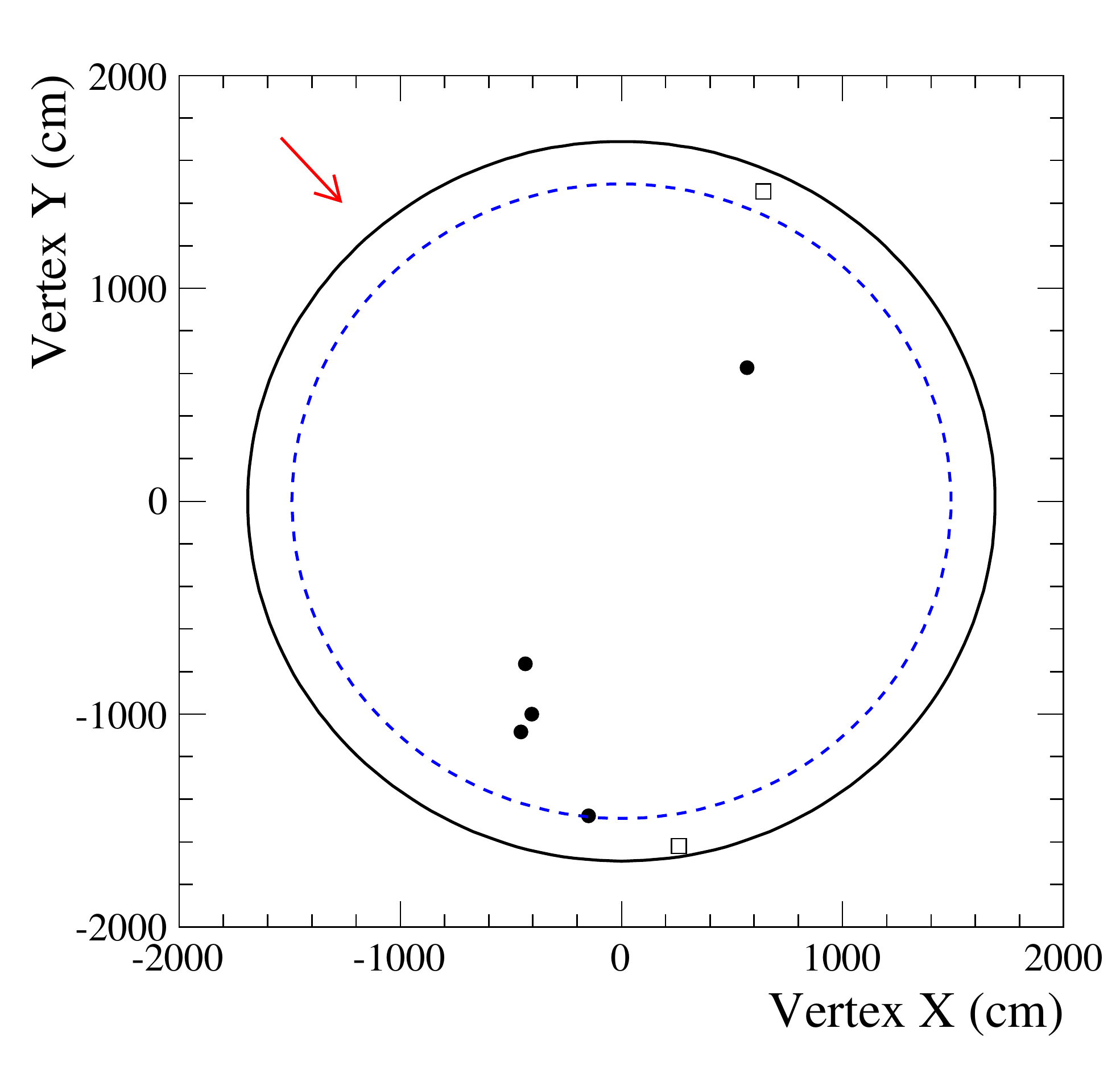}
  \includegraphics[width=0.48\textwidth]{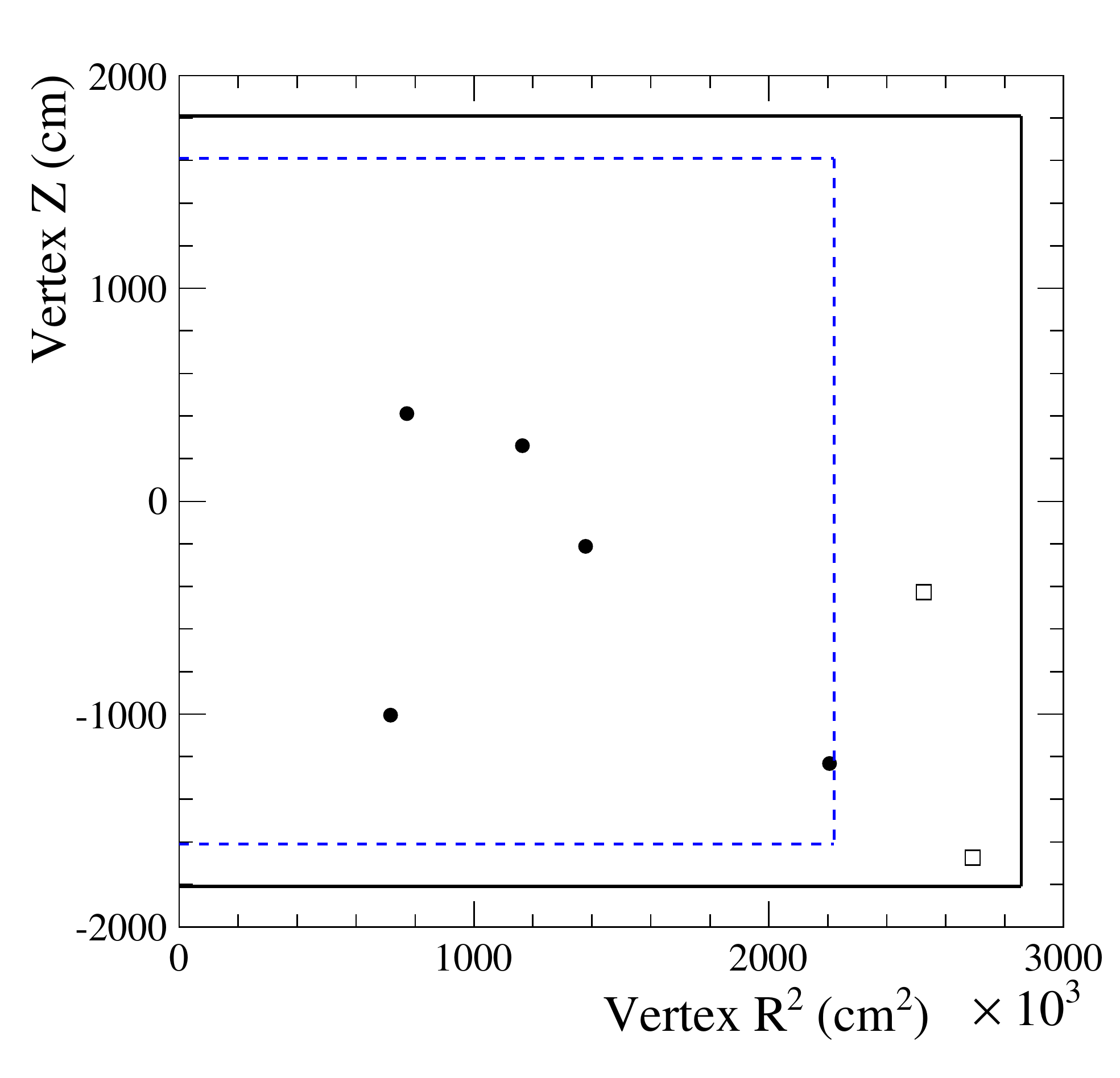}

\caption{Two-dimensional vertex distributions of the observed \nue \ccpip candidate events in 
$(X,Y)$ and $(R^2,Z)$. 
%and $(r^2 = x^2 + y^2,z)$.  
The arrow indicates the neutrino beam direction and 
the dashed line indicates the fiducial volume boundary.  
Events indicated 
by open square markers passed all of the \nue \ccpip selection cuts except for the fiducial 
volume cut.}
\label{fig:SK_nueCC1pi_vtx}
\end{center}
\end{figure*}

Fig.~\ref{fig:SK_nueCC1pi_vtx} 
shows the vertex distribution of the \nue \ccpip candidate events 
in the SK tank coordinate system.

\subsection{SK detector systematic uncertainties}
\label{sec:sksyst}

This section discusses the estimation of the uncertainty in the selection efficiency
and background for the oscillation samples that result from modeling of the SK detector.
This topic has been covered in detail in previous publications \cite{Abe:2015awa} but
there have been a number of updates, particularly related to the addition of the
\nue \ccpip sample.

Control samples unrelated to the T2K beam are used to assess the uncertainties.
%FC, FV, Michel electrons
Cosmic-ray muon samples are used to estimate uncertainties related to the
FC, fiducial-volume and decay-electron requirements, for the selections of both 
\nueoreb and \numormb CC candidates.
The error from the initial FC event selection is negligible.
%note: it used to be 1%, all coming from an event-by-event flasher cut that is no
%longer applied.
The uncertainty in the fiducial volume
is estimated to be 1\% using the vertex distribution of cosmic ray muons which 
have been independently determined to have stopped inside the ID.
The uncertainty due to the Michel electron tagging efficiency is estimated by comparing cosmic-ray stopped muon
data with MC.  The rate of falsely identified Michel electrons is estimated from MC and a 100\% uncertainty on that rate
is assumed.

\begin{table}[!h]
 \centering
 \caption{Criteria for categorization of simulated events by final state topology for systematic studies.
 $N_x$ is the number of particles of type $x$ and
 the number of charged pions ($N_{\pi^{\pm}}$) and protons ($N_{p}$)
 only includes those particles produced with momentum above Cherenkov threshold set at 156.0\,\mevc
 and 1051.0\,\mevc respectively.}
 \begin{tabular}{lp{5.5cm}}
   \hline
   \hline
   Event type & MC truth selection criteria \\
   \hline
   CC 1$e$  & $\nu_{e}$ CC and $N_{\pi^{0}}=0$ and $N_{\pi^{\pm}}=0$ and $N_{P}=0$ \\
   CC $e$ other            & $\nu_{e}$ CC and not $\nu_{e}$ CC1$e$ \\
   CC 1$\mu$    & $\nu_{\mu}$ CC and $N_{\pi^{0}}=0$  and $N_{\pi^{\pm}}=0$ and $N_{P}=0$\\
   CC $\mu$ other                & $\nu_{\mu}$ CC and $N_{\pi^{0}}=0$ \\
   CC $\mu$ $\pi^{0}$ other\ \ \ \ & $\nu_{\mu}$ CC and $N_{\pi^{0}}>0$ \\
   NC 1$\pi^{0}$                 & NC and not NC 1$\gamma$ and $N_{\pi^{0}}=1$ and $N_{\pi^{\pm}}=0$ and $N_{P}=0$\\
   NC $\pi^{0}$ other            & NC and not NC 1$\gamma$ and $N_{\pi^{0}}\geq1$ and not NC 1$\pi^{0}$\\
   NC 1$\gamma$                 & NC 1$\gamma$ in the MC generator\\
   NC 1$\pi^{\pm}$               & NC and not NC 1$\gamma$ and $N_{\pi^{0}}=0$ and $N_{\pi^{\pm}}=1$ and $N_{P}=0$\\
   NC other                     & NC and not NC 1$\gamma$ and not NC 1$\pi^{0}$ and not NC 1$\pi^{\pm}$
                                and not NC $\pi^{0}$ other \\
   \hline
   \hline
 \end{tabular}
 \label{tab:FSmodes}
\end{table}

Other studies of systematic uncertainty in SK modeling divide simulated events into
categories according to their final state topologies, with the criteria shown in Tab.~\ref{tab:FSmodes}.
These topologies do not correspond exactly with true interaction modes due to
subsequent interactions 
within the nucleus or with neighboring nuclei or because one or more particles 
are produced below Cherenkov threshold.
The dominant uncertainties are described in the following paragraphs.
Atmospheric neutrino data are used to assess possible
mismodeling of the ring counting (RC), particle identification (PID), and \piz rejection for the first four topologies shown in
Tab.~\ref{tab:FSmodes}.
In addition to flux and cross-section uncertainties from the atmospheric neutrino analysis, additional parameters are included
in the fit to alter the cut values applied to the MC for the three classifiers above, thereby allowing for possible mismodeling of the events.
Separate parameters are used for each of the four final state topologies and for different ranges of visible energy.  A likelihood is defined
comparing the data and MC which is then marginalized over flux and cross section parameters using a Markov chain Monte Carlo to estimate corrected efficiencies for the four final state topologies in bins of $E_{vis}$ and their covariance.  The measured shifts between the nominal and fitted efficiencies are included in the
final uncertainty assuming full correlation between samples.

To estimate the uncertainty in modeling $\pi^{0}$'s we construct a set of hybrid data-MC control samples.  Events in these samples are
created by overlaying a single electron-like ring from the SK data with a simulated photon ring.
The kinematics of the photon ring are chosen such that when combined, the two rings follow the decay kinematics of $\pi^{0}$ events from the T2K MC.  Hybrid samples with both rings from the SK MC are also produced for comparison with the hybrid data.
The difference in the selection efficiency when using the oscillation analysis sample candidate criteria is used to determine the systematic error.
This procedure is performed twice, with the data electron as the higher energy and lower energy ring, and the uncertainties are combined.

In previous analyses, the CC1$\mu$ events in the \nue appearance sample that did not involve
muon decay-in-flight (81\% of  CC1$\mu$) were assigned a conservative 150\% error due to the ring-counting, PID and
$\pi^0$ rejection cuts.  However while these events
form only $\sim$1\% of the \nue CCQE-like candidate sample, they are $\sim$10\% of the \nue \ccpip candidate sample
and thus the uncertainty was re-evaluated with a dedicated analysis using stopped cosmic-ray
muons and atmospheric neutrinos.  Based upon this work, this uncertainty has been reduced
to 63.2\%, where the error on the particle identification is the dominant source of uncertainty. For CC1$\mu$ events involving a muon decay-in-flight, a 16\% uncertainty is assigned, unchanged with respect to the previous analysis.

All aspects of the SK detector simulation that can affect the modeling of the candidate event selection 
described above are propagated using a vector of systematic uncertainty parameters, $\vec{s}$,
which scale the nominal expected number of events in bins of
the observable kinematic variables \erec or $p_{l}$ for 
the true neutrino interaction mode categories.  

\section{Neutrino oscillation framework}
\label{sec:oscprob}
%Describe oscillation probabilities in neutrino and antineutrino mode, justify the choice of taking data in RHC mode.  
%
%Effect of \dcp, sin(\dcp), cos(\dcp). 
%
%Strategy of the joint fit to treat correlation among oscillation parameters. 
%
%External inputs. 
%
%Expected length: 2.5 pages.
%
%Editor: Davide. 

%
% TODO:
% - replace fig ellipse_dcp with consistent param values
% - add another plot with oscillation probability ??? 
% 
The previous sections have described how the near detector samples are used to reduce the uncertainties on the neutrino fluxes before oscillation, and on the neutrino interaction model parameters.
%the analysis of the near detector data,
%performed to infer the unoscillated neutrino flux,
%has been shown.
The best-fit values of the neutrino and antineutrino flux and cross-section
parameters together with their correlations are extrapolated 
to SK, where the oscillation probabilities are measured.
In this section, the general model that describes the oscillations
between the three standard neutrino flavors,
$\nu_e$, $\nu_{\mu}$ and $\nu_{\tau}$,
are described, with a focus on the relevant oscillation channels for the T2K experiment. Finally, external measurements of neutrino oscillation parameters, which are used as prior constraints in some fits, are summarized.

\subsection{Oscillation probabilities in the PMNS framework}
\label{sec:pmnsoscprob}

As anticipated by Pontecorvo,
the neutrino flavor eigenstates do not correspond to mass eigenstates,
but are linear superpositions of them:  

\begin{align}
| \nu_{\alpha} \rangle  = \sum_k U_{\alpha k} | \nu_k \rangle \hspace{1cm} \left( \alpha = e, \mu, \tau \right)
\label{eq:flavstate}
\end{align}

\noindent where only the three active neutrinos, $k= 1,2,3$ are considered.
$U_{\alpha k}$ is an element of the $3 \times 3$ unitary matrix, 
called the Pontecorvo-Maki-Nakagawa-Sakata (PMNS) matrix~\cite{Maki:1962mu,Pontecorvo:1967fh},
which can be parametrized as:

\begin{widetext}
\begin{align}
&U = 
& %\hspace{-0.7cm}
\left( \begin{array}{ccc}  
 1  &0           &0 \\
 0  &c_{23}   &s_{23} \\
 0  &-s_{23}  &c_{23} 
 \end{array} 
 \right)
 \left( \begin{array}{ccc}  
 c_{13}  &0           &s_{13} e^{-i \delta_{CP}} \\
 0  & 1   & 0 \\
 -s_{13} e^{i \delta_{CP}}  &0  &c_{13} 
 \end{array} 
 \right)
\left( \begin{array}{ccc}  
 c_{12}  &s_{12}   &0            \\
 - s_{12}  & c_{12}   & 0 \\
 0 & 0 & 1 
 \end{array} 
 \right)
%\left( \begin{array}{ccc}  
%1 &0   &0            \\
%0 & e^{i \lambda_2}   & 0 \\
% 0 & 0 & e^{i \lambda_3}  
% \end{array} 
% \right) 
\nonumber \\
 %
% &=
% \left( \begin{array}{ccc}  
% c_{12} c_	{13} 		    								& s_{12} c_{13}           &s_{13} e^{- i \delta_{CP}} \\
% -s_{12} c_{23} - c_{12} s_{23} s_{13} e^{i \delta_{CP}}		 &c_{12} c_{23} - s_{12} s_{23} s_{13} e^{i\delta_{CP}}   &s_{23} c_{13} \\
% s_{12} s_{23} - c_{12}c_{23} s_{13} e^{i \delta_{CP}}		 & - c_{12} s_{23} - s_{12} c_{23} s_{13} e^{i \delta_{CP}} &c_{23} c_{13} 
% \end{array} 
% \right)
 \label{eq:pmns}
\end{align}
\end{widetext}

\noindent where $s_{ij} \equiv \sin \theta_{ij}$ and $c_{ij} \equiv \cos \theta_{ij}$, $\theta_{ij}$ are the three mixing angles
and $\delta_{CP}$ is the CP-violating phase.  
The Majorana phases are neglected here as the three-flavor oscillation probability is invariant under their rotation.

The neutrino oscillation probability is determined by six parameters:
the mixing angles $\theta_{12}$, $\theta_{13}$, $\theta_{23}$,
which define the amplitude %and the shape 
of the oscillation probability;
the differences in the squared masses of the eigenstates,
$\Delta m^2_{21} = m^2_2 - m^2_1$ and $\Delta m^2_{32} = m^2_3 - m^2_2$, 
that define the oscillation frequency and the position of the oscillation maxima as a function of $L/E$;
and the CP-violating phase, $\delta_{CP}$.
%contained by the imaginary part of the oscillation probability.
While it is known that 
%$m_2 > m_1$, 
%i.e. 
$\Delta m^2_{21}>0$, 
%is positive,
the neutrino mass ordering (MO) and hence the sign of $\Delta m^2_{32}$ has not yet been determined.

Effects due to charged-current coherent scattering of $\nu_e$ with the electrons in the Earth matter 
can affect the neutrino event rate.
The matter effect potential is proportional to 
$+N_e$ for $\nu_e$ 
and
$-N_e$ for $\bar{\nu}_e$,
where $N_e$ is the matter electron density. 
Neglecting matter effects, the \numormb survival probability can be written as
%Eq.~\ref{eq:posc_nu_nubar} can be approximated 
%and the $\nu_{\mu}$ ($\bar{\nu}_{\mu}$) survival probability can be written as:
%
\begin{widetext}
\begin{align}
& \hspace{-0.5cm} P(\;\nu^{\bracketbar}_{\mu} \rightarrow \nu^{\bracketbar}_{\mu}) = 1  - 4 \left( s^2_{12} c^2_{23} + s^2_{13} s^2_{23} c^2_{12} + 2 s_{12} s_{13} s_{23} c_{12} c_{23} \cos \delta_{CP} \right) s^2_{23} c^2_{13} \sin^2 \phi_{31} \nonumber \\
& \hspace{-0.5cm} - 4 \left( c^2_{12} c^2_{23} + s^2_{13} s^2_{23} s^2_{12} - 2 s_{12} s_{13} s_{23} c_{12} c_{23} \cos \delta_{CP} \right) s^2_{23} c^2_{13} \sin^2 \phi_{32}  - 4 \left( s^2_{12} c^2_{23} + s^2_{13} s^2_{23} c^2_{12} + 2 s_{12} s_{13} s_{23} c_{12} c_{23} \cos \delta_{CP} \right) \nonumber \\
& \hspace{-0.5cm} \times \left( c^2_{12} c^2_{23} + s^2_{13} s^2_{23} s^2_{12} - 2 s_{12} s_{13} s_{23} c_{12} c_{23} \cos \delta_{CP} \right) \sin^2 \phi_{21},
\label{eq:psurv_numu}
\end{align} 
\end{widetext}

\noindent where 
\begin{align}
\phi_{kj} = \frac{ \Delta m^2_{kj} L}{4E} .
\end{align}

$|\Delta m^2_{32}|$ and $\sin^2 2 \theta_{23}$ can be determined from Eq.~\ref{eq:psurv_numu},
but the $\theta_{23}$ octant, 
i.e., whether $\theta_{23} > \pi/4$ or $\theta_{23} < \pi/4$, cannot be distinguished.
At the T2K baseline and peak neutrino energy, the impact of the matter effect is small, and a full calculation including them would only modify the $P(\nu_{\mu} \rightarrow \nu_{\mu})$ and $P(\bar{\nu}_{\mu} \rightarrow \bar{\nu}_{\mu})$ probabilities by $\sim$0.1\%.

The $\nu^{\bracketbar}_{e}$ appearance probability, approximated to first order in the matter effect (see~\cite{Arafune:1997hd}), is
\begin{widetext}

\begin{align}
& \hspace{-0.3cm} P(\;\nu^{\bracketbar}_{\mu} \rightarrow \nu^{\bracketbar}_{e}) = 4c^2_{13} s^2_{13} s^2_{23} \sin^2 \phi_{31} \left( 1 + \frac{2\alpha}{\Delta m^2_{31}} \left( 1 - 2 s^2_{13} \right)  \right)  + 8c^2_{13} s_{12} s_{13} s_{23} \left( c_{12} c_{23} \cos \delta_{CP} - s_{12} s_{13} s_{23} \right) \cos \phi_{23} \sin \phi_{31} \sin \phi_{21} \nonumber \\
& \hspace{-0.3cm} \bplus{-} 8c^2_{13} c_{12} c_{23} s_{12} s_{13} s_{23} \sin \delta_{CP} \sin \phi_{32} \sin \phi_{31} \sin \phi_{21} \ + 4s^2_{12} c^2_{13} \left( c^2_{12} c^2_{23} + s^2_{12} s^2_{23} s^2_{13} - 2c_{12} c_{23} s_{12} s_{23} s_{13} \cos \delta_{CP} \right) \sin^2 \phi_{21} \nonumber \\ 
& \hspace{-0.3cm} - 8c^2_{13} s^2_{13} s^2_{23} \left( 1 - 2s^2_{13} \right) \frac{ {\bminus{+}} \alpha L}{4 E} \cos \phi_{32} \sin \phi_{31}, \label{eq:psurv_nue}
\end{align}

%\begin{align}
%& \hspace{-0.3cm} P(\bbar{\nu}_{\mu} \rightarrow \bbar{\nu}_{e}) = \nonumber \\
%& \hspace{-0.3cm} 4c^2_{13} s^2_{13} s^2_{23} \sin^2 \phi_{31} \left( 1 + \frac{2\alpha}{\Delta m^2_{31}} \left( 1 - 2 s^2_{13} \right)  \right)  \nonumber \\
%& \hspace{-0.3cm} + 8c^2_{13} s_{12} s_{13} s_{23} \left( c_{12} c_{23} \cos \delta_{CP} - s_{12} s_{13} s_{23} \right) \cos \phi_{23} \sin \phi_{31} \sin \phi_{21} \nonumber \\
%& \hspace{-0.3cm} \bplus{-} 8c^2_{13} c_{12} c_{23} s_{12} s_{13} s_{23} \sin \delta_{CP} \sin \phi_{32} \sin \phi_{31} \sin \phi_{21} \nonumber \\
%& \hspace{-0.3cm} + 4s^2_{12} c^2_{13} \left( c^2_{12} c^2_{23} + s^2_{12} s^2_{23} s^2_{13} - 2c_{12} c_{23} s_{12} s_{23} s_{13} \cos \delta_{CP} \right) \sin^2 \phi_{21} \nonumber \\ 
%& \hspace{-0.3cm} - 8c^2_{13} s^2_{13} s^2_{23} \left( 1 - 2s^2_{13} \right) \frac{ {\bminus{+}} \alpha L}{4 E} \cos \phi_{32} \sin \phi_{31} \label{eq:psurv_nue}
%\end{align} 

\end{widetext}
where $``(-)$'' corresponds to the change of sign required for antineutrino oscillations.
$\rho$ is the average density of the Earth matter through which the neutrinos travel
and $\alpha \left[ \text{eV}^2 / \text{c}^4 \right] = 7.56 \times 10^{-5} \rho \left[ \text{g}/\text{cm}^3 \right] E_{\nu} \left[ \text{GeV} \right]$.
In all oscillation analyses described here, the exact three-flavor oscillation probabilities are used, rather than the approximations shown in this section.

As well as being proportional to $\sin^2 \theta_{13}$, the first, leading, term of Eq.~\ref{eq:psurv_nue} is proportional to $\sin^2 \theta_{23}$, which makes the appearance probability (in contrast to the disappearance probability) sensitive to the octant of $\theta_{23}$. The precise determination of $\theta_{23}$ and the solution of the octant degeneracy are important for the determination of \dcp. Because of the precise measurement of $\theta_{13}$ from reactor experiments (see Tab.~\ref{tab:pdg}), the largest uncertainty in the oscillation model which affects the determination of \dcp is due to the uncertainties on the value of $\theta_{23}$.

By looking at the other terms in Eq.~\ref{eq:psurv_nue}, it is easy to understand why the $\nu^{\bracketbar}_e$ appearance transition
%, discovered by T2K with a significance of more than 7 standard deviations
%to a nonzero $\theta_{13}$ 
is the golden channel in the search for CP violation.
The term containing $\sin \delta_{CP}$ in Eq.~\ref{eq:psurv_nue}
%, due to the imaginary part of Eq.~\ref{eq:posc_nu_nubar}
has an opposite sign for neutrinos and antineutrinos,
and is the only term that can violate CP symmetry.
Since all the mixing angles have been measured as nonzero,
the requirement for violating CP symmetry in neutrino oscillations is
$\sin \delta_{CP} \neq 0$, 
i.e. $\delta_{CP} \neq 0, \pi$.

In T2K, the term proportional to $\sin \delta_{CP}$ can change the appearance probability by as much as $\pm30\%$. This means that an extreme value of \dcp such as $\dcp=-\pipi/2$ would increase (decrease) the \nue (\nueb) by 30\% with respect to $\dcp = 0$.
In Eq.~\ref{eq:psurv_nue} there are also other terms containing $\cos \delta_{CP}$. While these terms do not violate CP symmetry, they change the shape of the appearance spectrum and are important for a precise determination of the value of \dcp.

%However these terms cannot violate the CP symmetry, though they may be important
%for a more precise measurement of $\delta_{CP}$. 

%Furthermore, thanks to a relatively small value of $\theta_{13}$,
%the $\sin \delta_{CP}$ term has a non negligible effect on the T2K appearance probabilities.
%In Eq.~\ref{eq:psurv_nue} there are also other terms containing $\delta_{CP}$ as argument of a cosine function.
%However these terms cannot violate the CP symmetry, though they may be important
%for a more precise measurement of $\delta_{CP}$.

%Furthermore, thanks relatively small value of $\theta_{13}$,
%the $\sin \delta_{CP}$ term has a non negligible effect on the appearance probabilities.

%In Eq.~\ref{eq:psurv_nue} there are also terms containing $\delta_{CP}$ as argument of a $\cos$ function.
%However these terms do not violate the CP symmetry, though they may be important
%for a more precise determination of \dcp.

In summary, at first order, the term in $\sin \delta_{CP}$ defines the normalization of the 
$\nu^{\bracketbar}_{\mu} \rightarrow \nu^{\bracketbar}_e$ spectrum, 
while the term in $\cos \delta_{CP}$ provides information on the shape
and allows the discrimination of the CP-conserving phases $\delta_{CP} = 0$ and $\pi$.

As mentioned above the appearance probability is also affected by the matter effects. 
Matter effects can enhance either 
${\nu}_{\mu} \rightarrow {\nu}_{e}$,
or 
$\bar{\nu}_{\mu} \rightarrow \bar{\nu}_{e}$ 
depending on the sign of $\Delta m^2_{32}$.
For instance if the mass ordering is normal (inverted)
${\nu}_{\mu} \rightarrow {\nu}_{e}$ oscillations are enhanced (suppressed)
while $\bar{\nu}_{\mu} \rightarrow \bar{\nu}_{e}$ oscillations are suppressed (enhanced).
%$\nu^{\bracketbar}_{\mu}$ survival probability is not sensitive to the MO 
%because the matter effects are small.

\begin{figure}[htbp]
 \includegraphics[width=0.5\textwidth]{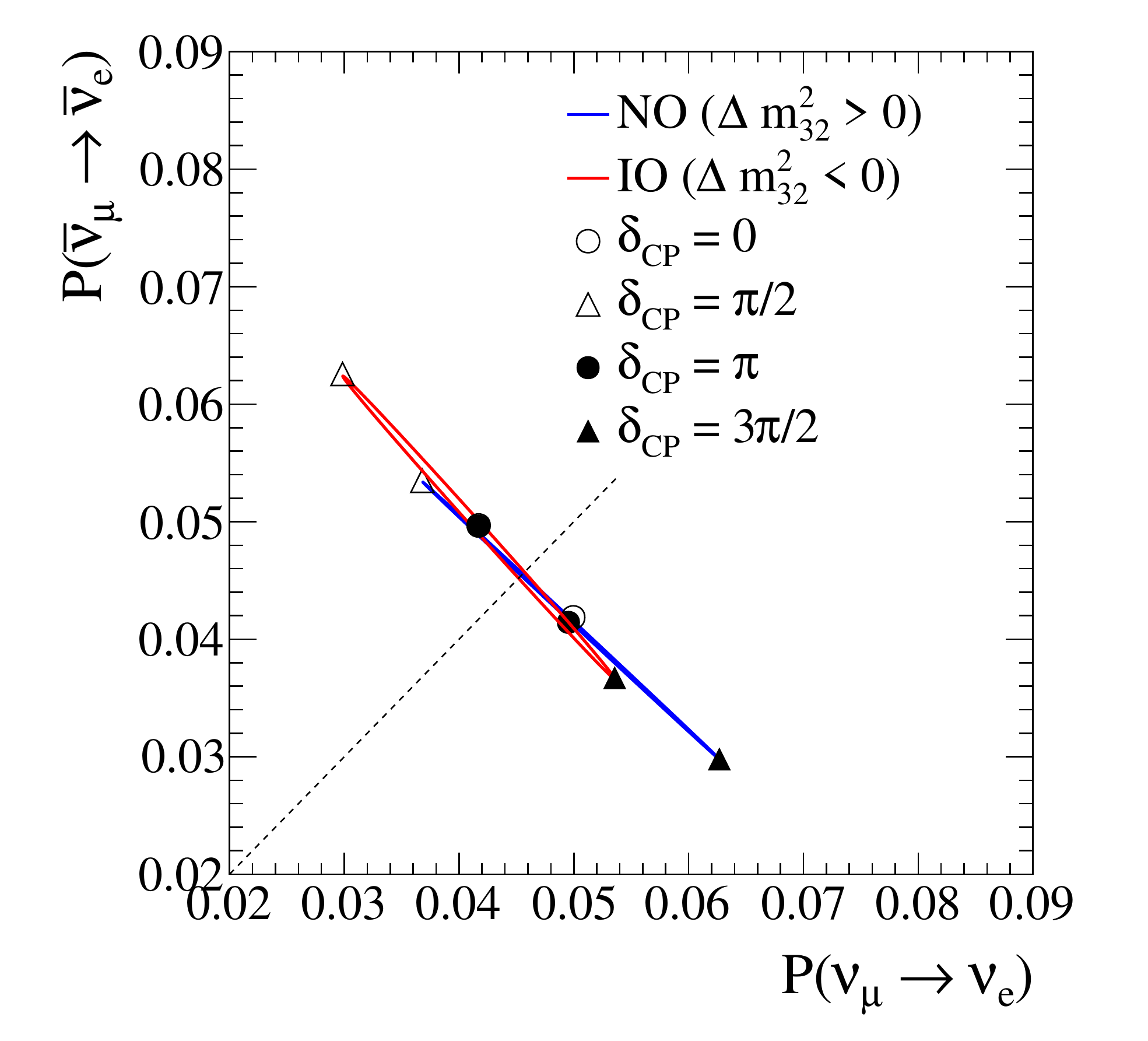}
 \caption{The oscillation probability for ${\nu}_{\mu} \rightarrow {\nu}_{e}$ and $\bar{\nu}_{\mu} \rightarrow \bar{\nu}_{e}$ 
 transitions as a function of $\delta_{CP}$ and MO 
 for $E_\nu = 0.6$~GeV and a baseline of 295~km.
The other oscillation parameters are fixed to
the values shown in Tab.~\ref{tab:nomosc_par}. The dashed line indicates $P({\nu}_{\mu} \rightarrow {\nu}_{e})=P(\bar{\nu}_{\mu} \rightarrow \bar{\nu}_{e})$.  }
\label{fig:oscprob_ellipse}
\end{figure}

With its low neutrino energy and baseline of 295 km, the T2K experiment is not sensitive to the matter effect. The two possible solutions for the mass ordering (MO) change the appearance probability by approximately $\pm10\%$ and a definitive measurement of the MO cannot be made by exploiting the matter effects in T2K beam. The interplay between effects due to \dcp and effects due to the MO on the \nueoreb-appearance probabilities are shown in Fig.~\ref{fig:oscprob_ellipse}. Each ellipse shows the effect due to \dcp on the two appearance probabilities and the two ellipses represent the two possible solution for the mass ordering. 
The two regions overlap for more than half of the \dcp range, indicating the large degeneracy among the MO and \dcp at the T2K baseline.

%where the two ellipses,
%that show the relation between $\nu_e$ and $\bar{\nu}_e$ appearance probabilities
%for NO and IO,
%The fact that the two ellipses overlaps for a large combination of the values of \dcp indicates that, except for some lucky combinations of the values of \dcp and MO,  set of are overlapped for more than half of the $\delta_{CP}$ range.

%In order to measure the MO exploiting the matter effects a very long baseline (at least $O(1000 \text{~km})$) 
%would be ideal. 
%Consequently T2K, with a baseline of only 295 km, cannot provide a precise measurement of the MO.
%Nevertheless the oscillations in matter are neither CP- nor CPT- invariant,
%so the MO gives effects similar to $\delta_{CP}$.
%This is clearly shown in Fig.~\ref{fig:oscprob_ellipse},
%where the two ellipses,
%that show the relation between $\nu_e$ and $\bar{\nu}_e$ appearance probabilities
%for NO and IO,
%are overlapped for more than half of the $\delta_{CP}$ range.
%Solving the degeneracy between the two parameters becomes 
%very important for the measurement of $\delta_{CP}$.

%Since T2K has good sensitivity to the absolute value of $\Delta m^2_{32}$ 
%but not to its sign, we will consider $|\Delta m^2_{32}|$
%and the MO as two separate parameters.
%T2K  is not sensitive to
%the solar neutrino oscillation parameters, 
%$\Delta m^2_{21}$ and $\sin^2 \theta_{12}$,
%which have a completely negligible impact on the oscillation probability.

% Goal of T2K
The goal of T2K is the precise measurement of
$\theta_{13}$, 
$\theta_{23}$, 
$|\Delta m^2_{32}|$ 
and to search for evidence of CP violation.
Since T2K has excellent sensitivity to the absolute value of $\Delta m^2_{32}$ 
but not to its sign,
the absolute value of $\Delta m^2_{32}$
and the MO are considered as two separate parameters.
As mentioned above, the violation of CP symmetry would manifest itself as a
difference between neutrino and antineutrino oscillation probabilities.
In order to find a hint of this violation it is necessary to directly compare 
${\nu}_{\mu} \rightarrow {\nu}_{e}$ and
$\bar{\nu}_{\mu} \rightarrow \bar{\nu}_{e}$
oscillation probabilities, 
as also shown in Fig.~\ref{fig:oscprob_ellipse},
where the anti-correlation as a function 
of both $\delta_{CP}$ and MO is visible.
The measurement of the 
$\nu^{\bracketbar}_{\mu}$ survival probabilities is 
important for determining 
$|\Delta m^2_{32}|$, 
i.e. the position of the first oscillation maximum,
and $\sin^2 2 \theta_{23}$,
drastically reducing the uncertainty on $\delta_{CP}$.
The degeneracy between $\delta_{CP}$ and the $\theta_{23}$ octant
can be solved by comparing $\nu_e$ and $\bar{\nu}_e$ appearance probabilities.
Indeed, in contrast to $\delta_{CP}$, 
$\sin^2 \theta_{23}$ 
has the same effect on both the 
${\nu}_{\mu} \rightarrow {\nu}_{e}$ and the
$\bar{\nu}_{\mu} \rightarrow \bar{\nu}_{e}$
oscillation probabilities.
For this reason, T2K measures 
the $\nu_{\mu} \rightarrow \nu_{\mu}$ ($\bar{\nu}_{\mu} \rightarrow \bar{\nu}_{\mu}$) disappearance
and the 
$\nu_{\mu} \rightarrow \nu_e$ ($\bar{\nu}_{\mu} \rightarrow \bar{\nu}_e$) appearance 
probabilities in
\nunu- and \nub-mode simultaneously in the analyses described in this paper.
%For this reason, a joint analysis of 
%$\nu_{\mu}$, $\bar{\nu}_{\mu}$, $\nu_e$ and $\bar{\nu}_e$
%candidate samples is extremely important and has been pursued in the analyses described in this paper.
The joint analysis also allows all of the correlations between the oscillation parameters to be properly taken into account.

% Write oscillation probability for both nu and nubar (DONE)
% Describe matter effects and their effect on MH (DONE)
% Majorana phases have no effect because (DONE) 
% Neutrino oscillations --> m!=0 (DONE)

% Describe the role of deltaCP: 
% - sinsqth13 not zero, but not large (constrained from reactors) --> sensitive to deltaCP (DONE)
% - why antineutrino data taking is important (DONE)
% - deltaCP parameter, sin_deltaCP and cos_deltaCP (second maximum) 

% Describe correlations between the oscillation parameters:
% - impact of MH and degeneracy with deltaCP (DONE)
% - impact of sinsqth23 (DONE)
% - theta23 octant (DONE)
% - importance of theta23 (and octant) for deltaCP (DONE)
% - add sentence about sensi to deltaCP as a function of sinsqth23 (DONE)

% plot of effect of oscillation probability for different values of deltaCP and MH (NOT NEEDED)
% ellipse plot with oscillation probability (DONE)

\subsection{External constraints on the oscillation parameters}
\label{sec:pmnspriors}
For some of the oscillation parameters, Gaussian priors, based on external experiments, are used in the oscillation analyses. These parameters are the solar oscillation parameters, to which T2K is insensitive, and, in some cases, the value of \sto as measured by reactor experiments. In the latter case, the weighted average of Daya Bay, RENO and Double Chooz as reported in Ref.~\cite{PDG2015} is used. The Gaussian priors used in the analysis are shown in Tab.~\ref{tab:pdg}.
For all the results shown in this paper, it will be specified whether reactor constraints on \sto are used.

All the parameters that are measured by T2K are included in the oscillation analyses with flat priors which extend beyond the allowed regions for those parameters. The boundaries on the flat priors used in the analysis are $0.3<\stt<0.7$, 2$<$\dmsq($\times10^{-3}~eV^2/c^4$)$<$3, $-\pipi<\dcp<\pipi$, and $0<\stot<0.4$ (when no reactor measurements are used).

%T2K is insensitive to the solar oscillation parameters 

% External inputs:
% - table of PDG
% - short description of reactor measurements of theta13

%Some of the neutrino oscillation parameters, 
%for instance the solar parameters and $\sin^2 \theta_{13}$
%can be constrained by using external measurements.
%The gaussian priors used for the oscillation parameters are shown in tab.~\ref{tab:pdg}. 

\begin{table}
  \centering      
\caption{Gaussian priors used for the neutrino oscillation parameters, taken from Ref.~\cite{PDG2015}.}\label{tab:pdg}
      {\renewcommand{\arraystretch}{1.2}
        \begin{tabular}{>{\centering}p{2.5cm}|p{3cm}<{\centering}}
          \hline \hline
          Oscillation parameter & Mean and standard deviation\\
	         
          \hline
          $\sin^2 2 \theta_{12}$ & $0.846 \pm 0.021$   \\
          $\Delta m^2_{21}$   & $(7.53 \pm 0.18) \times 10^{-5}~\text{eV}^2/\text{c}^4$   \\
          $\sin^2 2 \theta_{13}$  & $0.085 \pm 0.005$   \\
          \hline \hline
      \end{tabular}}
\end{table}

\section{Oscillation analysis method}
\label{sec:oamethod}

In this section, the oscillation analyses which have been developed to 
%which use a combination of T2K and external data to 
provide oscillation parameter estimations are described: three different analyses have been developed within frequentist and Bayesian frameworks. 

For all analyses, the appearance and disappearance channels in both $\nu$- and $\nub$-modes are analysed simultaneously, incorporating constraints from several T2K samples and external experiments. The oscillation analyses compare the rate and distribution of events in analysis bins defined by combinations of the reconstructed neutrino energy, \erec (see Equation~\ref{eq:SK_Erec}), the reconstructed lepton momentum, and the reconstructed angle of the outgoing lepton with model predictions. %The far detector selection procedure and uncertainties are described in Sec.~\ref{sec:sk} and the model prediction in Sec.~\ref{sec:niwg} and \ref{sec:oscprob}.
Systematic uncertainties are characterized by systematic model parameters, which tune the model prediction, and by a covariance matrix, which determines the sizes and correlations of their uncertainties. 
The fit to the near detector data is applied as a multivariate Gaussian penalty term to constrain flux and cross-section uncertainties common to the near and far detectors.
%A full description of the systematic uncertainties is given in Sec.~\ref{sec:oasyst}.

For a given set of systematic and oscillation parameters, the expected distribution is computed in the following way. The expected SK MC samples are produced assuming nominal systematic parameters without oscillation. The effect of oscillations are then included by applying the neutrino oscillation
probability %multiplicatively 
to events defined by neutrino energy and interacting flavor. 
In order to do this all the analyses use the full three-flavor oscillation probabilities.
%The full three-flavor oscillation probabilities is used in all the analyses, not the approximated ones shown in Sec.~\ref{sec:oscprob},
The number of events is weighted by the oscillation probability and for the systematic parameters, as will be described in Sec.~\ref{sec:oasyst}.
%After reweighting the number of events for the oscillation probability, 
%also the event reweighting coming from the systematic parameters is applied, as it will be described in Sec.~\ref{sec:oasyst}.
% This is followed by the application of systematic parameters as described in Sec.~\ref{sec:oasyst}.
Then the expected number of events in each bin of the far detector samples are extracted from a Poisson distribution, leading to a likelihood of the form

%\begin{equation} 
%\label{eq:sk_likelihood}
%%\displaystyle
%%\chi^{2} 
%-2 \; \textrm{ln} \; \lambda (\vec{\theta},{\bf a}) = 2 \cdot \sum_{i=0}^{N-1} 
%   \Big( 
%     n^{obs}_{i} \cdot \textrm{ln} (n^{obs}_{i}/n^{exp}_{i}) +
%     (n^{exp}_{i}-n^{obs}_{i}) 
%   \Big) 
%   + ({\bf a}-{\bf a_{0}})^{T} \cdot {\bf C}^{-1} \cdot ({\bf a}-{\bf a_{0}})
%\end{equation}where ${\bf a}$ is a vector which defines the systematic parameters, ${\bf a_{0}}$ is the vector of the systematic parameters at their nominal values, ${\bf C}$ is the covariance matrix, and $n^{obs}_{i}$ and $n^{exp}_{i}$ are the number of observed and expected events in the $ith$ analysis bin respectively. 
%$\vec{\theta}$ is the vector of oscillation model parameters described in \ref{sec:oscprob}. 
%Finally, the sum runs over all N bins in the far detector samples. 

\begin{equation}  \label{eq:sklikelihood}
%\displaystyle
%\chi^{2} 
-2 \; \textrm{ln} \; \mathcal{L} (\vec{o},\vec{f}) = 2\sum_{i=0}^{N-1} 
   \Big[ 
     n^{obs}_{i} \cdot \textrm{ln} (n^{obs}_{i}/n^{exp}_{i}) +
     (n^{exp}_{i}-n^{obs}_{i}) 
   \Big],
   %+ ({\bf a}-{\bf a_{0}})^{T} \cdot {\bf C}^{-1} \cdot ({\bf a}-{\bf a_{0}})
\end{equation}

\noindent where 
$\vec{o}$ and $\vec{f}$ are particular choices of the vector 
of  
the neutrino oscillation parameters (free in the fit),
and of the nuisance parameters.
$n^{obs}_i$ is the observed number of events in the $i$th analysis template bin,
and
$n^{exp}_i = n^{exp}_i (\vec{o},\vec{f})$
is the expected number of events in the $i$th bin.

The method described in Ref.~\cite{Cousins:1991qz} is used to project out the nuisance parameters.
The high dimensionality of the likelihood function is solved by computing the marginal likelihood. This marginal likelihood for the parameters of interest is found by integrating the product of the likelihood function and the priors over all the parameters that are not of interest, and defined as:

\begin{align}
\mathcal{L}_{marg} (\vec{o} ;x) &= \int_{F} \mathcal{L} (\vec{o},\vec{f} ; x) \pi{(\vec{f})} d \vec{f}  \nonumber \\
	&= \frac{1}{n} \sum_{i=1}^{n} \mathcal{L} (\vec{o}, \vec{f}_i ;x).
\label{eq:likemarg}
\end{align}

%
% NOT SURE WHETHER "data" SHOULD GO BEFORE OR AFTER \vec{o}
% IT SHOULD BE 
%

This defines the likelihood which is a function only of the parameters of interest $\vec{o}$,
 given a dataset, $x$, 
and a model prediction defined by a set of systematic and oscillation parameters. 
%A major modification, from previous T2K results, is the treatment of systematic uncertainties. 
The marginal likelihood of Eq.~\ref{eq:likemarg} can be used to perform both a hybrid frequentist-Bayesian analysis 
as well as a full Bayesian analysis.
In both cases the prior p.d.f. for flux, cross section and SK detector parameters
is a multivariate Gaussian,
with mean $\mu$ and covariance $V$.
All the analyses have been validated with simulated datasets using different neutrino interaction models and different sets of oscillation parameters for the generation of the simulated samples. Examples of the validations will be shown in Sect.~\ref{sec:fds}.
%The mean values and covariance
%of the beam flux and neutrino cross-section parameters 
%were obtained from the fit with the near detector data,
%as described in Sec.~\ref{sec:banff}.
%The ones corresponding to the Super-K detector systematic parameters
%are described in Sec.~\ref{sec:sksyst}.

%Previously, when there was a need to move to a likelihood defined by x parameters from one defined by N parameters, N-x parameters were often profiled. In that case, the new x dimensional likelihood was maximised with respect to the N-x eliminated parameters at each point in the space. This method was frequently used in the past to move from a likelihood which depended on both systematic and oscillation parameters to one that only depended on oscillation parameters. The systematic parameters were "profiled out" of the likelihood.
%
%In all analyses presented here systematic parameters have been marginalised. In this case to move from an N dimensional likelihood to an N-x dimensional likelihood, the N dimensional likelihood is integrated with respect to the N-x parameters to produce the marginal likelihood, a function of only x parameters. 

%Define likelihood used in the OA.
%
%Describe the Frequentist analysis and the main changes with respect to previous publications.
%
%Describe the Bayesian analysis and the main changes with respect to previous publications.
%
%We can refer to the joint paper published in 2015.
%
%Effect of the systematic uncertainties.

\subsection{Frequentist analyses}
\label{sec:oafrequentist}
%Define likelihood used and method to treat systematic uncertainties and other oscillation parameters.
%Marginalization of systematics.
%2D informations for the e-like samples (2 analyses p-theta or Erec-theta). \\
%Expected length: 1.5 pages.
%Editors: Davide, and Raj. \\

Two analyses have been developed in the frequentist framework. The main difference between the two analysis is that one uses 2D templates in \{$\erec$, $\theta_{lep}$\} for the e-like samples and the other uses  \{$p_{lep}$, $\theta_{lep}$\} templates. The reasons for these choices will be described in Sec.~\ref{sec:oatemplates}.
Both analyses construct confidence intervals or regions 
by fitting the SK dataset,
using the likelihood ratio
$-2 \Delta \ln \mathcal{L}(\vec{o}) = -2 \ln [ \mathcal{L}(\vec{o};x) / \mathcal{L}_{max}(x) ]$
as the test statistic, 
where $\mathcal{L}(\vec{o};x)$ is as defined in Eq.~\ref{eq:likemarg}.
If the Gaussian approximation holds, Eq.~\ref{eq:likemarg} converges to a $\chi^2$ distribution,
and the negative log-likelihood ratio is sometimes denoted as $\Delta \chi^2 (\vec{o})$.
%The flux, systematic and detector parameters are marginalized as described above.

Due to the difficulties in providing confidence intervals with high dimensionality, a subset of the oscillation parameters is projected out by using the marginalization method 
described above, with the priors shown in Tab.~\ref{tab:pdg}. 

Two approaches can be used in order to calculate the confidence intervals.
%In the Gaussian approximation, i.e., where the parameter dependence is linear, the Neyman construction~\cite{PDG2015}
%provides the correct coverage:
%the confidence region consists of the set of oscillation parameter values
%that satisfies the condition 
%$-2 \Delta \ln \mathcal{L}(\vec{o}) \leq X_{crit}(\vec{o})$.
%$\Delta \chi^2 (\vec{o}) \geq X_{crit}(\vec{o})$.
%The precomputed values of $X_{crit} (\vec{o})$, that define the $\alpha \%$ confidence level for different dimensionality, can be found in Ref.~\cite{PDG2015}.
%This method is sometimes called as the ``constant $\Delta \chi^2$ method'',
%because $X_{crit} (\vec{o})$ does not depend on the values of $\vec{o}$.
In the Gaussian approximation, where the parameter dependence is linear, the so-called ``constant $\Delta \chi^2$ method'' provides the correct coverage: the confidence region consists of the set of oscillation parameter values that satisfies the condition  $-2 \Delta \ln \mathcal{L}(\vec{o}) \leq X_{crit}(\vec{o})$. The precomputed values of $X_{crit} (\vec{o})$~\cite{PDG2015}, that define the $\alpha \%$ confidence level for different dimensionality, do not depend on the value of $\vec{o}$.

Since in the neutrino oscillation analysis the gaussian approximation condition does not always hold, a toy MC method,
recommended by Feldman and Cousins \cite{feldmancousins}, 
should be used to define the values of $X_{crit}(\vec{o})$ 
as a function of $\vec{o}$, in order to provide the correct coverage. An ensemble of %at least 
10,000 toy datasets was produced for each point of the oscillation parameter grid,
fine enough to describe the variation of $X_{crit}(\vec{o})$ as a function of the free parameters.
The flux, cross-section and detector systematic parameters were sampled from the same multivariate Gaussian with mean $\mu$ and covariance $V$
used as prior p.d.f. in the marginalization method, as described above.
The oscillation parameters constrained by external measurements, such as $\sin^2 \theta_{12}$, $\Delta m^2_{21}$ and $\sin^2 \theta_{13}$,
were sampled by using the p.d.f.'s with parameters given in Tab.~\ref{tab:pdg}.
The remaining oscillation parameters 
%treated as nuisance but not measured by any other experiment 
were sampled using the distribution of $-2 \Delta \ln \mathcal{L}(\vec{o})$ 
obtained from the fit of the T2K data.
This method will be referred as ``posterior-predictive".

%Each toy dataset had a set of values of the systematic parameters sampled from a multi-dimensional Gaussian having means at the nominal values, and covariances V. Each oscillation parameter is sampled from a Gaussian with mean and sigma values listed in Sec. VIIB. 
%The values of  CP are sampled uniformly. 
%The systematic parameters and these additional oscillation parameters are removed from the likelihood function by profiling. 
%In order to calculate an interval of just one oscillation parameter, we determine the critical values by marginalizing over the second oscillation parameter. 
%The marginalization assumes that the probability is proportional to the likelihood using T2K data.

Since the Feldman-Cousins method is computationally intensive, it is used only for the most important result of the oscillation analysis,
i.e. the confidence intervals as a function of $\delta_{CP}$ and MO.
For all the other results the constant $\Delta\chi^2$ method, which is more practical and provides good coverage to a first order approximation, is used.

%%%%%%%%%%%%%%%%%%%%

\subsection{Bayesian analysis}
\label{sec:oabayesian}

The Bayesian analysis performs the integration of Eq.~\ref{eq:likemarg} using a Markov Chain Monte Carlo (MCMC) method.
The Metropolis-Hastings algorithm~\cite{Hastings:1970a} is used to populate the space of the oscillation and nuisance parameters, 
distributed according to the posterior probability density function.

This algorithm uses a weighted random walk to explore the parameter space with a chain of many points, $X_i$. A new proposed step to $X_{i+1}$ is accepted with a probability equal to the minimum between 1 and $P(X_{i+1})/P(X_i)$, i.e., it is accepted with 100\% probability when $P(X_{i+1})>P(X_i)$ and with a probability equal to $P(X_{i+1})/P(X_i)$ when $P(X_{i+1}) < P(X_i)$. 
When a step to $X_{i+1}$ is rejected, the point $X_i$ is repeated in the chain, and a a new step is proposed again from the previous step $X_i$. Typically, a chain consists of $10^6 - 10^7$ steps. The population of accepted points in parameter space thus constructed is distributed according to the posterior probability density.

%This algorithm uses a weighted random walk to explore the parameter space with a chain of many steps, $X_i$. A new proposed step $X_{i+1}$ is accepted with a probability equal to $P(X_{i+1}) / P(X_{i})$, i.e., it is accepted with 100\% probability when $P(X_{i+1}) \geq P(X_i)$ and with a probability equal to $P(X_{i+1})/P(X_i)$ when $P(X_{i+1})<P(X_i)$. When a step $X_{i+1}$ is rejected a new step $X_{i+1}'$ is proposed again from the previous step $X_i$.  Typically, a chain consists of $10^6 - 10^7$ steps. The population of accepted step points in the parameter space is proportional to the posterior probability density. 

Credible intervals are built by selecting the highest density region in the space of the parameters of interest,
typically one or two, that contains $\alpha \%$ of the points. A kernel density estimator (KDE)~\cite{Mills:KDE,Cranmer:KDE},
is used to extract the most probable values of the 
%four 
oscillation parameters of interest, 
%for instance $\sin^2 \theta_{23}$, $\Delta m_{32}^2$, $\sin^2 \theta_{13}$ and $\delta_{CP}$.
A KDE PDF is built by smearing the step points with a Gaussian function which has a width inversely proportional to the number of points.
The most probable values of the parameters of interest were found by maximizing the 
KDE PDF using MINUIT~\cite{minuit}.

\subsection{Analysis templates}
\label{sec:oatemplates}

The three analyses use different far detector templates.
% and different statistical approaches. 
All of them use the reconstructed neutrino energy, \erec, for the $\nu_{\mu}$ and $\bar{\nu}_{\mu}$ candidate samples,
while different templates are used for the $\nu_{e}$ and $\bar{\nu}_{e}$ candidate samples.
One of the two frequentist analyses and the MCMC Bayesian analysis 
use \erec and the angle between the lepton and the neutrino beam direction, $\theta_{lep}$.
The choice of these two parameters is due to the fact that \erec is important to infer the oscillation probability function, 
which depends on the true neutrino energy.
$\theta_{lep}$ is used because it provides additional separation between $\nu$ and $\bar{\nu}$ candidate events.
The $\nu$ and $\bar{\nu}$ cross sections differ as a function of \qq, and leptons produced in $\nu$ interactions tend to be emitted at larger angles than those produced in \nub interactions.
%not only in the normalization, % OR 3???
%but also cover different regions of the phase space.
%The shape informations provided by the different kinematics of the outgoing lepton
%is used in the oscillation analysis in order to better constrain the wrong-sign flux background 
%as well as the $\nu$ and $\bar{\nu}$ cross section. 

\begin{figure}[htbp]
\begin{center}
\includegraphics[width=0.45\textwidth]{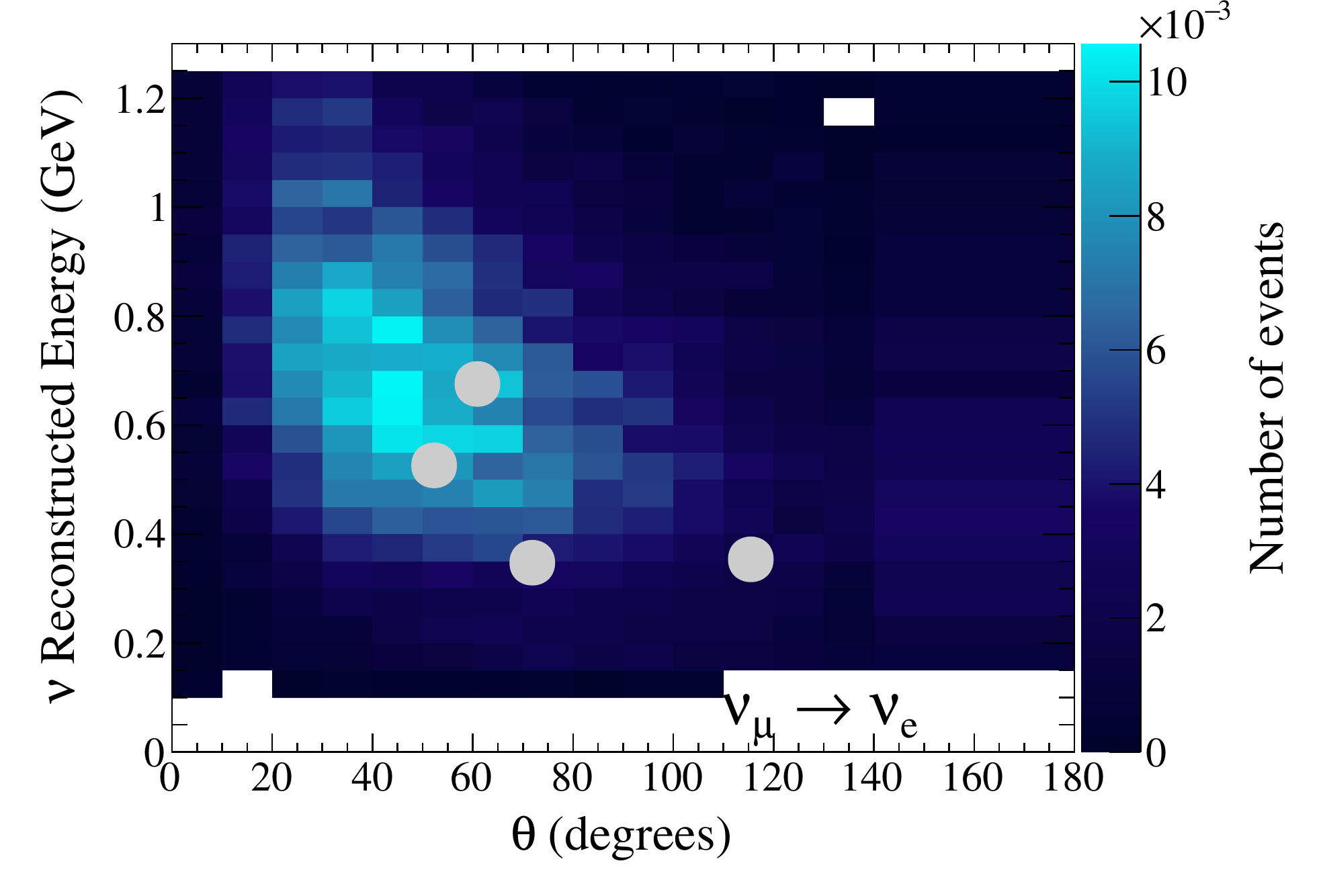}
\includegraphics[width=0.45\textwidth]{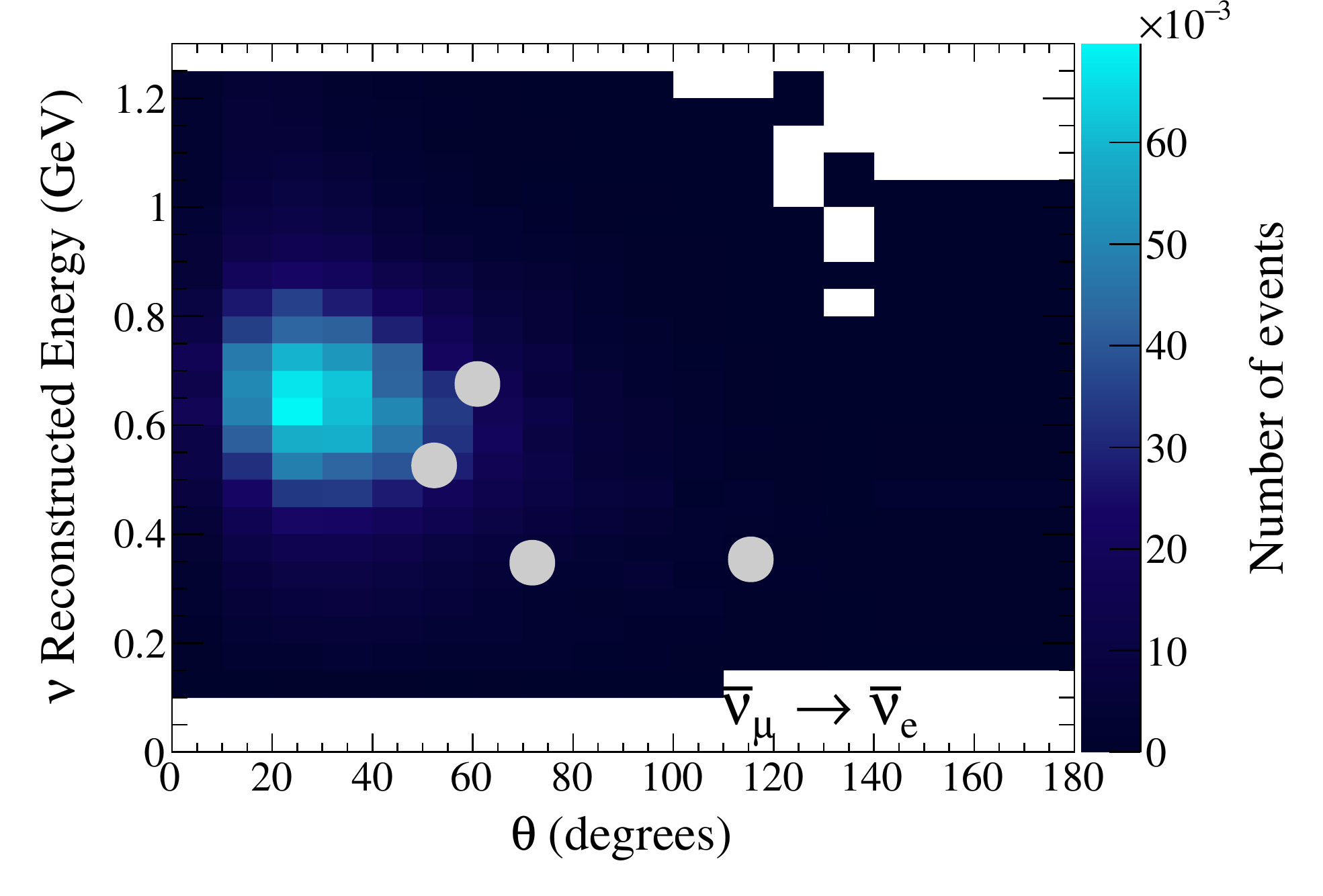}
\caption{
Expected \{\erec, $\theta_{lep}$\} distributions for the 
%signal (top) and background (bottom) events
signal $\bar{\nu}_{\mu} \rightarrow \bar{\nu}_{e}$ (top)
and 
wrong-sign background $\nu_{\mu} \rightarrow \nu_e$ (bottom) 
in the e-like CCQE-like selection in \nub-mode.
The superimposed grey dots correspond to the data. 
The expectation is based on the parameters of Tab.~\ref{tab:nomosc_par}.
}
\label{fig:2d_templ_fhc_ccqe_sig_bkg}
\end{center}
\end{figure}

As was shown in Tab.~\ref{tab:SK_nue_events}, while the number of $\bar{\nu}_{\mu} \rightarrow \bar{\nu}_{e}$ in $\nu$-mode is negligible, in $\nub$-mode, the number of oscillated \nue is roughly 30\% of the number of oscillated \nueb and thus it is crucial to use templates which are able to distinguish between the two cases.
In Fig.~\ref{fig:2d_templ_fhc_ccqe_sig_bkg}, the expected \{\erec, $\theta_{lep}$\} distribution in the single ring e-like sample in $\nub$-mode for the signal ($\bar{\nu}_{\mu} \rightarrow \bar{\nu}_{e}$) and for the oscillated \nue background is shown.

%This is clearly shown in fig.~\ref{fig:2d_templ_fhc_ccqe_sig_bkg},
%where the MC signal $\bar{\nu}_e \rightarrow \bar{\nu}_{\mu}$ 
%and 
%wrong-sign background $\nu_e \rightarrow \nu_{\mu}$
%\{$\erec$,$\theta_{lep}$\} 
%distributions 
%of the $\bar{\nu}_e$ CCQE candidate sample are shown.
%Differently from the $\nu_{e}$ candidate samples,
%where the contamination from $\bar{\nu}_{\mu} \rightarrow \bar{\nu}_{e}$ is almost negligible,
%the wrong-sign background in the $\bar{\nu}_{e}$ candidate sample is about 17\%
%and the separation of $\nu$ and $\bar{\nu}$ spectra is very important for the search of CP violation, 
%which consists of a difference between the $\nu$ and $\bar{\nu}$ oscillation probabilities.

\begin{figure}[htbp]
\begin{center}
\includegraphics[width=0.45\textwidth]{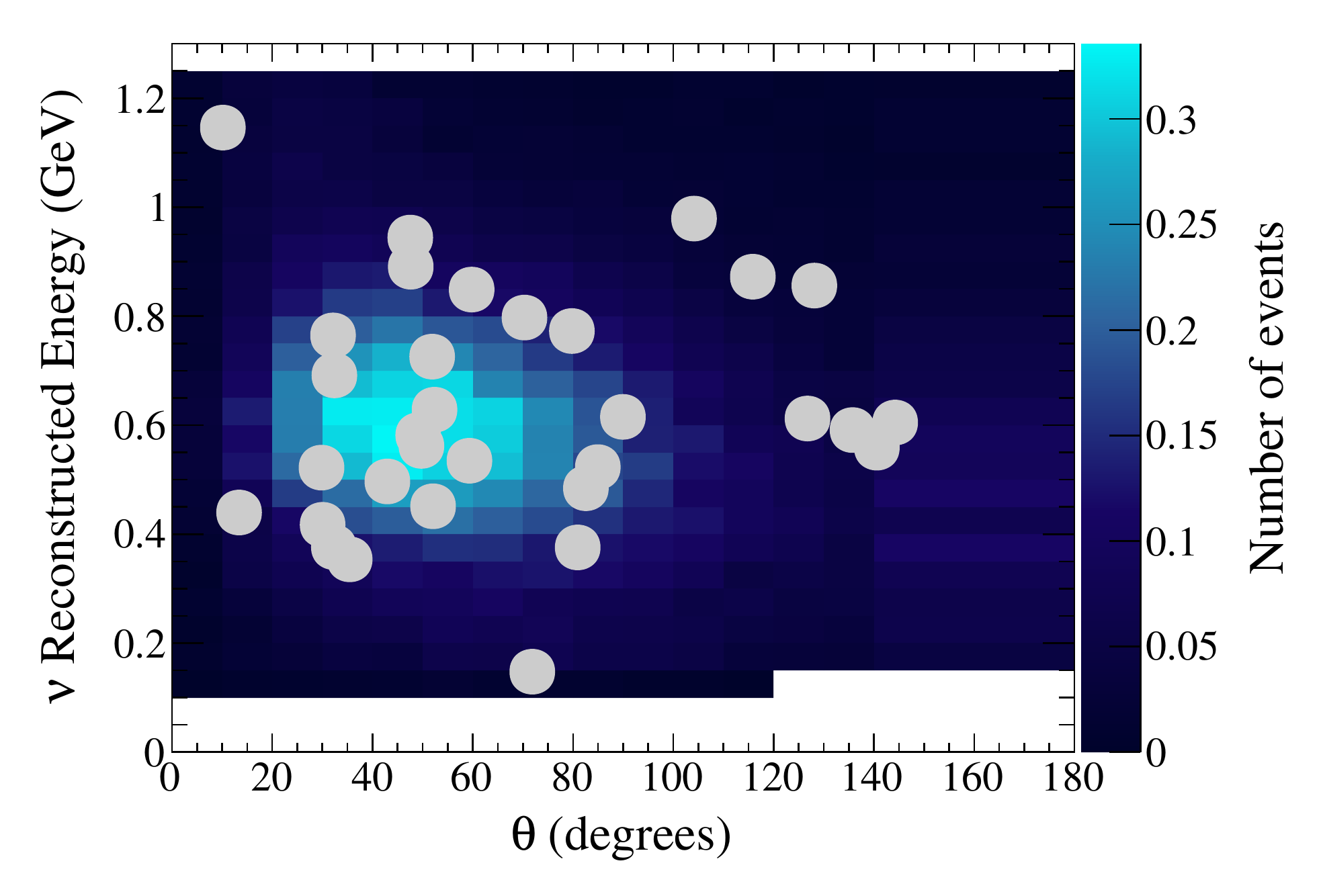}
\includegraphics[width=0.45\textwidth]{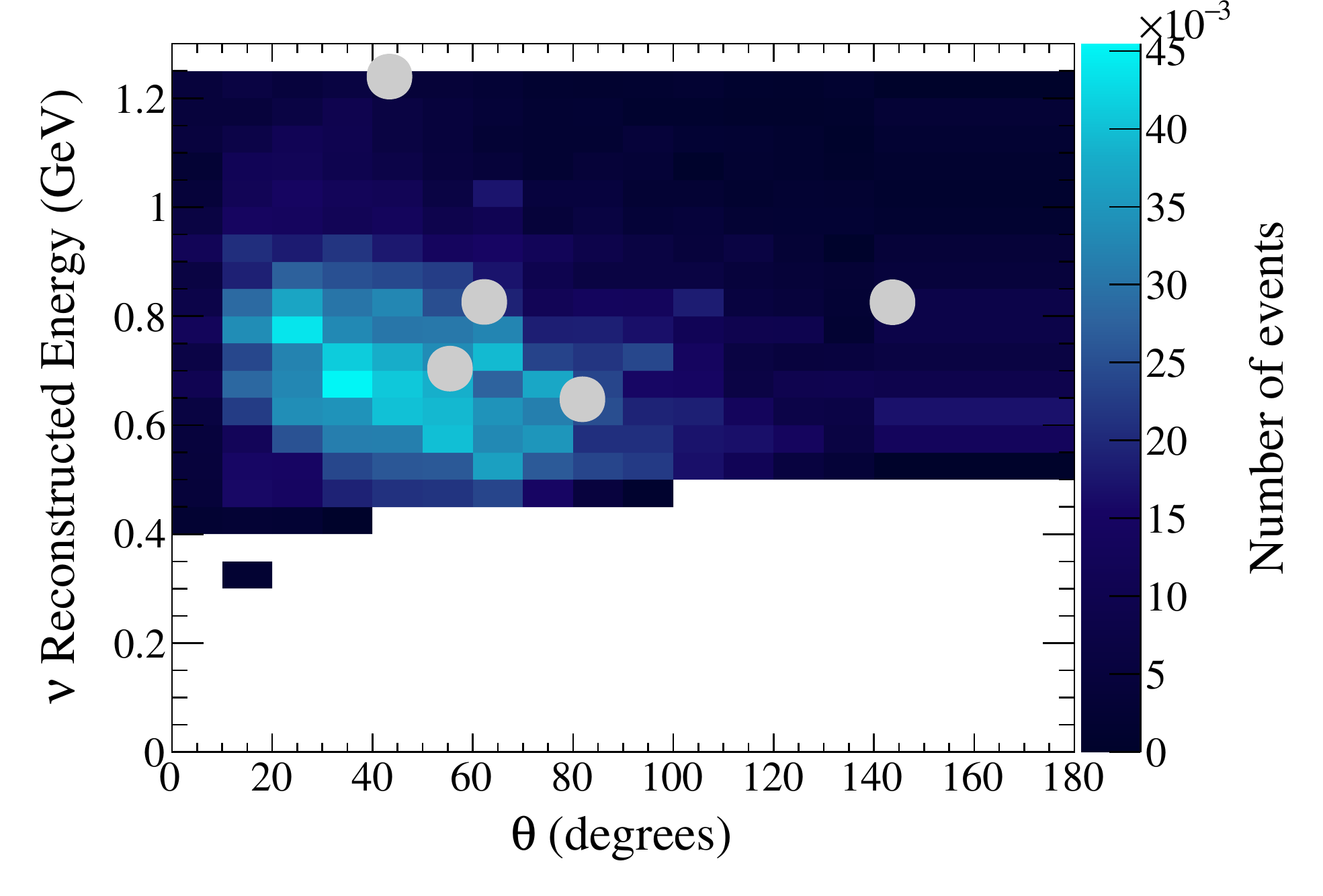}
\includegraphics[width=0.45\textwidth]{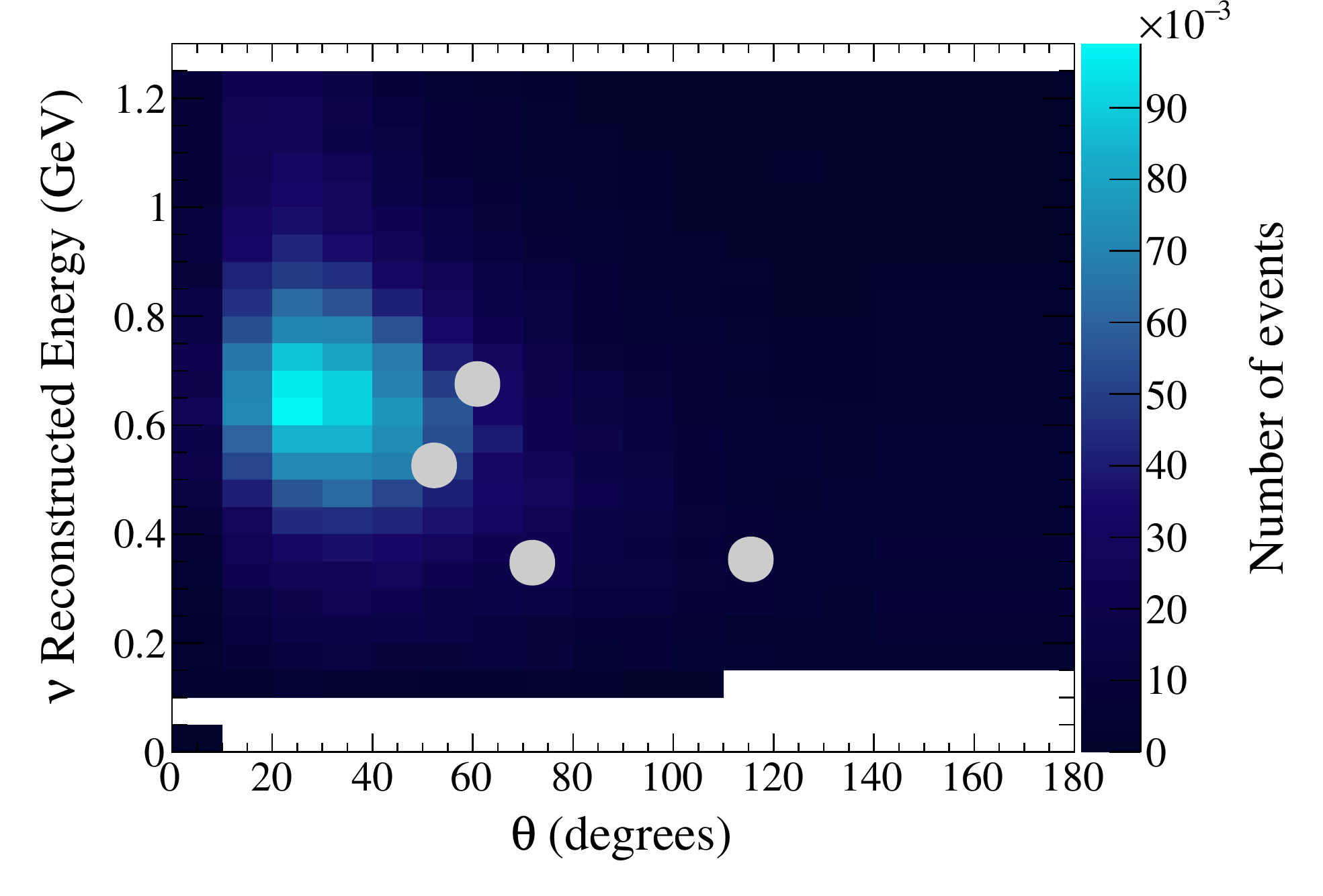}
\caption{
\{$\erec$,$\theta_{lep}$\} templates for the 
$\nu_e$ CCQE-like (top),
$\nu_e$ \ccpip (middle) in \nunu-mode
and 
$\bar{\nu}_e$ CCQE-like (bottom) in \nub-mode
candidate samples.
Both signal and background events are included in the expected distributions based on the oscillation parameters of Tab.~\ref{tab:nomosc_par}.
The superimposed grey dots correspond to the data. 
%~\ref{tab:nomosc_par}.}
}
\label{fig:2d_templ_ccqe_cc1pi}
\end{center}
\end{figure}

In Fig.~\ref{fig:2d_templ_ccqe_cc1pi} the 2D templates for the three $\nu_e$ candidate samples are shown. The $\nu_e$ \ccpip sample has no events with reconstructed energy below 400~MeV due to the $\Delta^{++}$ production requirement in Eq.~\ref{eq:SK_CC1pi_Erec} used to reconstruct the neutrino energy.

The second frequentist analysis also uses 2D templates but instead of using \erec, it uses lepton momentum, $p_{lep}$, and $\theta_{lep}$.
%which provides the same $\nu$ - $\bar{\nu}$ separation as described above.
This choice provides similar $\nunu$--$\nub$ separation as \{$p_{lep}$, $\theta_{lep}$\} templates are directly related to \erec.
Indeed, a particular value of \erec corresponds to a slice in the \{$p_{lep}$, $\theta_{lep}$\} phase space.
For this reason, \{\erec, $\theta_{lep}$\} and \{$p_{lep}$, $\theta_{lep}$\} analysis templates 
provide very similar sensitivities to the oscillation parameters.

Finally, another difference among the three analyses is that the Bayesian analysis performs a simultaneous fit of the ND280 and SK datasets
in order to validate the extrapolation of the flux, cross-section and detector systematic parameters from the near to far detector.
The other two analyses perform a fit of the far detector data using the priors obtained by the near detector data fit
for the flux and cross section systematic parameters.
The two methods show very good agreement in the estimation of flux and cross-section parameters as was shown in Sec.~\ref{sec:banff}.

\subsection{Systematic uncertainties on the oscillation analyses}
\label{sec:oasyst}
%Event rates and error envelops for flux, cross-section, detector systematics for the five samples
%Expected length: 1 page.
%Editor: Ko.\\
%%%
The systematic uncertainties of the oscillation analysis are separated into four different categories: flux, cross section, and SK detector and momentum scale parameters. The flux and many of the cross-section parameters (mentioned in Tab.~\ref{tab:xsec_parameters}) are correlated by the ND280 measurements as shown in Sec.~\ref{sec:banff}, effectively reducing the systematic uncertainties of the measurements at SK. %The methods of evaluating the systematic error inputs of each component are described in previous sections.

The flux parameters assign the systematic uncertainties on the neutrino and antineutrino flux. These parameters are applied as normalization factors for different neutrino flavor subcategories and true energy bins. The flux uncertainties at SK are constrained by the ND280 measurements.

The cross-section parameters are applied based on the true neutrino interaction category of each event. Most of these parameters are normalization factors, but some also contain shape information, modifying the kinematic distributions in $p_{lep}-\theta_{lep}$ and \erec probability density functions for corresponding oscillation analysis samples. 
%Some of the systematic uncertainties on the cross section are constrained by the ND280 measurements as described in Sec.~\ref{sec:banff}. 
%However, the parameters that are applied for specific nuclear targets, carbon ($^{12}$C) for ND280 and oxygen ($^{16}$O) for SK, are independent to the ND280 measurements and therefore not constrained.

There are several categories of SK uncertainty: detector efficiency, final state interaction (FSI), secondary interaction (SI), and photo-nuclear (PN) effect uncertainties. The update on the SK detector efficiency uncertainty related to the addition of the CC1$\pi$ e-like sample is described in Sec.~\ref{sec:sksyst}. In addition, the FSI, SI, and PN effect uncertainties on the \ccpip e-like sample are evaluated using the same method introduced in Ref.~\cite{Abe:2015awa}. The increased uncertainties for this sample is mainly due to the larger backgrounds affecting it, and to larger uncertainties on the pion FSI and SI as shown in Tab.~\ref{tab:systematics_numode}, where the SK systematic uncertainties are separated into the SK detector and the FSI+SI+PN contributions.

The SK energy scale uncertainty is applied independently from other parameters. The energy scale uncertainty is applied as a normalization of \erec for each event, which may vary the total event rate by shifting the events into the cut regions of the visible energy and reconstructed neutrino energy selection criteria. The SK energy scale uncertainty is estimated to be 2.4\%.

The effect of the systematic uncertainties on the predicted event rates of the $\nu$- and \nub-mode samples are summarized in Tab.~\ref{tab:systematics_numode} and Tab.~\ref{tab:systematics_anumode} respectively. The $1\sigma$ uncertainties are obtained by throwing large sets of toy experiments, varying only the selected systematic parameters for each experiment, and calculating the relative uncertainties from the distributions of the event rates. The anti-correlations between flux and cross section parameters reduce the systematic uncertainties when both these sources are taken into account.

%%%
\begin{table}[htb]
\caption{Effect of 1$\sigma$ variation of the systematic uncertainties on the predicted event rates of the $\nu$-mode samples.}
\begin{tabular}{l c c  c} \hline \hline
Source of uncertainty& $\nu_e$ CCQE-like & $\nu_{\mu}$& $\nu_{e}$ \ccpip \\ 
 &$\delta N/N$&$\delta N/N$&$\delta N/N$\\ \hline

Flux &3.7\%& 3.6\%&3.6\% \\ 
(w/ ND280 constraint) & &  & \\ \hline
Cross section &5.1\%& 4.0\%&4.9\% \\ 
(w/ ND280 constraint) & &  & \\ \hline
Flux+cross-section&&& \\
(w/o ND280 constraint)&11.3\% &10.8\%&16.4\%  \\
(w/ ND280 constraint)&4.2\% &2.9\%& 5.0\% \\ \hline
FSI+SI+PN at SK& 2.5\%& 1.5\%&10.5\% \\ \hline
SK detector & 2.4\%& 3.9\%&9.3\% \\ \hline
All & && \\
(w/o ND280 constraint)& 12.7\%& 12.0\%&21.9\% \\ 
(w/ ND280 constraint)& 5.5\%& 5.1\%&14.8\% \\ \hline \hline
\end{tabular}
\label{tab:systematics_numode}
\end{table}
%%%

%%%
\begin{table}[htb]
\caption{Effect of 1$\sigma$ variation of the systematic uncertainties on the predicted event rates of the \nub- mode samples.}
\begin{tabular}{l c c  } \hline \hline
Source of uncertainty& $\overline \nu_e$ CCQE-like & $\overline \nu_{\mu}$ \\ 
 &$\delta N/N$&$\delta N/N$\\ \hline
Flux&3.8\%& 3.8\% \\ 	 
(w/ ND280 constraint) &  & \\ \hline
 Cross section &5.5\%& 4.2\% \\ 
(w/ ND280 constraint) & &  \\ \hline
Flux+cross-section&& \\
(w/o ND280 constraint)&12.9\% &11.3\%  \\
(w/ ND280 constraint)&4.7\% &3.5\%\\ \hline
FSI+SI+PN at SK & 3.0\%& 2.1\% \\ \hline
SK detector & 2.5\%&3.4\% \\ \hline
All & &\\ \hline
(w/o ND280 constraint)& 14.5\%& 12.5\% \\ 
(w/ ND280 constraint)& 6.5\%& 5.3\% \\ \hline \hline
\end{tabular}
\label{tab:systematics_anumode}
\end{table}

\section{Sensitivity of oscillation parameters to neutrino interaction modeling}
%\section{Fake Data studies}
\label{sec:fds}
The neutrino interaction uncertainties are one of the main contributors to the systematic uncertainty on all oscillation measurements and there is a global effort underway to improve the understanding of neutrino cross sections.
This has lead to the creation of a number of interaction models which can, at least partially, describe existing cross-section data but which also produce different predictions for the oscillated event rates and spectra at SK.
This analysis uses the high statistics near detector data to constrain both the flux and cross-section model uncertainties, improving the prediction of the far detector event rate, and reducing the uncertainty on that prediction. However, the near and far detectors observe different neutrino energy spectra (mainly due to oscillations) so the underlying neutrino interactions they sample will be different.  The different design of the two detectors also means that they are sensitive to different regions of the lepton kinematic phase space.  The near detector fit therefore tunes the flux and cross-section models to a set of neutrino interactions and a lepton kinematic phase space that is not the same as that observed at the far detector.
%However, due to different detector designs and different neutrino energy spectra (mainly due to oscillation), the near and far detectors are sensitive to different parts of the neutrino interaction phase space. The near detector analysis mainly select leptons emitted with $\cos\theta_{beam}>0.3$ where $\theta_{beam}$ is the angle of the lepton with respect to the beam direction, while the selection efficiency at the far detector is flat with respect to $\theta_{beam}$.    
%The near detector fit therefore tunes the flux and cross-section models to a region of phase space that is not the same as that observed at the far detector.
These differences could then be incorrectly attributed to neutrino oscillation effects in the oscillation analysis, which would result in biased oscillation measurements.

This has been studied in a phenomenological context using a simplified oscillation analysis, summarized in Ref.~\cite{Ankowski:2016jdd}.
The results of the studies show that long-baseline experiments may be biased by cross-section model choices, as might be expected from the qualitative arguments above.
Given the potential impact of these model choices it is important to investigate these issues using the full T2K oscillation analysis framework.
This has been performed for a range of neutrino interaction model variations, which were discussed in Sec.~\ref{sec:niwgfake}.

\subsubsection{\label{sec:fd_procedure}Production and analysis of simulated datasets}
%\subsubsection{\label{sec:fd_procedure}Fake data generation and analysis}

\begin{figure*}[htbp]
%    \begin{subfigure}[t]{.48\textwidth}
        \includegraphics[height=5.5cm]{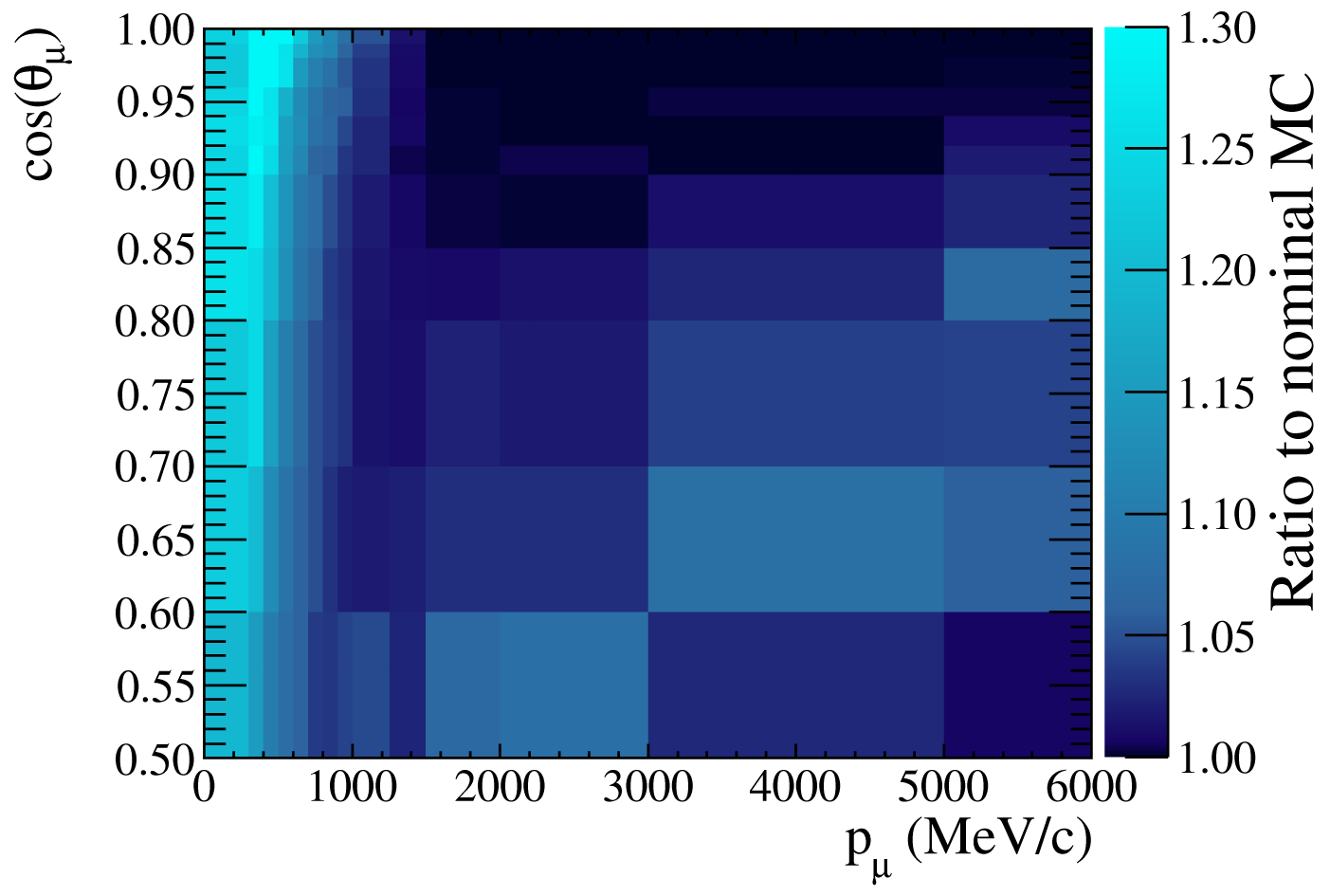}
%        \centering
%        \caption{Pion-less delta decay like event weights}
%        \label{fig:fig:nd80_fake_data_PDD}
%    \end{subfigure}
    \hfill
%    \begin{subfigure}[t]{.48\textwidth}
        \includegraphics[height=5.5cm]{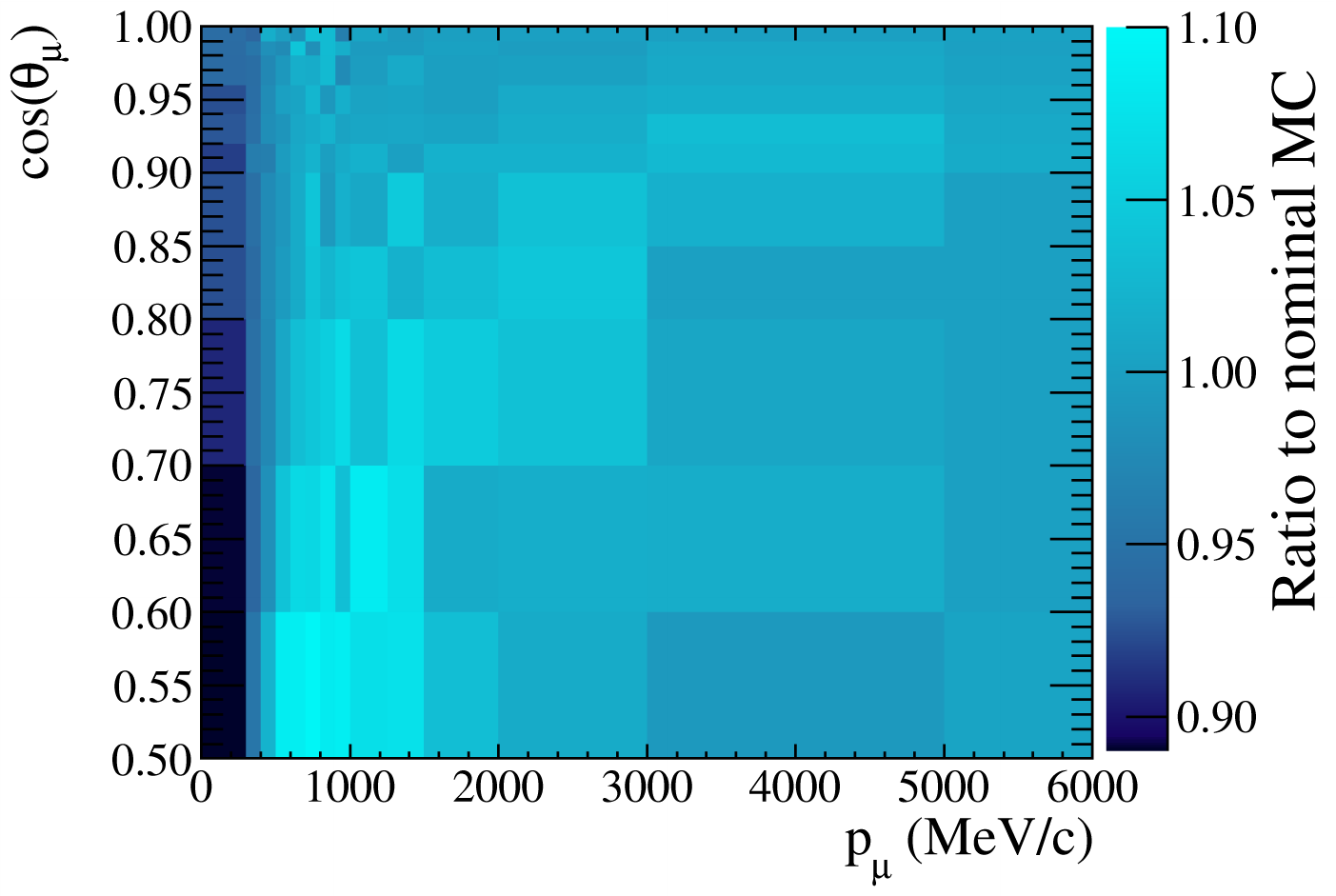}
    \caption{Ratio of the alternative 2p2h simulated data (left) and alternative 1p1h simulated data (right) to the nominal MC, shown as a function of reconstructed lepton kinematics for the FGD1 CC-0$\pi$ sample.}
    \label{fig:nd80_fake_data}
\end{figure*}

The analysis is performed using replica datasets created by changing one part of the nominal MC.
Simulated data are created by applying the event selection described in Sec.~\ref{sec:nd280} and Sec.~\ref{sec:sk} to the nominal T2K MC.
A weight is then applied to each selected event, calculated as the ratio of the altered interaction model to the nominal cross section model.
For the SK simulated data, the relevant oscillation probability is also applied. 
This produces event samples corresponding to the alternative interaction model.
%Finally, the events are weighted to reflect the current T2K POT.
An example of this is shown in Fig.~\ref{fig:nd80_fake_data} for the ND280 samples, where the left plot shows the ratio of the alternative 2p2h simulated data to the nominal MC for the FGD1 CC-0$\pi$ sample while the right shows the ratio for the alternative 1p1h model.

For each interaction model variation, simulated data are generated at both the ND280 and SK.
The nominal flux and interaction models are then fit to the ND280 simulated data using the procedure described in Sec.~\ref{sec:banff}.
The result is then used as the ND280 input to the oscillation analysis, described in Sec.~\ref{sec:oafrequentist}.
The SK simulated data are fit to produce a set of likelihood contours for the oscillation parameters.
These likelihood contours are then compared to the expected likelihood contour for fits where the nominal MC is used as data.
For each oscillation parameter a bias is calculated, defined as the difference in the parameter best-fit point between the simulated data fit and the nominal data fit, divided by the $1\sigma$ uncertainty on that parameter from the nominal fit.

\subsubsection{\label{sec:fd_results}Results}
\begin{figure*}[htbp]
%    \begin{subfigure}[t]{.48\textwidth}
        \includegraphics[height=6cm]{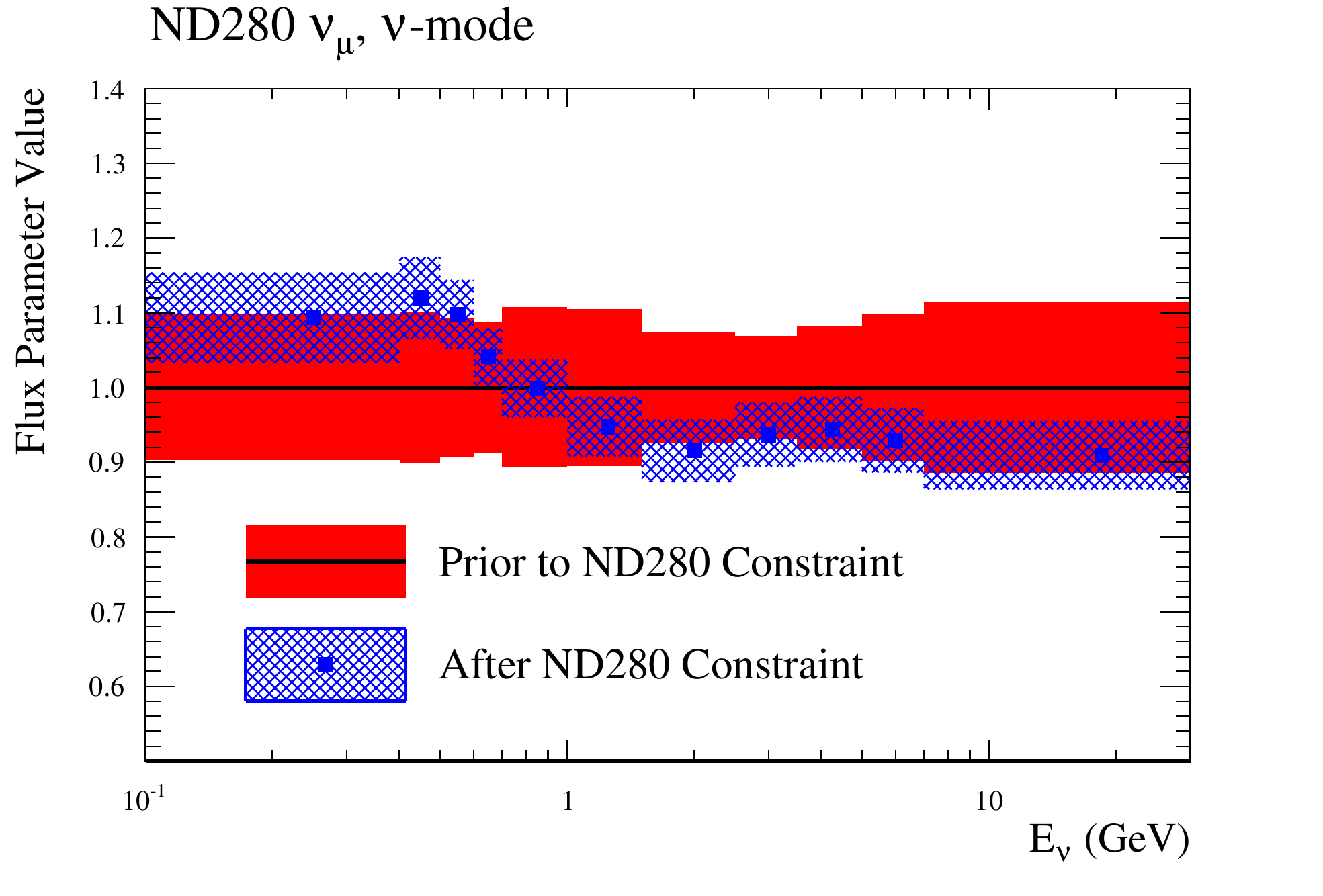}
%        \centering
%        \caption{ND280 flux parameters}
%        \label{fig:flux_martini}
%    \end{subfigure}
    \hfill
%    \begin{subfigure}[t]{.48\textwidth}
        \includegraphics[height=6cm]{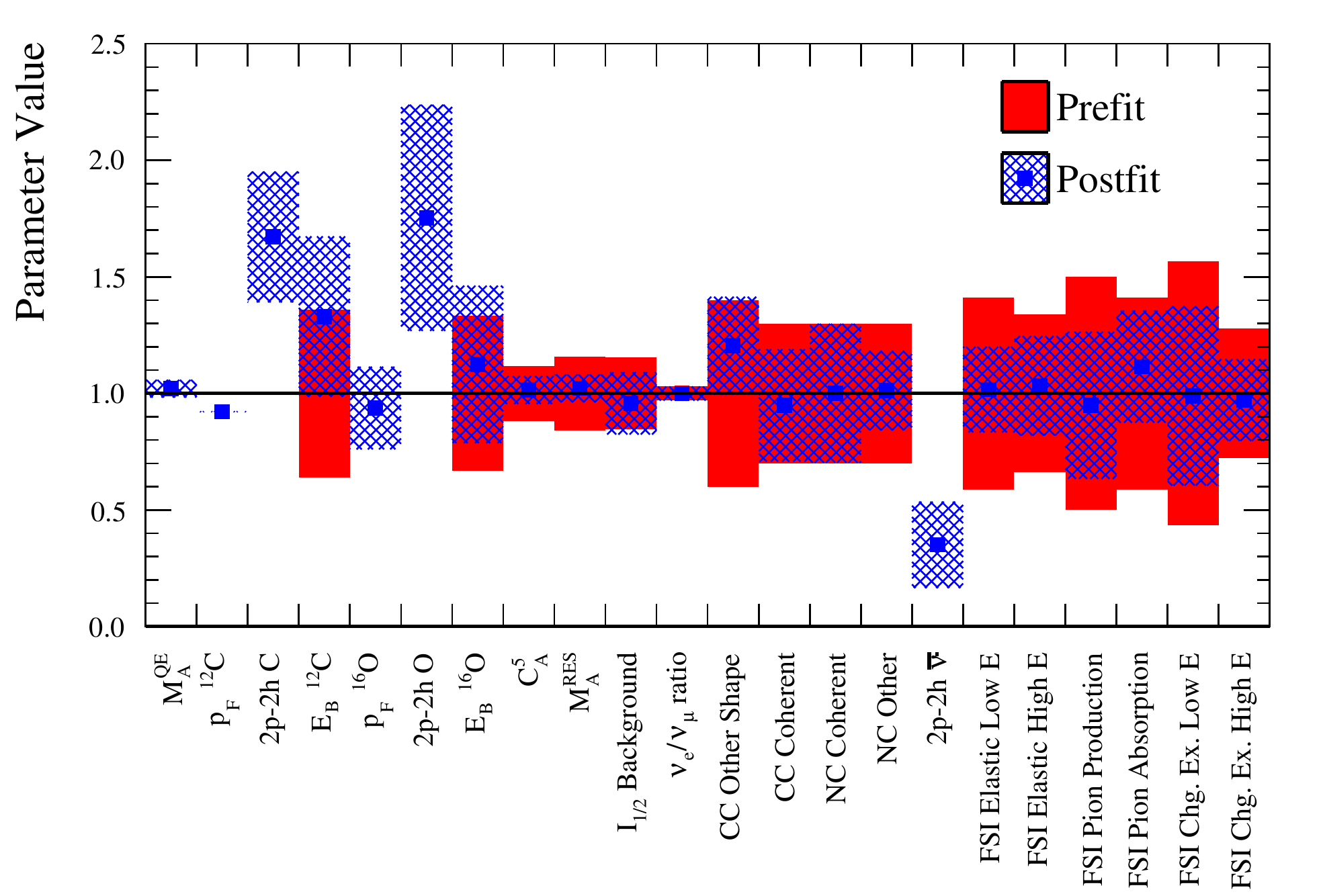}
%        \centering
%        \caption{Cross-section parameters}
%        \label{fig:xsec_martini}
%    \end{subfigure}
    \caption{The nominal values (red) and postfit values (blue) of the ND280 $\nu$-mode $\nu_{\mu}$ flux (left) and the cross-section (right) parameters, shown as a fraction of their nominal value, following the near detector fit to the alternative 2p2h ND280 simulated data.}
    \label{fig:martini_banff}
\end{figure*}

\begin{figure*}[htbp]
%    \begin{subfigure}[t]{.48\textwidth}
        \includegraphics[height=6cm]{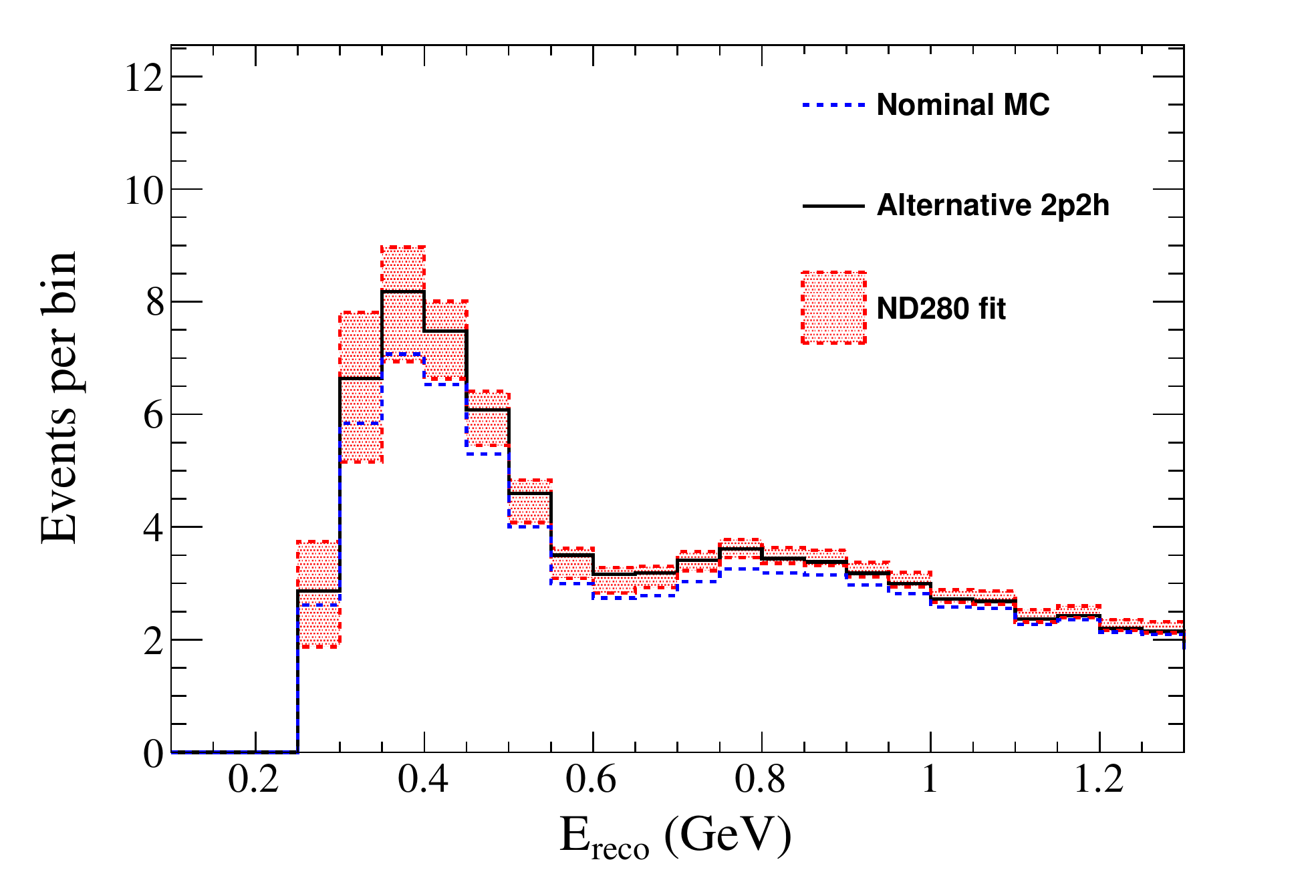}
%        \centering
%        \caption{Placeholder for NuMu spectrum}
%        \label{fig:numu_comp}
%    \end{subfigure}
    \hfill
%    \begin{subfigure}[t]{.48\textwidth}
        \includegraphics[height=6cm]{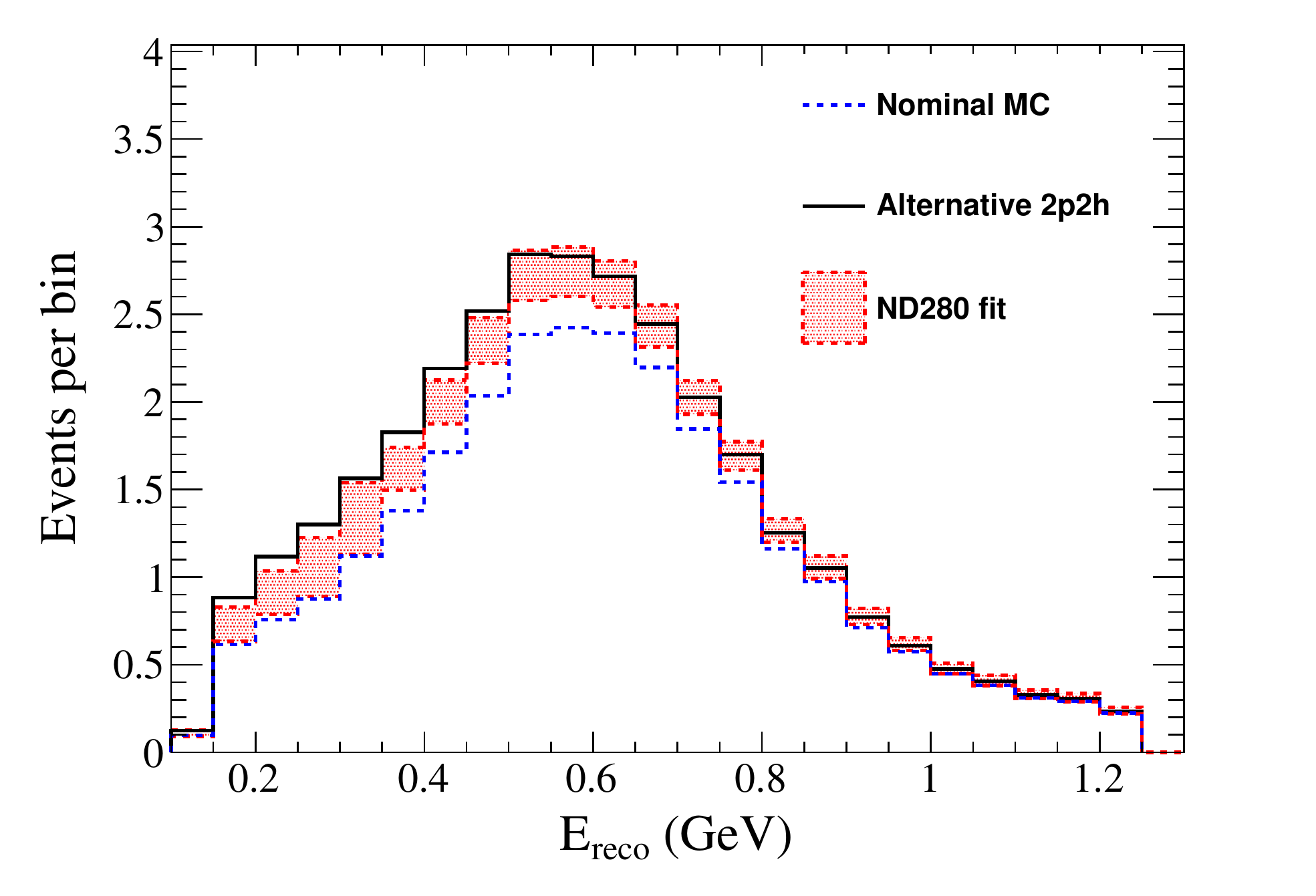}
%        \centering
%        \caption{Selected $\nu_{e}$ events}
%        \label{fig:nue_comp}
%    \end{subfigure}
    \caption{Single-ring expected event spectrum at SK as a function of \erec, assuming CCQE kinematics, for the selected $\mu$-like (left) and e-like (right) events in \nunu-mode.  The nominal SK prediction is shown by the dashed blue line, the SK simulated data by the solid black line and the prediction from the ND280 simulated data fit by the hashed red region.}
    \label{fig:sk_spectra}
\end{figure*}

Results from the ND280 fit are used for central values and uncertainties for the neutrino flux and cross-section model parameters.
%The ND280 fit alters the value and uncertainty of the T2K flux and cross-section model parameters.
When fitting the nominal MC, the parameter estimators are found to be unbiased while their uncertainties are reduced.
For the fits to simulated data generated from alternative models the values of both the flux and cross-section parameters change, as shown in Fig.~\ref{fig:martini_banff}. These show the results obtained by fitting the data generated using the alternative 2p2h model, rather than the nominal model.
Figure~\ref{fig:martini_banff} demonstrates how differences between the true and model interaction cross sections can be absorbed by both the flux and cross-section parameters.%that, if the true interaction cross section is not well represented by the T2K model, these differences can be absorbed by both the flux and cross-section parameters. 
The ND280 fit result is used to predict the unoscillated event spectra at SK, producing the \nunu-mode $\nu_{\mu}$ and $\nu_{e}$ event spectra shown in Fig.~\ref{fig:sk_spectra}.
%The nominal MC samples are clearly different from the modified MC samples, although it can be seen that, including the near detector fit, the uncertainty in the prediction largely accounts for this difference.
The predicted spectra falls between the nominal and 2p2h models and the uncertainty typically covers much of the difference between the models.

\begin{table}[!tbp]
    \centering
    \caption{Oscillation parameter values used as inputs for the studies of simulated data.}
    \label{tab:fd_osc_pars}
    %\begin{tabular}{c|c|c}
    \begin{tabular*}{0.9\columnwidth}{c@{\extracolsep{\fill}} c@{\extracolsep{\fill}} c}
        \hline
        \hline
        Parameter & Maximal  & Non-maximal  \\
         &  oscillation &  oscillation \\
        \hline
        $\sin^2(\theta_{23})$ & 0.523 &  0.6  \\
         $\delta_{CP}$ (rad) & -1.601 & -1.601 \\
          \hline
        $\sin^2(\theta_{13})$ & \multicolumn{2}{c}{0.0217} \\     
        $\sin^2(\theta_{12})$ & \multicolumn{2}{c}{0.304} \\
        $\Delta m^2_{21}$ $(10^{-5}$~eV$^2)$ & \multicolumn{2}{c}{7.53} \\
        $\Delta m^2_{32}$ $(10^{-3}$~eV$^2)$ & \multicolumn{2}{c}{2.51} \\
        \hline
        \hline
    \end{tabular*}
\end{table}

\begin{figure*}[htbp]
%    \begin{subfigure}[t]{.48\textwidth}
        \includegraphics[height=5cm]{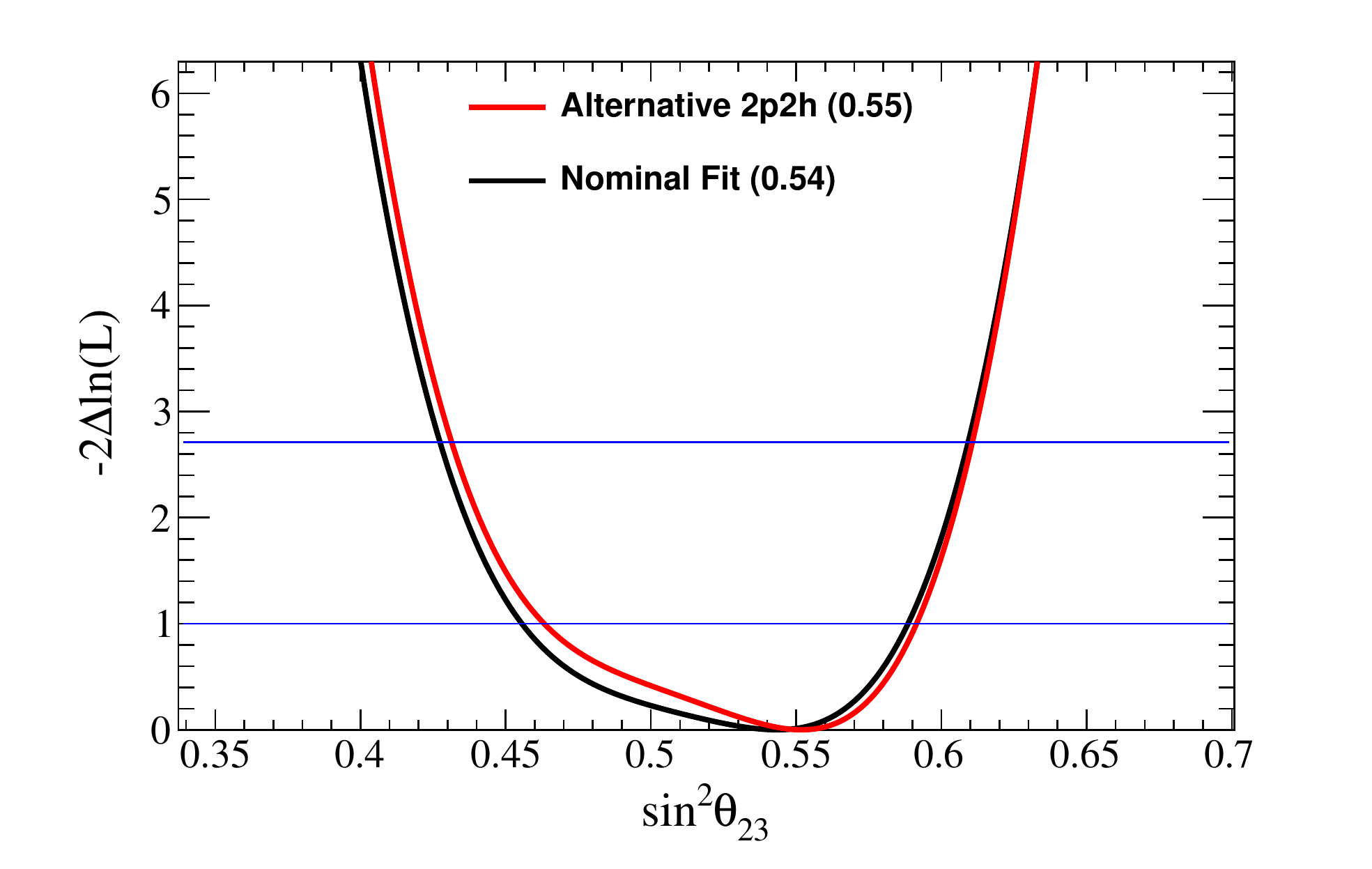}
%        \centering
%        \caption{$\sin^{2}\theta_{23}$}
%        \label{fig:martini_theta23}
%    \end{subfigure}
    \hfill
%    \begin{subfigure}[t]{.48\textwidth}
        \includegraphics[height=5cm]{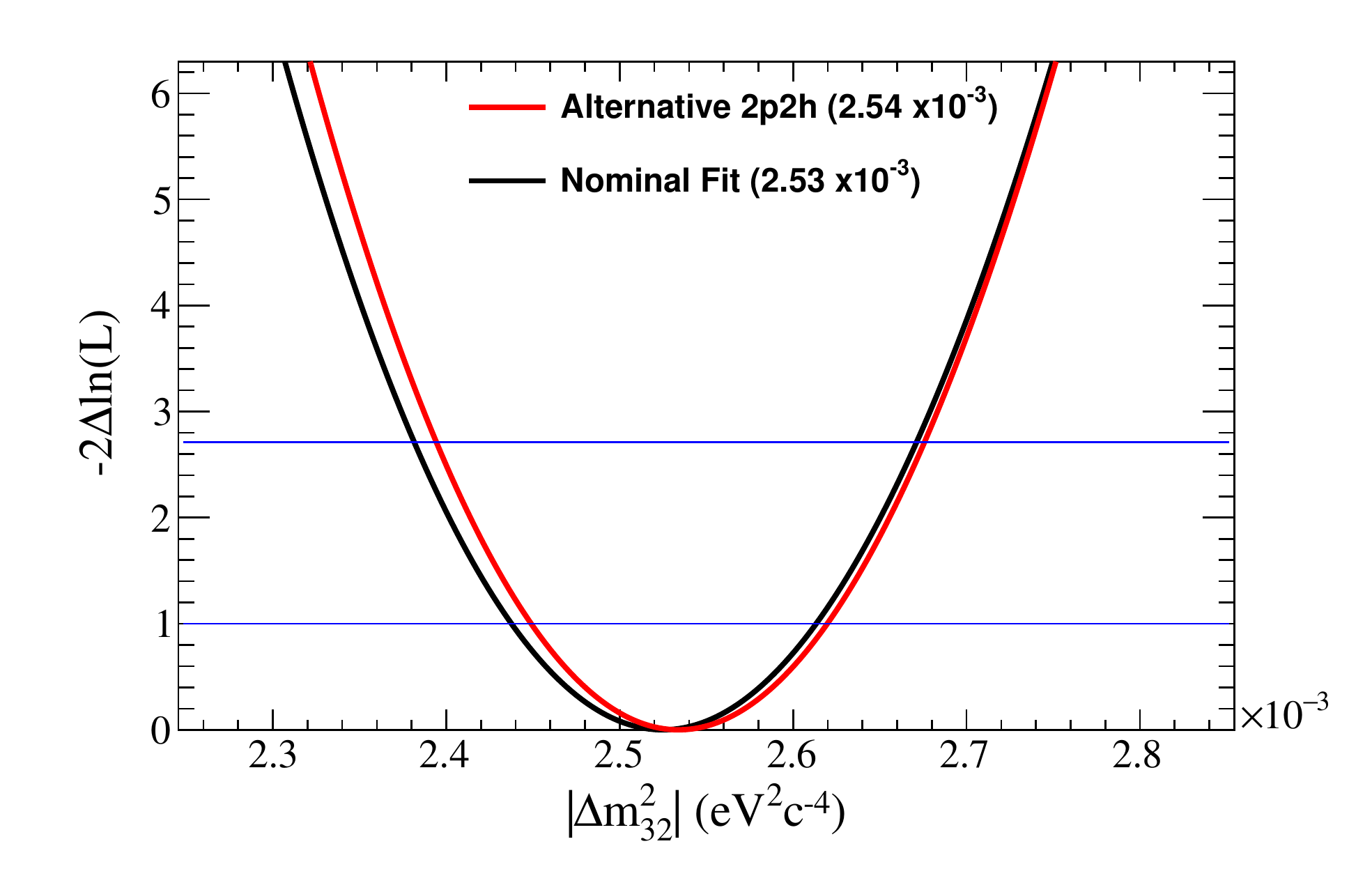}
%        \centering
%        \caption{$\Delta m^{2}_{32}$}
%        \label{fig:martini_dm2}
%    \end{subfigure}
%    \begin{subfigure}[t]{.48\textwidth}
        \includegraphics[height=5cm]{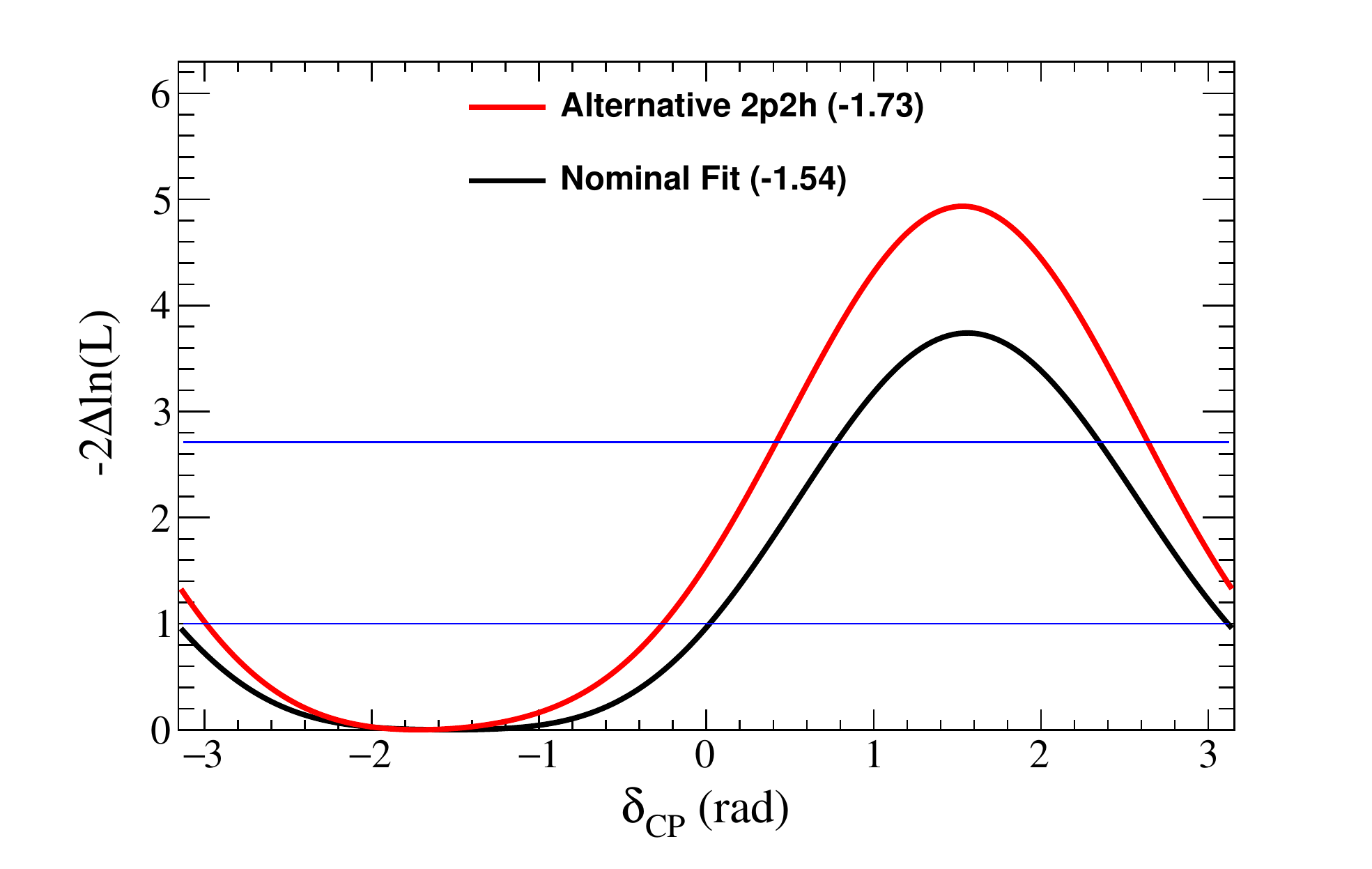}
%        \centering
%        \caption{$\delta_{CP}$}
%        \label{fig:martini_dcp}
%    \end{subfigure}
    \hfill
%    \begin{subfigure}[t]{.48\textwidth}
        \includegraphics[height=5cm]{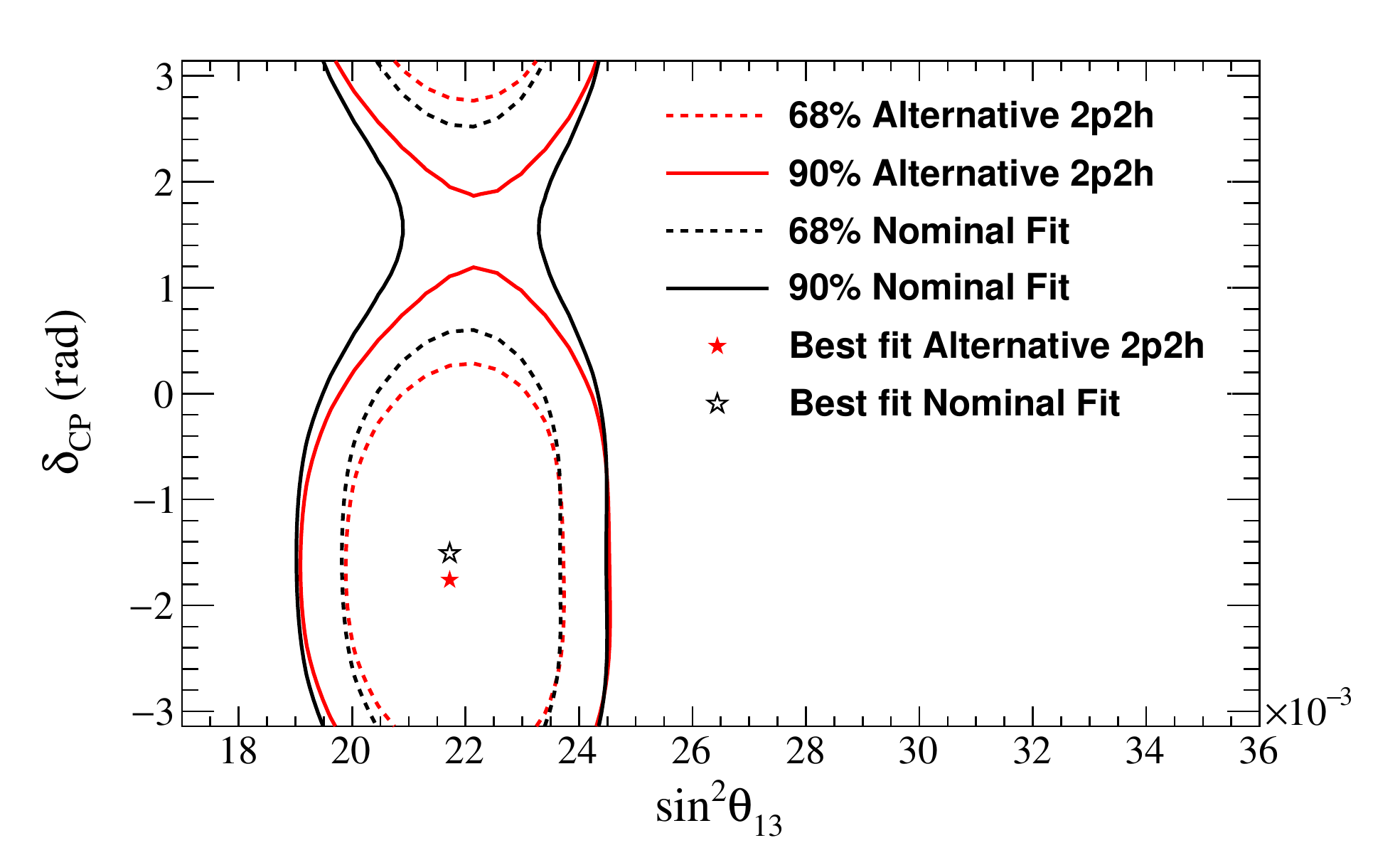}
%        \centering
%        \caption{$\delta_{CP}$--$\sin^{2}\theta_{13}$}
%        \label{fig:martini_2d}
%    \end{subfigure}
    \caption{The likelihood surfaces (red) from the oscillation fit of the alternative 2p2h simulated data at SK, using the result of the fit to the ND280 simulated data as an input. The oscillation parameters of Tab.~\ref{tab:fd_osc_pars} (maximal mixing) were used. The likelihood surfaces from the nominal fit are shown by the black line.  The contours for $\sin^{2}\theta_{23}$ (top left), $\Delta m^{2}_{32}$ (top right), $\delta_{CP}$ (bottom left) and $\delta_{CP}$ vs $\sin^{2}\theta_{13}$ (bottom right) are shown. For the one-dimensional likelihoods the point of minimum $\Delta \chi^2$ is given in each legend.}
    \label{fig:martini_likelihood}
\end{figure*}

%These predicted spectra are then fit, using the output uncertainties and central values from the ND280 fit result, to the SK simulated data in order to extract the oscillation parameters.
To evaluate possible bias in the oscillation parameter estimators, a fit is performed with SK simulated data.
The likelihood surfaces for the oscillation parameters of interest obtained by this fit are shown in Fig.~\ref{fig:martini_likelihood}.
These results use the true parameter values listed in the ``maximal oscillation'' column of Tab.~\ref{tab:fd_osc_pars}.
Fig.~\ref{fig:martini_likelihood} shows that if the alternative 2p2h model is correct, using the Nieves model in the nominal MC produces small ($\sim$2\%) biases in the measured values of $\sin^{2}\theta_{23}$ and $\Delta m^{2}_{32}$ ($<1\%$).
The study also shows a change in the $\delta_{CP}$ likelihood contour, increasing the exclusion around $\delta_{CP} = \pipi/2$ by 1.2 units of $\Delta\chi^{2}$.
Overall, as seen in the bottom right of Fig.~\ref{fig:martini_likelihood}, the effect on the appearance analysis is small, compared to the statistical uncertainty.

\begin{figure*}[htbp]
%    \begin{subfigure}[t]{.48\textwidth}
        \includegraphics[height=5cm]{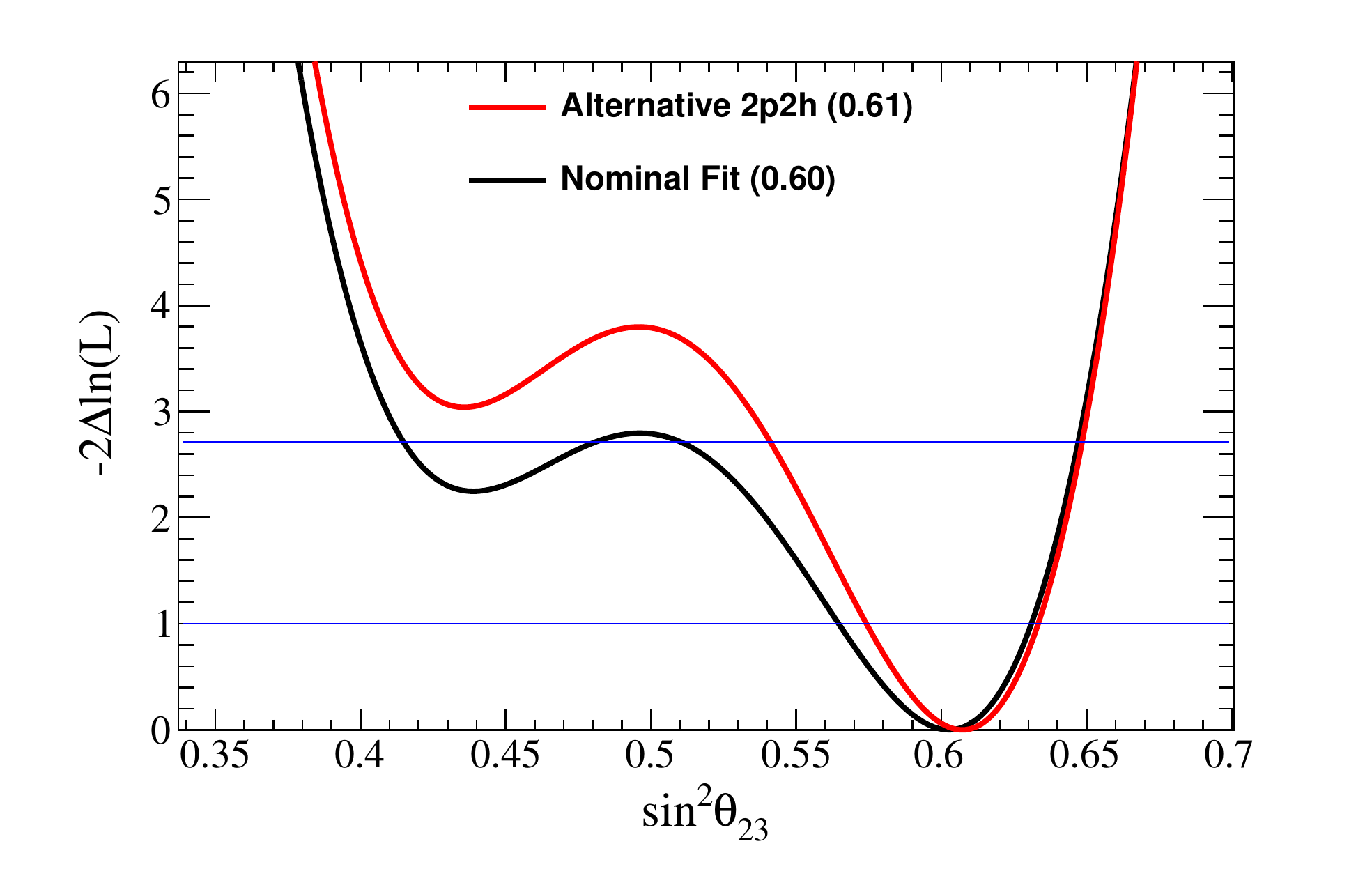}
%        \centering
%        \caption{$\sin^{2}\theta_{23}$}
%        \label{fig:martini_theta23}
%    \end{subfigure}
    \hfill
%    \begin{subfigure}[t]{.48\textwidth}
        \includegraphics[height=5cm]{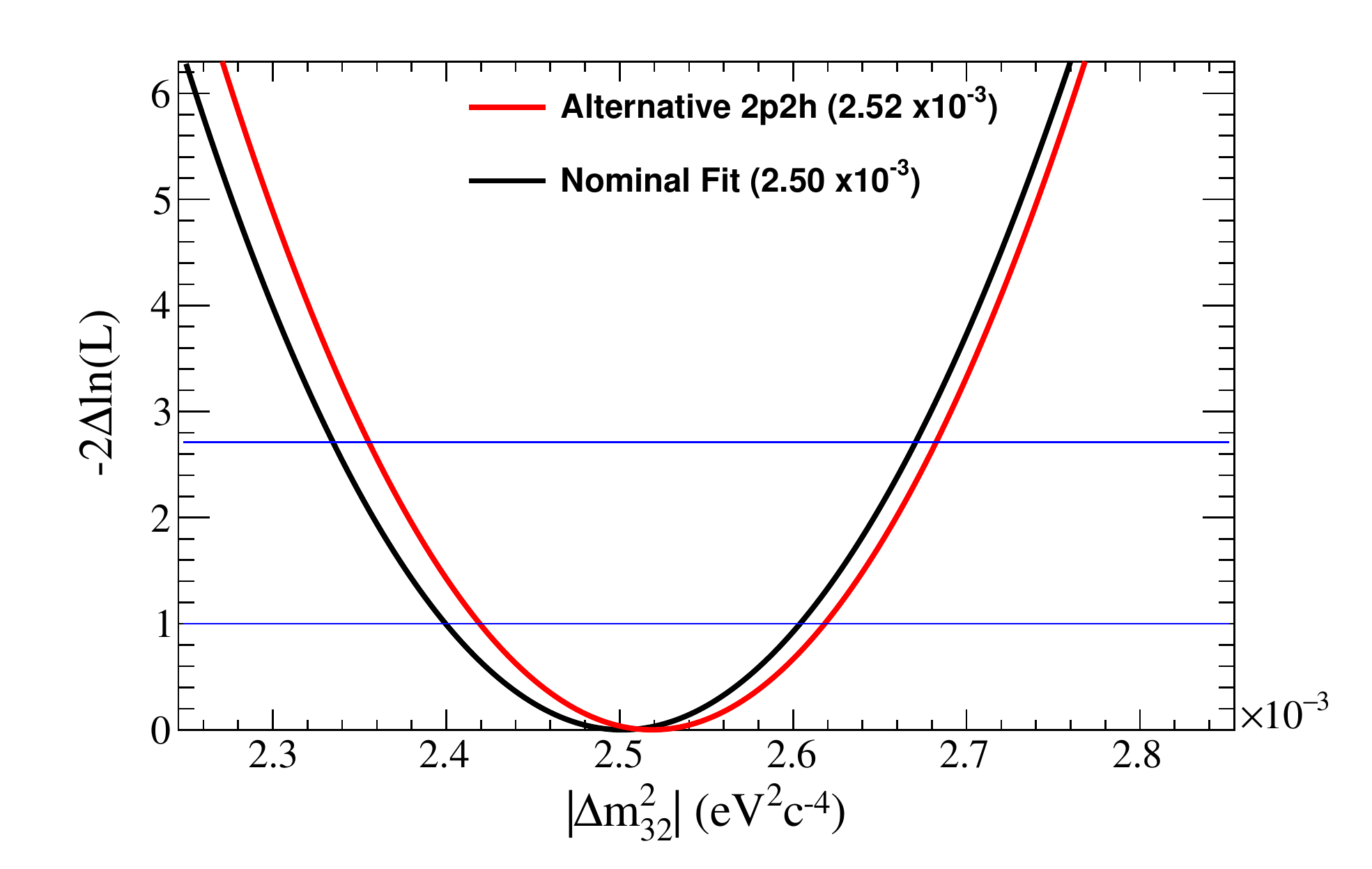}
%        \centering
%        \caption{$\Delta m^{2}_{32}$}
%        \label{fig:martini_dm2}
%    \end{subfigure}
%    \begin{subfigure}[t]{.48\textwidth}
        \includegraphics[height=5cm]{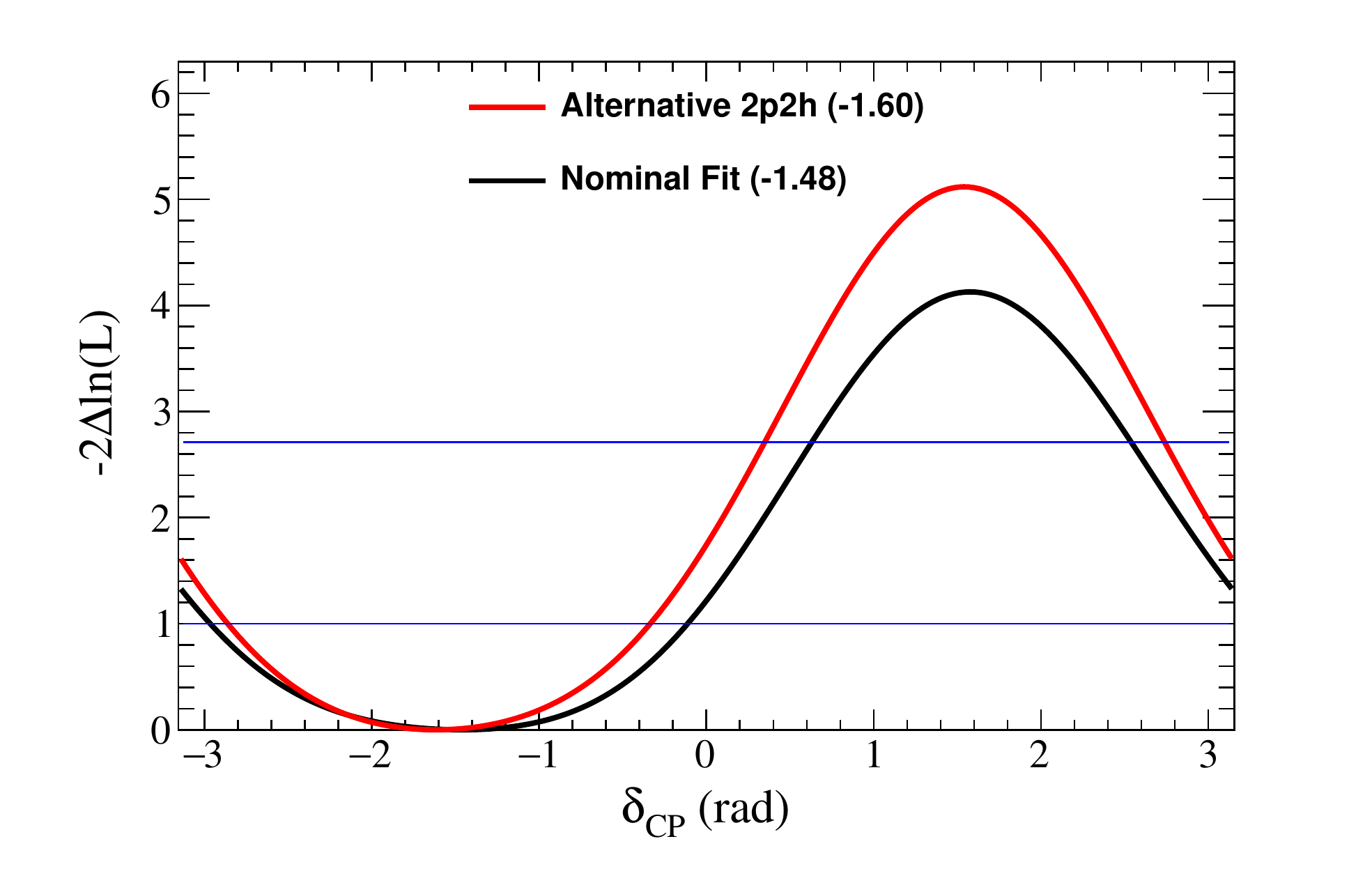}
%        \centering
%        \caption{$\delta_{CP}$}
%        \label{fig:martini_dcp}
%    \end{subfigure}
    \hfill
%    \begin{subfigure}[t]{.48\textwidth}
        \includegraphics[height=5cm]{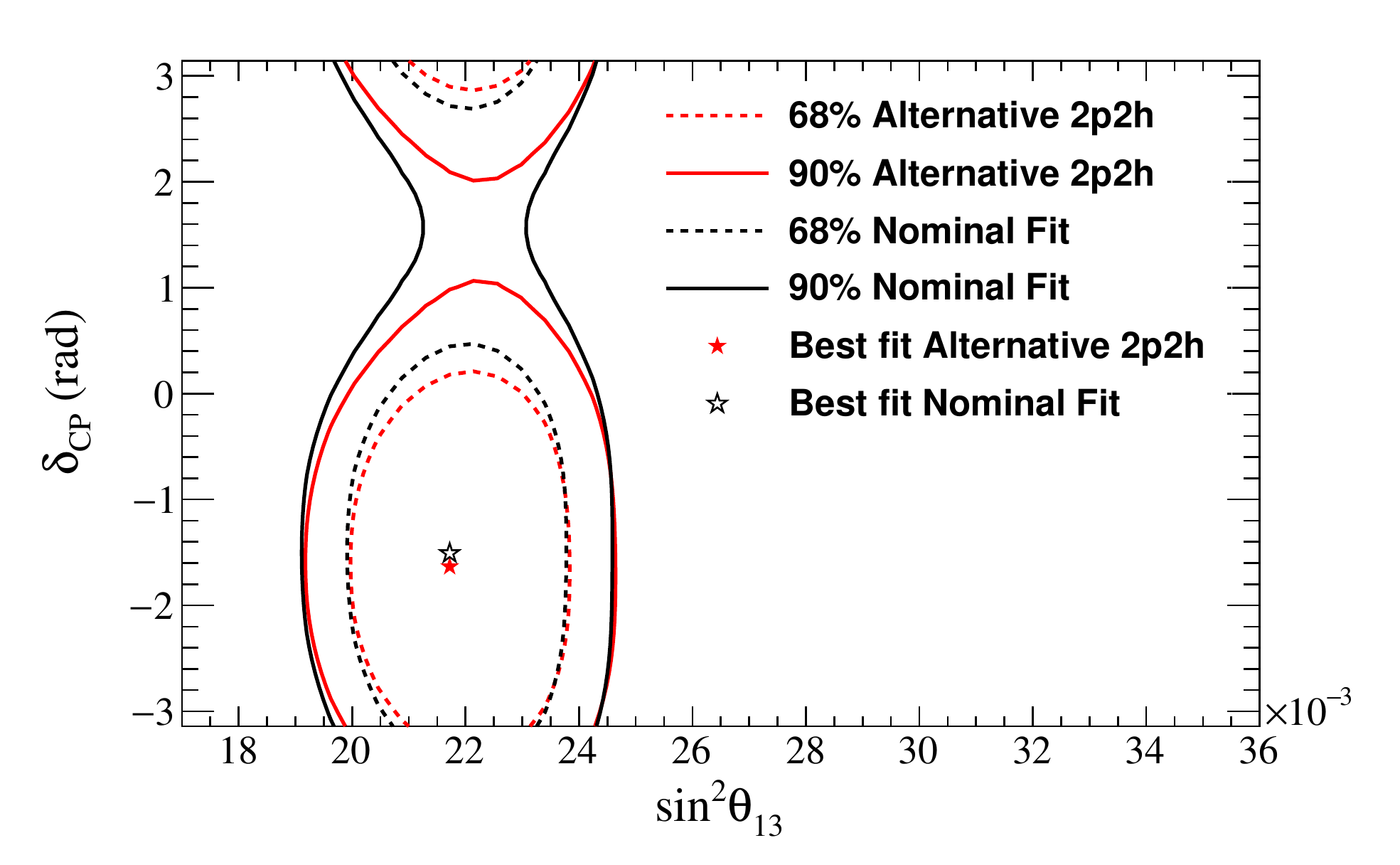}
%        \centering
%        \caption{$\delta_{CP}$--$\sin^{2}\theta_{23}$}
%        \label{fig:martini_2d}
%    \end{subfigure}
      \caption{The likelihood surfaces (red) from the oscillation fit of the alternative 2p2h simulated data at SK, using the result of the fit to the ND280 simulated data as an input. The oscillation parameters of Tab.~\ref{tab:fd_osc_pars} (non-maximal mixing) were used.  The likelihood surfaces from the nominal fit are shown by the black line.  The contours for $\sin^{2}\theta_{23}$ (top left), $\Delta m^{2}_{32}$ (top right), $\delta_{CP}$ (bottom left) and $\delta_{CP}$ vs $\sin^{2}\theta_{13}$ (bottom right) are shown. For the one-dimensional likelihoods the point of minimum $\Delta \chi^2$ is given in each legend.}
      \label{fig:martini_nonmax}
\end{figure*}

%The choice of true oscillation parameter values affects the simulated data fit result.
Figure~\ref{fig:martini_nonmax} shows fits to the same alternative 2p2h simulated data as in Fig.~\ref{fig:martini_likelihood}, but using the non-maximal oscillation parameters from Tab.~\ref{tab:fd_osc_pars}.
For non-maximal disappearance, a larger change is observed in the $\sin^{2}\theta_{23}$ likelihood contour.
%There is a small distortion of the likelihood at the 1$\sigma$ level, but minimal bias is seen in the fitted parameter values.
The effect on the point estimate and 68\% CL interval is relatively small.

\begin{figure*}[htbp]
%    \begin{subfigure}[t]{.48\textwidth}
        \includegraphics[height=5cm]{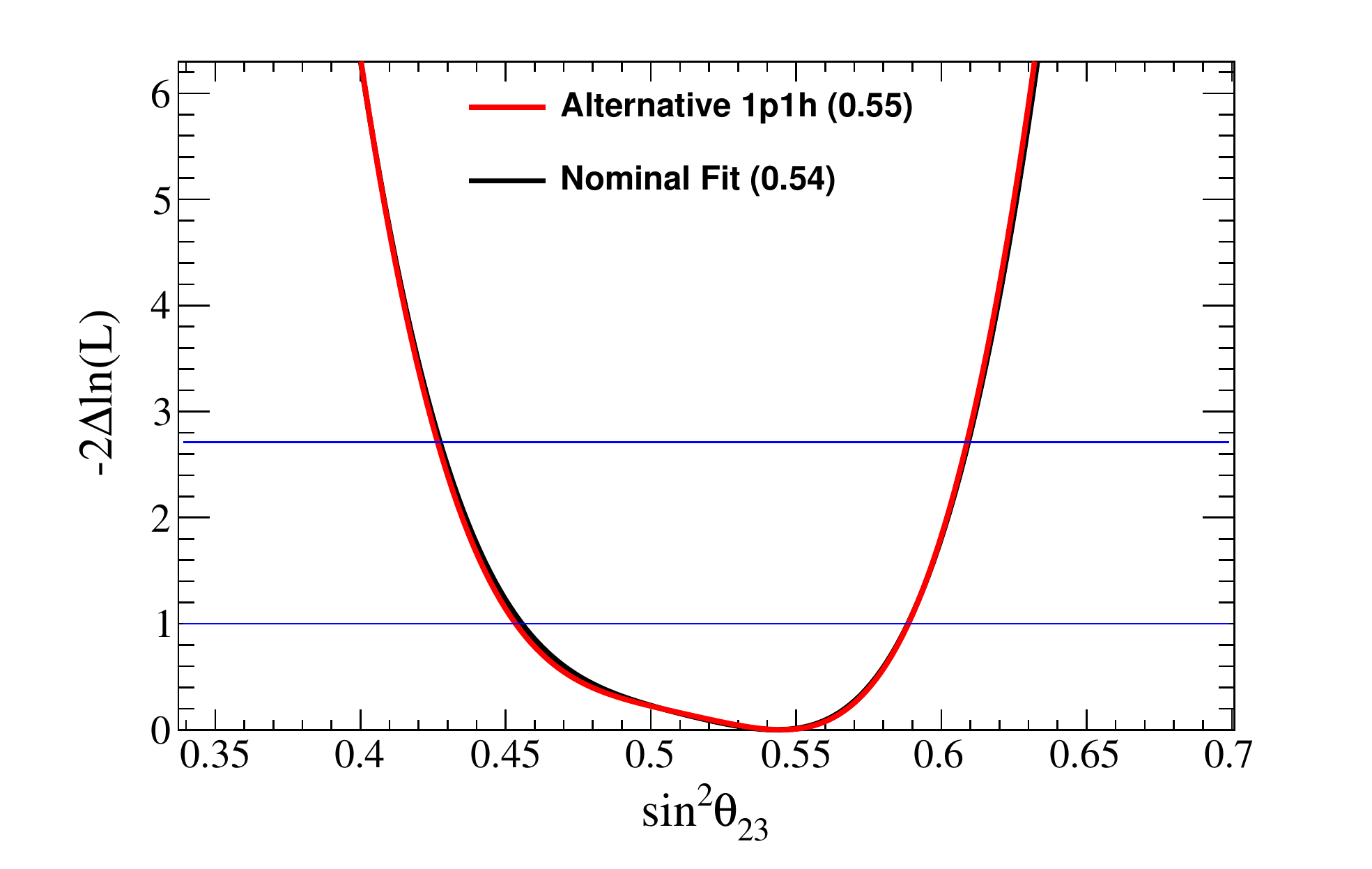}
%        \centering
%        \caption{$\sin^{2}\theta_{23}$}
%        \label{fig:1p1h_theta23}
%    \end{subfigure}
    \hfill
%    \begin{subfigure}[t]{.48\textwidth}
        \includegraphics[height=5cm]{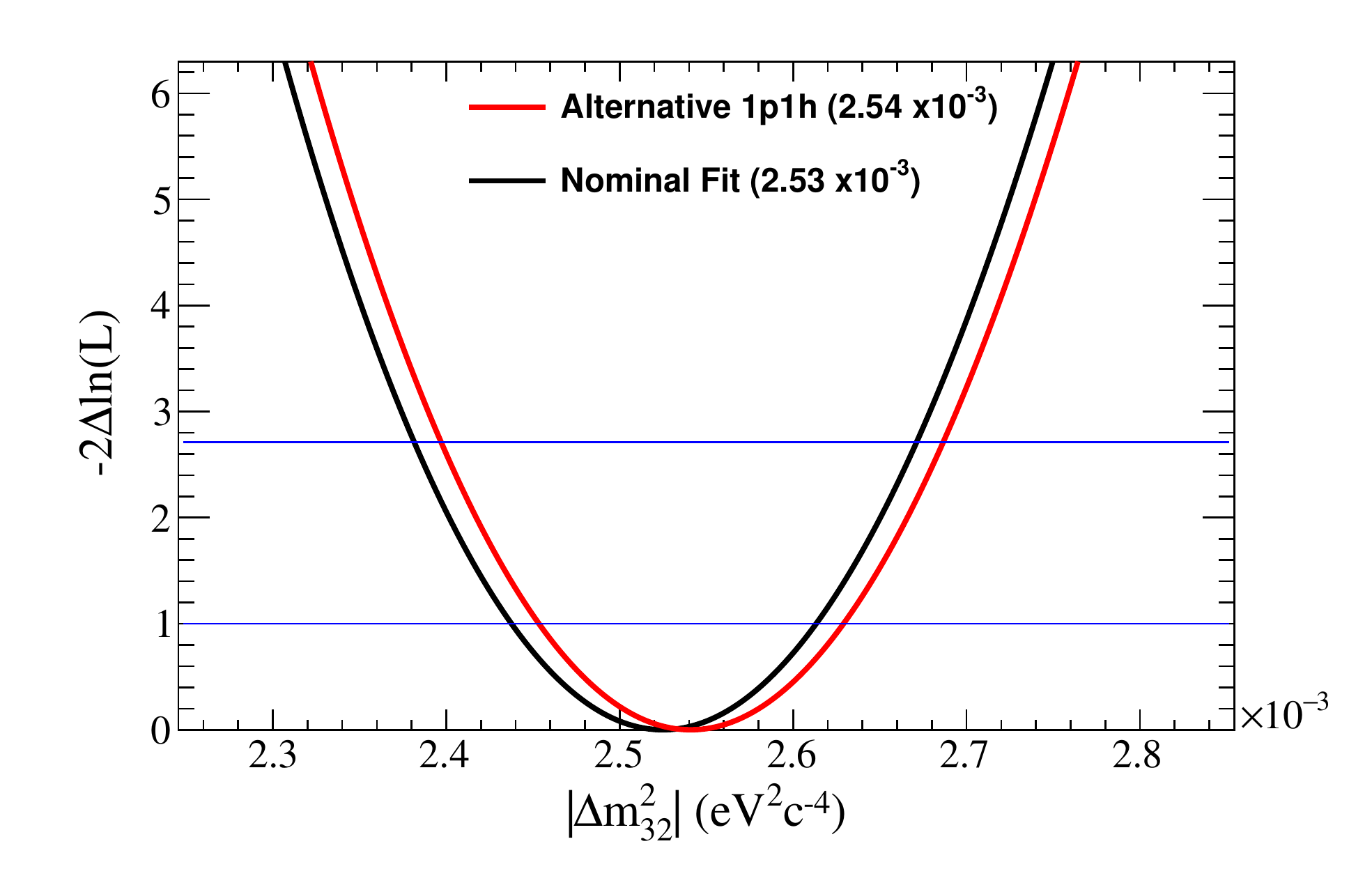}
%        \centering
%        \caption{$\Delta m^{2}_{32}$}
%        \label{fig:1p1h_dm2}
%    \end{subfigure}
%    \begin{subfigure}[t]{.48\textwidth}
        \includegraphics[height=5cm]{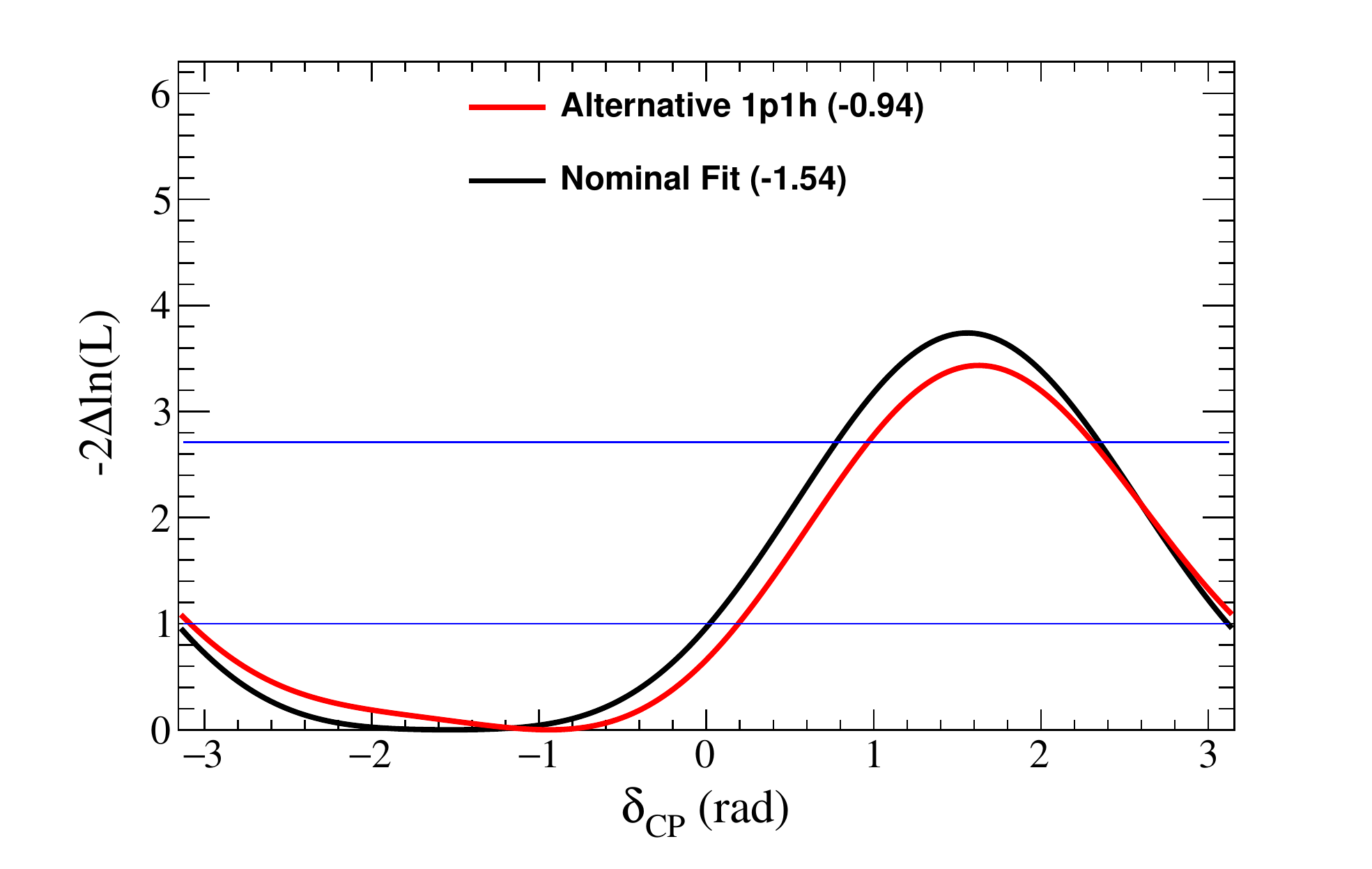}
%        \centering
%        \caption{$\delta_{CP}$}
%        \label{fig:1p1h_dcp}
%    \end{subfigure}
    \hfill
%    \begin{subfigure}[t]{.48\textwidth}
        \includegraphics[height=5cm]{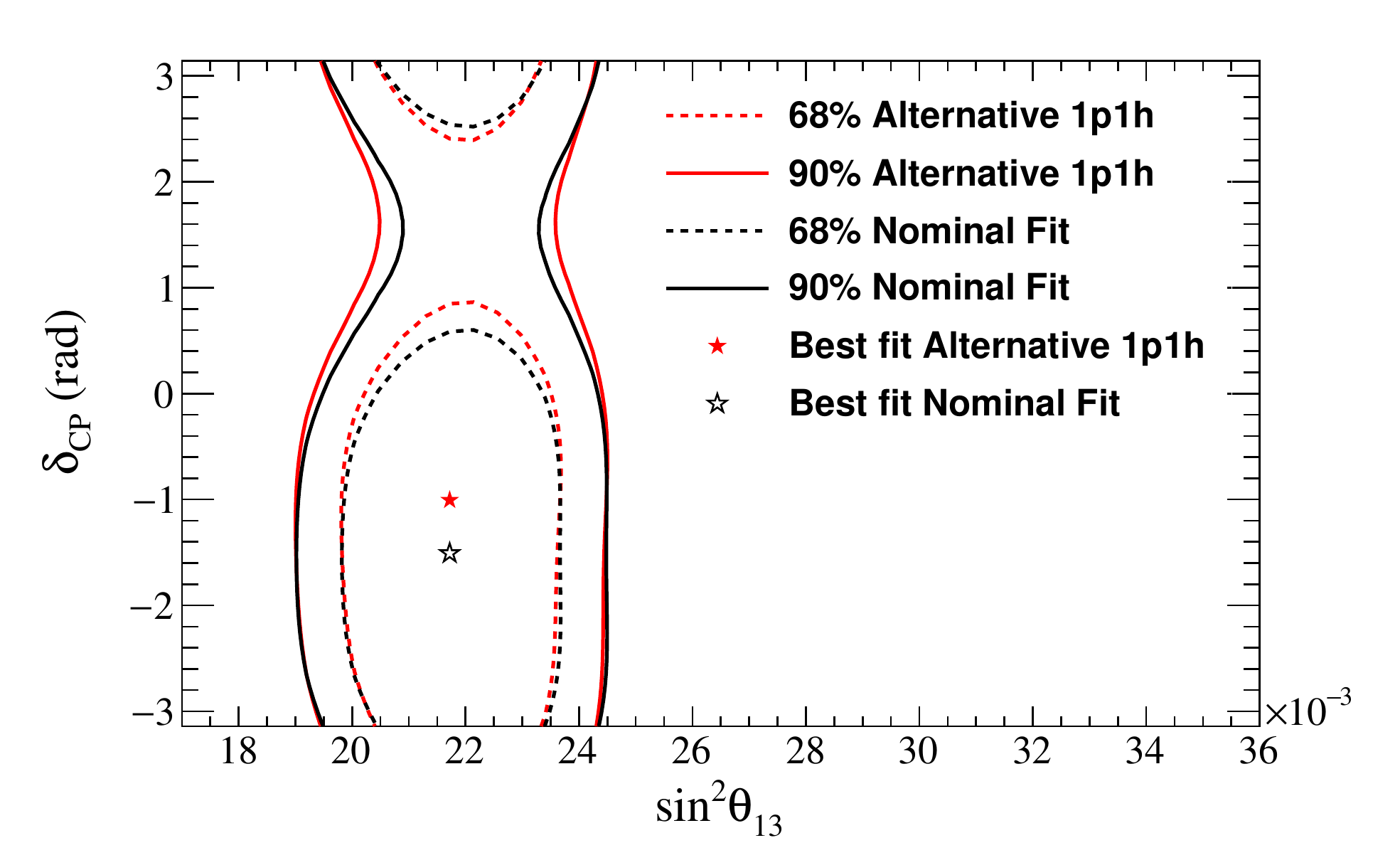}
%        \centering
%        \caption{$\delta_{CP}$--$\sin^{2}\theta_{13}$}
%        \label{fig:1p1h_2d}
%    \end{subfigure}
    \caption{The likelihood surfaces (red) from the oscillation fit of the alternative 1p1h simulated data at SK, using the result of the fit to the ND280 simulated data as an input. The oscillation parameters of Tab.~\ref{tab:fd_osc_pars} (maximal mixing) were used. The likelihood surfaces from the nominal fit are shown by the black line.  The contours for $\sin^{2}\theta_{23}$ (top left), $\Delta m^{2}_{32}$ (top right), $\delta_{CP}$ (bottom left) and $\delta_{CP}$ vs $\sin^{2}\theta_{13}$ (bottom right) are shown. For the one-dimensional likelihoods the point of minimum $\Delta \chi^2$ is given in each legend.}
     \label{fig:1p1h_results}
\end{figure*}

Fig.~\ref{fig:1p1h_results} shows the results of the fits to alternative 1p1h simulated dataset described in Sec.~\ref{sec:niwgfake}.
The alternative 1p1h model changes the T2K MC as a function of both neutrino energy and the angle at which the lepton is produced relative to the neutrino direction.
Similarly to the alternative 2p2h case, this model variation introduces a small bias in the measurement of $\Delta m^{2}_{32}$ and has a small effect on the $\delta_{CP}$ likelihood contour.

\subsubsection{\label{sec:fd_summary}Summary}

%Detailed studies have been performed to understand how the choice of neutrino interaction model could have an effect on the T2K oscillation analysis.
%The near--far extrapolation procedure tunes the neutrino flux and interaction models to the near detector simulated data, reducing the differences between the predicted spectra from ND280 fit and the simulated spectra at SK obtained using alternative neutrino interaction models.
Detailed studies have been performed to evaluate the sensitivity of the T2K oscillation analysis to neutrino interaction modeling. By using the near detector data to estimate flux and interaction model parameters, the oscillation parameter estimates are found to be essentially unaffected by changes to the interaction model.
The largest observed effect on the oscillation parameter likelihood contours is shown in Fig.~\ref{fig:martini_nonmax}, while Fig.~\ref{fig:1p1h_results} is more representative of the majority of the model variations discussed in Sec.~\ref{sec:niwgfake}.

A summary of the maximum bias observed for all of the alternative models studied is shown in Tab.~\ref{tab:summary_current_stats}.
The bias is presented as a fraction of the expected 1$\sigma$ uncertainty on each oscillation parameter, and is the maximum bias seen from all true oscillation parameter values tested.
T2K does not expect a significant constraint on $\delta_{CP}$ given the integrated POT available for this analysis.
As a result, it is difficult to quantify the effect on $\delta_{CP}$ in a single number, and so this parameter is not included in the summary table.
For reference, the largest effect observed for $\delta_{CP}$ is shown in Fig.~\ref{fig:martini_nonmax}.

\begin{table}[htb]
\caption{Summary of the maximum bias seen in oscillation parameters when fitting simulated datasets (defined in Sect.~\ref{sec:niwgfake}), presented as a fraction of the expected 1$\sigma$ uncertainty on each parameter.  Fits were performed for a number of true oscillation parameter assumptions. The numbers shown here are the maximum bias found for a given parameter and alternative model across all true oscillation parameter values tested.}
\centering
\begin{tabular}[t]{m{2cm} >{\centering\arraybackslash}p{2cm} >{\centering\arraybackslash}p{2cm} >{\centering\arraybackslash}p{2cm}}
\hline
\hline
Alternative model                    & \multicolumn{3}{c}{Maximum bias on parameter ($\sigma$)} \\
                                     & $\Delta m^{2}_{32}$ & $\sin^{2}\theta_{23}$ & $\sin^{2}\theta_{13}$ \\ \hline
SF                                   & 0.09                & 0.17                 & 0.17 \\
\hline
Effective RPA                        & 0.00                & 0.06                 & 0.00 \\
\hline
Alternative 1p1h          & 0.20                & 0.09                 & 0.07 \\
\hline
Alternative 2p2h          & 0.20                & 0.21                 & 0.18 \\
\hline
Delta-enhanced 2p2h       & 0.10                & 0.10                 & 0.00 \\
\hline
Not-Delta 2p2h            & 0.11                & 0.07                 & 0.05 \\
\hline
\end{tabular}
\label{tab:summary_current_stats}
\end{table}

\section{\nueb appearance analysis}
\label{sec:nuebar}
%%%%%%%%%%%%%%%%%%%%%%%%%%%%%%%%%%%
%%Nuebar appearance at T2K%%%%%%%%%
%%%%%%%%%%%%%%%%%%%%%%%%%%%%%%%%%%%

%%%%%%%%%%%%%%%%%%%%%%%%%%%%%%%%%%%
%%%%%%%%%%Plan%%%%%%%%%%%%%%%%%%%%%
%1. Motivations for the nuebar:%%%% 
%nue observed at T2K but nuebar%%%%
%never observed%%%%%%%%%%%%%%%%%%%%
%2. Method%%%%%%%%%%%%%%%%%%%%%%%%%
%3. Asimov + Toy results%%%%%%%%%%%
%4. Data results%%%%%%%%%%%%%%%%%%%
%5. PMNS testing results%%%%%%%%%%%
%%%%%%%%%%%%%%%%%%%%%%%%%%%%%%%%%%%
%Before moving to the results of the full joint analysis, in this section the results of an analysis designed to search only for \nueb apperance are described.

T2K has already observed $\nue$-appearance in a \num beam~\cite{Abe:2013hdq}, but no direct evidence of \nueb-appearance has been reported so far. 
A search for this process, with the data collected by T2K in \nub-mode, has been performed using the analysis tools described in Sec.~\ref{sec:oamethod}.

 %is investigated not only using the $\nueb$ sample but a joint fit on the different samples defined in Sec.~\ref{sec:sk}. 
%The fit extracts the model parameters from observable samples. In this analysis, the fit parameters are the oscillation and systematic uncertainty parameters. They are both treated as nuisance parameters, and marginalized over. The joint fit method allows to maximize the predictive power of the datasets and avoid any bias that may arise from neglected correlations. The full T2K dataset is used.

%\subsection{Methodology}
%\subsection{The hypothesis test}
In order to look for \nueb-appearance independently from \nue-appearance, a parameter $\beta$ is introduced which multiplies the \nueb-appearance probability
%The $\nueb$ appearance is tested independently from the $\nue$ appearance. This is done by defining a parameter $\beta$ that multiply the \nueb appearance probability:
% the $\nueb$ appearance probability is multiplied by a factor $\beta$:
\begin{equation}
P(\numb \rightarrow \nueb) \rightarrow \beta \times P(\numb \rightarrow \nueb)
\end{equation}
In this analysis, $\beta$ can have two values: $\beta=0$, which corresponds to no $\numb\rightarrow\nueb$ oscillations; and $\beta=1$, which corresponds to the appearance probability as predicted by the PMNS framework. 
%Both hypotheses are tested by estimating their p-values%, in the first and final parts of this section respectively.

%\subsection{The test statistics}
The two hypotheses are tested using rate-only and rate-plus-shape analyses. The test statistic is different in the two cases. The test statistic used in the rate-only analysis is the number of e-like candidates observed at SK in \nub-mode, 
%while for the rate-plus-shape analysis, a model comparison between the no-$\nueb$ appearance and the PMNS predicted $\nueb$ appearances is used in order to maximize the predictive power
while for the rate-plus-shape analysis, the likelihood ratio for the two hypotheses, assumining PMNS oscillations and no $\numb \rightarrow \nueb$ is used:
\begin{align}
\label{Eq:dchi2}
  \Delta \chi^{2} &= \chi^{2}(\beta=0)-\chi^{2}(\beta=1)  \nonumber \\
  &= 2 \Big[ -\; \textrm{ln} \; \mathcal{L}(\beta=0;x) + \; \textrm{ln} \; \mathcal{L}(\beta=1;x) \Big],
\end{align}
%\noindent which is equivalent to the ratio between the likelihood assuming PMNS oscillations and the likelihood assuming no $\numb \rightarrow \nueb$ appearance. 
%The Neyman-Pearson Lemma ensure that the likelihood ratio is the most powerful discriminant between two models.

%\subsection{The toy ensemble}
A toy ensemble is generated according to the prior knowledge on the oscillation parameters defined in Sec.~\ref{sec:pmnspriors} for both values of $\beta$. 
%A value of 0 of the $\beta$ parameter is chosen to test the no-$\nueb$ appearance hypothesis, and a value of 1 is chosen for the test of the PMNS model hypothesis.
The T2K information from the $\num$-disappearance, $\numb$-disappearance and $\nue$-appearance~\footnote{Note that the e-like \ccpip sample was not included in this analysis, only CCQE-like samples were used.} channels are taken into account by using a posterior predictive method~\cite{demortier2008p}. In this method, the data from these channels are used to constrain the prior parameter space in generating the toy ensemble for the \nueb sample while preserving their correlations. 

%A joint fit method is used between the $\num$, $\numb$, $\nue$ and $\nueb$ samples. In the rate-only and rate+shape analyses, a posterior predictive method~\cite{demortier2008p} is used. The data from the $\num$, $\numb$ and $\nue$ channels are used to constrain the prior parameter space, on top of the other experiments. Therefore, the toy ensemble as well as the marginalization space are constrained by the $\num$, $\numb$ and $\nue$ samples.

The sensitivity of the analysis is computed by producing a simulated sample without statistical fluctuations using the values of the oscillation parameters defined in Tab.~\ref{tab:nomosc_par}.
The test statistics for this particular dataset are 6.28 expected \nueb candidates for the rate-only analysis, corresponding to the total number of expected events in \nub-mode in Tab.~\ref{tab:SK_nue_events}, and $\Delta \chi^{2} = 2.54$ for the rate-plus-shape analysis. When compared to the generated toy ensembles for the no-\nueb appearance hypothesis the p-values are 0.047 for both, the rate-only and rate-plus-shape analyses. When compared with the toy ensemble generated with $\beta=1$, p-values of 0.52 and 0.41 are found for the rate-only and rate-plus-shape analyses respectively.

The same analysis is then applied to the $e$-like data selected in \nub-mode. The test statistics in data for the rate-only and rate-plus-shape analyses are $N^{obs}=4$ $\nueb$ candidates and $\Delta \chi^{2} = -2.51$, obtained from the \{\erec, $\theta_{lep}$\} distribution of the $e$-like candidates in \nub-mode shown in Fig.~\ref{fig:2d_templ_fhc_ccqe_sig_bkg}. 
As shown in Tab.~\ref{Tab:NUEB:DataPvalue} this test statistics return a p-value of 0.41 (0.46) for the $\beta=0$ hypothesis for the rate-only (rate-plus-shape) analysis, while the p-value for $\beta=1$ is 0.21 (0.07). This analysis shows that, with the available data, the rate-plus-shape analysis excludes $\beta=1$ at 90\% CL. 

The observed p-values for the $\beta=0$ hypothesis are larger than expectation due to the lower than expected number of observed events in the e-like \nub-mode sample.
The lower p-value for the rate $\beta=1$ hypothesis is driven by the larger discrepancy between the selected $\nue$ and $\nueb$ rate (32 and 4 events respectively) than predicted by the simulated dataset generated, even under the assumption of a maximal CP-violation $\delta=-\pi/2$ (28.55 and 6.28 events respectively). The p-value in the rate-plus-shape analysis for the $\beta=1$ hypothesis is reduced further, due to the distribution of the \nueb-appearance candidates in electron momentum and angle. Their distribution is more compatible with the background expected distribution, as shown in Fig.~\ref{fig:SK_enurec_nue_plots}. %leading to a mild tension between the T2K data and the PMNS model.
%Since the $\nueb$ event repartition shown on Figure~\ref{fig:SK_enurec_nue_plots} is more compatible with background than oscillated signal, the p-value obtained for the rate and shape analysis is further reduced to $0.07$. It leads to a mild tension between the T2K data and the PMNS model.

\begin{table}[htb]
  \begin{center}
   \caption{Expected and observed (data) p-values for the $\beta=0$ and $\beta=1$ hypotheses, for both rate-only and rate-plus-shape analyses. The expected p-values are estimated for the simulated dataset defined in 
  %Table~\ref{tab:fd_osc_pars}.
  Tab.~\ref{tab:nomosc_par}.}
     \begin{tabular}{c|c|c}
      \hline
      \hline
       & Rate-only & Rate-plus-shape \\
      \hline
      Simulated sample & & \\
     $\beta=0$ & 0.047 & 0.047 \\
      $\beta=1$ & 0.52 & 0.41 \\
      \hline
      Data & & \\ 
      $\beta=0$ & 0.41 & 0.46 \\
      $\beta=1$ & 0.21 & 0.07 \\
      \hline
      \hline
    \end{tabular}
  \label{Tab:NUEB:DataPvalue} 
  \end{center}
\end{table}

\section{Joint neutrino and antineutrino oscillation analysis results}
\label{sec:oaresults}
%Describe OA results including the CC1pi sample with run1-7c.
%Expected plots for appearance sector: 1D \dcp (including FC only for results with reactor constraints), 2D \dcp vs \sto with and without reactor constraints.
%It would be nice to have one plot with the FHC and RHC data fit separately.
%1D plot for \dcp should show the 2 hierarchies with respect to the global minimum, 2D \dcp vs \sto should have the two hierarchies separated.
%Comparison between joint fit and joint fit + CC1pi sample
%Expected plots for disappearance sector: , 2D \stt vs \dmsq (one plot also showing the comparison with other experiments). We won't do FC for the disappearance sector so just show the chi2 curves corresponding to 68 and 90 CL.
%Include also the table with the posterior probability for hierarchy and octant with and without reactor constraints.
%Also show the postfit spectra for the 5 samples.
%Expected length: 6 pages.
In this section, joint three-flavor oscillation analyses performed with both the frequentist and the Bayesian approaches are presented. The five SK samples introduced in Sec.~\ref{sec:sk} are used, which allows the simultaneous study of the \nue and \nueb appearance channels and the \num and \numb disappearance channels. The oscillation parameters $|\dmsq|$, \stt, \sto, \dcp, and the mass ordering are determined with and without using the measurement of \sto from reactor experiments as a constraint.

\subsection{Frequentist analysis}
\label{sec:oaresfreq}
Although two frequentist analyses were introduced in Sec.~\ref{sec:oafrequentist}, the results are similar, so detailed results are only presented for the analysis which uses \{$\erec$, $\theta_{lep}$\} templates in this section. Comparisons between both frequentist, and the Bayesian analysis are shown in Sec.~\ref{sec:oarescomparison}.

\subsubsection{Results without reactor constraints}
\label{sec:oareswithoutreac}
\begin{figure*}[htbp]
\centering
%\begin{subfigure}[b]{0.49\textwidth}
\includegraphics[width=0.49\textwidth]{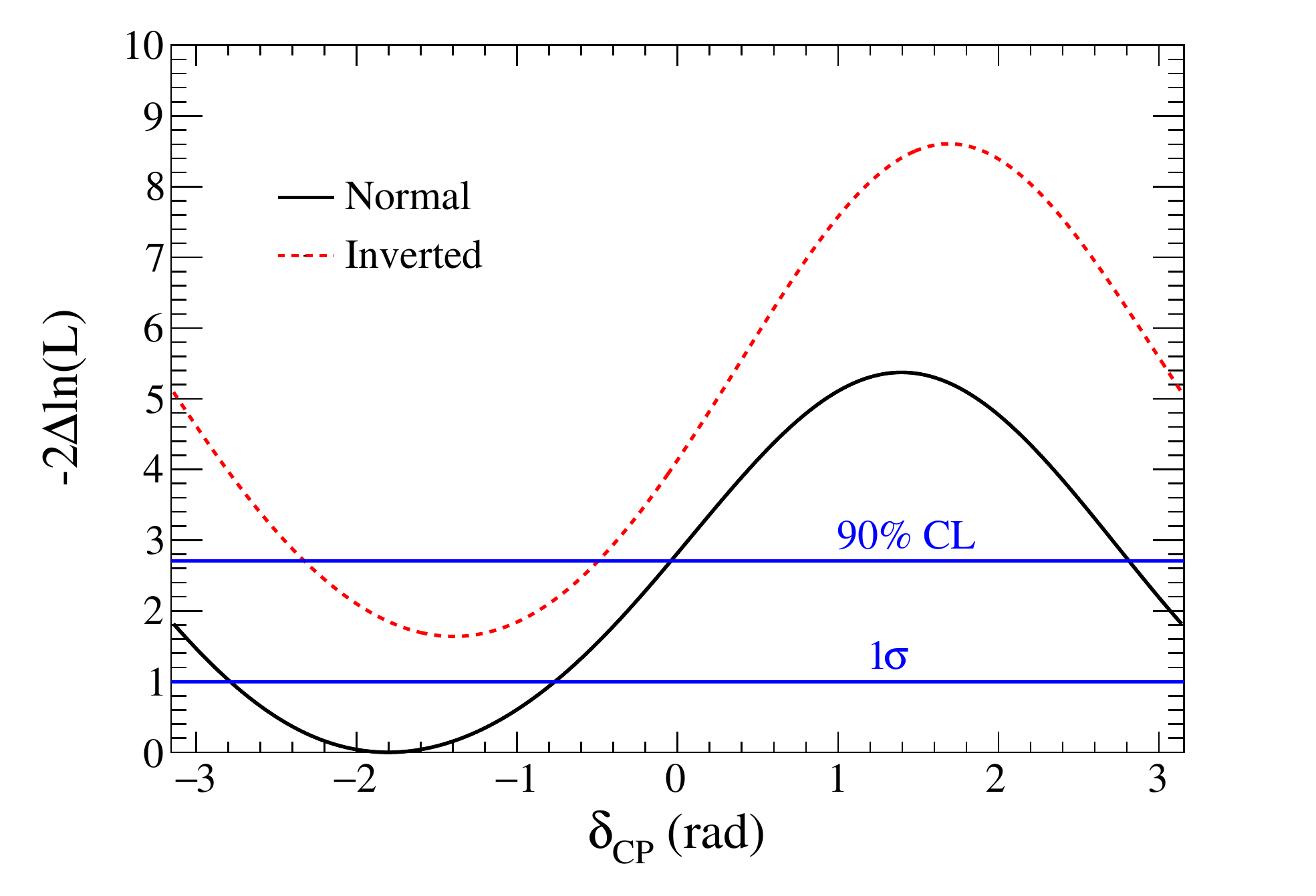}
%\caption{$\delta_{CP}$}
%\end{subfigure}
%\begin{subfigure}[b]{0.49\textwidth}
\includegraphics[width=0.49\textwidth]{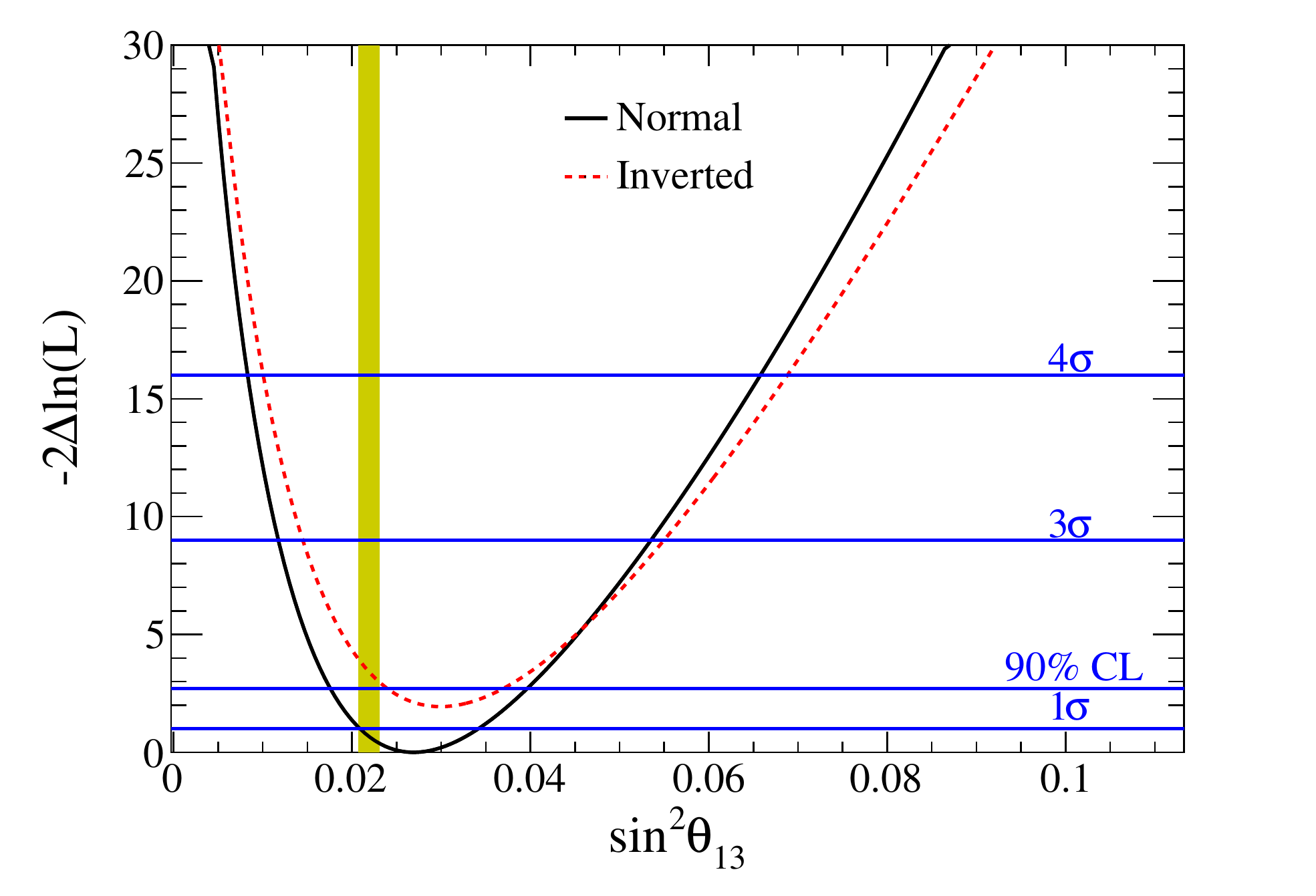}
%\caption{$\sin^2 \theta_{13}$}
%\end{subfigure}
\caption{
One-dimensional $\Delta\chi^2$ surfaces for oscillation parameters $\delta_{CP}$ and $\sin^2\theta_{13}$ using T2K-only data. The yellow band on the right plot corresponds to the reactor value on $\sin^2\theta_{13}$ from the PDG 2015~\cite{PDG2015}.
%The $\Delta\chi^2$ surfaces are produced by marginalising the likelihood with respect to all parameters other than the parameter of interest. %Contours produced with the ERec-Theta analysis. 
}
\label{fig:dcp_th13_t2konly}
\end{figure*}

This section describes the results obtained by the frequentist analysis when only T2K data are used to estimate the oscillation parameters. 
The main parameters of interest in this case are \dcp and $\sin^2\theta_{13}$ that can be directly compared to the reactor measurements.
The point estimates for these oscillation parameters  and the constant $\Delta\chi^2=1$ intervals are given for normal and inverted ordering in Tab.~\ref{tab:dcp_th13_t2konly}. The $\Delta\chi^2$ surfaces are shown in Fig.~\ref{fig:dcp_th13_t2konly}. These intervals have been produced via marginalization of all nuisance and oscillation parameters which are not of interest, as described in Sec.~\ref{sec:oafrequentist}.  

%\begin{table}[htb]
%\centering
%\begin{tabular}{| c | c || c | c | c || c | c | c |}
%\hline
%									& \multicolumn{3}{c||}{\bf Normal Ordering} & \multicolumn{3}{c|}{\bf Inverted Ordering} \\
%\hline										
%\bf Parameter									& Best-fit				& $\pm 1\sigma$ 	  	 	& Best-fit		 			& $\pm 1\sigma$ 	 \\
%\hline
%$\delta_{CP}$            				        			& -1.791  				& [-2.789; -0.764]			& -1.382					& [-2.296;-0.524]						\\
%\hline
%$\sin^2 \theta_{13}$    				             		& 0.0271 				& [0.0209; 0.0342]			& 0.0299					& [0.0232; 0.0380]							\\
%\hline
%\end{tabular}
%\caption{
%Point estimates and 1$\sigma$ confidence intervals under the constant $\Delta\chi^2$ approximation from an analysis considering T2K oscillation data only.
%}                                                                                                                                                                                          
%\label{tab:dcp_th13_t2konly}
%\end{table}

\begin{table}[!tbp]
\centering
\caption{Point estimates and 1$\sigma$ confidence intervals under the constant $\Delta\chi^2$ approximation from an analysis considering T2K oscillation data only.}
\begin{tabular}{ ccccc}\hline \hline
\multirow{2}{*}{Parameter}& \multicolumn{2}{c}{Normal ordering} & \multicolumn{2}{c}{Inverted ordering} \\ 
			 		& Best-fit& $\pm1\sigma$ & Best-fit & $\pm1\sigma$ \\ \hline 
 $\delta_{CP}$	&-1.791& [-2.789; -0.764]&-1.382&[-2.296;-0.524]\\  
  $\sin^{2}\theta_{13}$ 	&0.0271&[0.0209; 0.0342]&0.0299&[0.0232; 0.0380]\\ 
   \hline
  \hline
\end{tabular}
\label{tab:dcp_th13_t2konly}
\end{table}

\begin{figure}[htbp]
\center
\includegraphics[width=0.49\textwidth]{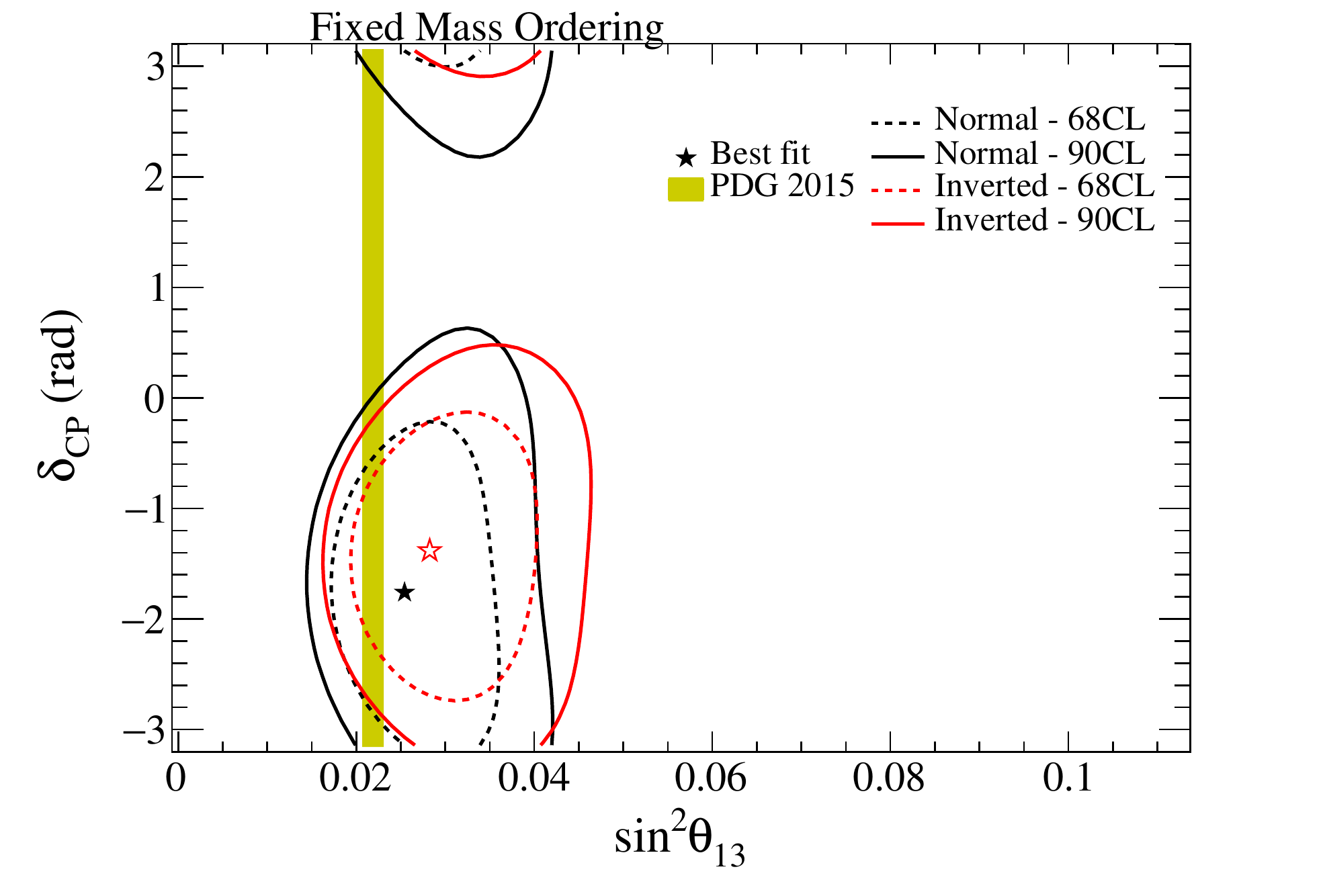}
\caption{
Two-dimensional constant $\Delta\chi^2$ contours for oscillation parameters $\delta_{CP}$ and $\sin^2\theta_{13}$ using T2K data only. The yellow band corresponds to the reactor value on $\sin^2\theta_{13}$ from the PDG 2015~\cite{PDG2015}.
%Produced by marginalising the likelihood with respect to all parameters other than the parameters of interest. %Contours produced with the ERec-Theta analysis.
}
\label{fig:dcp_th13_t2konly_2d}
\end{figure}

Two-dimensional contours of constant $\Delta\chi^2$ for the parameters $\delta_{CP}$ and $\sin^2\theta_{13}$, along with a comparison with the constraint on $\sin^2\theta_{13}$ from reactor experiments, are shown in Fig.~\ref{fig:dcp_th13_t2konly_2d}. The point estimate and constant $\Delta\chi^2$ confidence intervals of $\sin^2\theta_{13}$ from T2K data only are slightly larger than what is found by the reactor experiments. However, the T2K-only measurement of $\sin^2\theta_{13}$ is still consistent with the reactor measurement at the 68\% confidence level.

\begin{figure}[h]
\center
\includegraphics[width=0.49\textwidth]{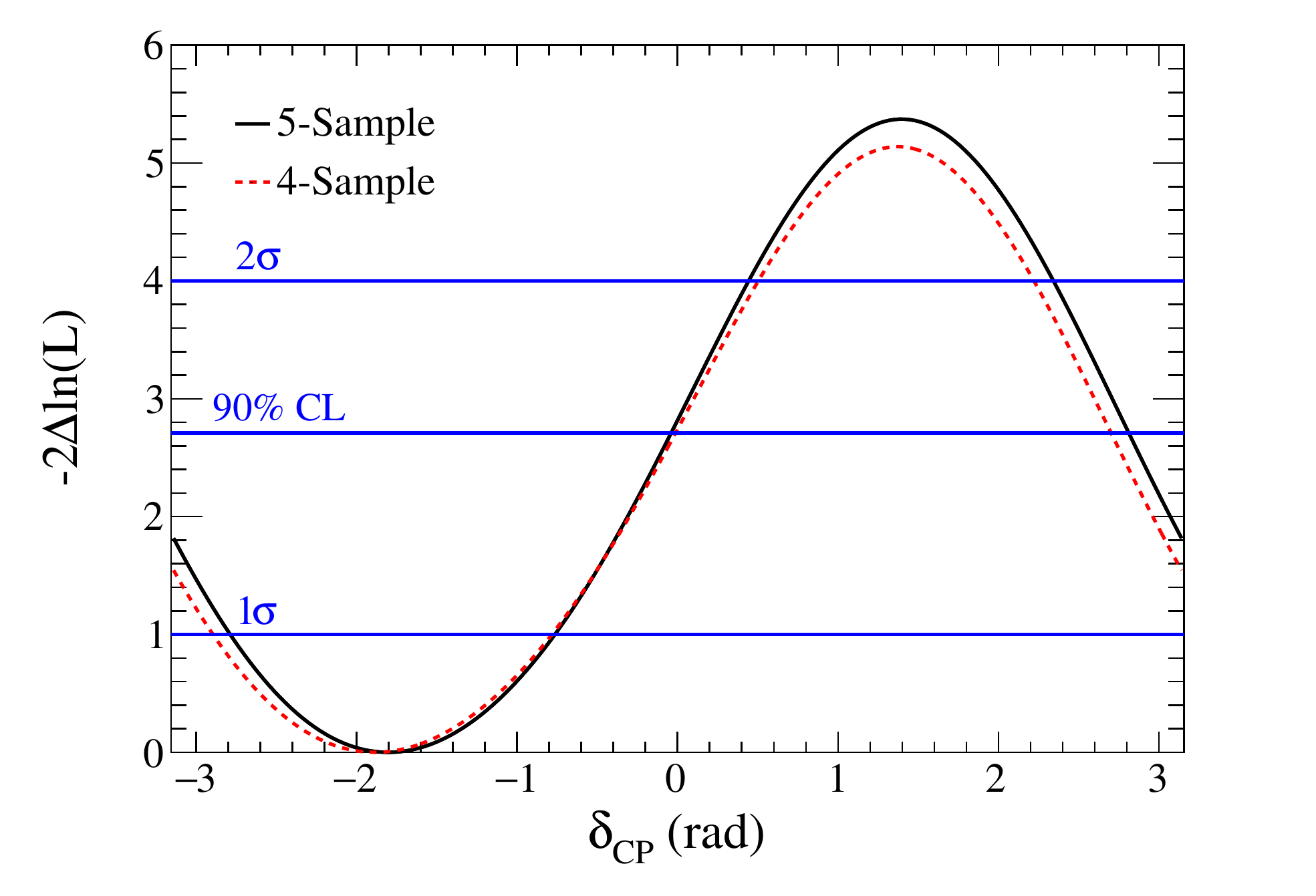}
%\begin{subfigure}[b]{0.49\textwidth}
%\includegraphics[width=0.49\textwidth]{figures/oa_fits/data_contours/2d/cp_13_nh_5sample_vs_4sample.pdf}
%\caption{Normal ordering}
%\end{subfigure}
%\begin{subfigure}[b]{0.49\textwidth}
%\includegraphics[width=0.49\textwidth]{figures/oa_fits/data_contours/2d/cp_13_ih_5sample_vs_4sample.pdf}
%\caption{Inverted ordering}
%\end{subfigure}
\caption{
A comparison of one-dimensional constant $\Delta\chi^2$ contours for normal ordering for $\delta_{CP}$ using T2K-only data for the four- and five-sample fits. %The contour is produced by marginalising the likelihood with respect to all parameters other than the parameter of interest.% Contours produced with the ERec-Theta analysis for the normal ordering hypothesis.
}
\label{fig:dcp_th13_t2konly_2d_4sample_vs_5sample}
\end{figure}

As mentioned above, in this analysis a fifth sample selecting \nue candidates at SK with one delayed Michel electron in the final state has been added for the first time. A comparison of the $\Delta\chi^2$ surface for \dcp only including the four single-ring samples used in previous analyses and the results obtained with the inclusion of the fifth sample is shown in Fig.~\ref{fig:dcp_th13_t2konly_2d_4sample_vs_5sample}. 
As expected, a small improvement is observed when the new sample is included. % with the Asimov dataset. 
%It is also clear that the charged current single pion dataset prefers a slightly larger value of $\sin^2\theta_{13}$ which has an impact on the $\delta_{CP}$ constraint when the dataset is combined with the reactor measurements.

%\todo DAVIDE: IS THE CCQE / CCQE+CC1PI COMPARISON NEEDED? IF YES WE SHOULD SHOW THE 1D DISTRIBUTIONS WHERE WE SHOW THE DELTACHISQ INCREASED BY THE NEW SAMPLE DATA. FROM 2D CONTOURS IT LOOKS LIKE WE ONLY GET A BIAS.

\begin{figure*}
\center
%\begin{subfigure}[b]{0.49\textwidth}
%\includegraphics[width=0.49\textwidth]{./joint_plots/data_contours/Run17c/ccpi/cp_13_nu_vs_nubar_nh_excluison.pdf} 
%\caption{Only 90\% CL - Normal ordering}
%\end{subfigure}
%\begin{subfigure}[b]{0.49\textwidth}
\includegraphics[width=0.49\textwidth]{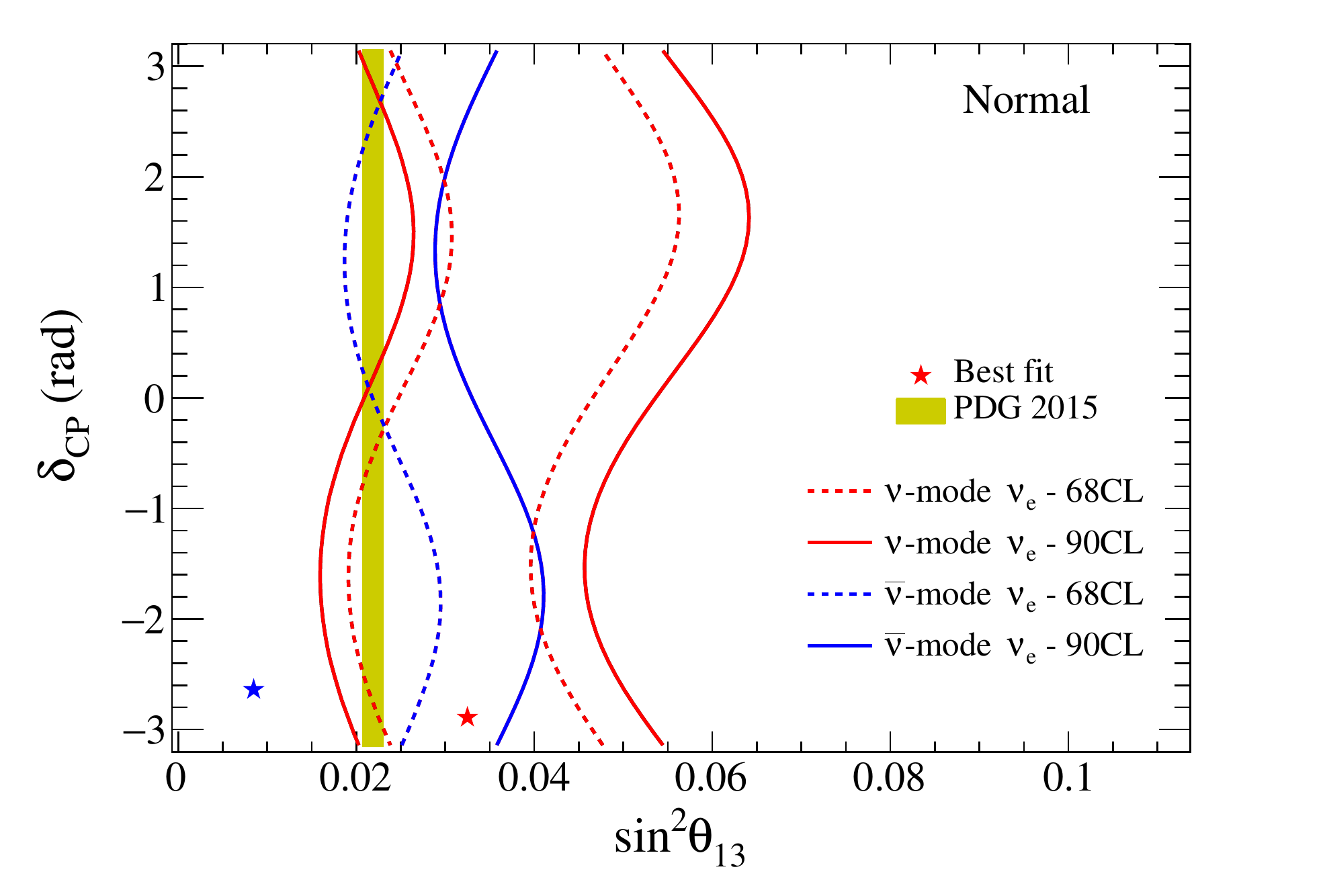} 
%\caption{Normal ordering}
%\end{subfigure}
%\begin{subfigure}[b]{0.49\textwidth}
%\includegraphics[width=0.49\textwidth]{./joint_plots/data_contours/Run17c/ccpi/cp_13_nu_vs_nubar_ih_excluison.pdf}
%\caption{Only 90\% CL - Inverted ordering}
%\end{subfigure}
%\begin{subfigure}[b]{0.49\textwidth}
\includegraphics[width=0.49\textwidth]{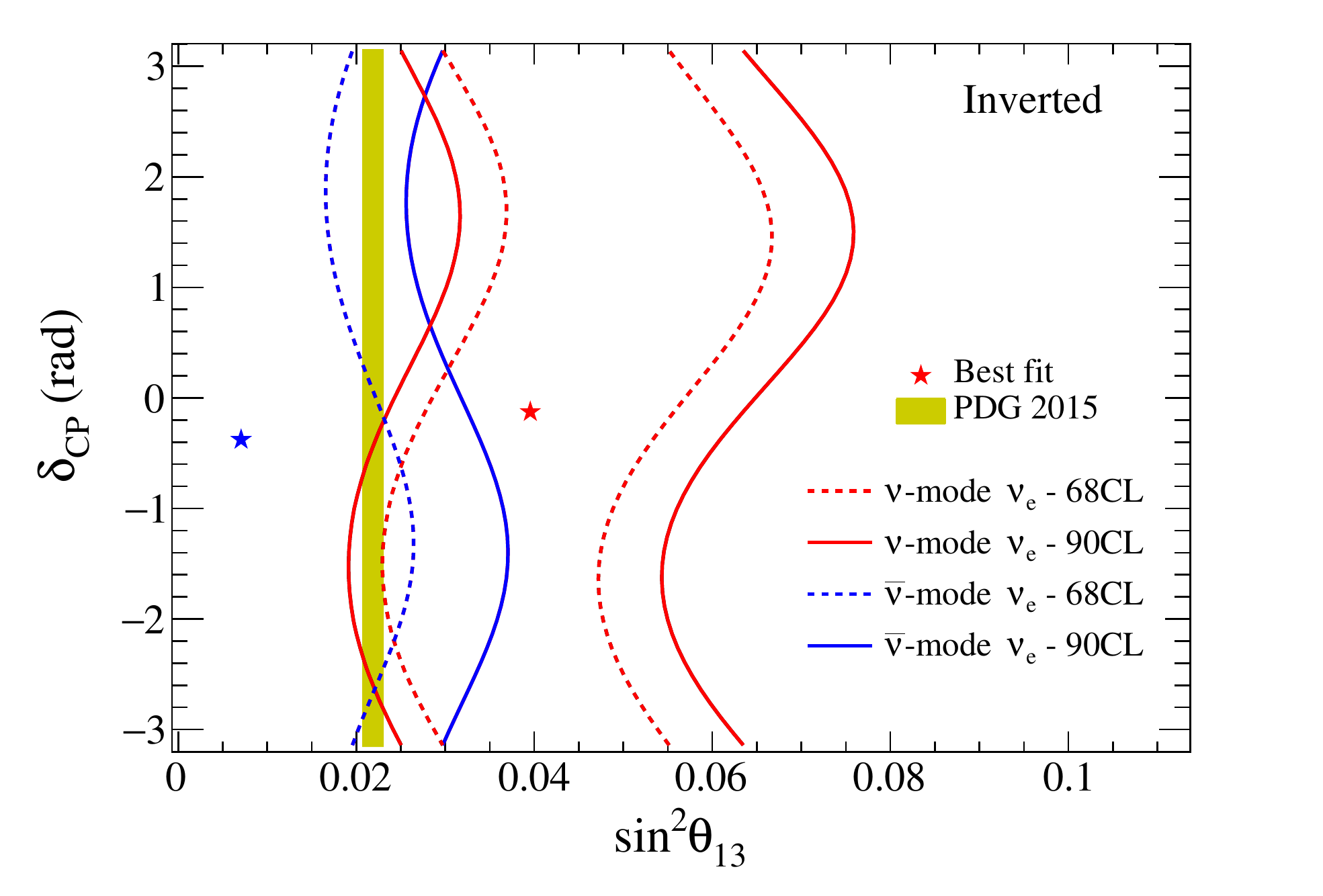}
%\caption{Inverted ordering}
%\end{subfigure}
\caption {
Contours in the $\sin^2 \theta_{13}$--$\delta_{CP}$ plane using T2K-only data,
obtained by analysing either the $\nu$- or \nub-mode appearance datasets are compared
for both orderings. Both $\nu$- and $\bar{\nu}$-mode disappearance datasets were used in all fits.
The yellow band corresponds to the reactor value on $\sin^2\theta_{13}$ from the PDG 2015~\cite{PDG2015}. %Contours produced with the constant $\Delta\chi^2$ approximation.
}
\label{fig:nu_only_vs_nubar_only}
\end{figure*}

Fig.~\ref{fig:nu_only_vs_nubar_only} shows a comparison of the constraints in the $\delta_{CP}$--$\sin^2\theta_{13}$ plane when appearance channels taken in \nunu-mode and in \nub-mode are considered independently. Both \nunu- and \nub-mode disappearance channels are used in both fits. The \nunu- and \nub-mode datasets alone prefer different values of $\sin^2\theta_{13}$, which is driven by the absolute appearance rate. It is clear that the \nub-mode appearance sample does not have the power to exclude a zero value of $\sin^2\theta_{13}$ by itself. In either case, the reactor measurement of $\sin^2\theta_{13}$ sits near the upper and lower 68\% confidence contours for the \nub-mode and \nunu-mode samples respectively.

\subsubsection{Results with reactor constraints}
\label{sec:oareswithreac}
%Editor: Ko.\\
Here, the oscillation parameters obtained by the T2K data fit 
where $\sin^2 \theta_{13}$ is marginalized using the reactor constraint given in Tab.~\ref{tab:pdg}. 
%the PDG 2015: $\sin^2 (2\theta_{13}) =0.085 \pm 0.005$. 
The best fit values of the T2K data with the reactor constraint are summarized in Tab.~\ref{tab:bestfit_rc_run1-7c}. 

\begin{figure}[h]
\center
\includegraphics[width=.49\textwidth]{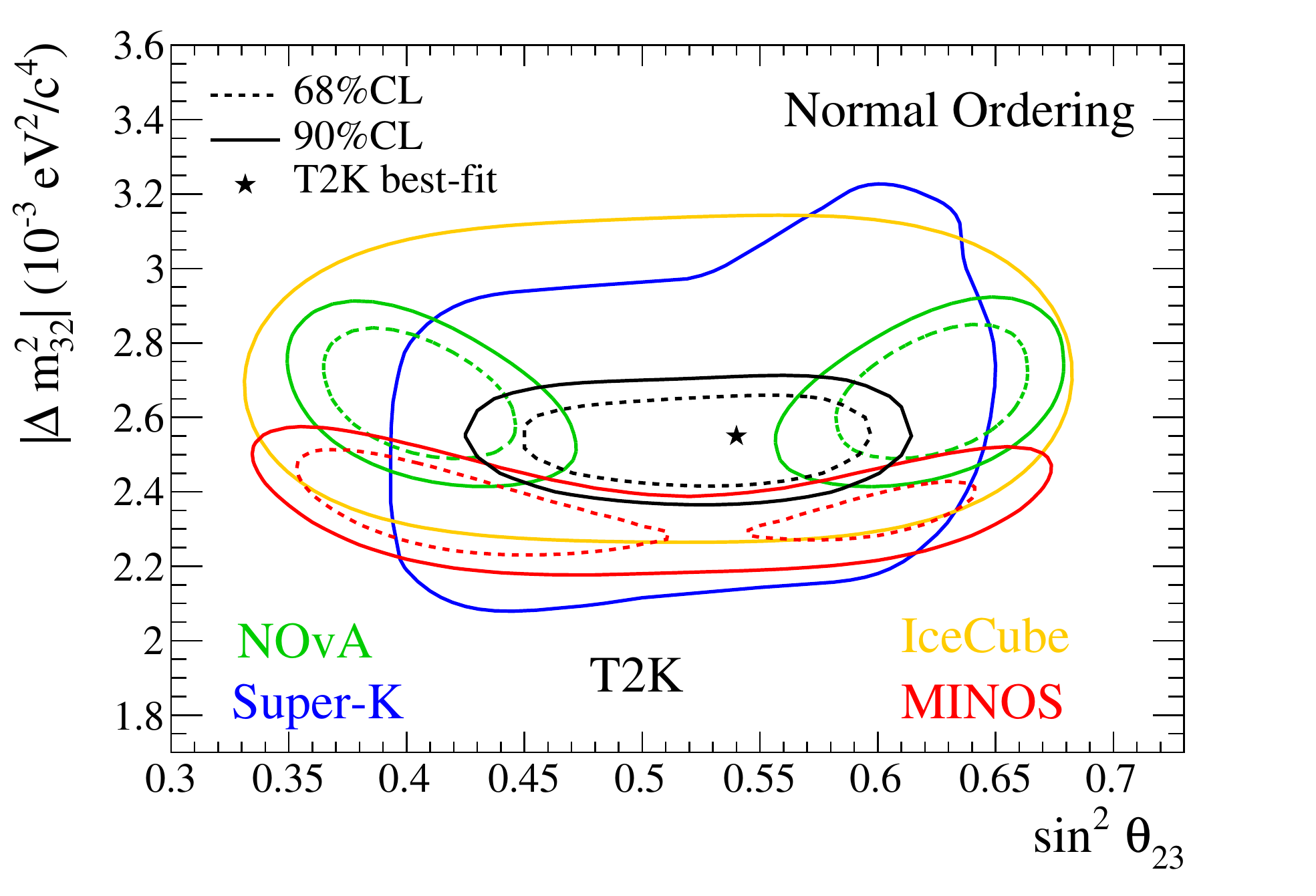}
\caption{Allowed region at 90\% confidence level for oscillation parameters $\sin^2\theta_{23}$ and $\Delta m_{32}^{2} $ using T2K data with the reactor constraint ($\sin^2 (2\theta_{13}) =0.085 \pm 0.005$). 
The normal mass ordering is assumed and the T2K results are compared with NO$\nu$A~\cite{Adamson:2017qqn}, MINOS~\cite{PhysRevLett.110.251801}, Super-K~\cite{Wendell:2015onk}, and IceCube~\cite{Aartsen:2016psd}.
}
\label{fig:sth23dm32}
\end{figure}

Fig.~\ref{fig:sth23dm32} shows the 90\% constant $\Delta\chi^2$ surface in the $\sin^2\theta_{23}$--$\Delta m_{32}^{2}$ plane, assuming normal mass ordering.
The interval is compared with other experiments, showing good agreement with IceCube and SK and some tension with MINOS and NO$\nu$A.

\begin{figure*}[htbp]
\center
\includegraphics[width=.49\textwidth]{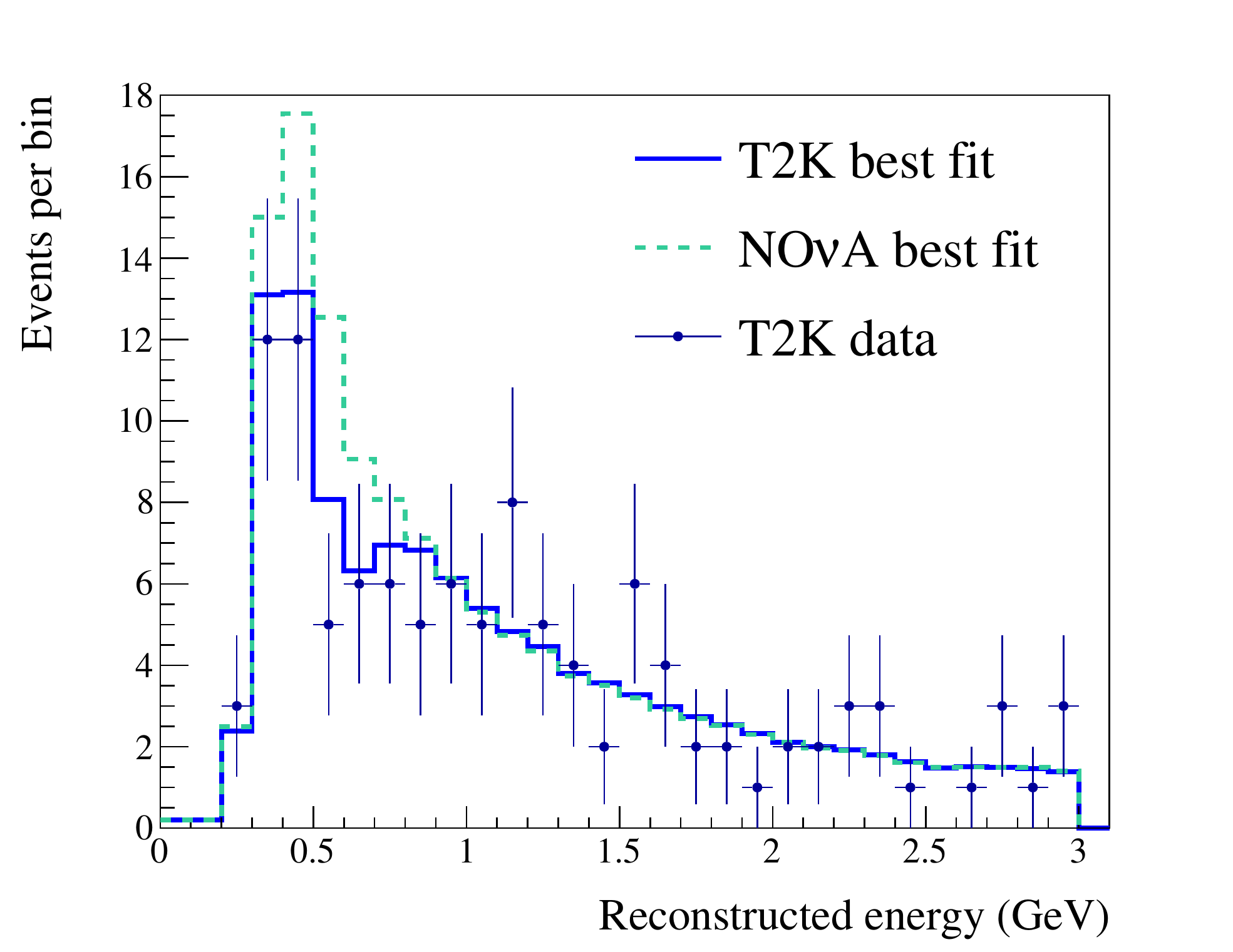}
\includegraphics[width=.49\textwidth]{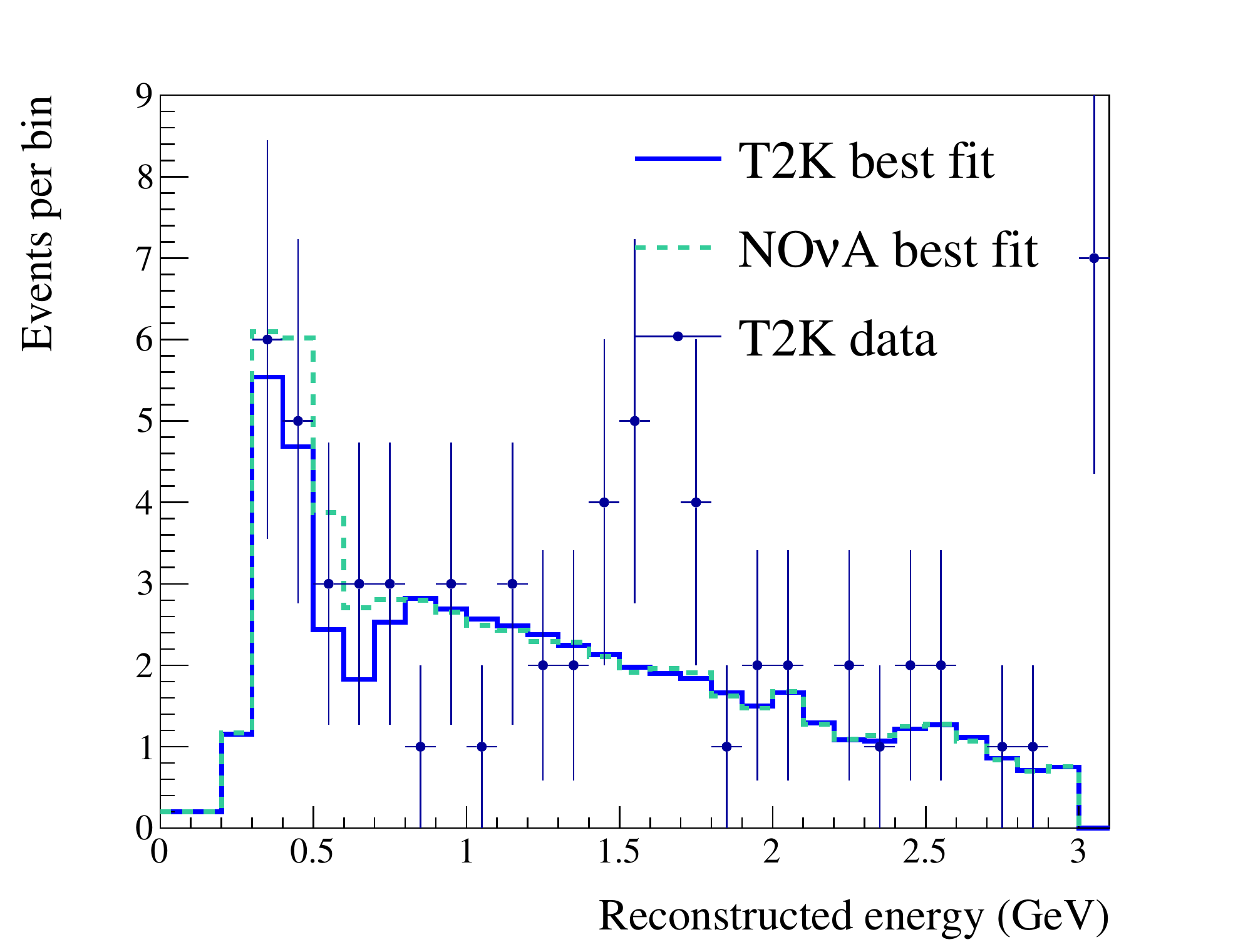}
\caption{Comparison of the T2K data in \num (left) and \numb (right) disappearance channels with the expected spectra obtained with the T2K most probable values of the oscillation parameters and using the NO$\nu$A most probable values for \stt (higher octant) and \dmsq taken from Ref.~\cite{Adamson:2017qqn}. }
\label{fig:novat2kbestfit}
\end{figure*}

%\todo ADD SPECTRA WITH T2K AND NOVA BEST FIT
The NO$\nu$A collaboration published \num-disappearance results disfavoring maximal mixing for \stt at 2.6$\sigma$~\cite{Adamson:2017qqn}. The T2K data in the \num- and \numb-disappearance channels, together with the T2K best fit and the expected spectrum produced using the NO$\nu$A best fit value for \stt (higher octant) and \dmsq, are shown in Fig.~\ref{fig:novat2kbestfit}. 
%The spectrum are generated using a posterior predictive method sampling 2000 steps from the Markov Chain and using the mean of the events distribution in each bin as the expected value.
% The $\Delta\chi^2$ between the two fits is 7.78.
% $\chi^2/d.o.f.$ for the T2K best fit is XXX (YYY) for \num (\numb) disappearance while by taking the NO$\nu$A best fit the $\chi^2/d.o.f.$ is XXX (YYY) confirming the tensions between the two results.

\begin{figure*}[htb]
\centering
\includegraphics[width=0.49\textwidth]{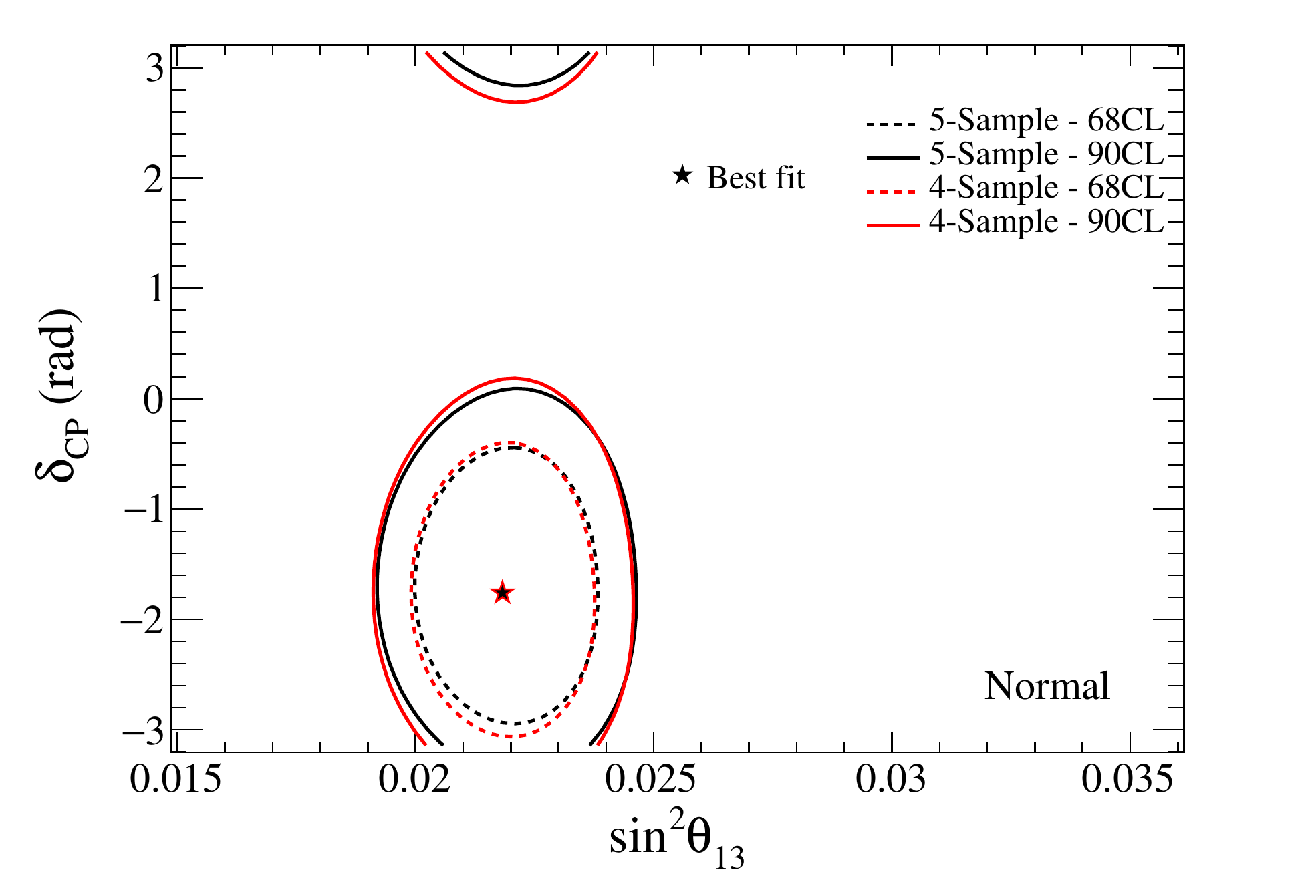}
\includegraphics[width=0.49\textwidth]{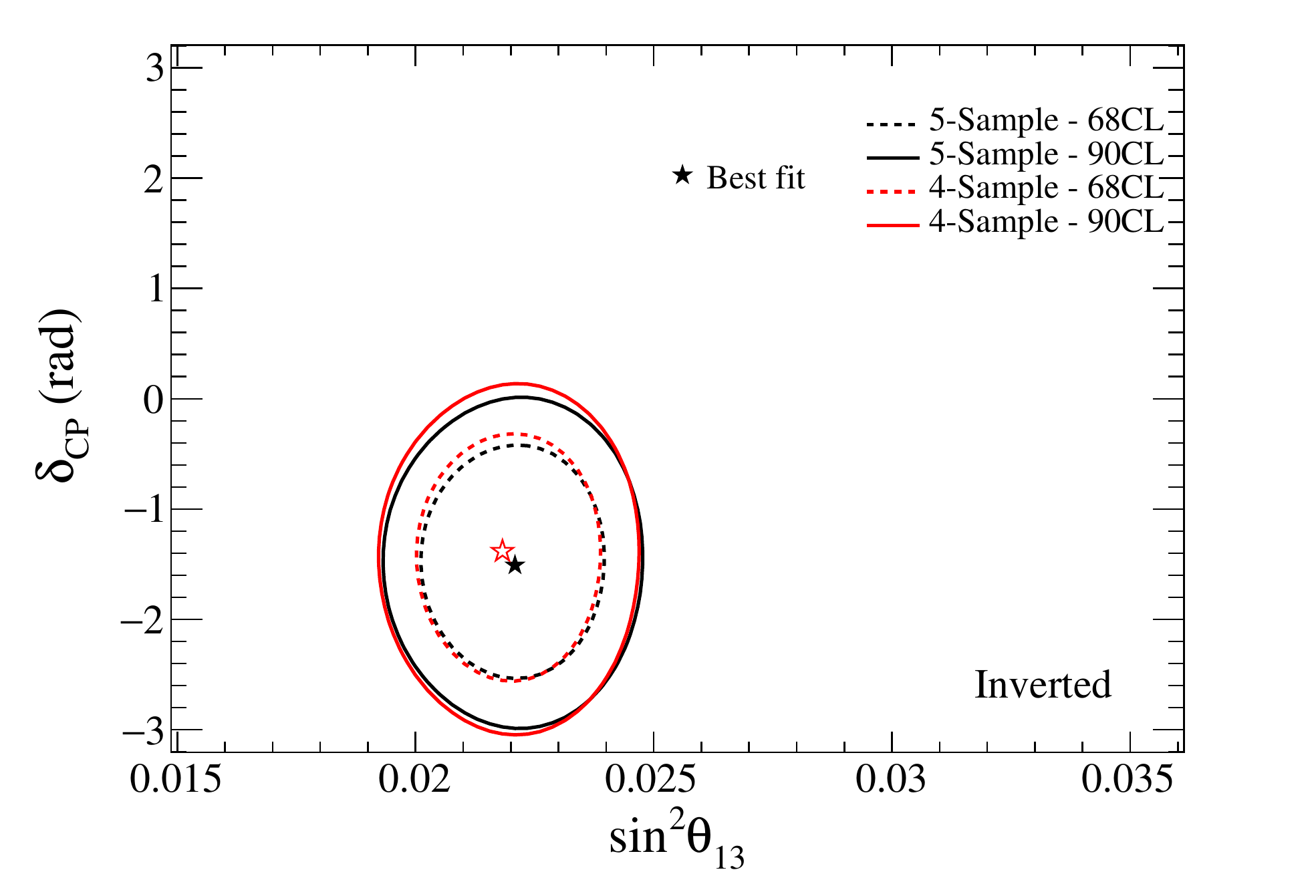}
\caption{
A comparison of two-dimensional constant $\Delta\chi^2$ contours in the $\delta_{CP}$--$\sin^2\theta_{13}$ plane using T2K data with the reactor constraint, for both four-sample (red) and five-sample (black) analyses with normal (left) and inverted (right) mass ordering hypotheses. The contours are produced by marginalizing the likelihood with respect to all parameters other than the parameters of interest. %P-Theta analysis plot; should it be updated with Erec-Theta plot from VALOR?
}
\label{fig:dcpt13_rc}
\end{figure*}

Fig.~\ref{fig:dcpt13_rc} shows the 2D $\sin^2\theta_{13}$--$\delta_{CP}$ confidence level contours for the data fit including the reactor constraint. The comparisons with the four-sample joint fit are also shown to demonstrate the effect of the inclusion of the \ccpip e-like sample in the appearance analysis. Compared to  the best-fit results obtained with the T2K-only data fit in Sec.~\ref{sec:oareswithoutreac}, the inclusion of the \ccpip e-like sample in the data fit with the reactor constraint results in a shift of best-fit value for the $\delta_{CP}$ phase towards the maximally violating phase of $-\pi/2$.

\begin{table}[htbp]
\centering
\caption{Best-fit results and the 1$\sigma$ confidence interval of the T2K data fit with the reactor constraint with normal and inverted hypotheses.}
\begin{tabular}{ ccccc }
\hline
\hline
	\multirow{2}{*}{Parameter}	& \multicolumn{2}{c}{Normal ordering} & \multicolumn{2}{c}{Inverted ordering} \\ 
 							& Best-fit& $\pm1\sigma$ & Best-fit & $\pm1\sigma$ \\ 
 \hline 
 $\delta_{CP}$&-1.728&[-2.538;-0.877]&-1.445&[-2.170;-0.768]\\  
   $\sin^{2}\theta_{23}$ &0.550&[0.465;0.601]&0.5525&[0.470;0.601]\\  

   $\Delta m^{2}_{32}$ &\multirow{2}{*}{2.54}&\multirow{2}{*}{[2.460;2.621]}&\multirow{2}{*}{2.51}&\multirow{2}{*}{[2.429;2.588]} \\
   	($10^{-3}$ eV$^2 / c^4$)	& & & & \\     
	\hline
\hline

%  $\sin^{2}2\theta_{13}$ &0.0855&[0.0807;0.0904]&0.0860&[0.0812;0.0909]\\  \hline
\end{tabular}
\label{tab:bestfit_rc_run1-7c}
\end{table}

%\begin{figure}[htb]
%\includegraphics[width=0.49\textwidth]{figures/oa_fits/data_contours/with_rc/Data_dcp_rc_run1-7c_5sample}
%\caption{
%1D Marginal $\Delta\chi^2$ surfaces for oscillation parameters $\delta_{CP}$ and $\sin^2\theta_{13}$ using T2K data with the reactor constraint ($\sin^2 (2\theta_{13}) =0.085 \pm 0.005$). The contour is produced by marginalizing the likelihood with respect to all parameters other than the parameter of interest. %P-Theta analysis plot; should it be updated with Erec-Theta plot from VALOR?
%}
%\label{fig:dcp_rc}
%\end{figure}

%move to the Bayesian analysis part
%\par Posterior probabilities of different hypotheses are studied by considering the mass ordering and the octant of $\theta_{23}$. The posterior probability of a hypothesis is defined by calculating the marginal likelihood for different hypotheses:
%\begin{eqnarray}
%p(H_{i} \;| \;N^{obs.},\mathbf{x}) & = & \frac{\mathcal{L}_{marg}(N^{obs.},\mathbf{x},H_{i}) P(H_i)}{\sum_{j} \mathcal{L}_{marg}(N^{obs.},\mathbf{x},H_{j}) P(H_j)}
%\end{eqnarray}
%with $\mathbf{x}$ the kinematic bins and $H_j$ the hypothesis which consists of four combinations with the two mass hierarchies and the two octants. The prior probabilities for each hypothesis are set to be equal (0.25).
%\par The posterior probabilities for four combinations of hypothesis with the reactor constraint is provided in \ref{tab:run1-7c_posterior_rc}. The results obtained in this study suggests that the T2K data continues to favor the normal ordering hypothesis.

%\begin{table}[htbp]
%\centering
%\begin{tabular}{| c | c   c | c | }\hline
%&  $\sin^2 \theta_{23} < 0.5$ &$\sin^2 \theta_{23}>0.5$  &Line Total\\ \hline 
%Inverted ordering&0.055  &0.150  & 0.205\\ 
%Normal ordering&0.232  &0.563  & 0.795\\ \hline
%Column Total& 0.287 & 0.713 &1\\ \hline
%\end{tabular}
%\caption{Posterior probabilities for different hypothesis with reactor constraint considered.}
%\label{tab:run1-7c_posterior_rc}
%\end{table}

\begin{figure}[htb]
\center
\includegraphics[width=.49\textwidth]{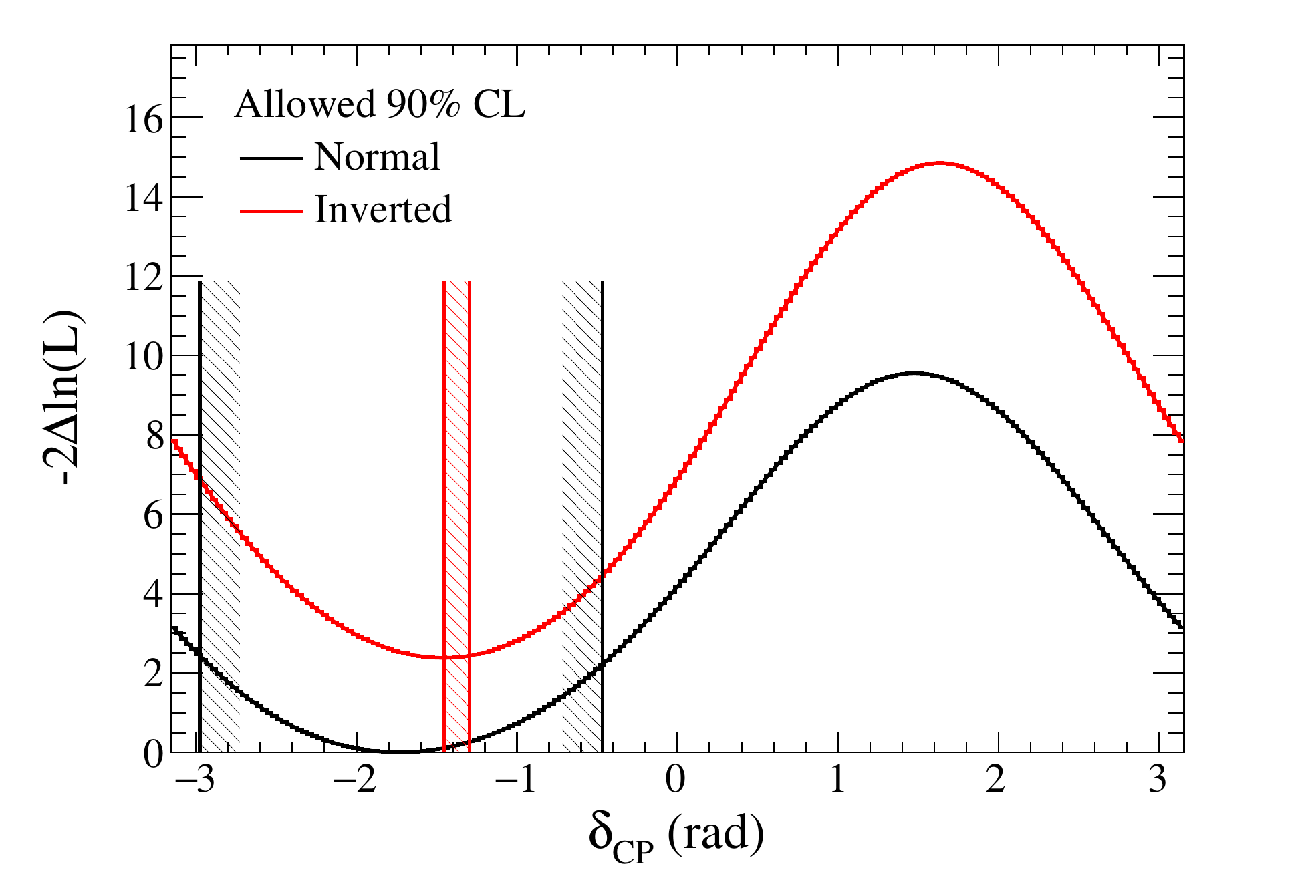}
\caption{One dimensional $\Delta\chi^2$ surfaces for oscillation parameter $\delta_{CP}$ using T2K data with the reactor constraint. 
%The contour is produced by marginalizing the likelihood with respect to all parameters other than the parameter of interest. 
The critical $\Delta \chi^2$ values obtained with the Feldman-Cousins method are used to evaluate the 90\% confidence level with the proper coverage.}
\label{fig:fc_run1-7c}
\end{figure}

Since there is a physical boundary at $\delta_{CP}=\pm \frac{\pi}{2}$, calculating the coverage near the boundary using a Gaussian 
approximation may not produce accurate results. 
To solve this problem, the coverage of the 1D $\Delta \chi^2$ distribution as a function of $\delta_{CP}$ is computed using the Feldman-Cousins approach, described in Sec.~\ref{sec:oamethod}.
In order to perform the study, 10,000 toy MC experiments were generated for different values of $\delta_{CP}$ and the mass ordering.
%Other oscillation parameters are randomly selected according to the posterior likelihood distributions
%,separately for normal and inverted ordering, 
%and the Gaussian distribution of $\sin ^2 2\theta_{13}$ with the reactor constraint. 
The 1D $\Delta \chi^2$ surface obtained with the Feldman-Cousins approach is used to evaluate the 90\% confidence intervals for $\delta_{CP}$ in both ordering cases, as shown in Fig.~\ref{fig:fc_run1-7c}. 
%and the critical values that define the 90\% confidence intervals are computed.
In this analysis, CP-conserving values of $\dcp~=~0,~\pipi$ are excluded at 90\% and 2$\sigma$ confidence levels respectively. Values of \dcp in the intervals  [-2.95,-0.44] ([-1.47, -1.27]) are allowed at 90\% confidence for normal (inverted) ordering.

%According to the proper coverage provided, the inclusion of the $\nu_e$ CC1$\pi^+$-like sample gives the $\delta_{CP}$ values of [-2.953,-0.440] and [-1.791, -1.100] with 90\% confidence level for normal and inverted ordering hypotheses, respectively.

%%%%%
% Discussion of the results

%A well known feature of confidence intervals constructed using the likelihood ratio as the test statistic,  as described in Sec.~\ref{sec:oamethod}, is that they do not show whether the obtained result is consistent or not with the physics model. A statistical fluctuation could favor an extreme case because the model is not allowed to move outside the physical boundaries. 
%
%It therefore becomes useful to compare the expected number of $\nu_e$ events (in both CCQE-like and \ccpip e-like samples) and $\bar{\nu}_e$ candidate events  for different values of $\delta_{CP}$, $\sin^2 \theta_{23}$, and mass ordering with those observed in the T2K dataset. 
%Fig.~\ref{fig:ellipse} shows that the T2K data have approximately a $1\sigma$ statistical fluctuation
%beyond the $\delta_{CP} = -\pi/2$ physical boundary. 
 
\begin{figure}[h]
\center
\includegraphics[width=.49\textwidth]{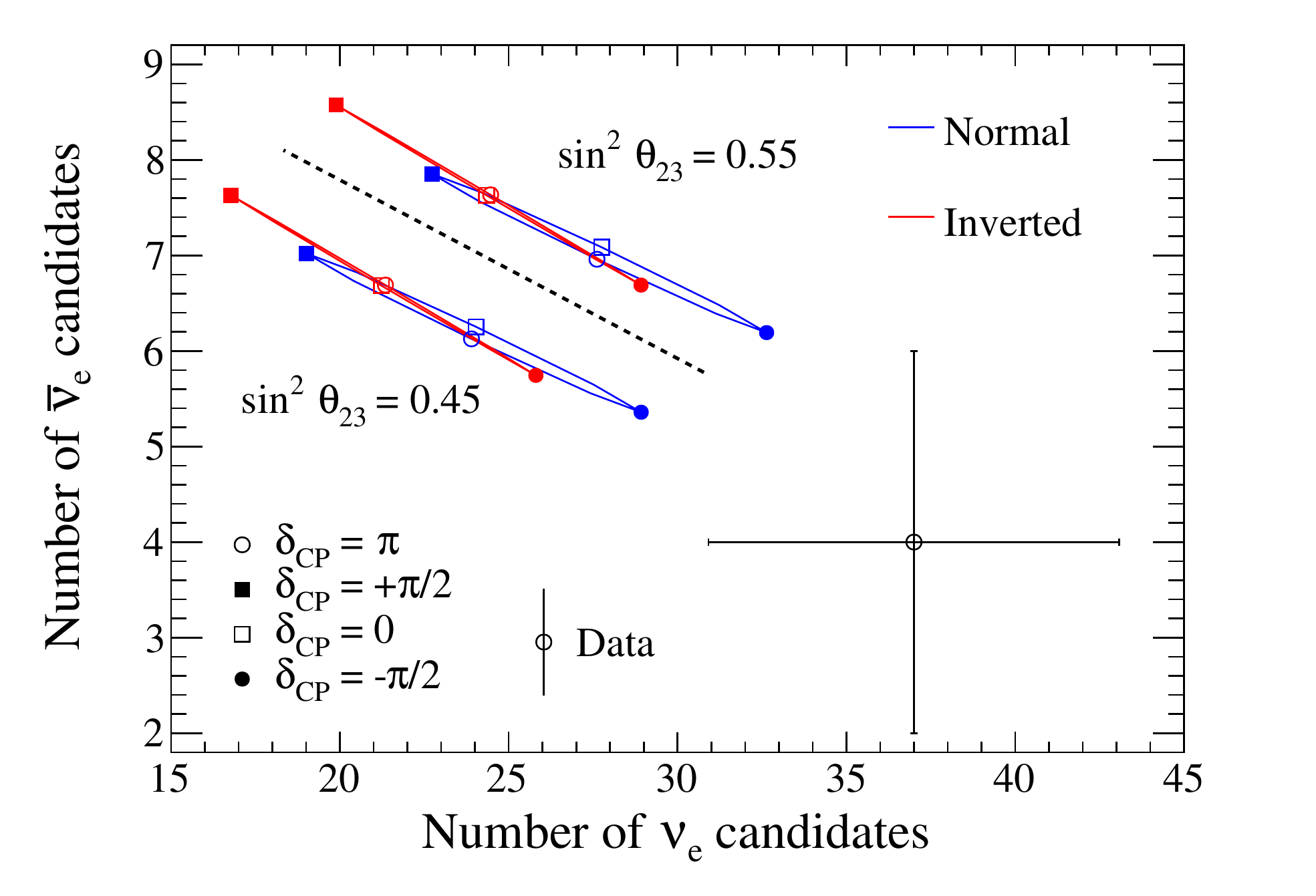}
\caption{Total predicted \nueoreb-appearance event rates in the $\nu$-mode samples and in the $\bar\nu$-mode sample as a function of $\delta_{CP}$ for different values of $\sin^2\theta_{23}$ and both mass orderings, compared to T2K data. The dashed line distinguishes the two solutions for the octant of $\theta_{23}$.}
\label{fig:ellipse}
\end{figure}

A useful way to visualize the results is to compare the observed number of events in the \nunu-mode (in both CCQE-like and \ccpip-like samples) and \nub-mode e-like samples with the expected events for different values of \dcp, $\sin^2 \theta_{23}$, and mass ordering. As it is shown in Fig.~\ref{fig:ellipse} the T2K data falls outside the physically allowed region.

%show the Figure~\ref{fig:ellipse} shows the observed and expected number of \nue (in both CCQE and \ccpip e-like samples) and \nueb candidate events for different values of \dcp, $\sin^2 \theta_{23}$, and mass ordering. 

\begin{figure}[htbp]
\center
\includegraphics[width=.49\textwidth]{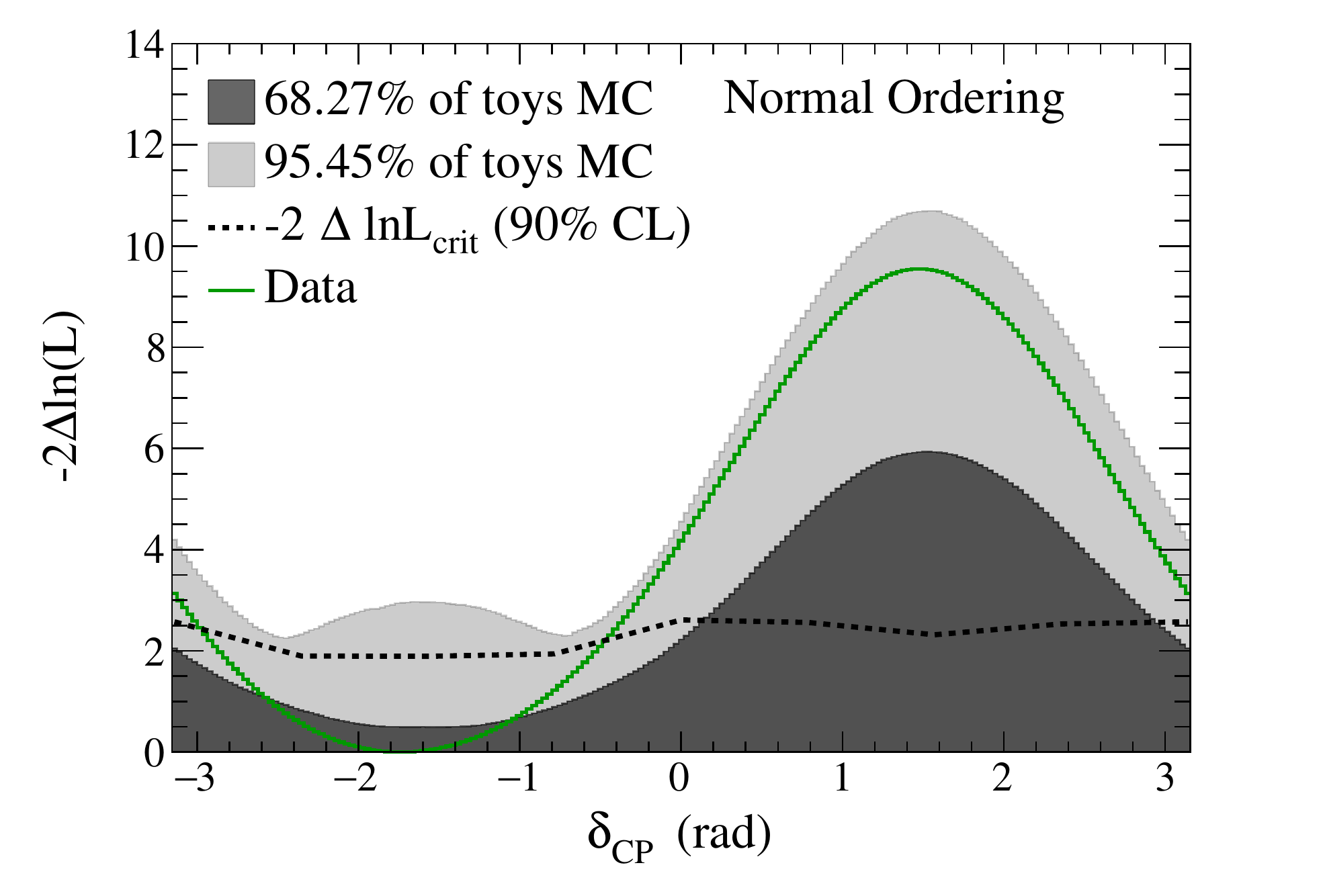}
\includegraphics[width=.49\textwidth]{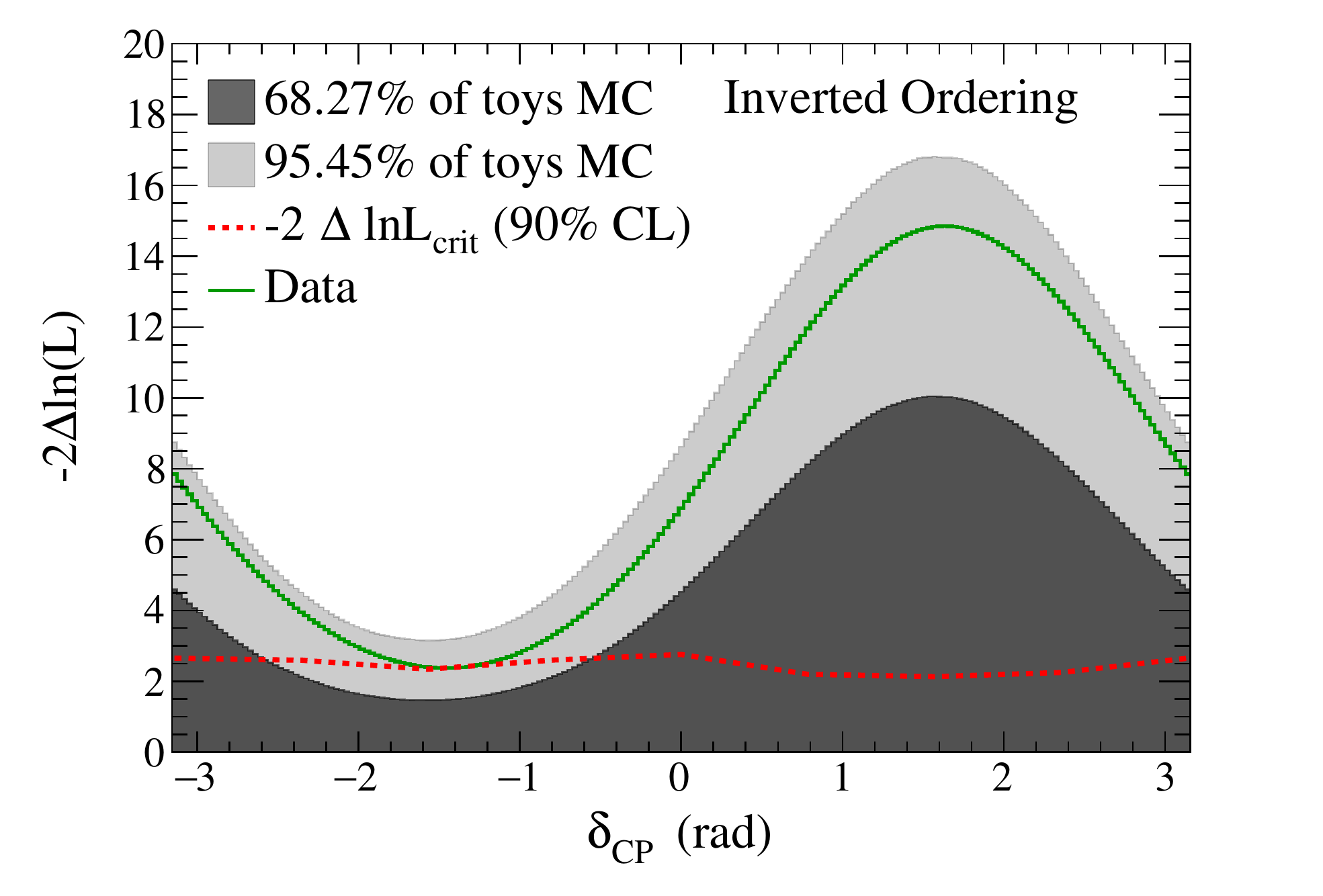}
\caption{One-dimensional marginal $\Delta\chi^2$ surfaces for oscillation parameters $\delta_{CP}$ and $\sin^2\theta_{13}$ using T2K data with the reactor constraint. The contour is produced by marginalizing the likelihood with respect to all parameters other than the parameter of interest. The red line shows the critical $\Delta \chi^2$ values obtained with the Feldman-Cousins method, used to evaluate the 90\% confidence level with the proper coverage. The green line show the $\Delta\chi^2$ obtained with the fit to the T2K data.}
\label{fig:brazilian}
\end{figure}

In order to quantify whether the T2K dataset is consistent with the PMNS framework in terms of significance, an additional toy MC study was performed.
%The goal was to cross-check the whole $-2 \Delta \ln L$ distribution.
An ensemble of 10,000 simulated datasets was obtained in the same way as described in Sec.~\ref{sec:oamethod} for the Feldman-Cousins method,
with $\delta_{CP}=-\pi/2$ and normal mass ordering.
The values of $-2 \Delta \ln L$ that contain 68.3\% and 95.5\% of the MC toys were computed and 
compared to the distribution obtained with the fit of the T2K dataset.
As shown in Fig.~\ref{fig:brazilian}, 
the T2K data $-2 \Delta \ln L$ distribution shows a less extreme fluctuation than at least 5\% of the toys MC 
for all the values of $\delta_{CP}$ and mass ordering,
i.e. if the experiment is repeated many times and the true value is $\delta_{CP} =-\pi/2$ with normal ordering, 
more than 5\% of the experiments are expected to show a more extreme statistical fluctuation
than the current T2K dataset over the whole range of $\delta_{CP}$ and mass ordering. 
%This means that, with the available statistics, T2K data are consistent with the PMNS framework.
From Fig.~\ref{fig:brazilian}, the fraction of experiments that would exclude $\delta_{CP} = 0, \pi$ at 90\% or $2\sigma$ confidence level can be estimated.
Assuming a true value of \dcp of -\pipi/2 and normal ordering, 24.3\% (21.3\%) of toy MC experiments exclude $\dcp=0$ (\pipi) at 90\% CL. The same can be repeated for different values of $\delta_{CP}$ and mass ordering as shown in Tab.~\ref{tab:sensi_fractoys_excluded}.  

%As shown in Tab.~\ref{tab:sensi_fractoys_excluded}, the fraction of toy MC experiments that exclude $\delta_{CP} = 0$ at $2\sigma$ confidence level
%or $\delta_{CP} = \pi$ at 90\%, for data corresponding to $\delta_{CP}$ = $-\pi/2$ with normal ordering, $\delta_{CP}$ = $0$ with normal ordering and $\delta_{CP}$ = $-\frac{\pi}{2}$ with inverted ordering. The 90\% and $2\sigma$ exclusion of $\delta_{CP} = 0$ observed the data is consistent with the exclusion in 24.3\% and 0.131\% of toy MC experiments with true $\delta_{CP} = -\pi/2$ and normal ordering.

\begin{table}[htb]
\centering
\caption{
The fraction of toy experiments for which $\delta_{CP} = 0,\pi$ and normal and inverted ordering
are excluded at 90\% and $2\sigma$ confidence is shown for different true values of $\delta_{CP}$ and mass ordering.
10,000 toy experiments are used for each set of values. 
}

\begin{tabular}{ccccc}
\hline
\hline
\multicolumn{4}{ c }{\textcolor{black}{True: $\delta_{CP} = -\pi/2$ --- normal ordering}}  \\
\bf $\delta_{CP}$	& Ordering	& 90\% CL       	& $2\sigma$ CL  \\ \hline

0 &  Normal & 0.243 & 0.131\\ 

$\pi$ &  Normal & 0.216 & 0.105\\ 

0 &  Inverted & 0.542 & 0.425\\ 

$\pi$ &  Inverted & 0.559 & 0.436\\ 
\hline
\hline
%\multicolumn{4}{| c |}{\textcolor{black}{True: $\delta_{CP} = -1.791$ - normal ordering}}  \\
%\hline
%$\delta_{CP}$		& ordering	& 90\% CL       	& $2\sigma$ CL  \\ \hline
%\hline
% $0$ 				& normal		& 0.191           	& 0.085  	\\ \hline
% $\pi$ 			& normal		& 0.157             & 0.065	\\ \hline
% $0$ 				& inverted		& 0.453             & 0.339 	\\ \hline
% $\pi$ 			& inverted		& 0.465             & 0.345	\\ \hline
%\hline
\multicolumn{4}{ c }{\textcolor{black}{True: $\delta_{CP} = 0$ --- normal ordering}}  \\
$\delta_{CP}$		& Ordering	& 90\% CL       	& $2\sigma$ CL  \\ \hline
0 &  Normal & 0.104 & 0.0490\\ 

$\pi$ &  Normal & 0.130 & 0.0591\\ 

0 &  Inverted & 0.229 & 0.137\\ 

$\pi$ &  Inverted & 0.205 & 0.122\\ 

\hline
\hline
\multicolumn{4}{ c }{\textcolor{black}{True: $\delta_{CP} = -\pi/2$ --- inverted ordering}}  \\
$\delta_{CP}$		& Ordering	& 90\% CL       	& $2\sigma$ CL  \\ \hline
0 &  Normal & 0.124 & 0.0515\\ 

$\pi$ &  Normal & 0.102 & 0.0413\\ 

0 &  Inverted & 0.290 & 0.194\\ 

$\pi$ &  Inverted & 0.308 & 0.207\\ 
 \hline
 \hline
\end{tabular}
\label{tab:sensi_fractoys_excluded}
\end{table}

%\todo Discuss the results with respect to the expected sensitivity (brazilian plots). Add some plots for dm2 and th32 with reactor constraints.

\subsection{Bayesian analysis}

\subsubsection{Results without reactor constraints}
\label{sec:oabreswithoreac}

This section describes the results obtained by the Bayesian analysis when using only T2K data to estimate the parameters $\sin^2 \theta_{23}$, $\Delta m_{32}^2$,  $\sin^2 \theta_{13}$ and $\delta_{CP}$ with the MCMC method described in Sec.~\ref{sec:oabayesian}.
In contrast with the frequentist analysis presented in Sec.~\ref{sec:oaresfreq}, the Markov chain walks in a parameter space where the sign of $\Delta m_{32}^2$ can flip, and results are presented for both mass orderings.
The best-fit point and $\pm1\sigma$ credible interval for each parameter, obtained with the KDE method, are summarized in Tab.~\ref{tab:bestfit_rc_run1-7c_mach3}. The best fit point is the mode of the four-dimensional histogram where the axes are the oscillation parameters.

\begin{table}[htbp]
\centering
\caption{Best-fit results and the 1$\sigma$ credible interval of the T2K data fit without the reactor constraint with the MCMC analyses including both mass orderings.}
\begin{tabular}{ l c c c c }\hline \hline 
 Parameter               & Best-fit                            &  $\pm1\sigma$     \\ \hline 
 $\delta_{CP}$           &  -1.815                             & [-2.275; -0.628]   \\  
 $\sin^{2}\theta_{13}$  &  0.0254                             & [0.0210; 0.0350]   \\  
 $\sin^2 \theta_{23}$    &  0.513                              & [0.460 ; 0.550]   \\  
 \multirow{2}{*}{$\Delta m_{32}^2$}       &  \multirow{2}{*}{$2.539\times 10^{-3}{eV^2/c^4}$}  & $[-2.628;-2.544] \times 10^{-3}{eV^2/c^4}$ \\
                         &                                     & $[2.436; 2.652]\times 10^{-3}{eV^2/c^4}$ \\  \hline\hline
\end{tabular}
\label{tab:bestfit_rc_run1-7c_mach3}
\end{table}

\begin{figure*}[htbp]
        \includegraphics[width=.9\textwidth]{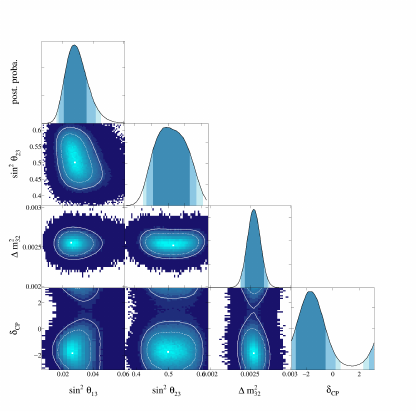}
%        \centering
%        \caption{$\delta_{CP}$--$\sin^{2}\theta_{13}$}
%        \label{fig:1p1h_2d}
%    \end{subfigure}
    \caption{The two-dimensional histograms represent the marginal posterior probability in the two-parameter space as a blue gradient. 
    The white solid (dashed) line is the 90\% ($1\sigma$) credible interval.
    The one-dimensional histograms represent the posterior probability density (post. proba.) of the oscillation parameter in the x-axis of the column marginalized over all other parameters.
    The blue areas are respectively the $1\sigma$ (dark), 90\% (medium), and 95\% (light) credible interval. }
    \label{fig:2dmach3_worc}
\end{figure*}

%The credible interval where the probability of containing the true value is 68.3\% and is in the column $\pm1\sigma$. 
%It has been computed from the posterior probability densities marginalised over all the other parameters as shown in Fig.~\ref{fig:2dmach3_worc}. 
The $\pm1\sigma$ credible intervals, which have a 68.3\% probability of containing the true value, are computed, for each parameter, from the posterior probability density marginalized over all the other parameters as shown in Fig.~\ref{fig:2dmach3_worc}.
Fig.~\ref{fig:2dmach3_worc} also shows the correlations between the oscillation parameters with the map of the marginal posterior density probability and the credible intervals in the space formed by two parameters.

% \begin{figure*}[htbp]
%         \includegraphics[width=.48\textwidth]{figures/oa_fits/data_contours_mach3/without_rc/jointFitNue1pi_data_woRC_1d_bothMH_dcp_ci.pdf}
%        \includegraphics[width=.48\textwidth]{figures/oa_fits/data_contours_mach3/without_rc/jointFitNue1pi_data_woRC_1d_bothMH_sth13_ci.pdf}
%        \includegraphics[width=.48\textwidth]{figures/oa_fits/data_contours_mach3/without_rc/jointFitNue1pi_data_woRC_1d_bothMH_sth23_ci.pdf}
%       \includegraphics[width=.48\textwidth]{figures/oa_fits/data_contours_mach3/without_rc/jointFitNue1pi_data_woRC_1d_bothMH_dm32_ci.pdf}
% 
%     \caption{The posterior probabilities for the four oscillation parameters fitted with the MCMC method.
%     The probabilities are marginalised over all parameters but the one shown on the axis, and take into account both mass hierarchies. 
%     The credible intervals for $\pm1\sigma$, $\pm 90 \%$ and $\pm 95 \%$ are also shown.}
%     \label{fig:1dmach3_worc}
% \end{figure*}

The proportion of the MCMC points with $\sin^2 \theta_{23} > 0.5$ or $< 0.5 $ gives the posterior probability of the octant. 
Similarly, the relative proportion of steps with $\Delta m_{32}^2 > $ or $< 0 $ gives the posterior probability of each mass ordering. 
They are shown in Tab.~\ref{tab:mhmach3_worc}. A Bayes factor can be computed as a ratio of the posterior probabilities~\cite{jeffreys1998}.
The Bayes factor for normal ordering is $B(\mathrm{NH}/\mathrm{IH}) = 2.28$; the Bayes factor for the upper octant is $B(\sin^2 \theta_{23} > 0.5 / \sin^2 \theta_{23} < 0.5) = 1.32$. 
Neither can be considered decisive.

\begin{table}[htbp]
\centering
\caption{Posterior probabilities for the mass orderings and $\sin^2 \theta_{23}$ when fitting T2K data only with an MCMC method.}
\begin{tabular}{l  c   c |  c }\hline \hline
                   &  $\sin^2 \theta_{23} < 0.5$ & $\sin^2 \theta_{23}>0.5$  & Line Total\\ \hline  
Inverted ordering & 0.137                       & 0.168                     & 0.305  \\ 
Normal ordering   & 0.294                       & 0.401                     & 0.695  \\  
\hline
Column total       & 0.431                       & 0.569                     & 1      \\ \hline\hline
\end{tabular}
\label{tab:mhmach3_worc}
\end{table}

\subsubsection{Results with reactor constraints}
\label{sec:oabreswithreac}

This section presents the results obtained with the MCMC analysis when adding a Gaussian prior on $\sin^2 \theta_{13}$ with the value given in Tab.~\ref{tab:pdg}.
The posterior mode marginalized over the nuisance parameters is given in Tab.~\ref{tab:bestfit_worc_run1-7c_mach3}. Including the reactor prior on $\sin^2 \theta_{13}$, the best-fit is closer to that obtained by the reactor experiments compared to the T2K-only results.
The $\delta_{CP}$ best-fit is closer to the maximum violating value of $-\pi / 2$ due to the correlations with $\sin^2 \theta_{13}$ shown in Fig.~\ref{fig:2dmach3_worc}.
\begin{table}[htbp]
\centering
\caption{Best-fit results and the 1$\sigma$ credible interval of the T2K data fit with the reactor constraint with the MCMC analyses including both mass orderings.}
\begin{tabular}{ l c c  c  c  }\hline \hline
 Parameter               & Best-fit                            &  $\pm1\sigma$     \\ \hline 
 $\delta_{CP}$           &  -1.789                             & [-2.450; -0.880]   \\  
 $\sin^{2}\theta_{13}$  &  0.0219                             & [0.0208; 0.0233]   \\  
 $\sin^2 \theta_{23}$    &  0.534                              & [0.490 ; 0.580]   \\  
 \multirow{2}{*}{$\Delta m_{32}^2$}       &  \multirow{2}{*}{$\unit[2.539\times 10^{-3}]{eV^2/c^4}$}  & [-3.000; -2.952] $\times 10^{-3} {eV^2/c^4}$ \\
                         &                                     & [2.424; 2.664]$\times 10^{-3}{eV^2/c^4}$ \\  \hline\hline
\end{tabular}
\label{tab:bestfit_worc_run1-7c_mach3}
\end{table}

\begin{figure}[htbp]
 \includegraphics[width=0.5\textwidth]{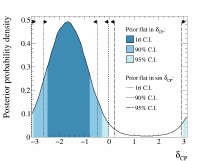}
\caption{\dcp marginal posterior probability as obtained with the MCMC method.
    The credible intervals for $\pm1\sigma$, $\pm 90 \%$ and $\pm 95 \%$ are shown when using a flat prior in $\delta_{CP}$ (usual fit), and when converting to a flat prior in $\sin\dcp$.    }
    \label{fig:1dmach3_wrc}
\end{figure}

%The $\pm1\sigma$ uncertainty on every parameters is the 68.3\% credible interval, shown on Fig.\ref{fig:1dmach3_wrc} for $\delta_{CP}$. 
The MCMC algorithm uses a flat prior on $\delta_{CP}$, but its dependence on this choice of prior has been tested by computing the credible intervals with a flat prior on $\sin\delta_{CP}$.
The two sets of intervals are in reasonable agreement as shown in Fig.~\ref{fig:1dmach3_wrc}.%eventhough the ones obtained with a flat prior on $\sin(\delta_{CP})$ are larger.

The Bayes factor for the mass ordering and the $\theta_{23}$ octant can be computed with the method described in Sec.~\ref{sec:oabreswithoreac}. 
Using the values from Tab.~\ref{tab:mhmach3_wrc}, they are found to be $B(\mathrm{NH}/\mathrm{IH}) = 3.71$ and $B(\sin^2 \theta_{23} > 0.5 / \sin^2 \theta_{23} < 0.5) = 2.39$ respectively. Also in this case, these cannot be considered decisive.
%Although adding the reactor constraint gives higher Bayes factor than fitting T2K data only, they are not still not decisive.

\begin{table}[htbp]
\centering
\caption{Posterior probabilities for the mass orderings and $\sin^2 \theta_{23}$ with an MCMC method when fitting T2K data with the reactor constraint.}
\begin{tabular}{l  c   c | c  }\hline \hline
                   &  $\sin^2 \theta_{23} < 0.5$ & $\sin^2 \theta_{23}>0.5$  & Line Total\\ \hline  
Inverted ordering & 0.060                       & 0.152                     & 0.212  \\ 
Normal ordering   & 0.235                       & 0.553                     & 0.788  \\
\hline
Column total       & 0.295                       & 0.705                     & 1      \\ \hline\hline
\end{tabular}
\label{tab:mhmach3_wrc}
\end{table}

\subsection{Comparison among the oscillation analyses}
\label{sec:oarescomparison}
%Editor: Leila.\\

\begin{figure}[htbp]
 \includegraphics[width=0.5\textwidth]{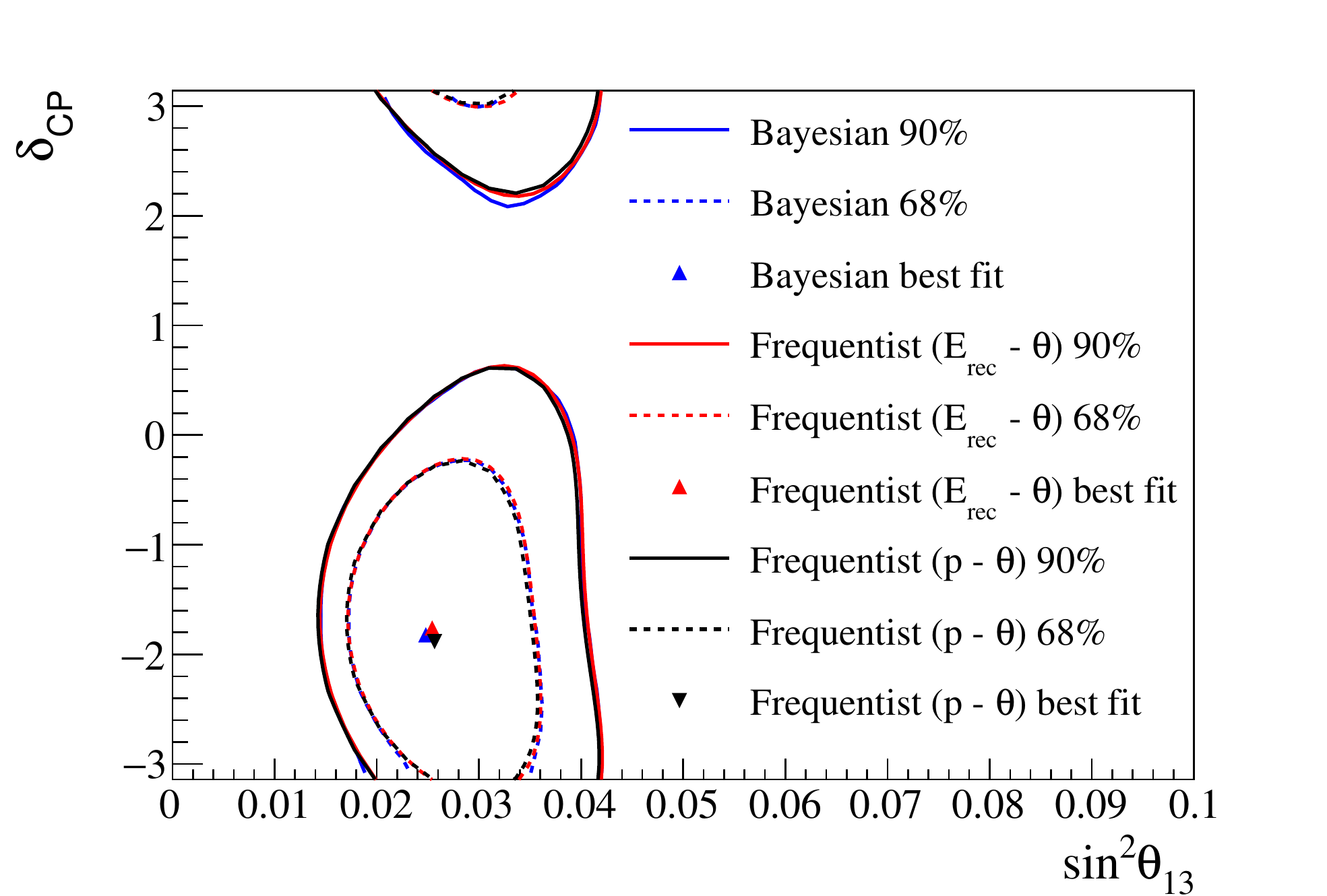}
\caption{Constant $\Delta \chi^2$ 68\% and 90\% intervals in the $\sin^2 \theta_{13}$--$\delta_{CP}$ plane for both frequentist analyses and the Bayesian fit.}
    \label{fig:comparefitters2d_worc}
\end{figure}

%Since the priors of the oscillation parameters are flat, the prior probability for normal and inverted orderings are the same, and the nuisance parameters are Gaussian, the frequentist and Bayesian fits can be directly compared.
The frequentist likelihood multiplied by Gaussian penalty terms for the nuisance parameters and uniform priors for the oscillation parameters is equivalent to the Bayesian posterior density, for the same priors. In order to compare the analyses, the posterior probability densities sampled by the Bayesian analyses are converted into a $\Delta \chi^2$ function and the intervals are recalculated to extract confidence intervals that are compared with the frequentist analyses. 
Fig.~\ref{fig:comparefitters2d_worc} shows the constant $\Delta \chi^2$ 68\% and 90\% intervals for all three oscillation analyses in the $\sin^2 \theta_{13}$--$\delta_{CP}$ plane, assuming normal ordering and only using T2K data. Differences exist among the three methods as the 2D templates fitted in the appearance samples are different and the Bayesian analyses does a combined fit of near and far detector samples but no major differences are found between the contours.
%We remind that the three methods fit the muon rings in reconstructed neutrino energy, and VALOR and MaCh3 fits the electron rings in reconstructed neutrino energy and lepton angle while p-theta fits them in lepton momentum and angle.

\begin{figure}[htbp]
        \includegraphics[width=.48\textwidth]{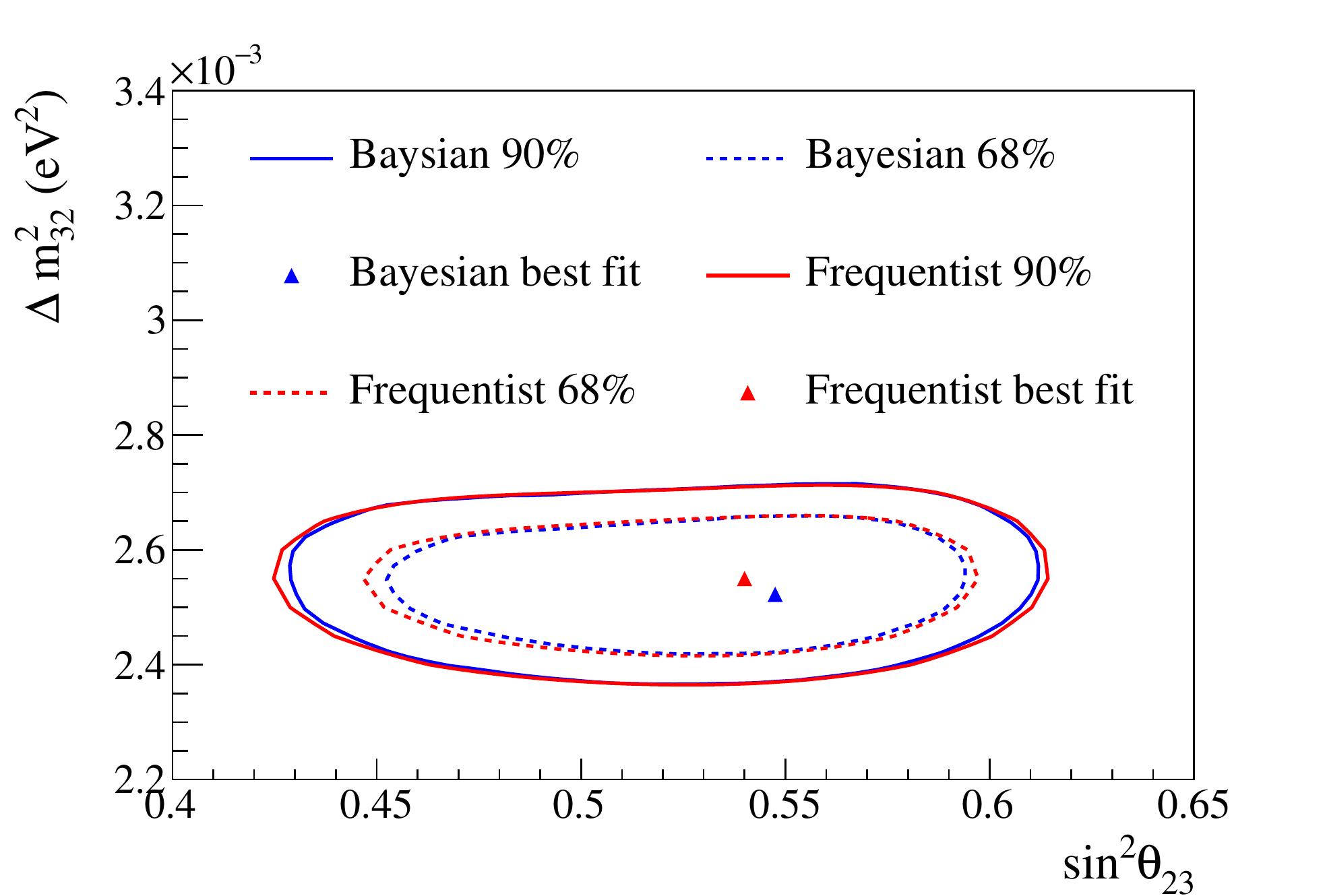}
    \caption{Constant $\Delta \chi^2$ 68\% and 90\% intervals in the $\sin^2 \theta_{23}$--$\Delta m_{32}^2$ plane for the frequentist fit which uses 2D \{$\erec$, $\theta_{lep}$\} templates and the Bayesian fit, assuming normal ordering. }
    \label{fig:comparefitters2d_wrc}
\end{figure}

Fig.~\ref{fig:comparefitters2d_wrc} shows the constant $\Delta \chi^2$ 68\% and 90\% intervals in the $\sin^2 \theta_{23}$--$\Delta m_{32}^2$ plane for both frequentist and Bayesian fits.
Both distributions and intervals agree between fitters, validating the extrapolation of the constraints on the nuisance parameters obtained in the near detector fit to SK.

\subsection{Best fit spectra}
\label{sec:oasum}
%Editor: Leila.\\

\begin{figure*}[!htbp]
     \includegraphics[width=.48\textwidth]{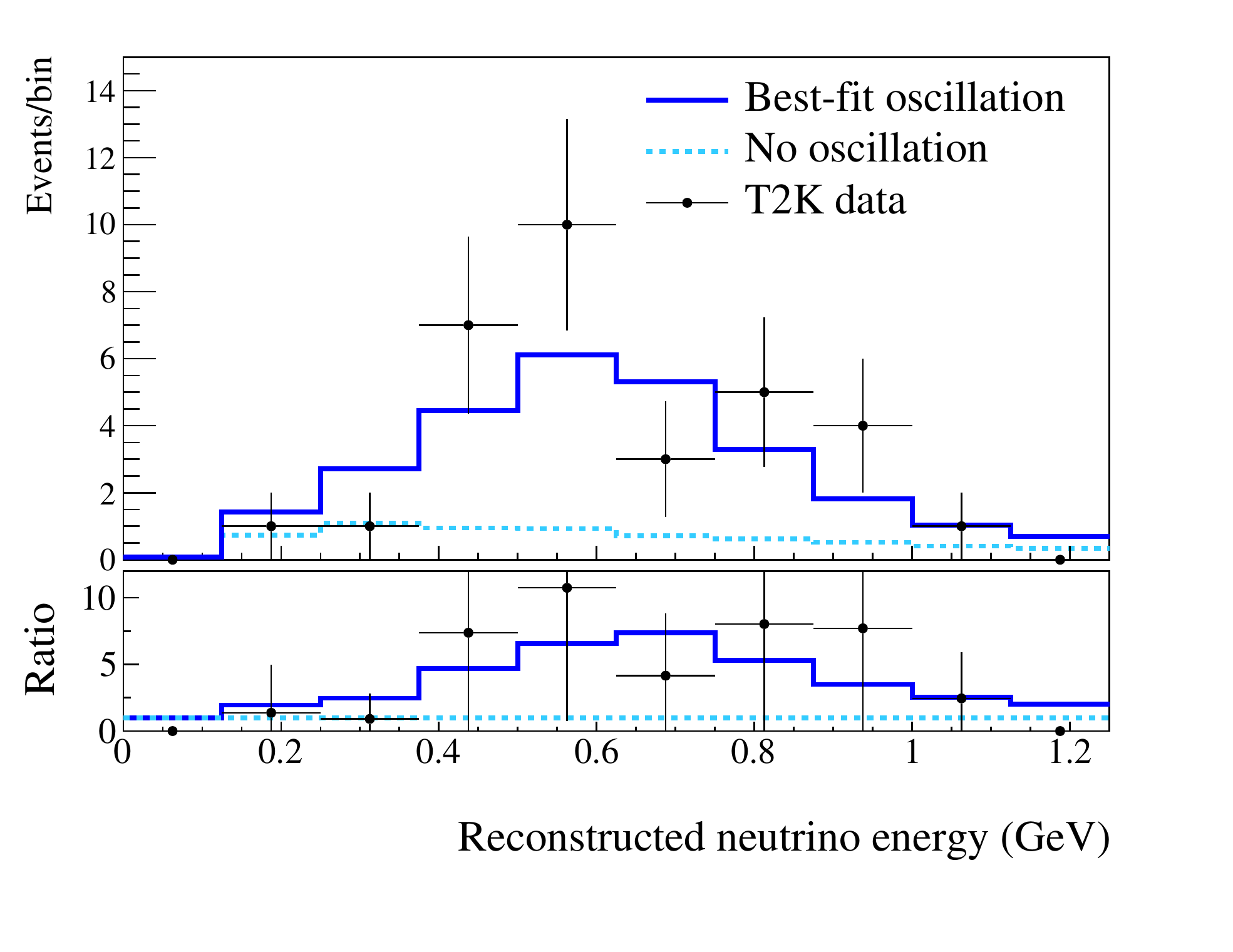}
     \includegraphics[width=.48\textwidth]{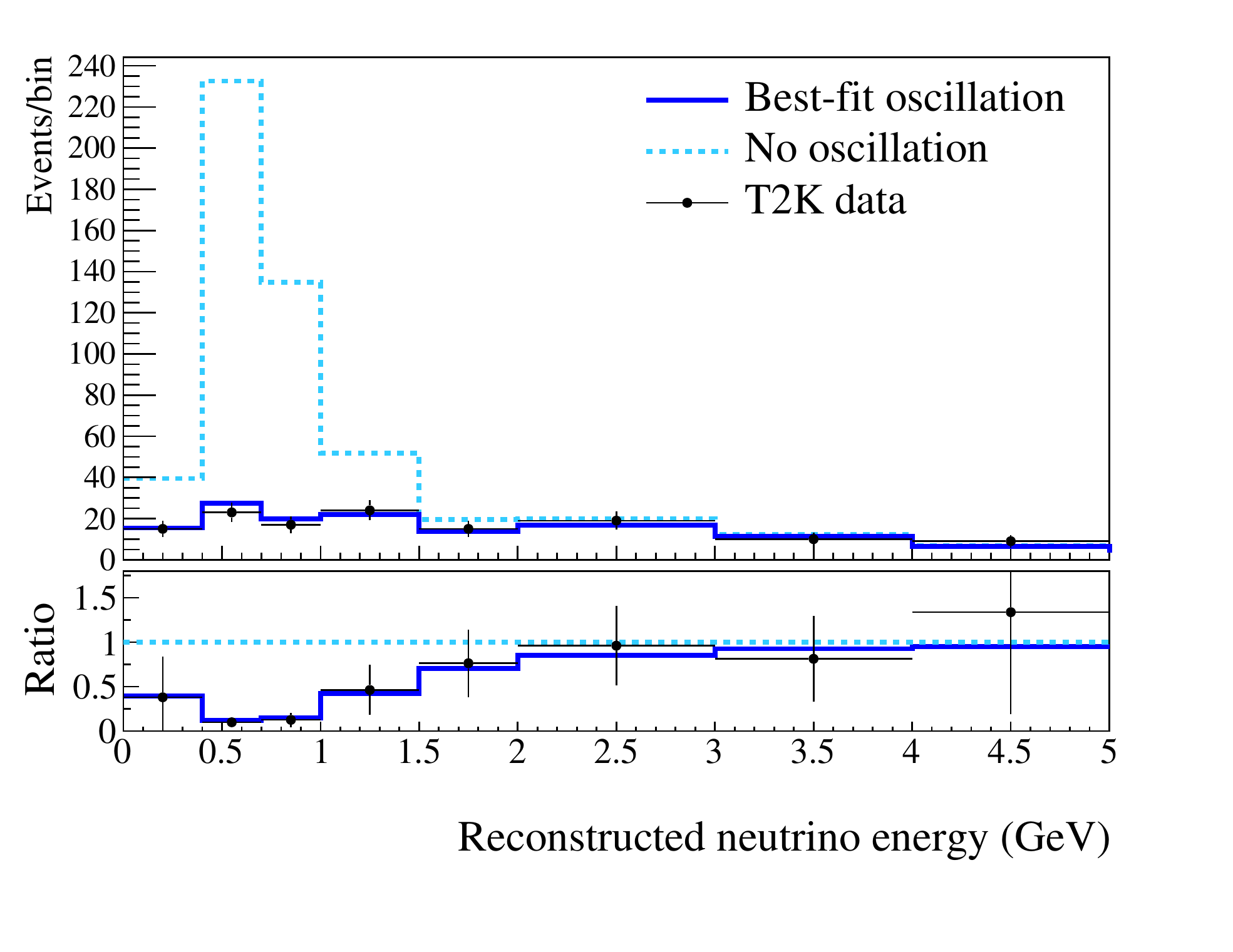}
     \includegraphics[width=.48\textwidth]{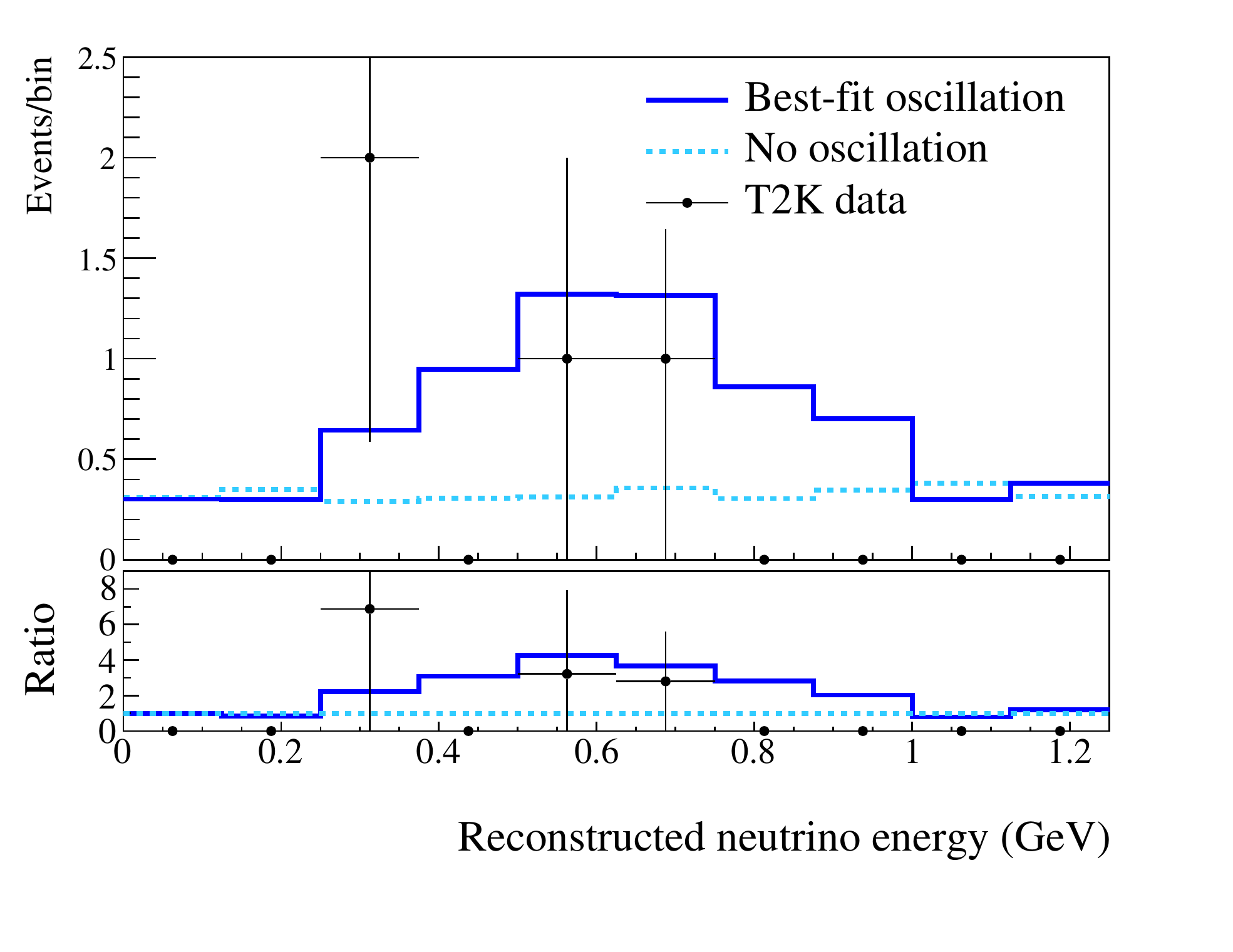}
     \includegraphics[width=.48\textwidth]{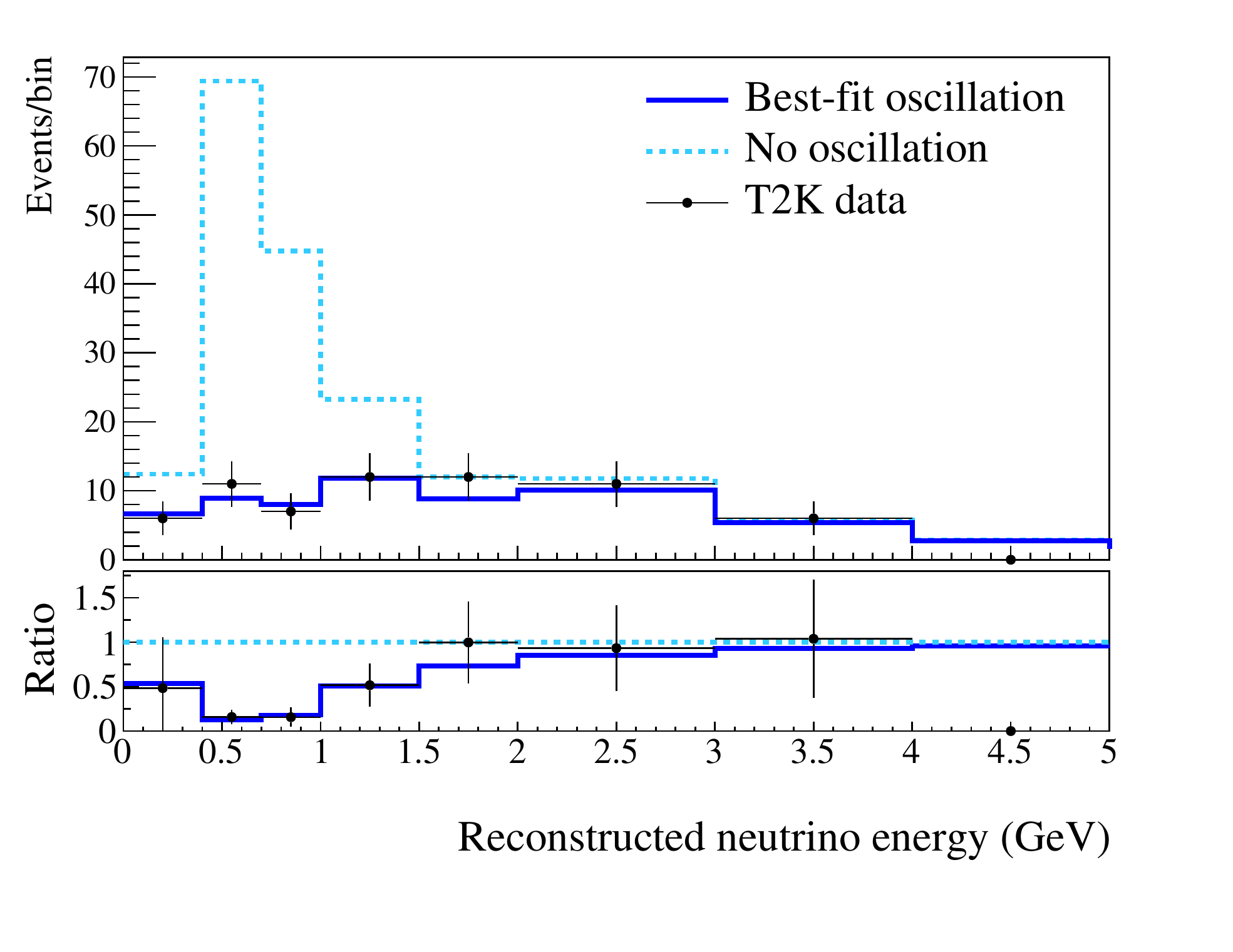}
     \includegraphics[width=.48\textwidth]{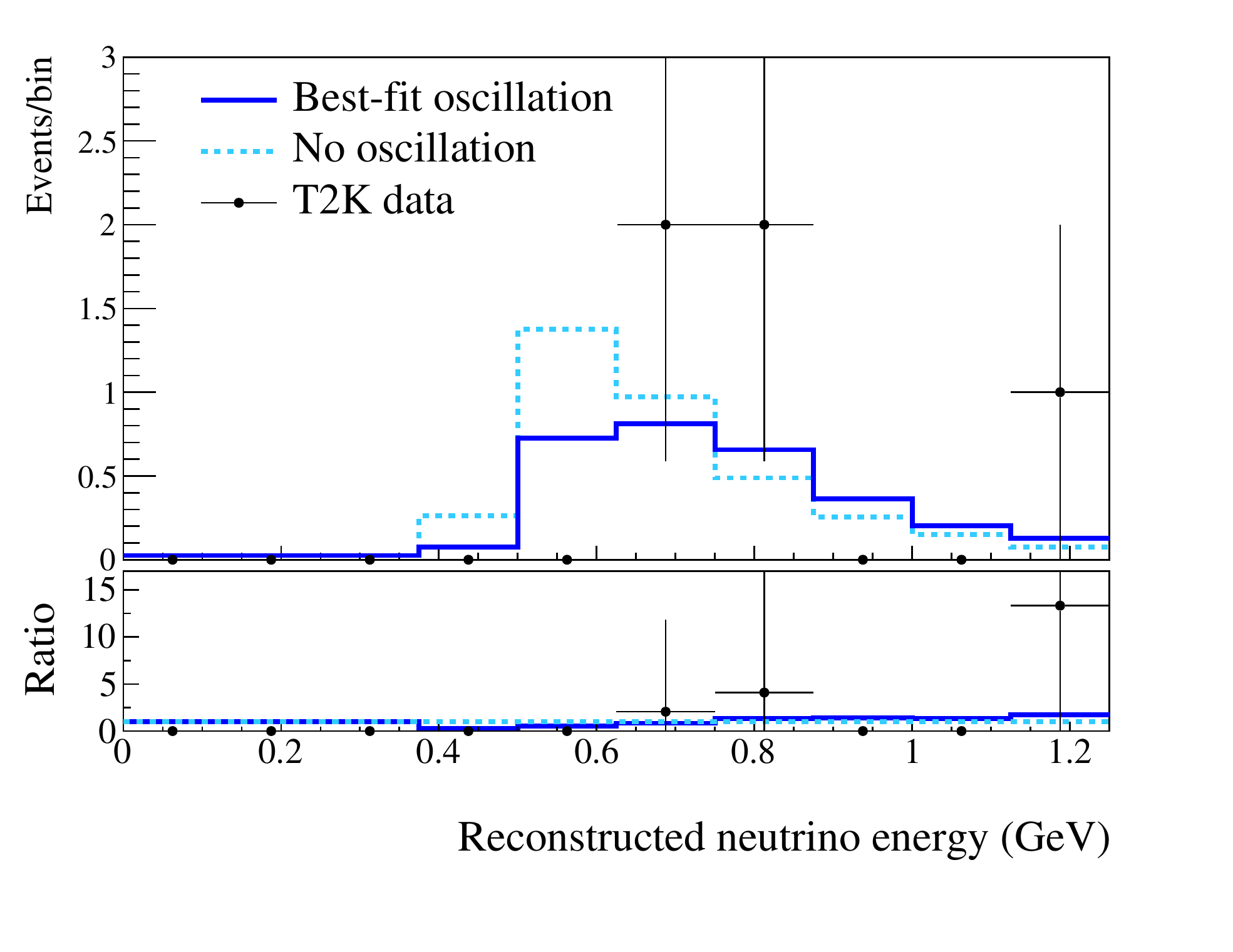}

    \caption{Comparison of the best-fit oscillated MC energy spectra, unoscillated spectra and T2K data for the five samples used in the fit: $\mu$-like sample in \nunu-mode and \nub-mode (top left and right), single ring e-like appearance sample in \nunu-mode and \nub-mode (middle left and right), \ccpip e-like appearance sample in \nunu-mode (bottom). 
    The larger unoscillated spectra in the \ccpip e-like sample compared to the single ring sample is due to the relatively large background of \num in the \ccpip sample, which does not disappear in the no-oscillation case. The ratio of the best fit to unoscillated spectra are also shown. }
    \label{fig:spectramach3_worc}
\end{figure*}

An estimate of the oscillation parameters $\Delta m_{32}^2$,  $\sin^2 \theta_{13}$ and $\delta_{CP}$ have been obtained with both frequentist and Bayesian analyses.
The agreement between the fit results and the data has been evaluated by comparing the expected spectra after the oscillation fit with the data points as shown in Fig.~\ref{fig:spectramach3_worc}.
%Each best-fit spectrum is formed by calculating the most probable value for the predicted number of events in each energy bin, using the MCMC points from the corresponding analysis. The fit spectrum for ?? CC events does not change appreciably when the reactor prior is included, but the ?e CC fit spectrum shows a noticeable reduction in the number of events.
The best-fit spectra are obtained by sampling 2000 points from the MCMC including the reactor constraint and 
fitting a Gaussian distribution to calculate the most probable value for the predicted number of events in each energy bin.

In order to extract the ratio of oscillated to unoscillated spectra, the expected spectra are also tuned to the no oscillation case.
%The MC is also tuned to the best fit oscillation parameters value, as well as no oscillation, in order to extract the ratio.
A coarser binning than the one used in the fit has been used for readability.

\section{Conclusions}
\label{sec:conclusions}
All data collected by the T2K experiment between 2010 and 2016 have been analyzed to estimate the oscillation parameters $|\dmsq|$, \stt, \sto, \dcp and the mass ordering. These parameters are estimated by doing a joint fit of the \num and \numb disappearance channels and \nue and \nueb appearance channels by using five samples selected at the far detector, thus, including the new additional \ccpip sample.  The data related to this measurement can be found in~\cite{datarelease}.
%A comprehensive set of studies of simulated data has been performed to estimate the impact that uncertainties arising from a poor understanding of neutrino interactions may have on the estimate of the oscillation parameters. 
A comprehensive study has been performed to evaluate the sensitivity of oscillation parameter estimates to neutrino interaction modeling, showing that the impact of these uncertainties is small compared to the total uncertainties on the measurement of all the oscillation parameters.

The general approach followed in this paper, which combines separate analyses of beamline, neutrino interactions, near and far detector selections through sets of systematic parameters and their covariances will be extended with additional data which will be collected by T2K in the coming years in both \nunu- and \nub-modes, and improved near and far detector samples. This is expected to greatly enhance the sensitivity of the T2K experiment to the measurement of the CP-violation phase \dcp as well as more precise measurements of the atmospheric parameters $|\dmsq|$ and \stt.

\begin{acknowledgments}

We thank the J-PARC staff for superb accelerator performance. We thank the 
CERN NA61/SHINE Collaboration for providing valuable particle production data.
We acknowledge the support of MEXT, Japan; 
NSERC (Grant No. SAPPJ-2014-00031), NRC and CFI, Canada;
CEA and CNRS/IN2P3, France;
DFG, Germany; 
INFN, Italy;
National Science Centre (NCN) and Ministry of Science and Higher Education, Poland;
RSF, RFBR, and MES, Russia; 
MINECO and ERDF funds, Spain;
SNSF and SERI, Switzerland;
STFC, UK; and 
DOE, USA.
We also thank CERN for the UA1/NOMAD magnet, 
DESY for the HERA-B magnet mover system, 
NII for SINET4, 
the WestGrid and SciNet consortia in Compute Canada, 
and GridPP and the Emerald High Performance Computing facility in the United Kingdom.
In addition, participation of individual researchers and institutions has been further 
supported by funds from ERC (FP7), H2020 Grant No. RISE-GA644294-JENNIFER, EU; 
JSPS, Japan; 
Royal Society, UK; 
the Alfred P. Sloan Foundation and the DOE Early Career program, USA.

\end{acknowledgments}

\bibliographystyle{unsrt}
\bibliography{long_paper}% Produces the bibliography via BibTeX.

\end{document}